\documentclass{jfm}

\usepackage{graphicx}
\usepackage{epstopdf, epsfig}

\usepackage{amsmath}
\usepackage{amssymb,amsbsy,bm}
\usepackage{comment} 
\usepackage[dvipsnames]{xcolor}

\usepackage[percent]{overpic} %earlier used \usepackage[percent]{overpic}
\usepackage{tikz}
\usetikzlibrary{arrows.meta}

\usepackage{subfiles}
\usepackage{subfig}

\usepackage{placeins}

\usepackage{mathtools,xparse}

\usepackage{hyperref}
\hypersetup{
    colorlinks=true,
    linkcolor=red,
    citecolor=blue,
    filecolor=cyan,
}
\usepackage{cleveref}

\crefformat{section}{\S#2#1#3} % see manual of cleveref, section 8.2.1
\crefformat{subsection}{\S#2#1#3}
\crefformat{subsubsection}{\S#2#1#3}

\DeclarePairedDelimiter{\abs}{\lvert}{\rvert}

\newcommand \avg[1]{\left\langle#1\right\rangle}
\newcommand \condavg[2]{\left\langle #1 \middle| #2 \right\rangle}
\newcommand \anisotr[1]{\overline{#1}}
\newcommand \orderof[1]{\textit{O}\left(#1\right)}
\newcommand \inlavg[1]{\langle#1\rangle}

\newcommand\de[0]{\textrm{d}}
\newcommand\Tr[0]{\textrm{Tr}}
\newcommand\eig[0]{*}

\newcommand\fu[0]{\widetilde{\bm{u}}}
\newcommand\fP[0]{\widetilde{P}}

\newcommand\fA[0]{\widetilde{\bm{A}}}
\newcommand\fS[0]{\widetilde{\bm{S}}}
\newcommand\fW[0]{\widetilde{\bm{W}}}
\newcommand\fH[0]{\widetilde{\bm{H}}}
\newcommand\fomega[0]{\widetilde{\bm{\omega}}}

\shorttitle{Filtered velocity gradient dynamics in the eigenframe}
\shortauthor{J. Tom, M. Carbone and A. D. Bragg}

\title{Exploring the turbulent velocity gradients at different scales from the perspective of the strain-rate eigenframe}

\author{Josin Tom\aff{1},
Maurizio Carbone\aff{1,2}\footnote{\small Present address: Max Planck Institute for Dynamics and Self-Organization,
Am Fa{\ss}berg 17, 37077 G{\"o}ttingen, Germany}
\and Andrew D. Bragg\aff{1}\corresp{\email{andrew.bragg@duke.edu}}}

\affiliation{\aff{1}Department of Civil and Environmental Engineering, Duke University, Durham, NC, USA
\aff{2}Dipartimento di Ingegneria Meccanica e Aerospaziale, Politecnico di Torino, Corso Duca degli Abruzzi 24, 10129 Torino, Italy
}

\begin{document}

\maketitle

\begin{abstract}
Expressing the evolution equations for the filtered velocity gradient tensor (FVGT) in the strain-rate eigenframe provides an insightful way to disentangle and understand various processes such as strain self-amplification, vortex stretching and tilting, and to consider their properties at different scales in the flow. Using data from Direct Numerical Simulation (DNS) of the forced Navier-Stokes equation, we consider the relative importance of local and non-local terms in the FVGT eigenframe equations across the scales using statistical analysis. The analysis of the eigenframe rotation-rate, that drives vorticity tilting, shows that the anisotropic pressure Hessian plays a key role, with the sub-grid stress making an important contribution outside the dissipation range, and the local spinning due to vorticity making a much smaller contribution. The results also show the striking behavior that the vorticity tilting term remains highly intermittent even at relatively large scales. We derive a generalization of the Lumley triangle that allows us to show that the pressure Hessian has a preference for two-component axisymmetric configurations at small scales, with a transition to a more isotropic state at larger scales. Correlations between the sub-grid stress and other terms in the eigenframe equations are considered, highlighting the coupling between the sub-grid and nonlinear amplification terms, with the sub-grid term playing an important role in regularizing the system. These results provide useful guidelines for improving Lagrangian models of the FVGT, since current models fail to capture a number of subtle features observed in our results.
\end{abstract}

\begin{keywords}
\end{keywords}

\section{Introduction}\label{sec:Intro}
%%%\subfile{Sections/Sec01_Introduction.tex}
%!%
The velocity gradient tensor provides an effective way to characterize the small-scale dynamics and kinematics of turbulent flows \citep{Meneveau2011}.
By filtering (coarse-graining) the velocity gradient on a length-scale $\ell$, one is able to analyze the properties of the velocity gradients at different scales in the flow by varying $\ell$ \citep{Borue1998}, providing insight into the multiscale dynamics of turbulence. The bare (un-filtered) velocity gradient provides insight into the local topology of the flow \citep{Chong1990}, and the structure of highly dissipative or vortical regions of the turbulence \citep{Buaria2019}, while the filtered gradient provides a way to characterize and understand the dynamics of the turbulent energy cascade \citep{Carbone2020a}. Some of the velocity gradient statistics are known to be qualitatively similar across the scales of the flow, i.e. for varying $\ell$. For example, the probability density function (PDF) of the second and third principe invariants of the velocity gradient has the well-known ``tear-drop'' shape not only for the bare velocity gradients, but also for the filtered ones \citep{Naso2005,Danish2018}. 

The dynamics of the velocity gradient can be analyzed effectively from a Lagrangian perspective, i.e.\ following a fluid particle trajectory \citep{Vieillefosse1984,Meneveau2011}. However, the pressure Hessian and viscous stress are non-local and unclosed in this frame of reference, requiring in-depth modelling. This work aims to enhance the understanding of the statistical properties of the velocity gradient dynamics at different scales using data from Direct Numerical Simulation (DNS) of the forced Navier-Stokes equations.

In \cite{Danish2018} the statistics of the filtered velocity gradients have been investigated, with a focus on how probability fluxes in the phase-space of the invariants of the filtered velocity gradients behave. In the present work, the multi-scale characterization of the velocity gradient is extended by analyzing it in the strain-rate eigenframe, formed by the eigenvectors of the symmetric strain-rate tensor. In this frame, the effect of the incompressibility constraint, the local strain self amplification/reduction and the centrifugal force due to rotation of the fluid element can be carefully untangled. Also, it has been recently shown that the description of the velocity gradient dynamics in the strain-rate eigenframe also allows for a dimensionality reduction of the non-local pressure Hessian when the single-time properties of the velocity gradients are considered \citep{Carbone2020b}, allowing for simpler modeling of the pressure Hessian.

An analysis of the velocity gradients in the strain-rate eigenframe has been employed in previous works for an effective description of the velocity gradient dynamics \citep{Vieillefosse1982,Dresselhaus1992,Nomura1998,Lawson2015}.
In the pioneering works by \cite{Vieillefosse1982,Vieillefosse1984} the so-called Restricted Euler (RE) model was introduced for an inviscid flow by neglecting the non-local part of the pressure Hessian, while retaining its local contribution. One of the consequences of setting the non-local part of the pressure Hessian to zero in the inviscid equations is conservation of the angular momentum of the fluid element. However, as Vieillefosse demonstrated, this localization of the pressure Hessian results in a model for the velocity gradients that exhibits a finite-time singularity. This singularity arises because although the rotation of the fluid element has a stabilizing effect on the dynamics, the strain self-amplification mechanism drives the system towards the finite-time singularity, in which the fluid element is flattened onto a plane \citep{Vieillefosse1984}.
Despite the finite-time singularity, which makes the system impractical for modelling, the RE model revealed many non-trivial geometrical features of the motion of an incompressible and inviscid flow \citep{Cantwell1992}.
For example, the RE system conserves several quantities and the presence of these first integrals is related to the onset of the singularity while the non-local pressure Hessian and viscous stress, which are key in a real flow, reduce the number of conserved quantities.
One of the conserved quantities is the determinant of the commutator between the symmetric and anti-symmetric parts of the velocity gradient tensor. That this quantity is conserved implies fundamental constraints on the eigenframe dynamics, namely, that the ordering of the (unordered) eigenvalues cannot change, and the vorticity components in the strain-rate eigenframe cannot change sign \citep{Vieillefosse1982}.
The presence of the non-local pressure Hessian and viscous stress in the real Navier-Stokes system can violate the conservation of this and other quantities that are conserved in RE. One of the objectives of the present paper is to explore this at different scales in the flow.

Given the crucial role played by the non-local pressure Hessian, as revealed through the RE model, several subsequent models have sought to derive closure models for this term, as well as the viscous term appearing the in Navier-Stokes system. Examples include an early stochastic model that is built on the assumption of a log-normal distribution of kinetic energy dissipation \citep{Girimaji1990a}, the Lagrangian tetrad model \citep{Chertkov1991,Naso2005}, the Recent Fluid Deformation Approximation (RFDA) model \citep{Chevillard2008}, the Gaussian Random Fields approximation \citep{Wilczek2014}, the Recent Deformation of Gaussian Fields (RDGF) model  \citep{Johnson2016}, and finally an extension of the RDGF model that captures the effects of multiple scales in the flow, allowing the velocity gradients to be predicted at arbitrary Reynolds numbers \citep{Johnson2017}. These models provide closures for the non-local pressure Hessian that are able to avoid finite-time singularities in the system, and to different degrees, they capture many of the important statistical properties of the velocity gradients. Work still needs to be done, however, to improve the accuracy of their predictions. Furthermore, these works focused on the bare velocity gradients, while equivalent models for the filtered counterpart are lacking.

All of the aforementioned closure models for the non-local pressure Hessian and viscous stress require detailed knowledge of the statistical geometry and invariants of the velocity gradient dynamics. For the filtered gradient, characterizing the statistical geometry and invariants of the sub-grid stress is also required to guide the development of Lagrangian models for the filtered velocity gradients. Moreover, as mentioned earlier, the RE model implies fundamental constraints on the eigenframe dynamics, and we wish to explore the extent to which these constraints are violated in Navier-Stokes turbulence, at different scales in the flow. These points motivate the present work.

%%%!
In the present work, the statistics of the dynamical terms in the filtered velocity gradient equations written in the strain-rate eigenframe, are characterized using results from Direct Numerical Simulation of statistically steady and isotropic incompressible turbulence.  The paper is organized as follows. In section \ref{sec:Theory}, the equations for the velocity gradient in the strain-rate principal basis are outlined. Details on the numerical simulations are in section \ref{sec:DNS_Details} and the numerical result are in section \ref{sec:DNS_Results}. In the numerical analysis, focus is put on the characterization of the non-local/unclosed dynamical terms conditioned on the local/closed dynamical terms.
A summary of the main results and the conclusions are in section \ref{sec:Conclusions}.
%!%
%%%

\section{Dynamical equations in the strain-rate eigenframe}\label{sec:Theory}
%%%\subfile{Sections/Sec02_Theory.tex}
%!%
In this section the equations for the filtered velocity gradient are presented and written in the eigenframe of the filtered strain-rate tenor. Since the equations for the velocity gradient in the strain-rate eigenframe are not often employed, and the formulation of these forms of the equations are only briefly presented in a few previous works \citep{Vieillefosse1982,Dresselhaus1992}, we will outline the key steps leading to these equations, as well as discuss the terms appearing in the equations which will be helpful for the results section.

\subsection{Equations for the filtered velocity gradient}

The filtered velocity field is governed by the incompressible, filtered, continuity and Navier-Stokes equations
\begin{subequations}
\label{theory_eq_NS}
\begin{align}
\bm{\nabla\cdot\fu} &= 0\\
\widetilde{D}_t\bm{\fu} \equiv \partial_t\bm{\fu}  + (\bm{\fu\cdot}\nabla) \bm{\fu} &= -\nabla \fP + \nu\nabla^2\bm{\fu} - \bm{\nabla\cdot\tau} %+ \bm{f}
\end{align}
\end{subequations}
where $\bm{\fu}(\bm{x},t)$, $\fP(\bm{x},t)$ are the filtered fluid velocity and pressure fields and $\nu$ is the kinematic viscosity. We use an isotropic filtering kernel $G_\ell$ with filtering length $\ell$, with which we define the filtering operation of an arbitrary field $\bm{\xi}$ as \citep{pope}
\begin{equation}
\widetilde{\bm{\xi}}(\bm{x},t) = \int G_\ell\left(\|\bm{x}-\bm{y}\|\right)\bm{\xi}(\bm{y},t)\de\bm{y},
\end{equation}
such that $\bm{\xi}$ is the bare (un-filtered) field, and $\widetilde{\bm{\xi}}$ the filtered field. The sub-grid stress is
\begin{equation}
\bm{\tau} \equiv \widetilde{\bm{uu}^\top} - \fu\fu^\top,
\end{equation}
where $\cdot^\top$ indicates transposition.

By taking the gradient of \eqref{theory_eq_NS}, the equations for the velocity gradient are obtained
\begin{subequations}
\label{theory_eq_grad}
\begin{align}
\Tr(\fA) &= 0,\\
\widetilde{D}_t\bm{\fA}  &=
-\bm{\fA\cdot\fA} - \bm{\fH}^P + \nu \nabla^2 \bm{\fA} - \bm{\nabla}\left(\bm{\nabla\cdot\tau}\right),
\end{align}
\end{subequations}
where the filtered velocity gradient and filtered pressure Hessian are
\begin{align}
\bm{\fA} &\equiv \partial_j \widetilde{u}_i \bm{e}_i\bm{e}_j^\top, \\
\bm{\fH}^P &\equiv \partial_j\partial_i \fP \bm{e}_i\bm{e}_j^\top.
\end{align}
In our notation, $\Tr(\cdot)$ denotes the trace operator, while ``$\bm{\cdot}$'' denotes an inner product (single contraction) between tensors, e.g. $\bm{\fA\cdot\fA} = \widetilde{A}_{ij}\widetilde{A}_{jk}\bm{e}_i\bm{e}_k^\top$. 
Here the tensors are represented with respect to the standard, right-oriented, orthonormal basis $\{\bm{e}_i\}$, which is constant in time and space.

The filtered velocity gradient is decomposed into its symmetric and anti-symmetric parts
\begin{align}
\bm{\fS} &\equiv \frac{1}{2}\left(\widetilde{A}_{ij}+\widetilde{A}_{ji}\right)\bm{e}_i\bm{e}_j^\top, \\
\bm{\fW} &\equiv \frac{1}{2}\left(\widetilde{A}_{ij} - \widetilde{A}_{ji}\right)\bm{e}_i\bm{e}_j^\top.
\end{align}
The symmetric part of the filtered velocity gradient is the filtered strain-rate while the antisymmetric part is associated with the filtered vorticity, $\bm{\fomega} = \bm{\nabla \times \fu}$. The vorticity components in the standard basis are $\widetilde{\omega}_i = \epsilon_{ikj} \widetilde{W}_{jk}$, where $\epsilon_{ijk}$ is the permutation symbol.

The equation for the filtered velocity gradient \eqref{theory_eq_grad} is decomposed into its symmetric and anti-symmetric part. The filtered strain-rate is governed by
\begin{subequations}
\label{theory_eq_strain}
\begin{align}
\Tr(\fS) &= 0\\
\widetilde{D}_t\bm{\fS} &= -\bm{\fS\cdot\fS} + \bm{\fW\cdot\fW}^\top - \bm{\fH}^P + \bm{\fH}^\nu - \bm{H}^\tau,
\end{align}
\end{subequations}
where the viscous and sub-grid stress contributions to the symmetric part of the gradient equation are
\begin{align}
\widetilde{\bm{H}}^\nu &\equiv \nu \partial_{k}\partial_{k} \widetilde{S}_{ij}\bm{e}_i\bm{e}_j^\top, \\
\bm{H}^\tau &\equiv \frac{1}{2}\partial_k\left(\partial_j\tau_{ik} + \partial_i\tau_{jk}\right)\bm{e}_i\bm{e}_j^\top.
\end{align}
The filtered vorticity is governed by the equation
\begin{align}
\widetilde{D}_t\bm{\fomega}  = \bm{\fS\cdot\fomega} + \widetilde{\bm{\Omega}}^\nu - \bm{\Omega}^\tau,
\label{theory_eq_vort}
\end{align}
to which the viscous and sub-grid stress directly contribute through their anti-symmetric parts
\begin{align}
\widetilde{\bm{\Omega}}^\nu  &\equiv \nu \partial_{k}\partial_{k} \widetilde{\omega}_{i} \bm{e}_i, \\
\bm{\Omega}^\tau &\equiv \epsilon_{ikj} \partial_k  \partial_m\tau_{jm} \bm{e}_i.
\end{align}

\subsection{Navier-Stokes equations in the strain-rate eigenframe}

The eigenvectors of the filtered strain-rate tensor $\{\bm{v}_i\}$, which are orthogonal and normalized to unit length, form a complete basis for the three-dimensional space. The basis $\{\bm{v}_i\}$ varies in space and time and will be referred to as strain-rate eigenframe. The bases $\{\bm{v}_i\}$ and $\{\bm{e}_j\}$ are related by the rotation matrix $V_{ji}$, the $i^{th}$ column of which contains the components of the $i^{th}$ strain-rate eigenvector with respect to the standard basis
\begin{equation}
V_{ji} \equiv \bm{v}_i\bm{\cdot} \bm{e}_j,
\label{theory_def_V}
\end{equation}
and $\bm{v}_i=V_{ji}\bm{e}_j$. The matrix $V_{ji}$ is orthonormal since $V_{ki} V_{kj} = \delta_{ij}$, where $\delta_{ij}$ is the Kronecker delta.
% (that is $(\bm{v}_1\bm{\times}\bm{v}_2)\bm{\cdot}\bm{v}_3=1$).
In the following, only right-oriented orthonormal bases are considered, with $\det\bm{V}=1 \; \forall \; \bm{x},t$.

The tensors are now represented with respect to the eigenframe in order to derive equations \eqref{theory_eq_grad} in that basis. For notation simplicity the tilde is suppressed in the following. The filtered strain-rate in its principal basis is
\begin{equation}
\bm{S} = S_{ij}\bm{e}_i\bm{e}_j^\top = V_{ik}S_{ij}V_{jm}\bm{v}_k\bm{v}_m^\top = S^\eig_{km}\bm{v}_k\bm{v}_m^\top,
\end{equation}
where $S^\eig_{ij}$ is a diagonal matrix containing the eigenvalues $\lambda_i$ on its diagonal.

The strain-rate eigenframe, formed by the principal basis, undergoes a rigid body rotation, since the eigenvectors $\{\bm{v}_i\}$ remain orthonormal and right-oriented for all times. Therefore, the angular velocity $\bm{\varpi}$ is the same for all the eigenvectors
\begin{equation}
D_t \bm{v}_i = \bm{\varpi\times} \bm{v}_i.
\label{theory_eq_S_ang_vel}
\end{equation}
Furthermore, the angular velocity of the principal basis is associated to the anti-symmetric tensor
\begin{equation}
\bm{\Pi} = \Pi_{ij} \bm{e}_i\bm{e}_j^\top = \epsilon_{ikj} \varpi_k \bm{e}_i\bm{e}_j^\top,
\end{equation}
whose components in the principal basis are
\begin{equation}
\Pi_{ij}^\eig  =
%\epsilon_{pkq} V_{pi} V_{qj}\varpi_k = 
%-(\bm{v}_i\bm{\times v}_j)\bm{\cdot e}_k \varpi_k = 
%-\epsilon_{ijk}\bm{v}_k\bm{\cdot \varpi} = 
\epsilon_{ikj}\varpi_k^\eig.
\label{theory_eq_rot_S}
\end{equation}
The rotation tensor $\bm{\Pi}$ represents the rate of rotation in the plane composed of two of the eigenvectors about the axis of the third, that is $\Pi_{ij}=\de_t\bm{v}_i\bm{\cdot}\bm{v}_j$ \citep{Nomura1998}. %%%%check sign

The dynamical equation for the strain-rate eigenvectors \eqref{theory_eq_S_ang_vel} in the eigenframe can be rewritten using \eqref{theory_eq_rot_S}
\begin{equation}
\widetilde{D}_t\bm{v}_i =
%\Pi_{pj} V_{ji} \bm{e}_p = 
%\epsilon_{pkj} V_{kn}\varpi_n^\eig V_{ji} \bm{e}_p =
%\bm{v}_n\bm{\times v}_i \varpi_n^\eig =
%-\epsilon_{inj}\varpi_n^\eig\bm{v}_j =
%\bm{v}_j \Pi_{ji}^\eig
\Pi_{kj} V_{ji} V_{km} \bm{v}_m =
\Pi_{mi}^\eig \bm{v}_m.
\label{theory_def_rot_basis}
\end{equation}
The time derivative of the strain-rate can be now expressed in the strain-rate eigenframe
\begin{equation}
D_t\bm{S} = D_t\left(  S^\eig_{ij}\bm{v}_i\bm{v}_j^\top  \right) = D_t S^\eig_{ij} \bm{v}_i\bm{v}_j^\top + S^\eig_{ij} \bm{v}_k \Pi_{ki}^\eig\bm{v}_j^\top + S^\eig_{ij} \bm{v}_i \Pi_{kj}^\eig \bm{v}_k^\top,
\end{equation}
that is, adjusting the indexes,
\begin{equation}
D_t\bm{S} =  \left( D_t S^\eig_{ij}  + \Pi_{ik}^\eig S^\eig_{kj}  - S^\eig_{ik} \Pi_{kj}^\eig  \right)\bm{v}_i\bm{v}_j^\top.
\end{equation}
The term due to the rotation of the eigenframe
%, has components 
%$\epsilon_{ikj}\varpi_k^\eig(\lambda_{(i)}-\lambda_{(j)})$
%with respect to $\{\bm{v}_i\}$, and it
is the commutator between the anti-symmetric tensor associated to the rotation of the strain-rate eigenbasis and the strain itself
\begin{equation}
[\bm{\Pi},\bm{S}] = \bm{\Pi \cdot S} - \bm{S \cdot \Pi} =
\left(\Pi_{ik}^\eig S^\eig_{kj}  - S^\eig_{ik} \Pi_{kj}^\eig\right)\bm{v}_i\bm{v}_j^\top.
\end{equation}
The equation for the strain-rate \eqref{theory_eq_strain} in the eigenframe reads
%by multiplying it by $\bm{v}_i^\top$ on the left and by $\bm{v}_j$ on the right
\begin{subequations}
\label{theory_eq_strain_princ}
\begin{align}
S_{ii}^\eig &= 0,\\
D_t S_{ij}^\eig + \Pi_{ik}^\eig S^\eig_{kj} - S^\eig_{ik} \Pi_{kj}^\eig &= -S^\eig_{ik}S^\eig_{jk} + \frac{1}{4}\left(\omega^2\delta_{ij}-\omega_i^\eig\omega_j^\eig\right) - H_{ij}^{P\eig} + H_{ij}^{\nu\eig} - H_{ij}^{\tau\eig},
\end{align}
\end{subequations}
where $H_{ij}^{P \eig} = \bm{v}_i^\top\bm{\cdot}\bm{H}^P\bm{\cdot}\bm{v}_j$, $H_{ij}^{\nu \eig} = \bm{v}_i^\top\bm{\cdot}\bm{H}^\nu\bm{\cdot}\bm{v}_j$ and $H_{ij}^{\tau \eig} = \bm{v}_i^\top\bm{\cdot}\bm{H}^\tau \bm{\cdot}\bm{v}_j$ are the components of the pressure, viscous and sub-grid symmetric contributions in the strain-rate eigenbasis.
%Since $S^\eig_{ij}$ is diagonal, i
It is convenient to split equation \eqref{theory_eq_strain_princ} into its diagonal and off-diagonal parts
\begin{subequations}
\label{theory_eq_strain_princ_split}
\begin{align}
\sum_{i=1}^3 \lambda_i &= 0,\label{theory_eq_strain_cont}\\
D_t \lambda_i  &= -\lambda_i^2 + \frac{1}{4}\left(\omega^2-\omega_i^{\eig 2}\right) - H_{i(i)}^{P\eig} + H_{i(i)}^{\nu\eig} - H_{i(i)}^{\tau\eig},\label{theory_eq_strain_lambda}\\
(\lambda_{(j)}-\lambda_{(i)})\Pi^\eig_{ij} &= -\frac{1}{4}\omega_i^\eig\omega_j^\eig 
- H_{ij}^{P\eig} + H_{ij}^{\nu\eig} - H_{ij}^{\tau\eig},\textrm{ for } i\neq j ,
\label{theory_eq_strain_rot}
\end{align}
\end{subequations}
where $\omega\equiv\|\bm{\omega}\|$ and indexes in parentheses are not contracted. The vorticity time derivative is expressed in the eigenframe using equation \eqref{theory_def_rot_basis}
\begin{equation}
D_t \bm{\omega} = D_t \left(\omega_i^\eig\bm{v}_i\right) = \left(D_t \omega_i^\eig + \Pi_{ij}\omega_j^\eig\right)\bm{v}_i,
\end{equation}
and then the vorticity equation \eqref{theory_eq_vort} in the eigenframe is obtained
\begin{equation}
D_t \omega_i^\eig = \lambda_{(i)}\omega_i^\eig - \Pi_{ij}^\eig\omega_j^\eig + \Omega_i^{\nu\eig} - \Omega_i^{\tau\eig},
\label{theory_eq_vort_princ}
\end{equation}
with $\Omega_{i}^{\nu \eig} = \bm{v}_i^\top\bm{\cdot}\bm{\Omega}^\nu$ and $\Omega_{i}^{\tau \eig} = \bm{v}_i^\top\bm{\cdot}\bm{\Omega}^\tau$.

%%%COMMENTS TO EQ
The first term on the right-hand side of \eqref{theory_eq_strain_lambda} is the strain-self interaction which acts to amplify $\lambda_3$ and suppress $\lambda_1$.
The second term represents a straining produced in the fluid due to the rotation of the fluid element and the associated centrifugal force. This term acts only in the plane orthogonal to the vorticity vector.
The third, fourth and fifth terms in \eqref{theory_eq_strain_lambda} are the symmetric contributions from the pressure Hessian, viscous stress and sub-grid stress. %\textcolor{red}{Add some comments about their roles?}
The local part of the pressure Hessian guarantees incompressibility. The anisotropic part of the pressure Hessian plays a major role in regularization of the dynamics generated by the local terms (which are expressible in terms of the gradient at the fluid particle position) and we will analyze its statistics in detail.
The viscous stress acts, on average, as a damping on both the strain rate and the vorticity. However, the statistical behaviour of the viscous stress differs from that of a simple linear damping and it also plays a relevant role in the transport of vorticity. The sub-grid stress represents the effect of the scales that have been filtered out on the filtered gradient dynamics. We will characterize its statistical properties across the scales and it will be shown how the sub-grid stress interacts with the pressure Hessian and viscous stress in a non-trivial way.

%%%%%%%%%%%%%%%%%%%%%%
The contributions to the eigenframe components of the rotation tensor $\bm{\Pi}$ are described by equation \eqref{theory_eq_strain_rot}, and correspond to contributions from the centrifugal force due to the rotation of the fluid element (which is retained in the RE model), the anisotropic pressure Hessian, viscous and sub-grid stresses are
\begin{align}
\Pi^{RE\eig}_{ij}  \equiv -\frac{1}{4}\frac{\omega_i^\eig\omega_j^\eig}{\lambda_{(j)}-\lambda_{(i)}}, &&
\Pi^{P\eig}_{ij}   \equiv -\frac{H_{ij}^{P\eig}}{\lambda_{(j)}-\lambda_{(i)}}, \nonumber\\
\Pi^{\nu\eig}_{ij} \equiv \frac{H_{ij}^{\nu\eig}}{\lambda_{(j)}-\lambda_{(i)}}, &&
\Pi^{\tau\eig}_{ij} \equiv -\frac{H_{ij}^{\tau\eig}}{\lambda_{(j)}-\lambda_{(i)}},
\label{theory_contr_strain_rot}
\end{align}
for $i\ne j$. The numerators in \eqref{theory_contr_strain_rot} may be interpreted as representing torques, which arise from local and non-local effects, while the denominator can be interpreted as the moment of inertia.

Pressure depends quadratically on the velocity gradient through its second invariant $Q\equiv -\Tr(\bm{A\cdot A})/2$
\begin{equation}
P(\bm{x},t) = -\frac{1}{2\pi} \int \frac{Q(\bm{y},t)}{\|\bm{y}-\bm{x}\|} \de \bm{y}.
\label{theory_eq_P}
\end{equation}
Since the kernel $\|\bm{y}-\bm{x}\|^{-1}$ decays slowly with distance from the fluid particle at $\bm{x}$, the local and non-local contributions from $P(\bm{x},t)$ to $\Pi_{ij}^{P\eig}$ may be of comparable magnitude. In fact, previous results for the bare velocity gradient dynamics show that the contribution from the non-local pressure Hessian to $\Pi_{ij}^{P\eig}$ dominates over the local contribution \citep{She1991,Dresselhaus1992}. We will consider whether this also is the case for finite filtering lengths $\ell_F>0$.
%%[[Tsinober looks at $\Pi_{ij}^{\nu\eig}$]]

The dynamics of the vorticity in the eigenframe is described by equation \eqref{theory_eq_vort_princ}.
The first term on the right-hand side of \eqref{theory_eq_vort_princ} is vortex stretching, that is particularly clear from this eigenframe perspective.
The second represents the reorientation (tilting) of the vorticity with respect to the eigenframe due to the rotation of the eigenframe. This term does not affect the evolution of the vorticity magnitude directly since $\Pi_{ij}^\eig\omega_j^\eig\omega_i^\eig=0$, although it indirectly contributes since the vortex stretching term depends on $\omega_j^\eig$.
Moreover, the angular velocity component along the vorticity direction does not affect the tilting of vorticity, and corresponds to a redundant degree of freedom with respect to the dynamical evolution of $\lambda_i$ and $\omega_j^\eig$ \citep{Carbone2020b}.
The Restricted Euler contribution to vorticity tilting in \eqref{theory_eq_vort_princ} is 
\begin{equation}
-\Pi^{RE\eig}_{ij}\omega_j^\eig = \frac{1}{4}\sum_{j \neq i}
\frac{\omega_j^{\eig2}}{\lambda_{j}-\lambda_{i}}\omega_i^\eig,
\end{equation}
and since the ordering of the eigenvalues cannot change in the RE model \citep{Nomura1998}, this contribution acts as a non-linear damping for $\omega_1^\eig$ and as a non-linear amplification for $\omega_3^\eig$ in the RE model. In real turbulence governed by the NSE, the eigenvalue ordering can change with time, such that the sign, and therefore the role of this term is not fixed with time. 

By substituting \eqref{theory_eq_strain_rot} into \eqref{theory_eq_vort_princ} it can be shown that the viscous stress contribution to vorticity tilting, $\Pi_{ij}^{\nu\eig}\omega_j$, is identically cancelled by part of the contribution coming from $\Omega_i^{\nu\eig}$ \citep{Dresselhaus1992,Nomura1998,Lawson2015}. However, we wish to consider the full viscous contribution, $\Omega_i^{\nu\eig}$, and therefore do not expand it into its subparts. The third and fourth terms on the right hand side of the vorticity equation \eqref{theory_eq_vort_princ} derive from the anti-symmetric part of the viscous and sub-grid stress. Since all the other terms in that equation are proportional to $\bm{\omega}$, these are the only terms that can generate vorticity from an initially irrotational state. 

\begin{comment}
\begin{itemize}
\item Start from $\lambda_2 = 0$ and $\omega = 0$
\item $\alpha_i$ (the torque) very small and $\bm{\omega}$ stretches along $\bm{v}_1$ (that is $\omega_1^\eig\uparrow$)
\item $\omega_3^2$ makes $\lambda_2>0$ increase since $\lambda_2$ is small vorticity dominates here
\item $\alpha_1<0$ increases and nullifies the growth in $\omega_1$
\item $\lambda_2$ still small compared to $\lambda_1$ so the stretching of $\omega_2$ lasts for longer time wrt initial stretching of $\omega_1$
\item $\omega_2^2$ grows and so $\alpha_3>0$ grows and $\omega_3$ becomes large
\item therefore $\alpha_2$ becomes negative and large and suppresses $\omega_2$
\item $\bm{\omega}$ temporarly aligns with $\bm{v}_3$ and $\omega$ is quickly destroyed
\end{itemize}
\end{comment}
%!%
%%%

\section{Direct Numerical Simulation}\label{sec:DNS_Details}
%%%\subfile{Sections/Sec03_DNSDetails.tex}
%!%

\begin{table}
\begin{center}
\begin{tabular}{c c}
Parameter                       & DNS Specification  \\
$R_{\lambda}$                   & 597                \\
$L$                   & $2 \pi$            \\
$\langle\epsilon\rangle$        & 0.228              \\
$\nu$                           & 0.00013            \\
$\mathcal{L}$                   & 1.43               \\
$\mathcal{L}/\eta$              & 812                \\
$\tau_{\mathcal{L}}$                        & 1.57               \\
$\tau_{\mathcal{L}}/\tau_{\eta}$            & 65.4               \\
$u'$                            & 0.915              \\
$u'/u_{\eta}$                   & 12.4               \\
$T/\tau_{\mathcal{L}}$                      & 5.75               \\
$N$                             & 2048               \\
$N_{\textrm{proc}}$                      & 16384              \\
$\kappa_{\textrm{max}}\eta$                   & 1.70             
\end{tabular}
%%%! modified
\caption{Flow parameters for the DNS study (all dimensional parameters are in arbitrary units). The simulation was performed in parallel on $N_{\textrm{proc}}$ processors and all statistics are averaged over $T$, the duration of the run. $R_{\lambda} \equiv u'\lambda/\nu \equiv 2K/\sqrt{5/3\nu \langle\epsilon\rangle} $ is the Taylor microscale Reynolds Number, $u' \equiv \sqrt{2K/3}$ is the root mean square of fluctuating fluid velocity, $K$ is the turbulent kinetic energy, $\lambda$ is the Taylor microscale, $\nu$ is the fluid kinematic viscosity, 
$\langle\epsilon\rangle \equiv 2 \nu \int\kappa^2 E(\kappa) \textrm{d}\kappa$ is the mean turbulent kinetic energy dissipation rate, $\kappa$ is the wavenumber in Fourier space, $E$ is the energy spectrum.
The integral length scale is defined as $\mathcal{L} \equiv (3\pi/2K) \int E(\kappa)/\kappa \textrm{d}\kappa$, $\eta \equiv (\nu^{3}/\langle\epsilon\rangle)^{1/4} $ is the Kolmogorov length scale, $\tau_{\eta} \equiv \sqrt{\nu/\langle\epsilon\rangle}$ is the Kolmogorov time scale, $u_{\eta} \equiv (\langle\epsilon\rangle \nu)^{3/4} $ is the Kolmogorov velocity scale, $\tau_{\mathcal{L}} \equiv \mathcal{L}/u'$ is the large-eddy turnover time.
The maximum resolved wavenumber is $\kappa_{\textrm{max}} = \sqrt{2N/3}$, $\kappa_{\textrm{max}}\eta$ is the small scale resolution, $L$ is the domain size and $N$ is the number of grid points in each direction.}
\label{Table01}
\end{center}
\end{table}

To analyze the dynamical properties of the filtered velocity gradients in the strain-rate eigenframe, we consider data from a Direct Numerical Simulation (DNS) of statistically stationary, isotropic turbulence. The data we use is from the DNS of \citet{Ireland2016a,Ireland2016b}, at a Taylor microscale Reynolds number $R_\lambda=597$. Incompressible Navier-Stokes equations were solved using a pseudo-spectral method on a three-dimensional, triperiodic cubic domain of length $2\pi$, discretized with $2048^3$ grid points. Deterministic forcing scheme kept the kinetic energy of the flow constant in time. The scale separation between the integral length scale $\mathcal{L}$ and the Kolmogorov scale $\eta$ in the DNS flow was $\mathcal{L}/\eta\simeq 812$. Further details on the numerical method used can be found in \citet{Ireland2013}. Details of the simulations are given in table \ref{Table01}.

We apply a sharp spectral cut-off at wavenumber $k_F$ to obtain the filtered field. In order to relate the spectral cut-off wavenumber $k_F$ to a physical space filtering scale, we define $\ell_F\equiv 2\pi/k_F$ \citep{eyink09}.  When constructing the pressure Hessian from the velocity field, there are some subtleties that must be carefully accounted for in order to ensure that the pressure Hessian computed has the correct properties. These issues are discussed in Appendix \ref{app_dealiasing}. The velocity field is filtered at scale $\ell_F$ and the resulting filtered velocity gradient, pressure Hessian, viscous stress and sub-grid stress are analyzed.
%!%
%%%

\section{Results and discussion}\label{sec:DNS_Results}
%%%\subfile{Sections/Sec04_DNSResults.tex}
%!%
We now turn to consider the role of the different terms appearing in the eigenframe dynamical equations for different filtering scales $\ell_F$, with
quantities normalized using the scale-dependent timescale
\begin{equation}
\widetilde{\tau}\equiv 1/\sqrt{2\langle\|\widetilde{\bm{S}}\|^2\rangle}.
\label{res_def_time_scale}
\end{equation}
Furthermore, while the eigenvalues $\lambda_1,\lambda_2,\lambda_3$ are not ordered in the dynamical equations discussed in \ref{sec:Theory}, it is standard and helpful to consider results in which the eigenvalues are ordered. Therefore, in the results that follow, the eigenvalues are ordered $\lambda_1\geq\lambda_2\geq\lambda_3$, with corresponding ordered eigenvectors $\bm{v}_1,\bm{v}_2,\bm{v}_3$. This ordering allows us to unambiguously interpret the significance of the sign of the local dynamical terms in equations \eqref{theory_eq_strain_princ_split} and \eqref{theory_eq_vort_princ}. 

We also remind the reader that for notational simplicity the tilde used earlier to denote filtered quantities has been dropped, and all variables correspond to filtered variables unless otherwise stated.

%%%%%contributions to lambda and omega
\subsection{Contributions to eigenvalue and vorticity component dynamics}
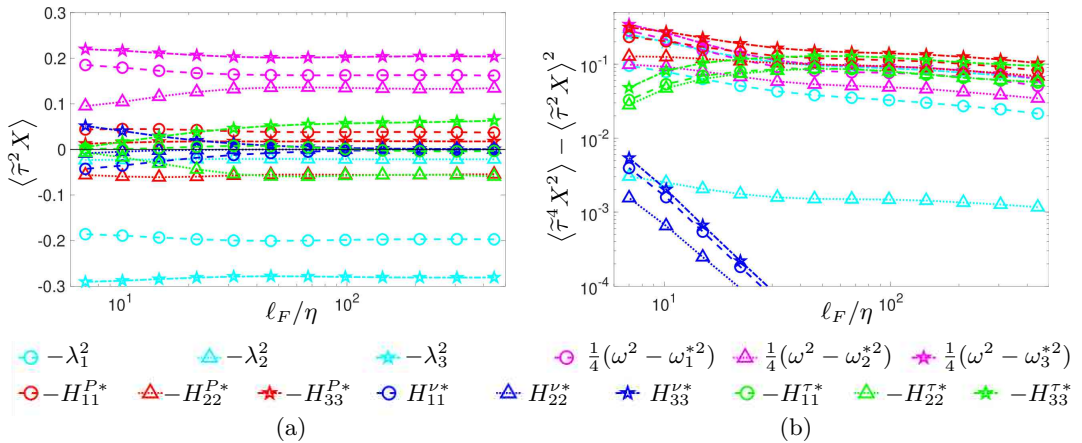
\begin{figure}
\centering
\vspace{0mm}			
    \subfloat[]	
	{\begin{overpic}
	[trim = 0mm -100mm -40mm -1mm,
	scale=0.125,clip,tics=20]{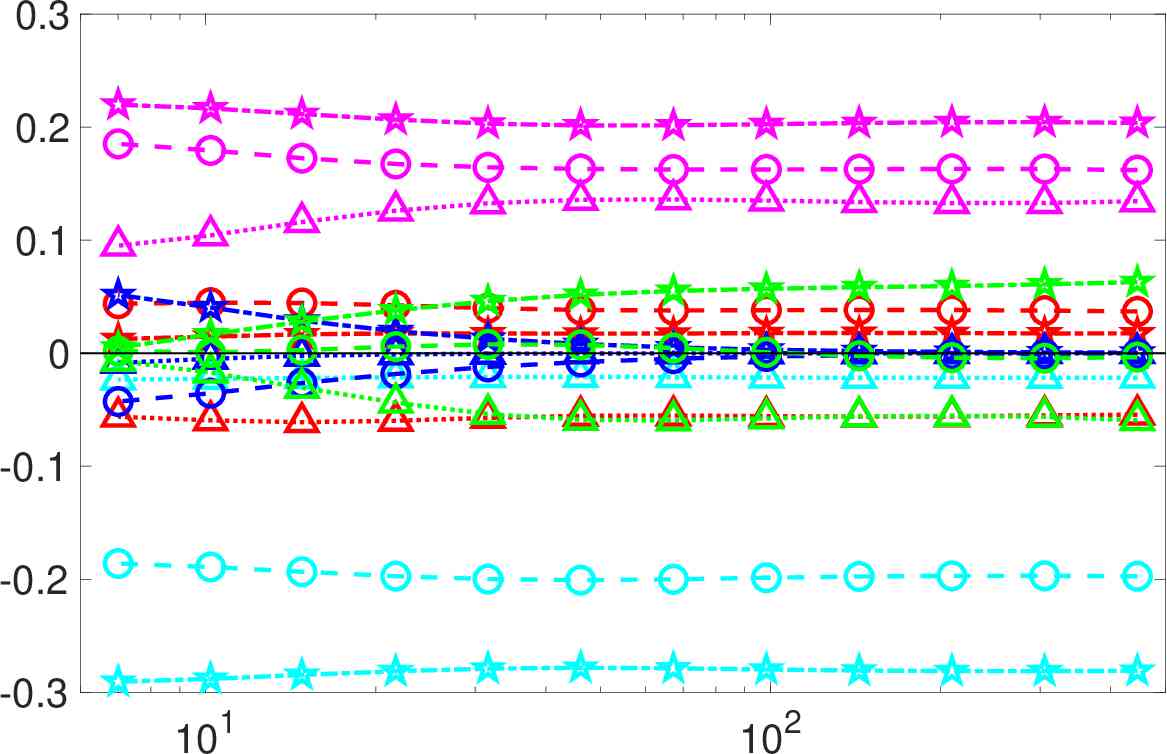}
			\put(45,17){$\ell_F/\eta$}
			% \put(-6,30){\rotatebox{90}{$\mathbb{E} (\widetilde{\tau}^2 X)$}}
			\put(-6,43){\rotatebox{90}{$\avg{\widetilde{\tau}^2 X}$}}
			
			\put(-7,6)
			{
			\begin{overpic}
			[trim = 35mm 272mm 30mm 27mm,
	        scale=0.30,clip,tics=20,]{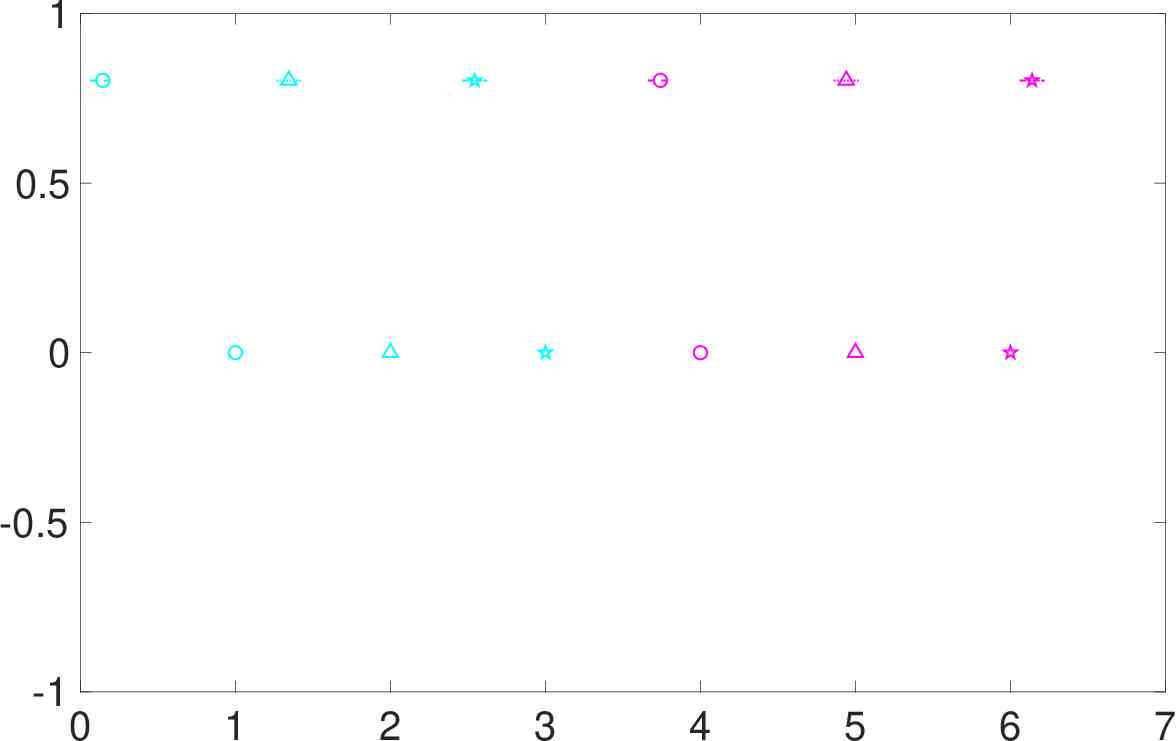}
			\put(3.5,1.25){\small{$-\lambda_1^2$}}
			\put(21.75,1.25){\small{$-\lambda_2^2$}}
			\put(40,1.25){\small{$-\lambda_3^2$}}
			\put(58.5,1.25){\small{$\frac{1}{4}(\omega^2-\omega_1^{\eig 2})$}}
			\put(76.5,1.25){\small{$\frac{1}{4}(\omega^2-\omega_2^{\eig 2})$}}
			\put(94.75,1.25){\small{$\frac{1}{4}(\omega^2-\omega_3^{\eig 2})$}}
			\end{overpic}
			}
			
			\put(-7,-2)
			{
			\begin{overpic}
			[trim = 35mm 280mm 30mm 15mm,
	        scale=0.30,clip,tics=20,]{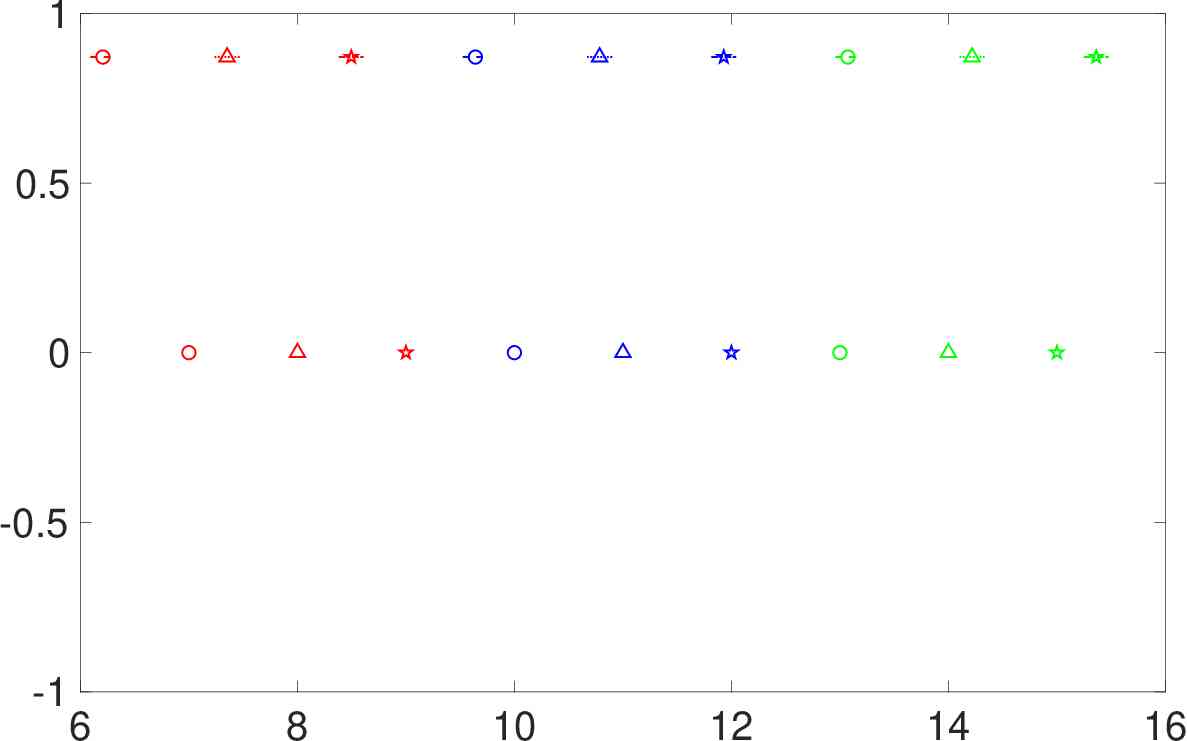}
			\put(3,1.5){\small{$-{H}_{11}^{P\eig}$}}
			\put(15.25,1.5){\small{$-{H}_{22}^{P\eig}$}}
			\put(27.5,1.5){\small{$-{H}_{33}^{P\eig}$}}
			\put(39.75,1.5){\small{${H}_{11}^{\nu\eig}$}}
			\put(52,1.5){\small{${H}_{22}^{\nu\eig}$}}
			\put(64.25,1.5){\small{${H}_{33}^{\nu\eig}$}}
			\put(75.5,1.5){\small{$-{H}_{11}^{\tau\eig}$}}
			\put(88,1.5){\small{$-{H}_{22}^{\tau\eig} $}}
			\put(100.5,1.5){\small{$-{H}_{33}^{\tau\eig}$}}
			\end{overpic}
			}
		\end{overpic}}
	\subfloat[]
	{\begin{overpic}
	[trim = -40mm -100mm 0mm -1mm,
	scale=0.125,clip,tics=20]{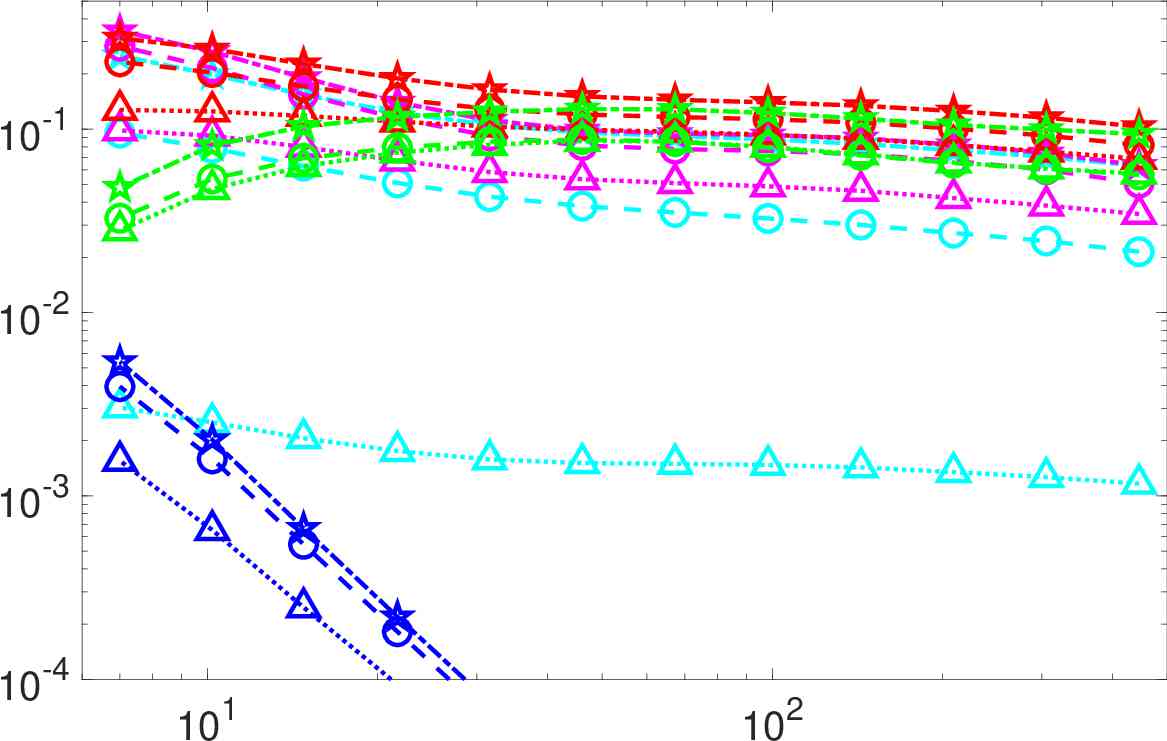}
			\put(55,17){$\ell_F/\eta$}
			% \put(0,18){\rotatebox{90}{$\mathbb{E} (\widetilde{\tau}^2X^2) - (\mathbb{E} (\widetilde{\tau}X))^2$}}
			\put(0,32){\rotatebox{90}{$\avg{ \widetilde{\tau}^4 X^2 }- {\avg{ {\widetilde{\tau}}^2 X}}^2$}}
	\end{overpic}}
	\caption{%\textcolor{red}{NewFig}
	(a) Average and (b) variance of the terms in \eqref{theory_eq_strain_lambda}, normalized by the timescale $\widetilde{\tau}$, and plotted as a function of the filtering scale $\ell_F/\eta$.
	Different colors distinguish the various terms, while different line types and symbols refer to different components of those terms.}
	\label{res_avg_dlambda}
\end{figure}

Figure \ref{res_avg_dlambda} shows the averages and second moments of the contributions to the evolution of the strain-rate eigenvalues, governed by equation \eqref{theory_eq_strain_lambda}.
Note that the average contributions need not be zero. For example, while $\langle{\bm{e}_i\bm{\cdot}\bm{H}^P\bm{\cdot}\bm{e}_j }\rangle=0$ for a homogeneous flow, $\langle H_{ij}^{P\eig}\rangle\equiv\langle{\bm{v}_i\bm{\cdot}\bm{H}^P\bm{\cdot}\bm{v}_j} \rangle$ need not be zero because $\bm{v}_i$ fluctuates and is correlated with $\bm{H}^P$.

The most negative strain-rate eigenvalue, $\lambda_3$, has on average the largest magnitude among the strain-rate eigenvalues. An implication of this is, for example, that its contribution dominates the strain self-amplification, $-\inlavg{\lambda_3^3}>\inlavg{\lambda_1^3}$ at all scales in the flow \citep{Tsinober2001,Carbone2020a}.
%%%!
The term $-\lambda_3^2$ drives the Restricted Euler system towards a finite-time singularity since it dominates the dynamics amplifying a negative $\lambda_3$. However, the rotation of the fluid element gives a strong stabilizing contribution through $\omega^2-\omega_3^{\eig 2}$, with magnitude that is comparable to that of the self-amplification of $\lambda_3$.
 The misalignment between $\bm{\omega}$ and $\bm{v}_3$ can be then traced back to the importance of the stabilizing effect of $\omega^2-\omega_3^{\eig 2}$. In particular, the vorticity component $\omega_3^\eig$ (and the corresponding alignment $\omega_3^\eig/\omega$) is small along the right Vieillefosse tail where the Restricted Euler system blows up, as shown in figure \ref{res_fig_omg_RQ}.
 The stabilizing effect of the rotation of the fluid element is exploited in reduced models for the velocity gradient dynamics to avoid the finite-time singularity thus producing steady-state statistics of the velocity gradient. Indeed, when the relative weight of the strain self-amplification $\alpha\lambda_i^2$ is reduced with respect to the magnitude of the rotation term $\beta(\omega-\omega_i^{\eig2})$, through the model coefficients $\alpha$ and $\beta$, then steady-state statistics can be obtained \citep{Wilczek2014,Lawson2015}.
This reduction of non-linearity is attributed mainly to the pressure Hessian.
%%%!
On the other hand, the term $-\lambda_1^2$ has by construction a stabilizing effect on $\lambda_1$ while the corresponding vorticity contribution $(\omega^2-\omega_1^{\eig 2})$ helps the growth of $\lambda_1$.

Since $-\lambda_3^2$ tends to make the dynamics unstable, it would be expected that, in order to prevent the velocity gradients becoming singular, the strongest effect of the pressure Hessian would be on $\lambda_3$. However, the average pressure Hessian contribution to $\lambda_3$, namely $\langle -H_{33}^{P\eig}\rangle$, is the smallest among the pressure components, although its variance is the largest, as shown in figure \ref{res_avg_dlambda}(b). Nevertheless, on average, this term does tend to hinder the growth of negative $\lambda_3$. 
The diagonal components of $\langle -H_{ij}^{P\eig}\rangle$, when normalized by $\widetilde{\tau}^2$, do not show significant variations as a function of $\ell_F$. Therefore, at least on average, $H_{ij}^{P\eig}$ scales approximately as $\sim\widetilde{\tau}^{-2}$, as expected from dimensional analysis, and consistent with models such as that of \cite{Wilczek2014}, which express the pressure Hessian as a linear combination of $\bm{S\cdot S}$ and $\bm{W\cdot W}^\top$. However, we would expect departures from this scaling for higher order moments of $-H_{ij}^{P\eig}$ due to intermittency.

The average effect of the pressure Hessian on $\lambda_1$ is $\langle-H_{11}^{P\eig}\rangle$ and is positive, indicating that on average the pressure Hessian helps the growth of $\lambda_1$. This also implies that the pressure Hessian indirectly contributes to the stretching of $\bm{\omega}$ along $\bm{v}_1$, opposing the preferential alignment of the vorticity with the intermediate eigenvector $\bm{v}_2$.
Interestingly, the largest average contribution from the pressure Hessian is for the intermediate eigenvalue, $\lambda_2$, and $\langle-H_{22}^{P\eig}\rangle$ is negative at all scales, driving $\lambda_2$ towards negative values. In this sense, the pressure Hessian hinders vortex stretching, suppressing the non-linear amplification of $\omega_2^\eig$ through $\lambda_2$ (see equation \eqref{theory_eq_vort_princ}).

The viscous term $\langle H_{i(i)}^{\nu\eig}\rangle$ tends to hinder all the eigenvalues, with $\langle H_{11}^{\nu\eig}\rangle<0$, $\langle H_{22}^{\nu\eig}\rangle<0$ and $\langle H_{33}^{\nu\eig}\rangle>0$. In the dissipation range, $\langle H_{i(i)}^{\nu\eig}\rangle$ is largest in magnitude for $i=3$, which is associated with $\lambda_3$ having the largest magnitude on average. Furthermore, $\langle H^{\nu\eig}_{22}\rangle $ is very small compared to the other components, and therefore because of imcompressibility, $\langle H_{11}^{\nu\eig}\rangle\approx -\langle H_{33}^{\nu\eig}\rangle$.
The term is related to the curvature of the strain field and the clear tendency for $\langle H_{22}^{\nu\eig}\rangle$ to be small can be a consequence of the moderate fluctuations of the intermediate eigenvalue. Indeed, the contribution from $\lambda_2$ to the strain self-amplification, namely $\inlavg{\lambda_2^3}$, is the smallest among the contributions of the eigenvalues \citep{Tsinober2001,Carbone2020a}.
Also, the sign of $\lambda_2$ fluctuates and the average $\lambda_2$ is small with respect to the average of the other eigenvalues.
The average viscous stress components vary considerably across the scales, as expected since by definition these terms play a sub-leading dynamical role outside of the dissipation range.

The role of the sub-grid stress has not been investigated much in the literature, especially from the perspective of the strain-rate eigenframe.
The sub-grid stress contribution to the dynamics increases with increasing $\ell_F$, and at the largest scales the sub-grid stress makes a leading order contribution to the eigenvalue dynamics. The sub-grid stress has strong variations across the scales even if normalized with a scale-dependent time scale.
However, across all the scales $\langle H^{\tau\eig}_{11}\rangle $ remains very small compared to the other components and, as a consequence, $\langle H^{\tau\eig}_{33}\rangle\simeq \langle H^{\tau\eig}_{22}\rangle$.
The sub-grid stress tends on average to drive $\lambda_2$ towards negative values and it hinders the growth of $|\lambda_3|$. In this sense, for the intermediate and most compressional principal directions, it acts similarly to the pressure Hessian, even if the quantitative trends of $H^{P\eig}_{i(i)}$ and $H^{\tau\eig}_{i(i)}$ across the scales are very different.

Figure \ref{res_avg_dlambda}(b) shows the variance of the contributions to the eigenvalue equation. The results show that the variance of $H_{i(i)}^{P\eig}$ is the largest of the contributions, with the variance of $H_{11}^{P\eig}$ and $H_{33}^{P\eig}$ decreasing as $\ell_F$ is increased, but becoming approximately constant in the inertial range when normalized by $\widetilde{\tau}$. On the other hand, the variance of $H_{22}^{P\eig}$ is almost independent of $\ell_F$ when normalized by $\widetilde{\tau}$, and this component is always the smallest. This is perhaps the reason why $\lambda_2$ undergoes smaller fluctuations in its time evolution than the other eigenvalues, which confirms the observation by \cite{Dresselhaus1992} concerning the relatively small magnitude and persistency of the intermediate eigenvalue. The variance of the viscous term, normalized by $\widetilde{\tau}$, becomes very small as $\ell_F$ is increased, decreasing as $\ell_F^{-\xi}$ with $\xi$ between 2 and 3.

Since there is a sign ambiguity in the definition of the eigenvectors $\bm{v}_i$ there is a corresponding ambiguity in the sign of $\omega^*_i$. When solving \eqref{theory_eq_vort_princ} in a Lagrangian frame this ambiguity is removed through the choice of a particular direction for $\bm{v}_i$ in the initial conditions. However, we are computing terms based on data in the Eulerian frame (i.e.\ at fixed grid points). Therefore, instead of considering the dynamical contributions to \eqref{theory_eq_vort_princ}, we will instead consider the dynamical contributions to the enstrophy equation (here we are calling $\omega_i^{\eig2}$ the enstrophy, although strictly speaking, the enstrophy is $\|\bm{\omega}\|^2=\sum_i \omega_i^{\eig2}$)
\begin{equation}
\frac{1}{2}D_t \omega_i^{\eig2} = \lambda_{(i)}\omega_i^{\eig2} - \Pi_{ij}^\eig\omega_j^\eig\omega_{(i)}^\eig + \Omega_i^{\nu\eig}\omega_{(i)}^\eig - \Omega_i^{\tau\eig}\omega_{(i)}^\eig.
\label{res_eq_vort2_princ}
\end{equation}

%%%%vorticity
\begin{figure}
\centering
\vspace{0mm}			
    \subfloat[]	
	{\begin{overpic}
	[trim = 0mm -100mm -40mm -1mm,
	scale=0.125,clip,tics=20]{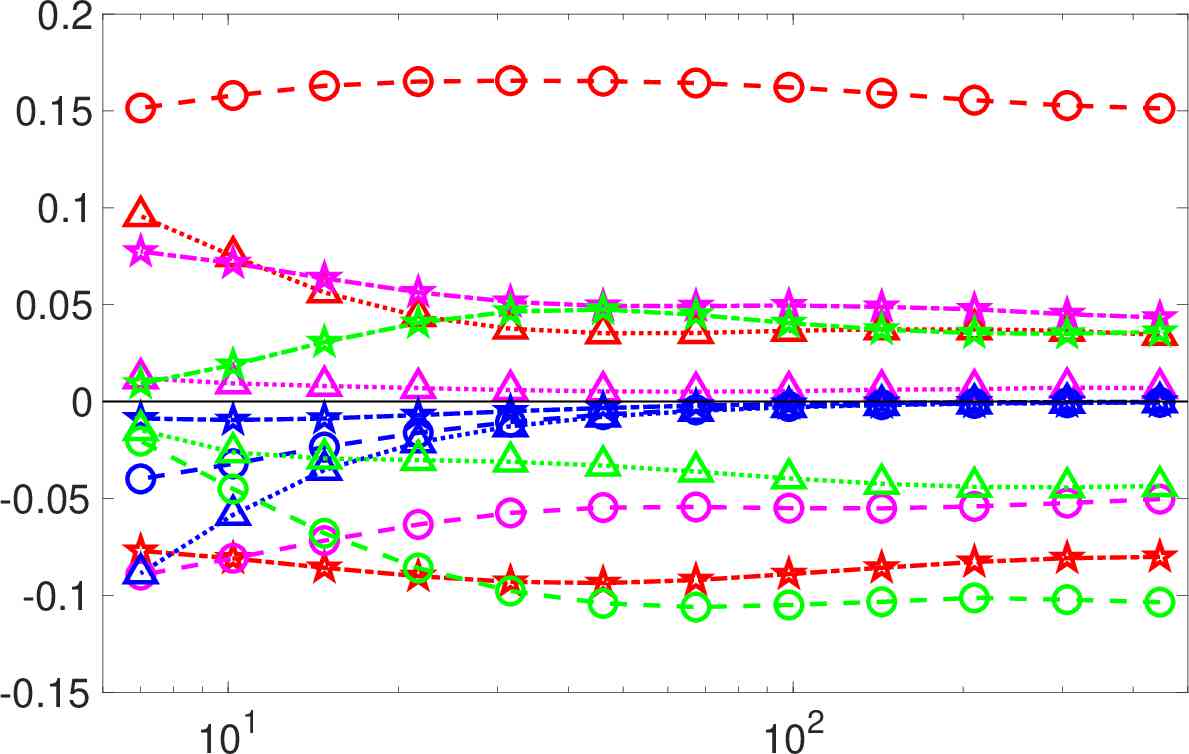}
			\put(45,17){$\ell_F/\eta$}
			% \put(-6,30){\rotatebox{90}{$\mathbb{E} (\widetilde{\tau}^2 X)$}}
			\put(-6,43){\rotatebox{90}{$\avg{\widetilde{\tau}^3 X}$}}
			
			\put(-5,6)
			{
			\begin{overpic}
			[trim = 35mm 272mm 30mm 27mm,
	        scale=0.30,clip,tics=20,]{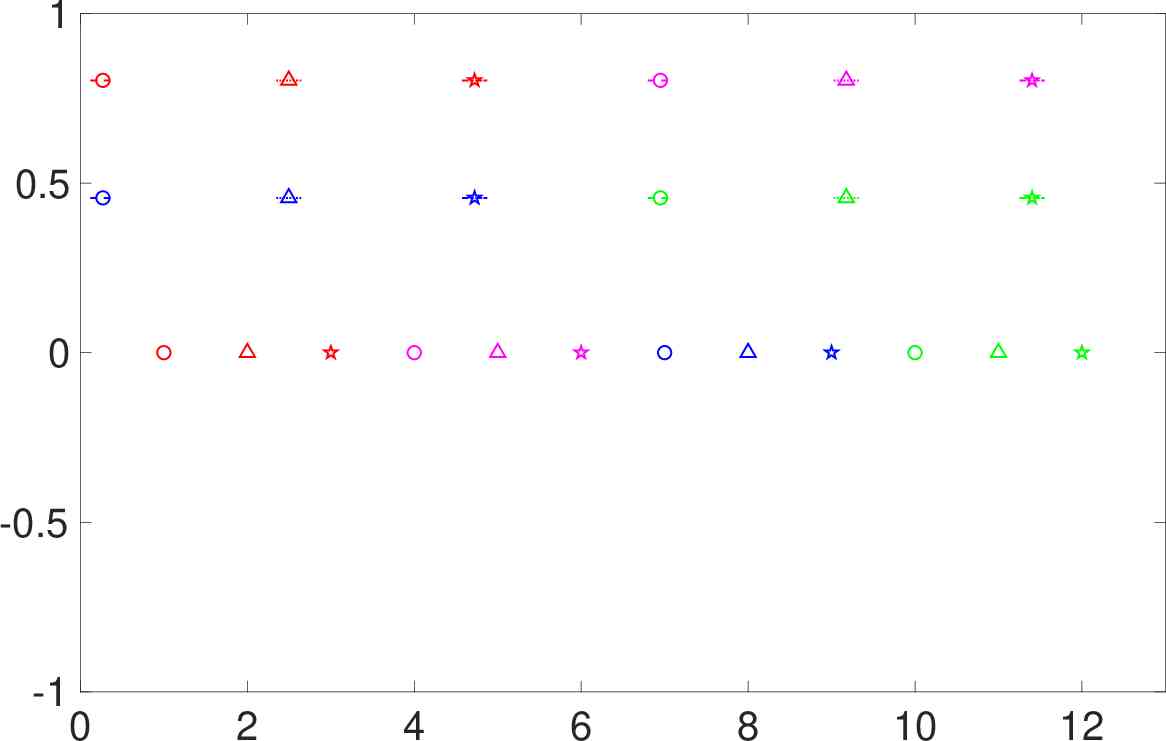}
			\put(3.5,1.5){\small{${\lambda_{1}{\omega_1^\eig}^2}$}}
			\put(22,1.5){\small{${\lambda_{2}{\omega_2^\eig}^2}$}}
			\put(40.5,1.5){\small{${\lambda_{3}{\omega_3^\eig}^2}$}}
			\put(59,1.5){\small{$- \Pi_{1j}^\eig\omega_j^\eig\omega_{1}^\eig$}}
			\put(77.5,1.5){\small{$- \Pi_{2j}^\eig\omega_j^\eig\omega_{2}^\eig$}}
			\put(96,1.5){\small{$- \Pi_{3j}^\eig\omega_j^\eig\omega_{3}^\eig$}}
			\end{overpic}
			}
			
			\put(-5,-2)
			{
			\begin{overpic}
			[trim = 35mm 222mm 30mm 80mm,
	        scale=0.30,clip,tics=20,]{JPEG_Figures/try_315/NA_avgvarterms_eqnwise_Fig09.jpg}
			\put(3.5,1.5){\small{$\bm{\Omega}_1^{\nu\eig} \omega_1^\eig$}}
			\put(22,1.5){\small{$\bm{\Omega}_2^{\nu\eig} \omega_2^\eig$}}
			\put(40.5,1.5){\small{$\bm{\Omega}_3^{\nu\eig} \omega_3^\eig$}}
			\put(59,1.5){\small{$-\bm{\Omega}_1^{\tau\eig} \omega_1^\eig$}}
			\put(77.5,1.5){\small{$-\bm{\Omega}_2^{\tau\eig} \omega_2^\eig$}}
			\put(96,1.5){\small{$-\bm{\Omega}_3^{\tau\eig} \omega_3^\eig$}}
			\end{overpic}
			}
		\end{overpic}}
	\subfloat[]
	{\begin{overpic}
	[trim = -40mm -100mm 0mm -1mm,
	scale=0.125,clip,tics=20]{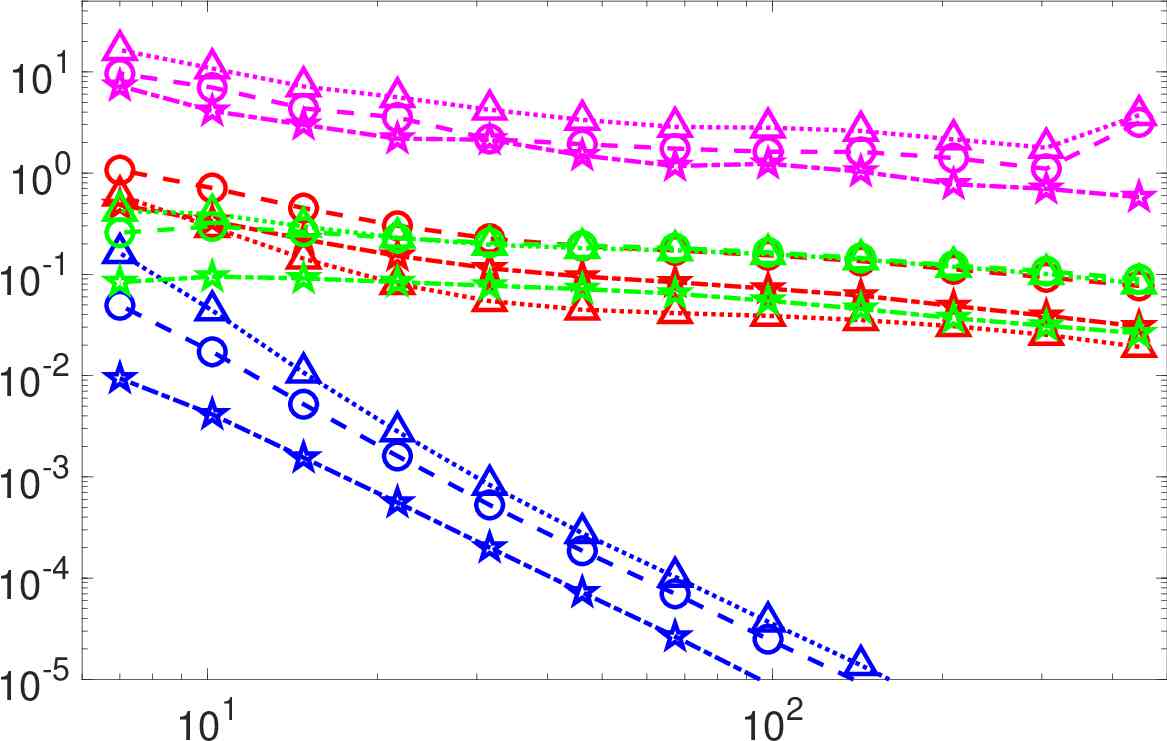}
			\put(55,17){$\ell_F/\eta$}
			% \put(0,18){\rotatebox{90}{$\mathbb{E} (\widetilde{\tau}^2X^2) - (\mathbb{E} (\widetilde{\tau}X))^2$}}
			\put(0,32){\rotatebox{90}{$\avg{ \widetilde{\tau}^6 X^2 }- {\avg{ {\widetilde{\tau}}^3 X}}^2$}}
	\end{overpic}}
	\caption{%\textcolor{red}{NewFig}
	(a) Average and (b) variance of the terms in \eqref{res_eq_vort2_princ} , normalized by the timescale $\widetilde{\tau}$, and plotted as a function of the filtering scale $\ell_F/\eta$.
	Different colors distinguish the various terms, while different line types and symbols refer to different components of those terms.}
    \label{res_avg_denstrophy}
\end{figure}
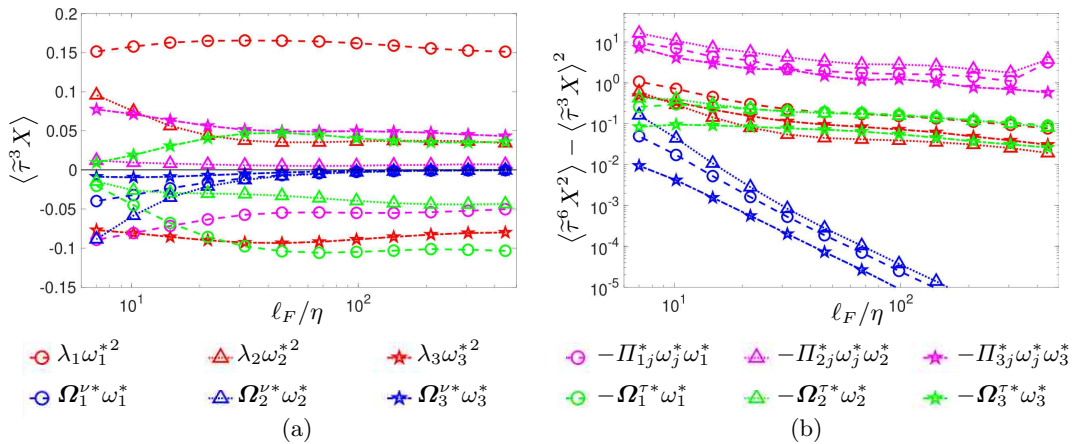

In figure \ref{res_avg_denstrophy}(a) we show the averages of the terms on the right hand side of \eqref{res_eq_vort2_princ}. The results show that the vortex stretching term $ \lambda_{(i)}\omega_i^{\eig2}$ acts on average to increase $\omega_1^{\eig 2}$ and $\omega_2^{\eig 2}$, and to reduce $\omega_3^{\eig 2}$. It is known that although $\bm{\omega}$ is preferentially aligned with $\bm{v}_2$ \cite{Ashurst1987,Meneveau2011}, $\lambda_1\omega_1^{\eig 2}$ gives on average the largest contribution to vortex stretching $\langle\bm{\omega}^\top\bm{\cdot S \cdot \omega}\rangle=\sum_{i}\langle\lambda_i\omega_i^{\eig 2}\rangle$ in the limit $\ell_F/\eta\to 0$ (i.e.\ for the unfiltered gradients) since $\lambda_1$ tends to be larger than $\lambda_2$ \cite{tsinober}. Our results show that for $\ell_F$ outside of the dissipation range, the contribution to vortex stretching from $\lambda_1\omega_1^{\eig 2}$ becomes increasingly dominant, with $\langle\lambda_2\omega_2^{\eig 2}\rangle\ll \langle\lambda_1\omega_1^{\eig 2}\rangle$ in the inertial range. It is also interesting to note that while the normalized average $\widetilde{\tau}^3\langle\lambda_i\omega_i^{\eig 2}\rangle$ changes substantially for $i=2$ as $\ell_F$ is increased from the dissipation to inertial range scales, it varies weakly with $\ell_F$ for $i=1,3$. This is in agreement with the weakening of the preferential alignment between the vorticity and the intermediate strain-rate eigenvector as $\ell_F$ is increased, while the statistical alignment of $\bm{\omega}$ with $\bm{v}_1$ and $\bm{v}_3$ depends very weakly on the filtering length  \citep{Danish2018}.

%%%!
The alignment between the vorticity and the strain-rate eigenvectors is in part governed by the vorticity tilting term $\Pi_{ij}^{\eig}\omega_j^\eig$. That term plays a central role in determining the preferential alignment between $\bm{\omega}$ and $\bm{v}_2$ in the Restricted Euler system, through $\Pi_{ij}^{RE\eig}\omega_j^\eig$, \citep{Dresselhaus1992,Nomura1998} and also in real turbulent flows due to the dominance of local over non-local contributions to the vorticity tilting \citep{Lawson2015}.
The vorticity tilting does not directly affect the total enstrophy $\|\bm{\omega}\|^2$, but it does contribute to the dynamics of the individual contributions $ \omega_i^{\eig2} $. In particular, the results in figure \ref{res_avg_denstrophy} indicate that on average $\Pi_{ij}^\eig\omega_i^\eig\omega_j^\eig$ tends to enhance $ \omega_3^{\eig2} $ while reducing $ \omega_1^{\eig2} $. Therefore, the tilting tends to oppose the stretching of vorticity along $\bm{v}_1$ and to oppose the compression of vorticity along $\bm{v}_3$. The results also show that the vorticity tilting makes a small positive contribution to the growth of $ \omega_2^{\eig2} $ on average. 

The viscous term acts on average to reduce the magnitude of all the components of enstrophy, with its effect strongest on $\omega_2^{\eig 2}$ and weakest on $\omega_3^{\eig 2}$. As expected, its average contribution markedly decreases with increasing filtering length as $\ell_F$ moves outside the dissipation range, with a $\ell_F^{-\xi}$ trend, with $\xi$ between $2$ and $3$. The average sub-grid stress contribution to $D_t \omega_i^{\eig2}$ in equation \eqref{res_eq_vort2_princ} increases as $\ell_F$ is increased and becomes approximately constant in the inertial range when normalized by $\widetilde{\tau}^3$, where its contribution becomes comparable in magnitude to the vortex-stretching terms for $i=2,3$. However, the sub-grid contribution is opposite in sign to the vortex stretching contributions, tending to hinder $\omega_1^{\eig2}$ and $\omega_2^{\eig2}$ and to enhance $\omega_3^{\eig2}$.

The variances of the terms on the right hand side of \eqref{res_eq_vort2_princ} are shown in figure \ref{res_avg_denstrophy}(b).
%%%!
The variance of the vorticity tilting terms is very large, which indicates sudden rotations of the vorticity vector with respect to the eigenframe.
Similar to the behavior of the averages in figure \ref{res_avg_denstrophy}(a), the results show that fluctuations in $\lambda_i\omega_i^{\eig 2}$ are greatest for $i=1$, and the contribution from $i=2$ becomes much smaller than that from $i=1$ for $\ell_F$ in the inertial range. Most interestingly, we find that the sub-grid contributions for $i=1,2$ are very similar (almost identical in the inertial range), while the contribution for $i=3$ is much smaller, a feature preserved across the scales. This indicates that there is not a simple relationship between the sub-grid and vortex stretching terms, which poses a challenge for Large Eddy Simulation (LES) modeling.

We have computed the moments in figure \ref{res_avg_dlambda} and \ref{res_avg_denstrophy} across
a range of  Reynolds numbers (not shown) and obtained the same trend for all the terms. This confirms that the statistics are converged and show that the moments depend weakly on the Reynolds number, at least in the range considered ($Re_{\lambda} \approx 200 - 600$).

%%%!
\subsection{Behavior of eigenvalues and vorticity in the $R$, $Q$ plane}

\begin{figure}
\centering
\vspace{0mm}			
    \subfloat[]	
	{\begin{overpic}
	[trim = 0mm -28mm -25mm -1mm,
	scale=0.085,clip,tics=20]{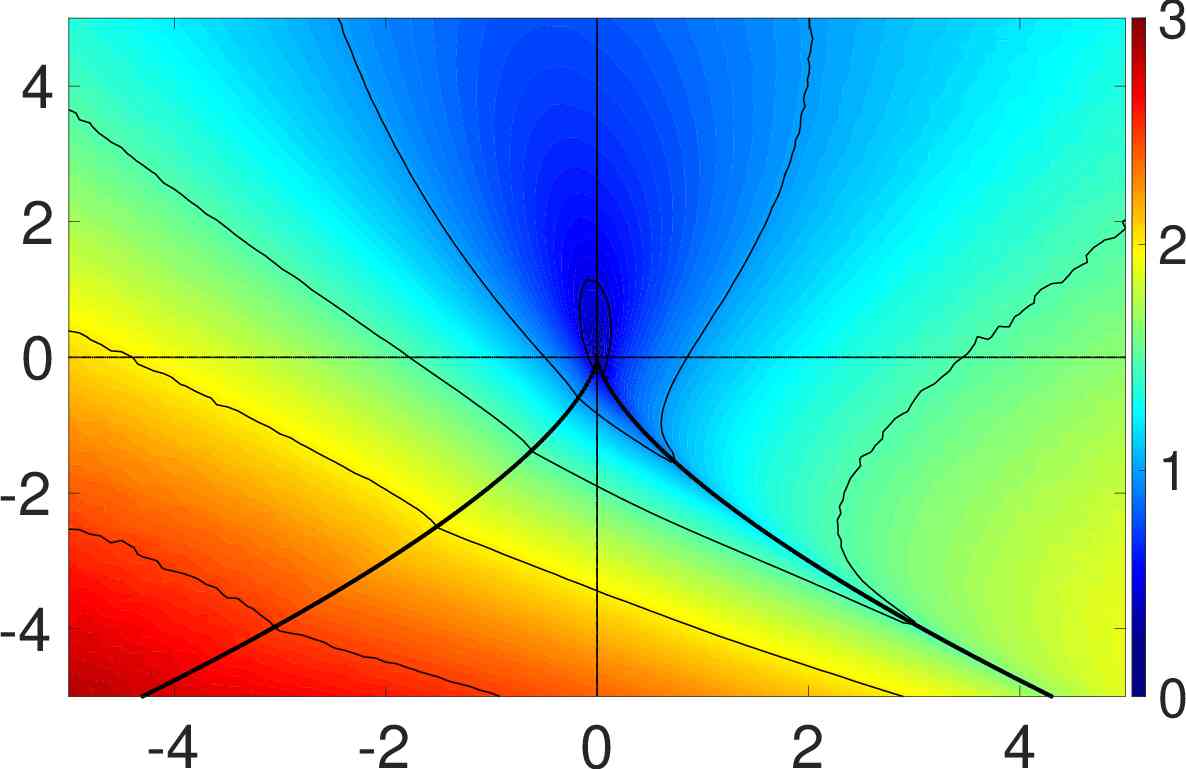}
	\put(-8,31){\rotatebox{90}{\small{${\widetilde{\tau}}^2 Q$}}}
    %\put(40,-2){\small{${\widetilde{\tau}}^3 R$}}
	\end{overpic}}
	\subfloat[]
	{\begin{overpic}
	[trim = -25mm -28mm -25mm -1mm,
	scale=0.085,clip,tics=20]{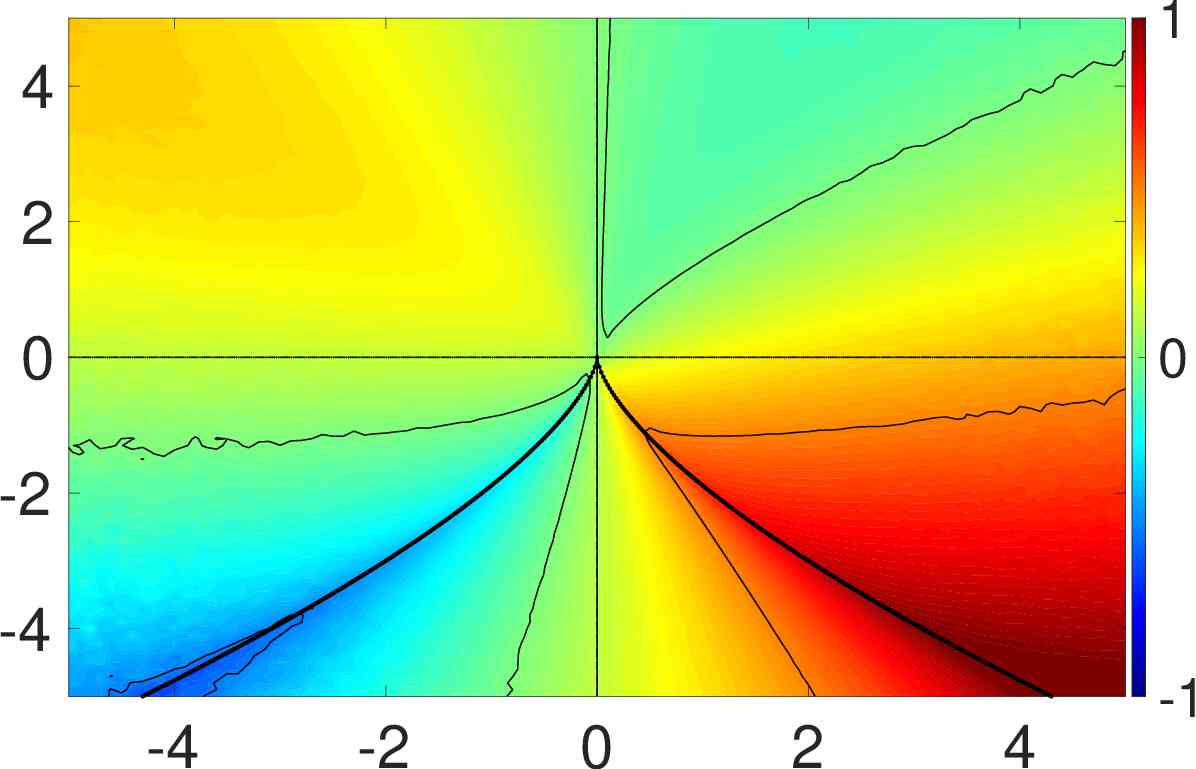}
	%\put(-3,31){\rotatebox{90}{\small{${\widetilde{\tau}}^2 Q$}}}
    %\put(43,-2){\small{${\widetilde{\tau}}^3 R$}}
	\end{overpic}}
	\subfloat[]
    {\begin{overpic}
	[trim = -25mm -28mm 0mm -1mm,
	scale=0.085,clip,tics=20]{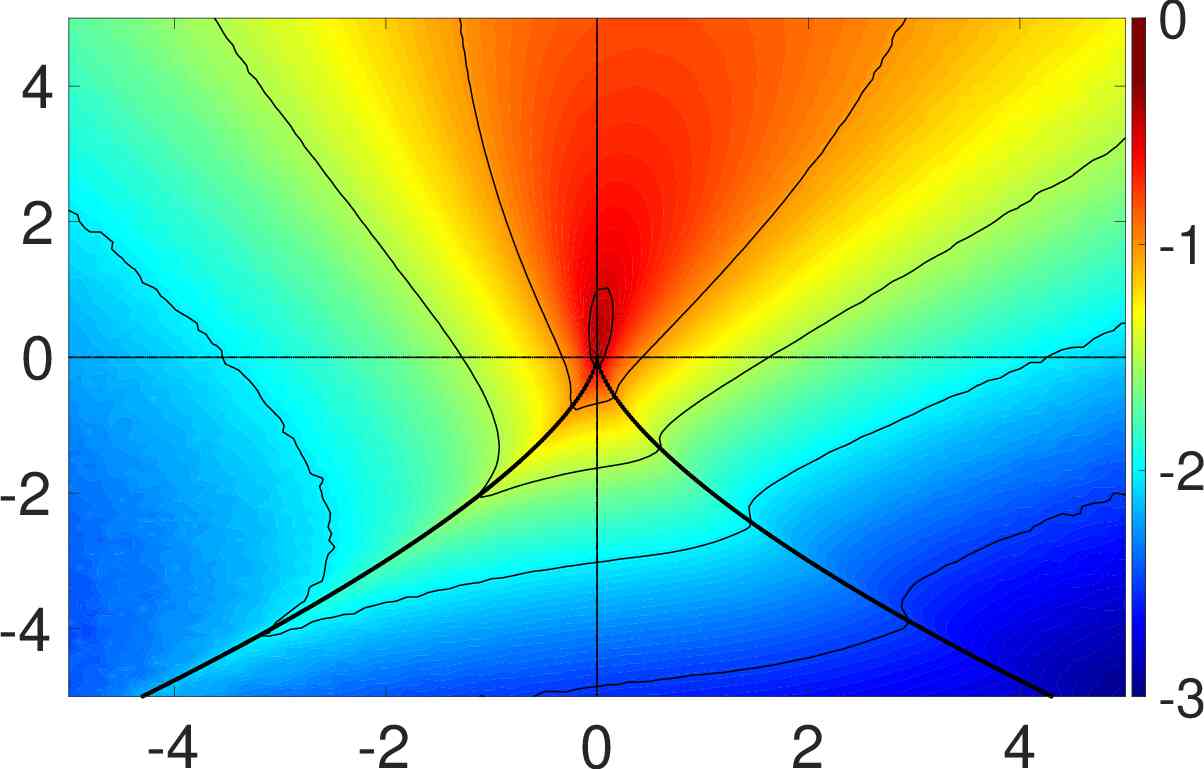}
	%\put(-3,31){\rotatebox{90}{\small{${\widetilde{\tau}}^2 Q$}}}
    %\put(45,-2){\small{${\widetilde{\tau}}^3 R$}}
	\end{overpic}} \\
   \subfloat[]	
	{\begin{overpic}
	[trim = 0mm -28mm -25mm -1mm,
	scale=0.085,clip,tics=20]{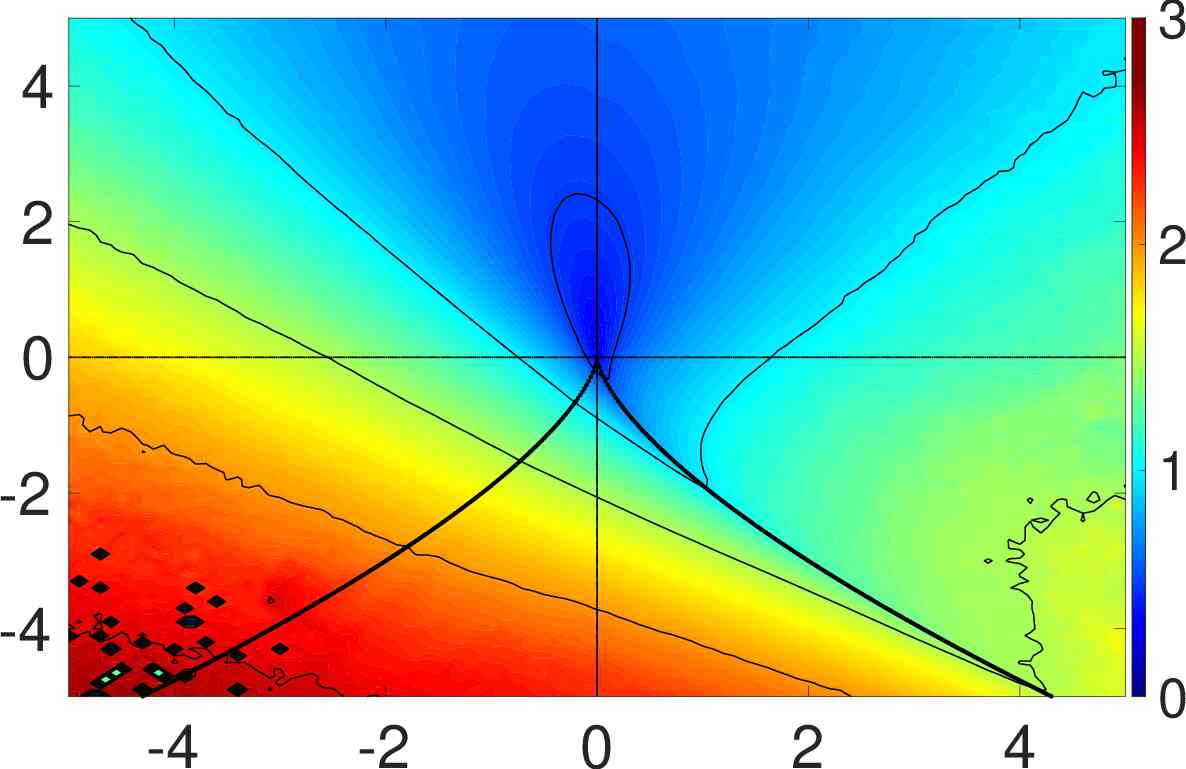}
	\put(-8,31){\rotatebox{90}{\small{${\widetilde{\tau}}^2 Q$}}}
    \put(40,-2){\small{${\widetilde{\tau}}^3 R$}}
	\end{overpic}}
	\subfloat[]
	{\begin{overpic}
	[trim = -25mm -28mm -25mm -1mm,
	scale=0.085,clip,tics=20]{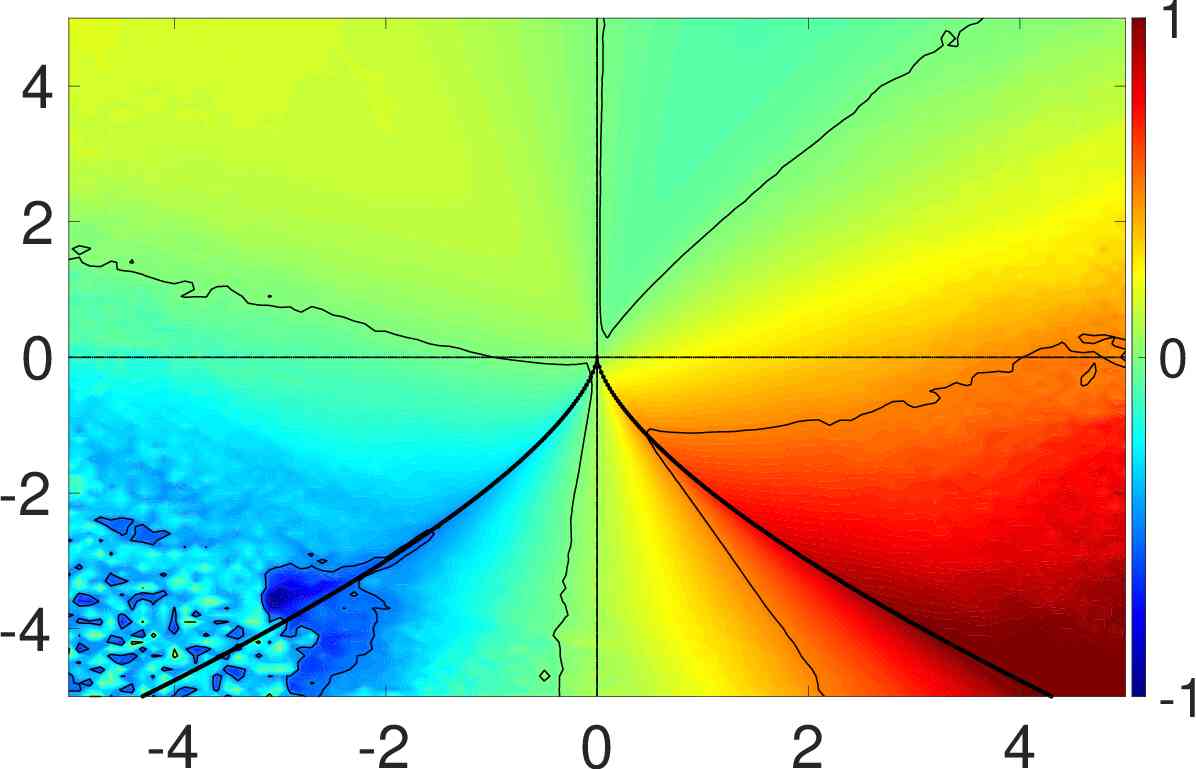}
	%\put(-3,31){\rotatebox{90}{\small{${\widetilde{\tau}}^2 Q$}}}
    \put(43,-2){\small{${\widetilde{\tau}}^3 R$}}
	\end{overpic}}
	\subfloat[]
	{\begin{overpic}
	[trim = -25mm -28mm 0mm -1mm,
	scale=0.085,clip,tics=20]{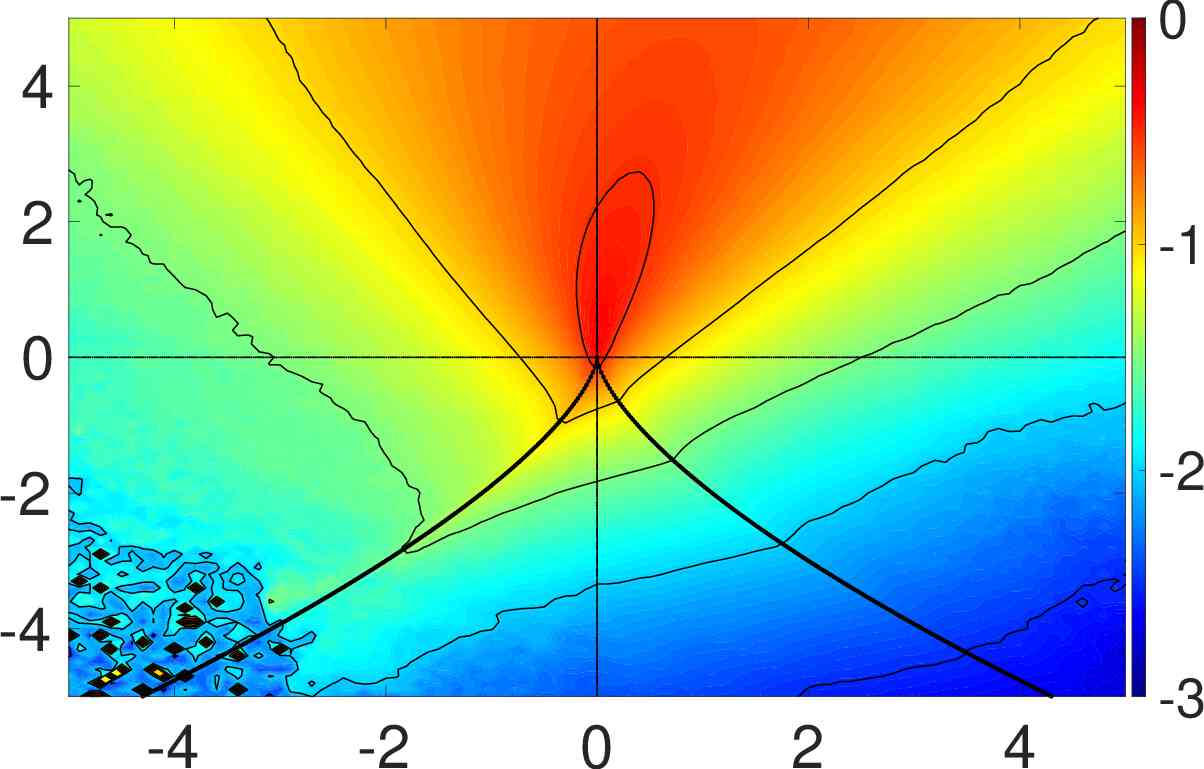}
	%\put(-3,31){\rotatebox{90}{\small{${\widetilde{\tau}}^2 Q$}}}
    \put(45,-2){\small{${\widetilde{\tau}}^3 R$}}
	\end{overpic}}
	\caption{Average of the filtered strain-rate eigenvalues conditioned on the second and third principal invariants of the filtered velocity gradient tensor.
	(a-b-c) show conditional averages $\inlavg{\widetilde{\tau}\lambda_{1}|R,Q}$, $\inlavg{\widetilde{\tau}\lambda_{2}|R,Q}$ and $\inlavg{\widetilde{\tau}\lambda_{3}|R,Q}$, respectively, for $\ell_F/\eta=7.0$ and (d-e-f) shows the corresponding quantities for $\ell_F/\eta=67.3$. The color map is in linear scale, and the thick black lines represent the Vieillefosse tails.}
	\label{res_fig_lam_RQ}
\end{figure}

%%%%%%lambda omega cond R Q

For further insight into the eigenvalue and vorticity dynamics, in figure \ref{res_fig_lam_RQ} we show the average of the filtered strain-rate eigenvalues conditioned on the principal invariants of the filtered velocity gradient $\inlavg{\widetilde{\tau}\lambda_i| {\widetilde{\tau}}^3 R, {\widetilde{\tau}}^2 Q}$, where $R=-\Tr(\bm{A}^3)/3$ and $Q=-\Tr(\bm{A}^2)/2$. The color map is in a linear scale, and the results refer to the filtering lengths $\ell_F/\eta=7.0$ and $\ell_F/\eta=67.3$.
The results show that $\inlavg{\lambda_2|R,Q}$ is relatively large and positive along the right Vieillefosse tail. Since the joint probability density function (PDF) of $R,Q$ is large along the right Vieillefosse tail \citep{Chong1998,Luthi2009,Elsinga2010,Meneveau2011} then that phase space region contributes strongly to the tendency for $\lambda_2$ to be positive, with $\inlavg{\lambda_2}>0$.
In contrast, $\inlavg{\lambda_1 | R,Q}$ is quite small along the right Vieillefosse tail, but is large along the left Vieillefosse tail, where $R<0$, corresponding to states of biaxial compression with $\lambda_2<0$. 
Both $\inlavg{\lambda_1|R,Q}$ and $\inlavg{\lambda_3|R,Q}$ tend to become relatively small in the quadrant $Q>0, R<0$, which corresponds to the vortex stretching quadrant \citep{Meneveau2011}. On the other hand, $\inlavg{\lambda_2|R,Q}$ is positive on average here, showing that it contributes to the stretching of vorticity along the direction $\bm{v}_2$ in this quadrant.

\begin{figure}
\centering
\vspace{0mm}			
    \subfloat[]	
	{\begin{overpic}
	[trim = 0mm -28mm -25mm -1mm,
	scale=0.085,clip,tics=20]{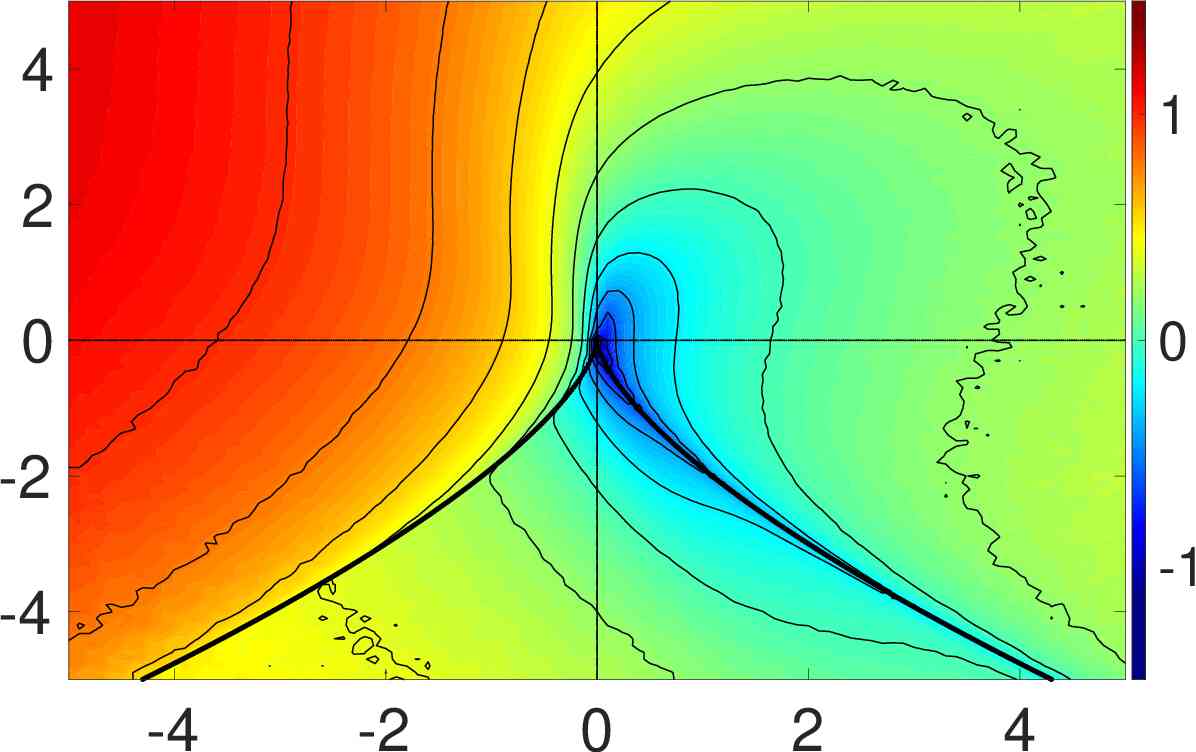}
	\put(-8,31){\rotatebox{90}{\small{${\widetilde{\tau}}^2 Q$}}}
    %\put(40,-2){\small{${\widetilde{\tau}}^3 R$}}
	\end{overpic}}
	\subfloat[]
	{\begin{overpic}
	[trim = -25mm -28mm -25mm -1mm,
	scale=0.085,clip,tics=20]{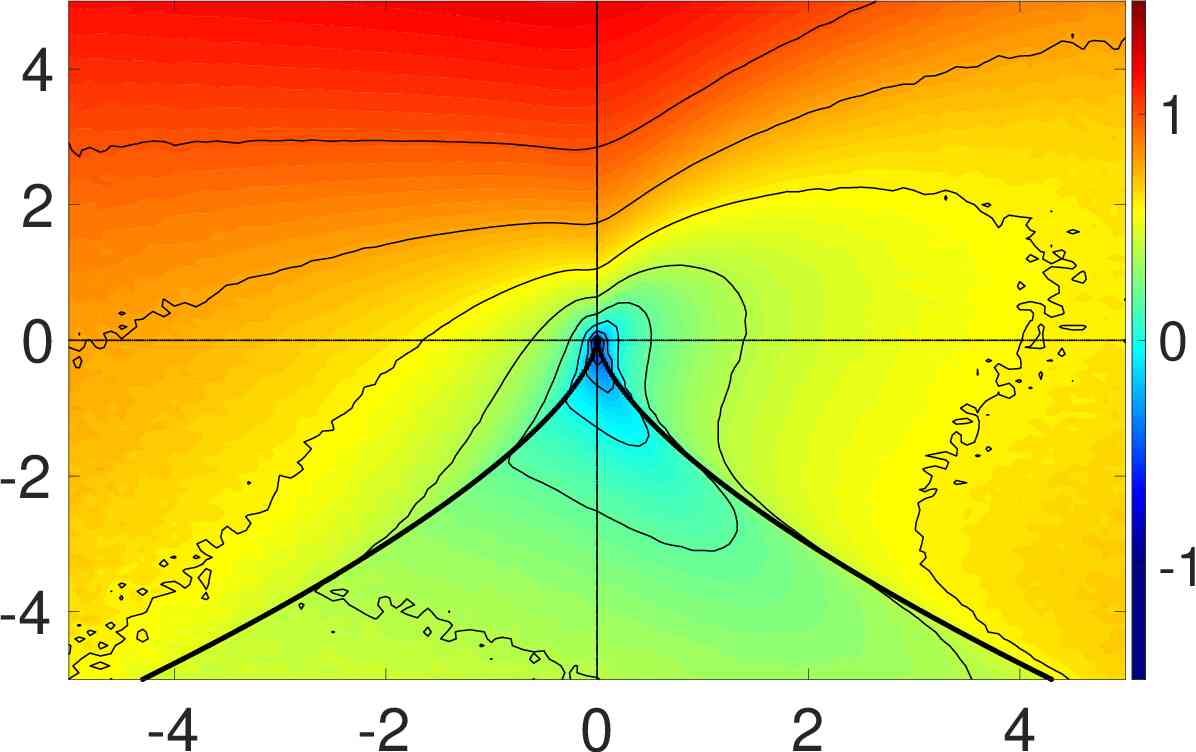}
	%\put(-3,31){\rotatebox{90}{\small{${\widetilde{\tau}}^2 Q$}}}
    %\put(43,-2){\small{${\widetilde{\tau}}^3 R$}}
	\end{overpic}}
	\subfloat[]
    {\begin{overpic}
	[trim = -25mm -28mm 0mm -1mm,
	scale=0.085,clip,tics=20]{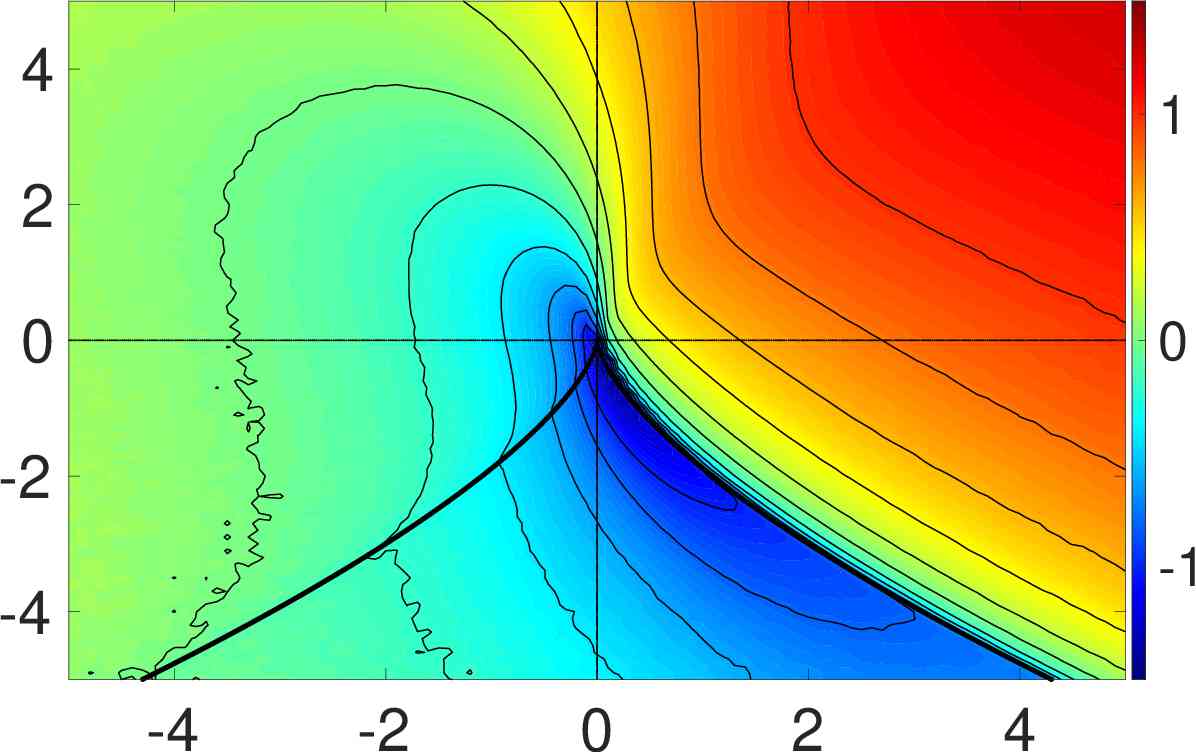}
	%\put(-3,31){\rotatebox{90}{\small{${\widetilde{\tau}}^2 Q$}}}
    %\put(45,-2){\small{${\widetilde{\tau}}^3 R$}}
	\end{overpic}} \\
   \subfloat[]	
	{\begin{overpic}
	[trim = 0mm -28mm -25mm -1mm,
	scale=0.085,clip,tics=20]{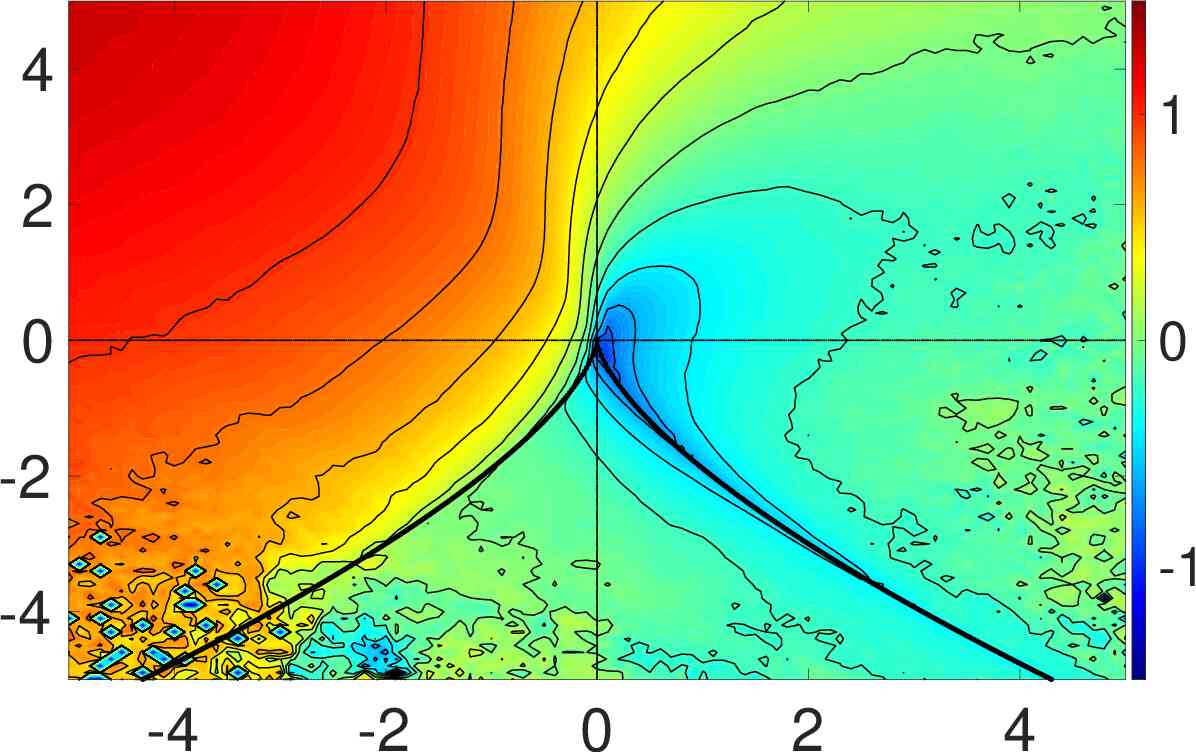}
	\put(-8,31){\rotatebox{90}{\small{${\widetilde{\tau}}^2 Q$}}}
    \put(40,-2){\small{${\widetilde{\tau}}^3 R$}}
	\end{overpic}}
	\subfloat[]
	{\begin{overpic}
	[trim = -25mm -28mm -25mm -1mm,
	scale=0.085,clip,tics=20]{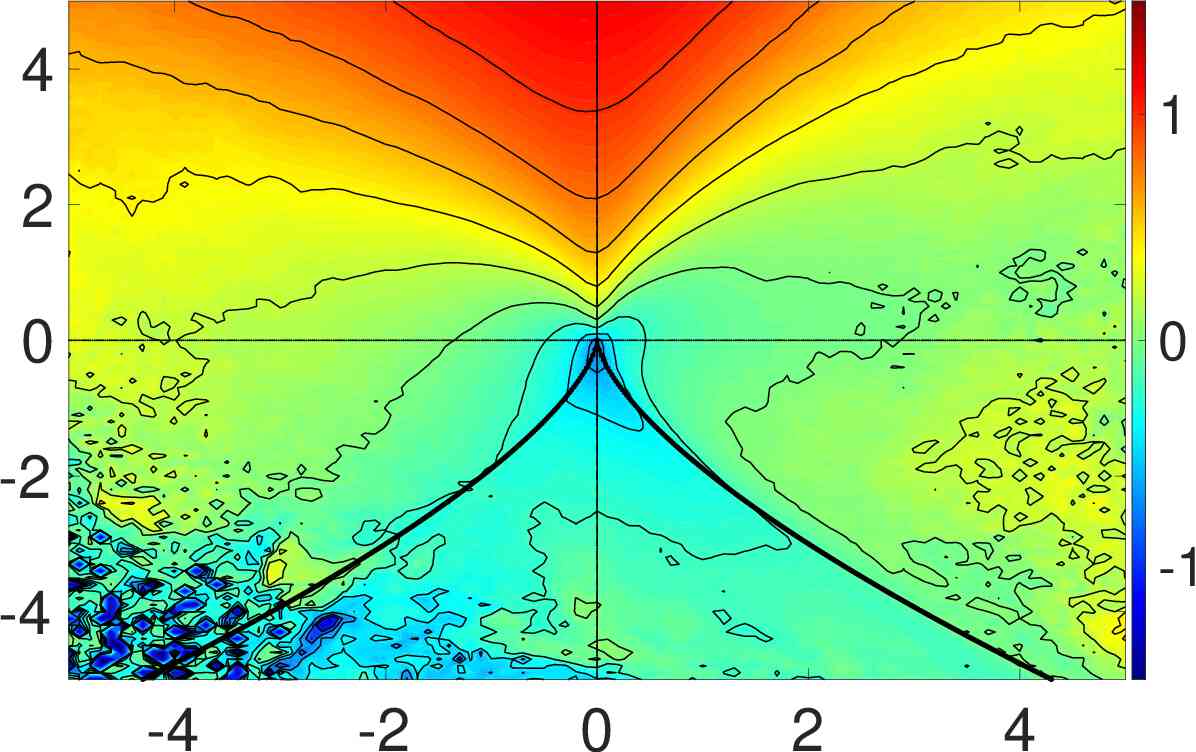}
	%\put(-3,31){\rotatebox{90}{\small{${\widetilde{\tau}}^2 Q$}}}
    \put(43,-2){\small{${\widetilde{\tau}}^3 R$}}
	\end{overpic}}
	\subfloat[]
	{\begin{overpic}
	[trim = -25mm -28mm 0mm -1mm,
	scale=0.085,clip,tics=20]{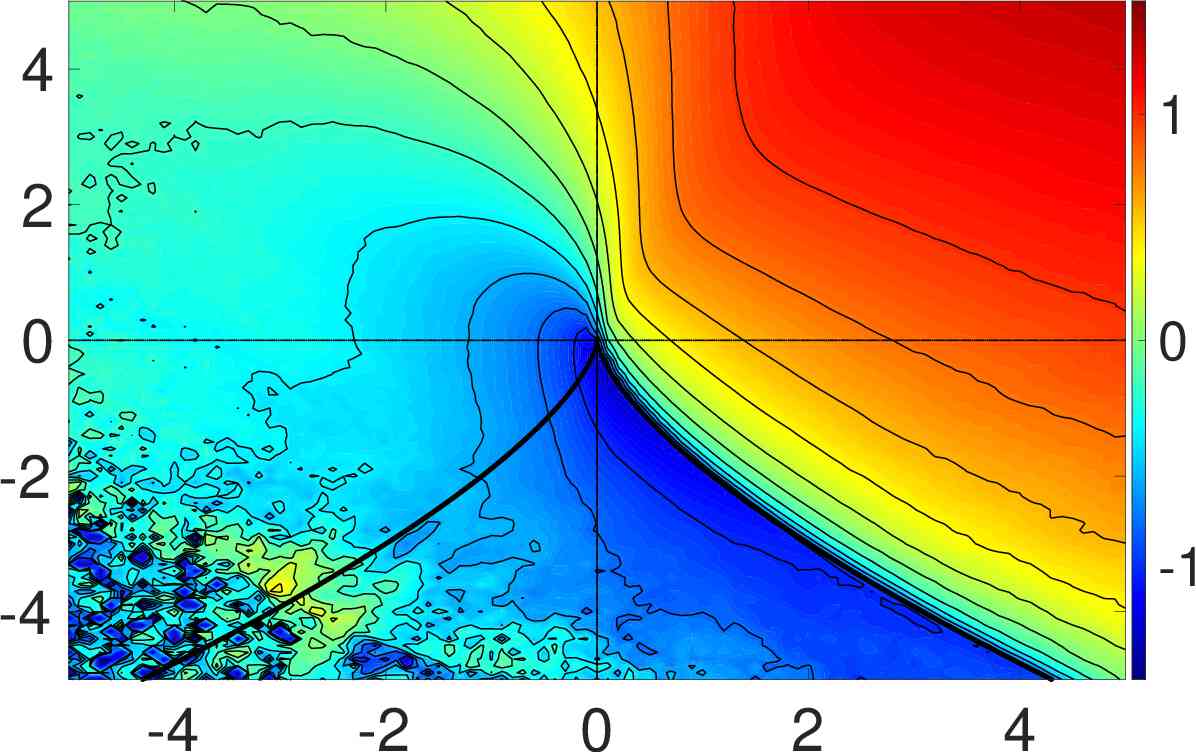}
	%\put(-3,31){\rotatebox{90}{\small{${\widetilde{\tau}}^2 Q$}}}
    \put(45,-2){\small{${\widetilde{\tau}}^3 R$}}
	\end{overpic}}
	\caption{Average of the square of the vorticity components conditioned on the second and third principal invariants of the filtered velocity gradient tensor.
	(a-b-c) show conditional averages $\inlavg{(\widetilde{\tau}\omega_1^\eig)^2|R,Q}$, $\inlavg{(\widetilde{\tau}\omega_2^\eig)^2|R,Q}$ and $\inlavg{(\widetilde{\tau}\omega_3^\eig)^2|R,Q}$, respectively, for $\ell_F/\eta=7.0$ and (d-e-f) shows the corresponding quantities at $\ell_F/\eta=67.3$. The color map is in $\log_{10}$ scale and the thick black lines represent the Vieillefosse tails.}
	\label{res_fig_omg_RQ}
\end{figure}

In figure \ref{res_fig_omg_RQ} we show the results for $\inlavg{(\widetilde{\tau}\omega_i^\eig)^2|{\widetilde{\tau}}^3 R, {\widetilde{\tau}}^2 Q}$, for the same filtering lengths $\ell_F/\eta=7.0$  and $\ell_F/\eta=67.3$. The quantity $(\omega_i^\eig)^2$ rather than $\omega_i^\eig$ has been employed to remove the ambiguity of the sign of $\omega_i^\eig$ that was discussed earlier, and the color map is in logarithmic scale. The quantity $\inlavg{(\widetilde{\tau}\omega_2^\eig)^2|R,Q}$ is large over a wide region of the upper plane $Q>0$ where rotational motion dominates the velocity gradients, and it also takes relatively large values close to the right Vieillefosse tail. The results for $\inlavg{(\widetilde{\tau}\omega_3^\eig)^2|R,Q}$ indicate that the vorticity tends to be orthogonal to $\bm{v}_3$ below and in the proximity of the right Vieillefosse tail. This can be related to the compression along the vorticity axis that arises due to incompressibility of the flow, described by the second term on the right hand side of \eqref{theory_eq_strain_lambda} together with the local pressure Hessian contribution. Since this term is non-negative, then it acts to suppress the growth of the compressional eigenvalue $\lambda_3$. However, if the vorticity was perfectly aligned with $\bm{v}_3$ then this term would vanish. Therefore, misalignment of the vorticity with $\bm{v}_3$ stabilizes the system by providing a mechanism to prevent the blow-up of $\lambda_3$ along the right Vieillefosse tail. Strong alignment between $\bm{\omega}$ and $\bm{v}_3$ does nevertheless take place in the quadrant $R>0$ and $Q>0$, which is expected since this is the quadrant associated with vortex compression \citep{tsinober}. %%%!

The conditional averages shown in figures \ref{res_fig_lam_RQ} and \ref{res_fig_omg_RQ} confirm that, qualitatively, the statistics are weakly dependent on the filtering length scale, as observed  e.g.\ for the probability density flux in the $R,Q$ plane \citep{Danish2018}. One exception is the behavior of $\omega_2^\eig$, for which comparing figures \ref{res_fig_omg_RQ}(b) and (e) reveals significant qualitative changes in the behavior of $\inlavg{(\widetilde{\tau}\omega_2^\eig)^2|R,Q}$ as $\ell_F$ is increased.

\subsection{Rotation of the strain-rate eigenframe and tilting of the vorticity vector}

The tilting of the vorticity vector with respect to the strain-rate  eigenframe plays a central role in determining the geometric alignments of the strain-rate and vorticity, and the rotation-rate of the eigenframe can take on very large values even in simple flows \citep{Dresselhaus1992}. This rotation-rate is determined by the interaction between vorticity, pressure gradient and viscous and sub-grid forces, and equation \eqref{theory_eq_strain_rot} suggests that the rotation can be very strong when the differences between the strain-rate eigenvalues becomes small. To understand this better, and the effect of filtering on these processes, we now turn to analyze in detail the statistics of the rotation-rate of the eigenframe and its dependence upon the local state of the velocity gradient.

\begin{figure}
\centering
\vspace{0mm}			
    \subfloat[]	
	{\begin{overpic}
	[trim = 0mm -28mm -15mm -1mm,
	scale=0.085,clip,tics=20]{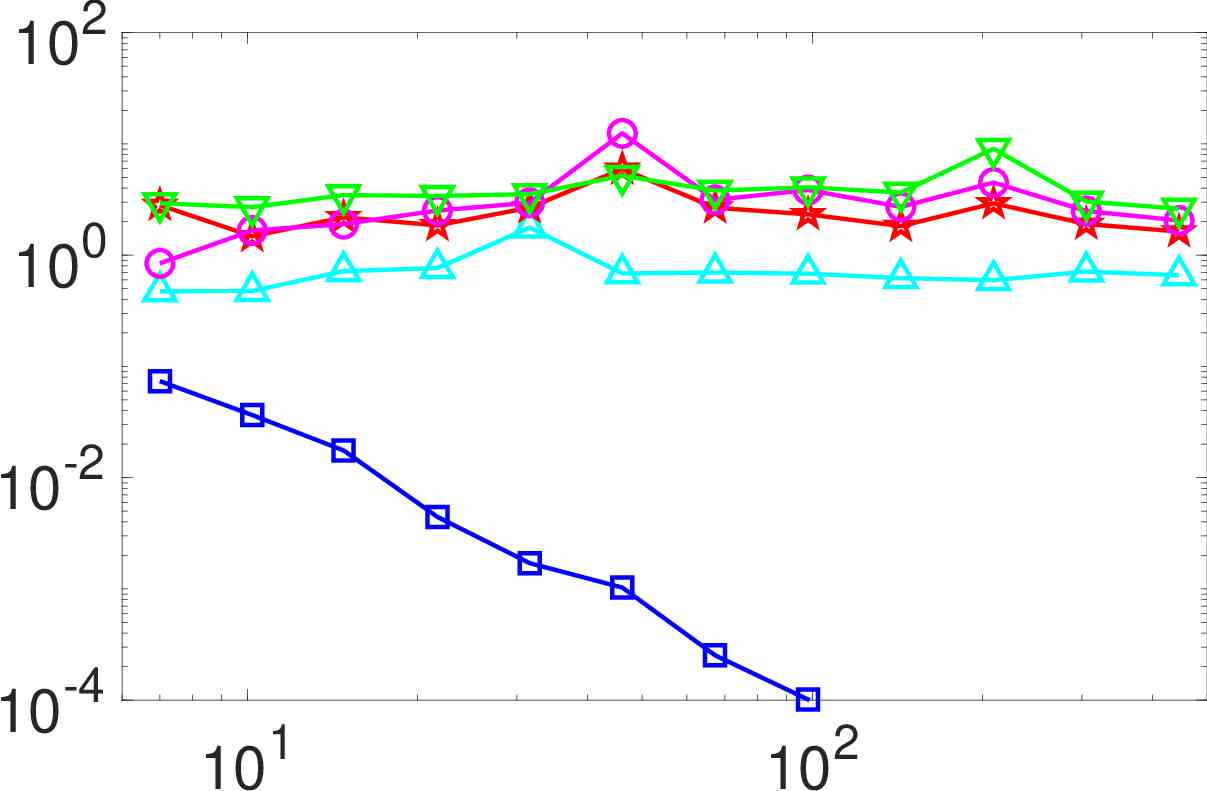}
	\put(-10,25){\rotatebox{90}{$\avg{{\widetilde{\tau}}^2 {X_{ij}}^2 }$}}
    \put(48,0){\small{$\ell_F / \eta$}}
        \begin{tikzpicture}[overlay]
            \node[draw,text width=0.70cm, black] at (3.5,1.25) {\scriptsize{{i = 3 j = 2}}};
        \end{tikzpicture}
	\end{overpic}}
	\subfloat[]
	{\begin{overpic}
	[trim = -15mm -28mm -15mm -1mm,
	scale=0.085,clip,tics=20]{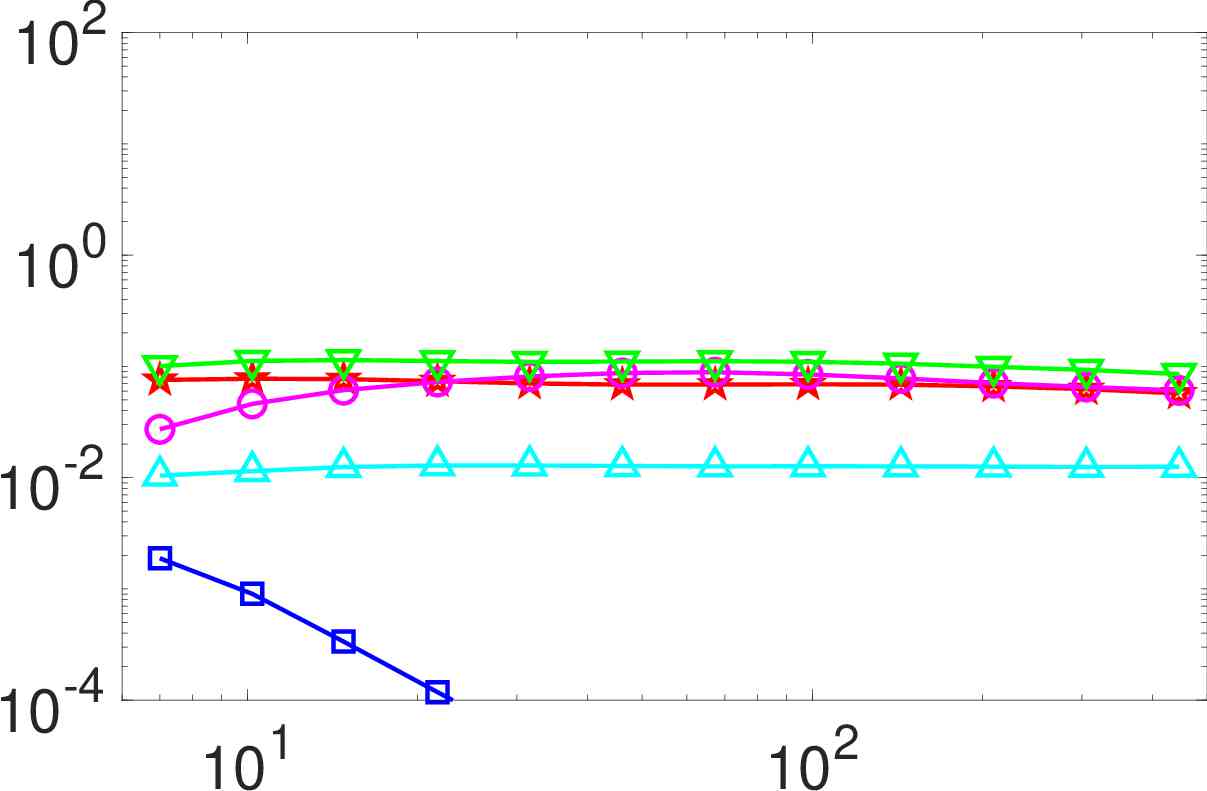}
	%\put(-3,31){\rotatebox{90}{$\avg{{\widetilde{\tau}}^2 {X_{31}}^2 }$}}
    \put(48,0){\small{$\ell_F / \eta$}}
        \begin{tikzpicture}[overlay]
            \node[draw,text width=0.70cm, black] at (3.75,2.35) {\scriptsize{{i = 3 j = 1}}};
        \end{tikzpicture} 
	\end{overpic}}
	\subfloat[]
    {\begin{overpic}
	[trim = -15mm -28mm 0mm -1mm,
	scale=0.085,clip,tics=20]{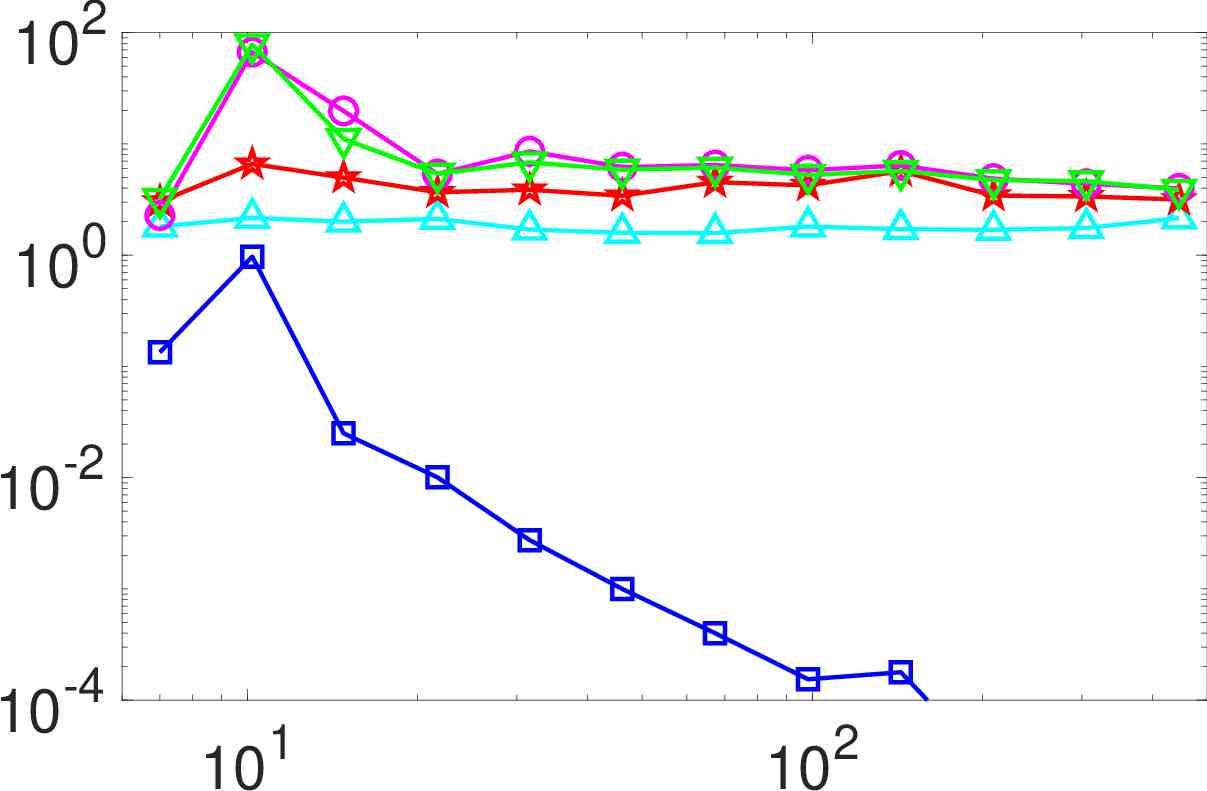}
	%\put(-3,31){\rotatebox{90}{$\avg{{\widetilde{\tau}}^2 {X_{21}}^2 }$}}
    \put(48,0){\small{$\ell_F / \eta$}}
        \begin{tikzpicture}[overlay]
            \node[draw,text width=0.70cm, black] at (3.65,1.25) {\scriptsize{{i = 2 j = 1}}};
        \end{tikzpicture}    
	\end{overpic}}\\
	{\begin{overpic}
	[trim = 35mm 275mm 150mm 20mm,
	scale=0.40,clip,tics=20]{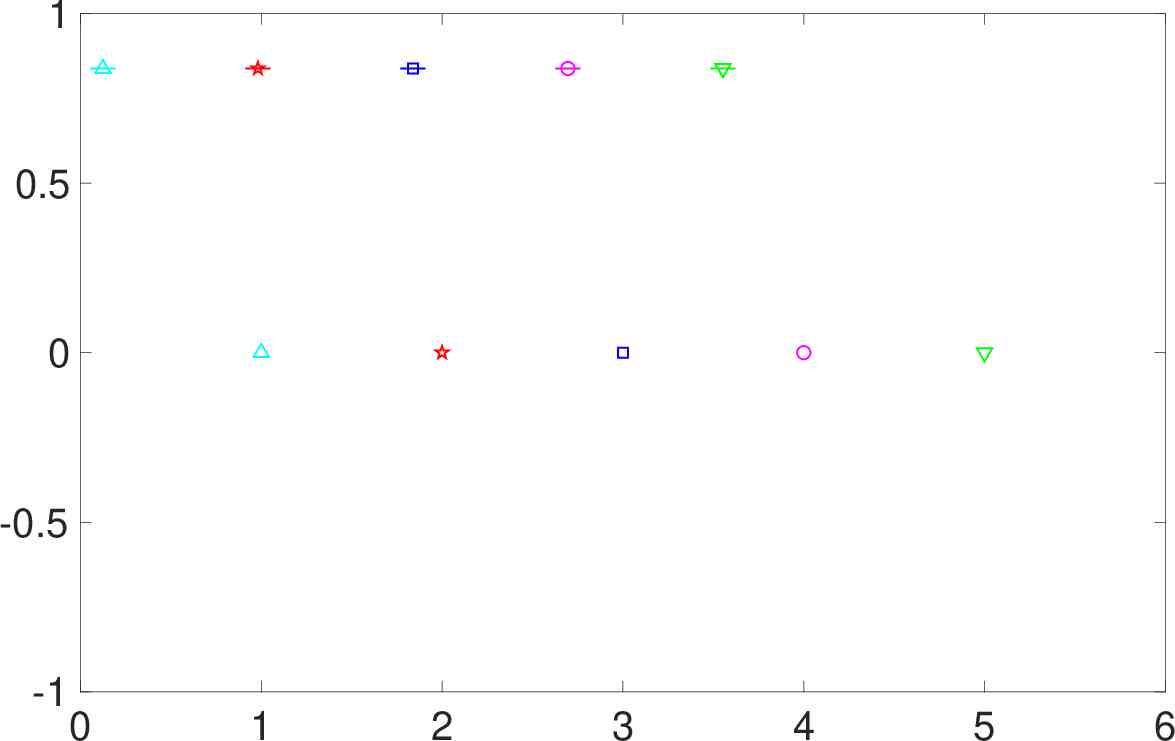}
		\put(5.5,2.5){$X = \Pi^{RE\eig}_{ij}$}
	    \put(26.5,2.5){$X = \Pi^{P\eig}_{ij}$}
	    \put(47.5,2.5){$X = \Pi^{\nu\eig}_{ij}$}
	    \put(68.5,2.5){$X = \Pi^{\tau\eig}_{ij}$}
	    \put(89.5,2.5){$X = \Pi^\eig_{ij}$}
	\end{overpic}}\\
	\subfloat[]
    {\begin{overpic}
	[trim = 0mm -28mm -10mm -16mm,
	scale=0.085,clip,tics=20]{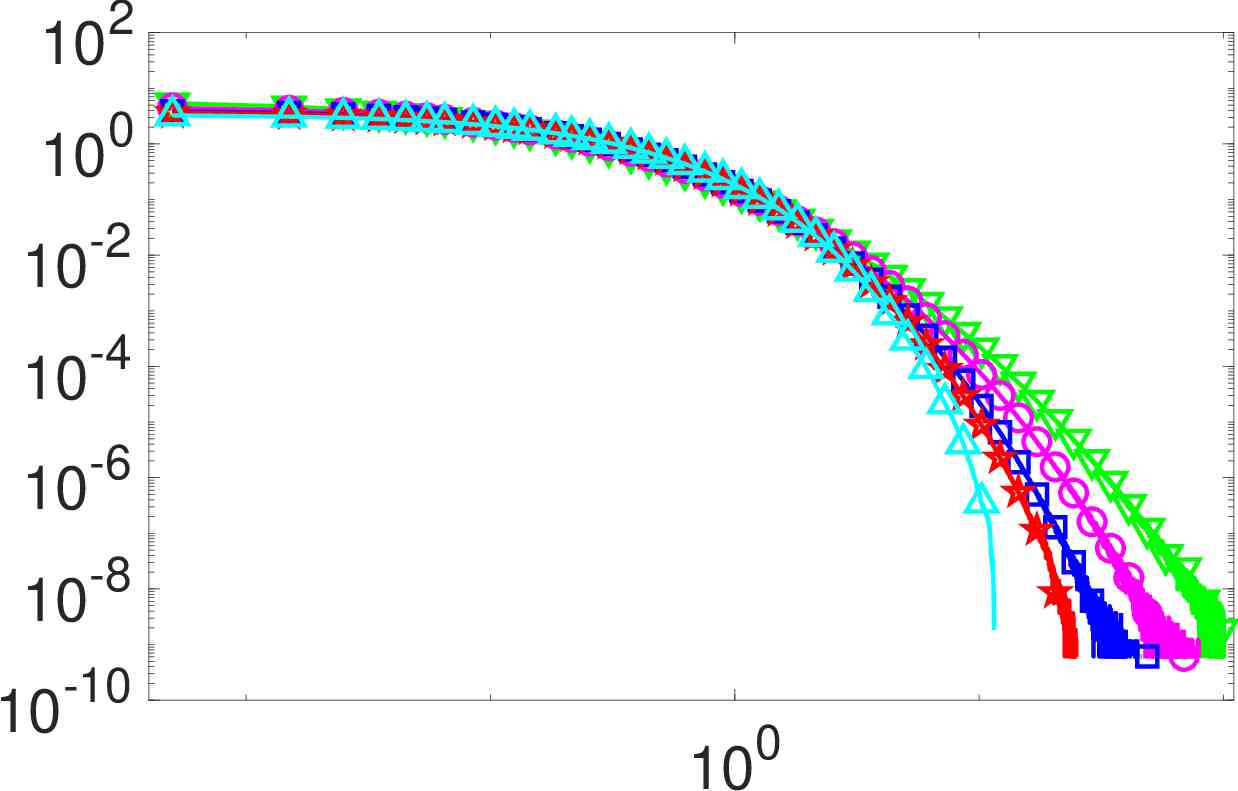}
	\put(-6,30){\rotatebox{90}{PDF}}
    \put(35,0){\small{$\widetilde{\tau}^2 (\lambda_2 - \lambda_3)^2$}}
	\end{overpic}}
	\subfloat[]
	{\begin{overpic}
	[trim = -10mm -28mm -10mm -16mm,
	scale=0.085,clip,tics=20]{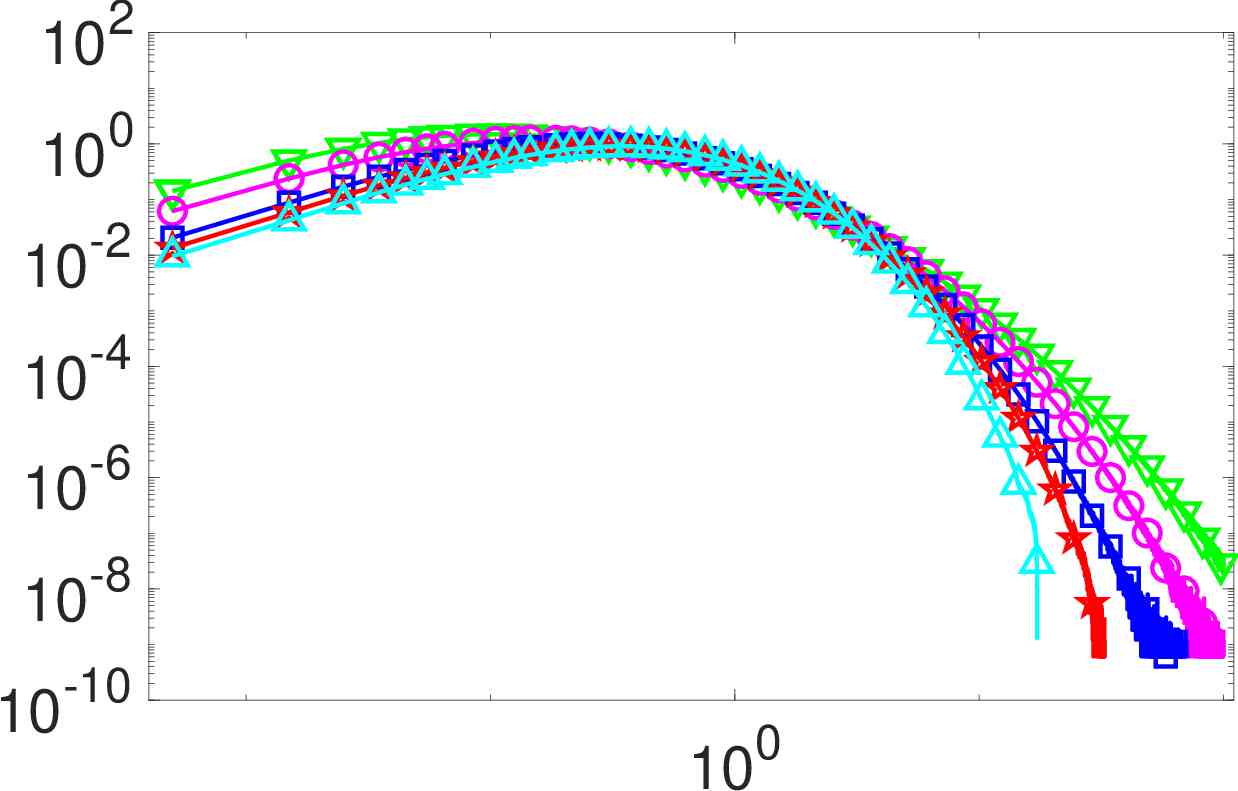}
	%\put(-3,31){\rotatebox{90}{$\avg{{\widetilde{\tau}}^2 {X_{31}}^2 }$}}
    \put(37,0){\small{$\widetilde{\tau}^2 (\lambda_1 - \lambda_3)^2$}}
	\end{overpic}}
    \subfloat[]	
	{\begin{overpic}
	[trim = -10mm -28mm 0mm -16mm,
	scale=0.085,clip,tics=20]{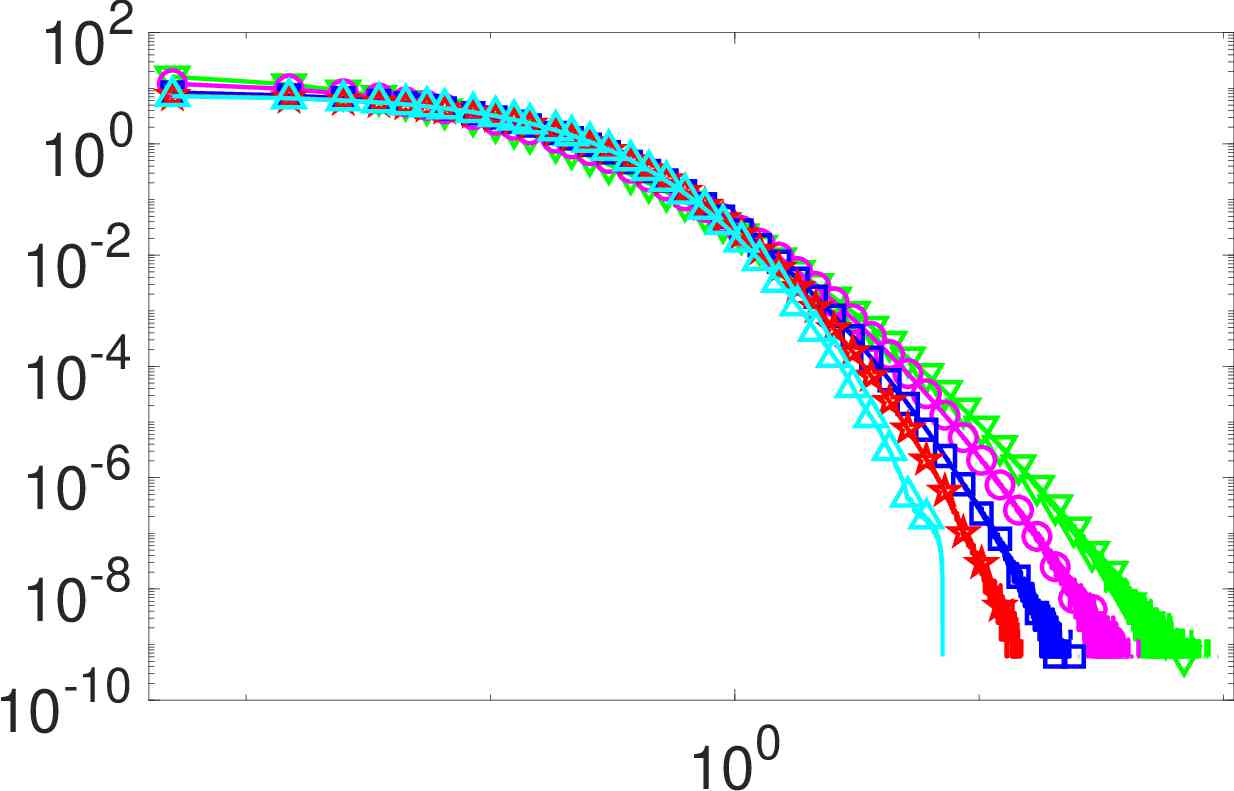}
    \put(37,0){\small{$\widetilde{\tau}^2 (\lambda_1 - \lambda_2)^2$}}
	\end{overpic}} \\
    {\begin{overpic}
	    [trim = 35mm 275mm 150mm 20mm,
	    scale=0.40,clip,tics=20]{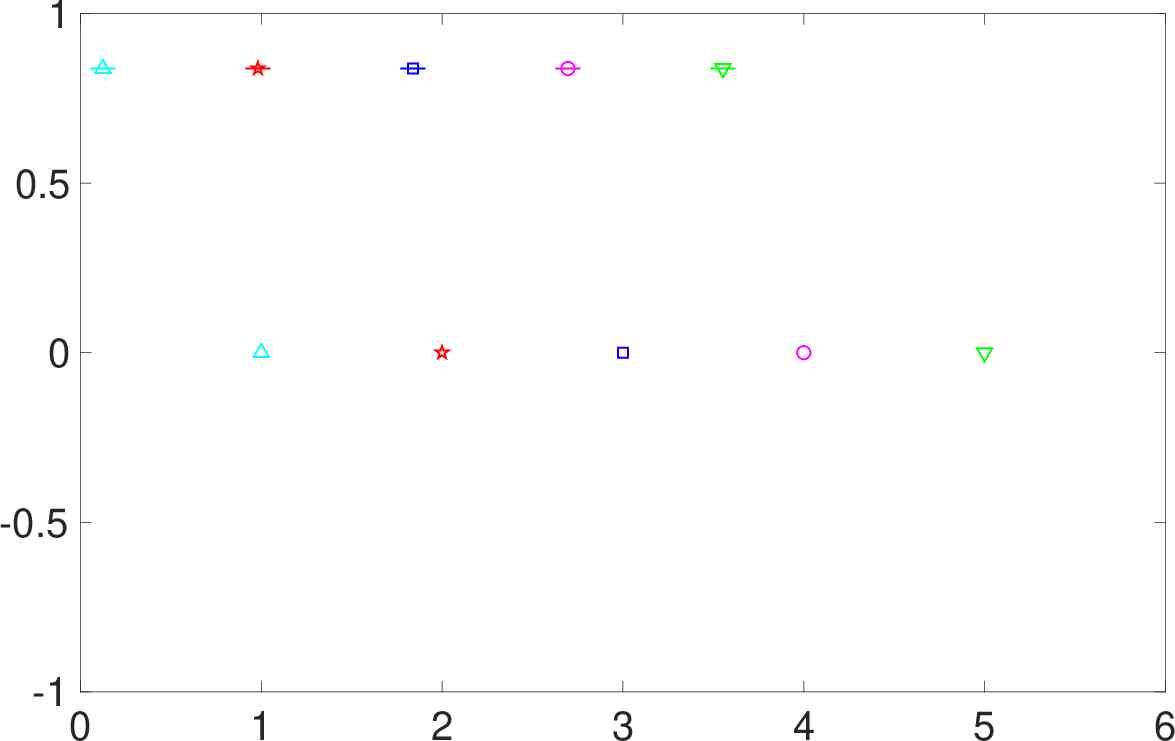}
		\put(5.5,2.5){$\ell_F / \eta = 209.4$}
	    \put(26.5,2.5){$\ell_F / \eta = 67.3$}
	    \put(47.5,2.5){$\ell_F / \eta = 31.6$}
	    \put(68.5,2.5){$\ell_F / \eta = 14.8$}
	    \put(89.5,2.5){$\ell_F / \eta = 7.0$}
    \end{overpic}}
	\caption{%\textcolor{red}{NewFig}
	(a-b-c) Second moment of the components of the Restricted Euler, pressure, viscous and sub-grid contributions to the strain-rate eigenframe rotation-rate, together with the overall rotation-rate.
	(d-e-f) PDF of the square of the differences in the eigenvalues, which appear in the equation for the rotation-rate of the eigenframe \eqref{theory_eq_strain_rot}.}
	\label{res_fig_S_rot}
\end{figure}
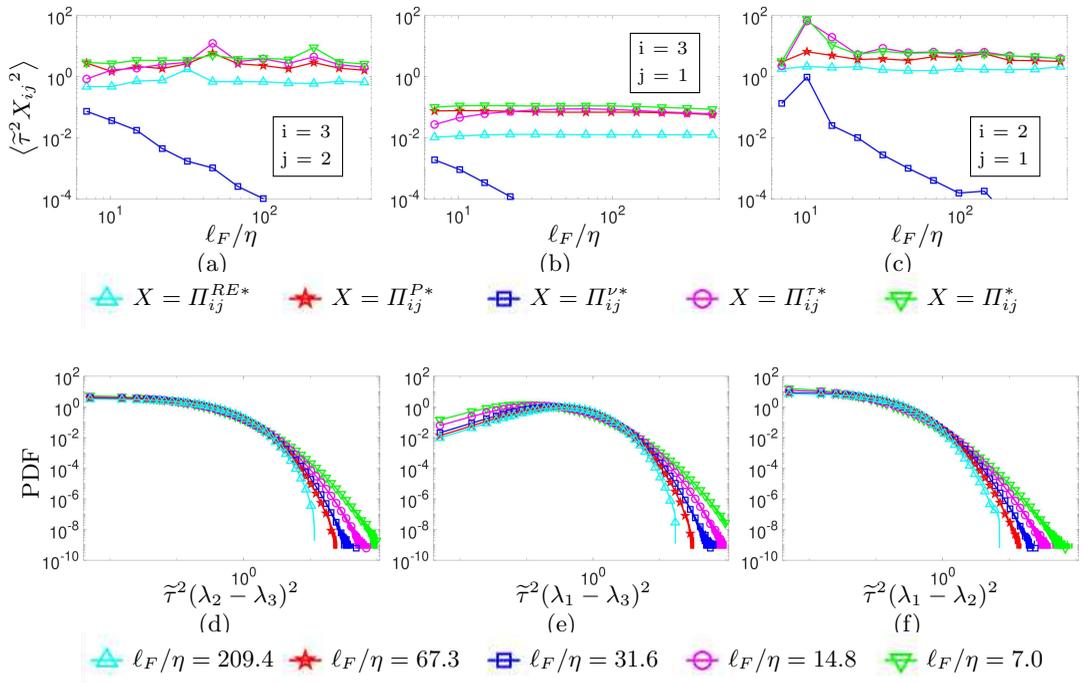

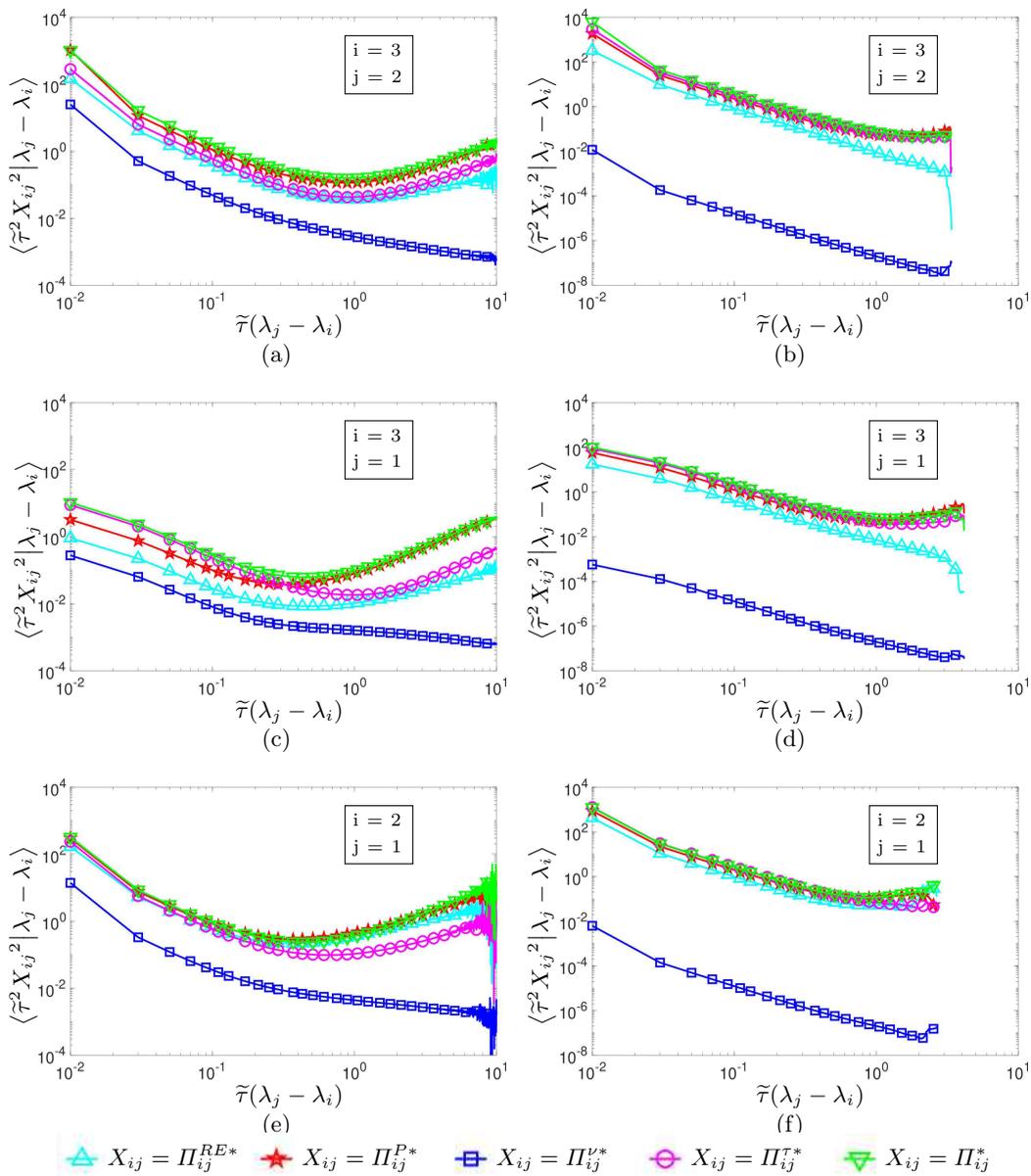
\begin{figure}
\centering
\vspace{0mm}
	\subfloat[]
	{\begin{overpic}
	[trim = -20mm -25mm -30mm -1mm,
	scale=0.125,clip,tics=20]{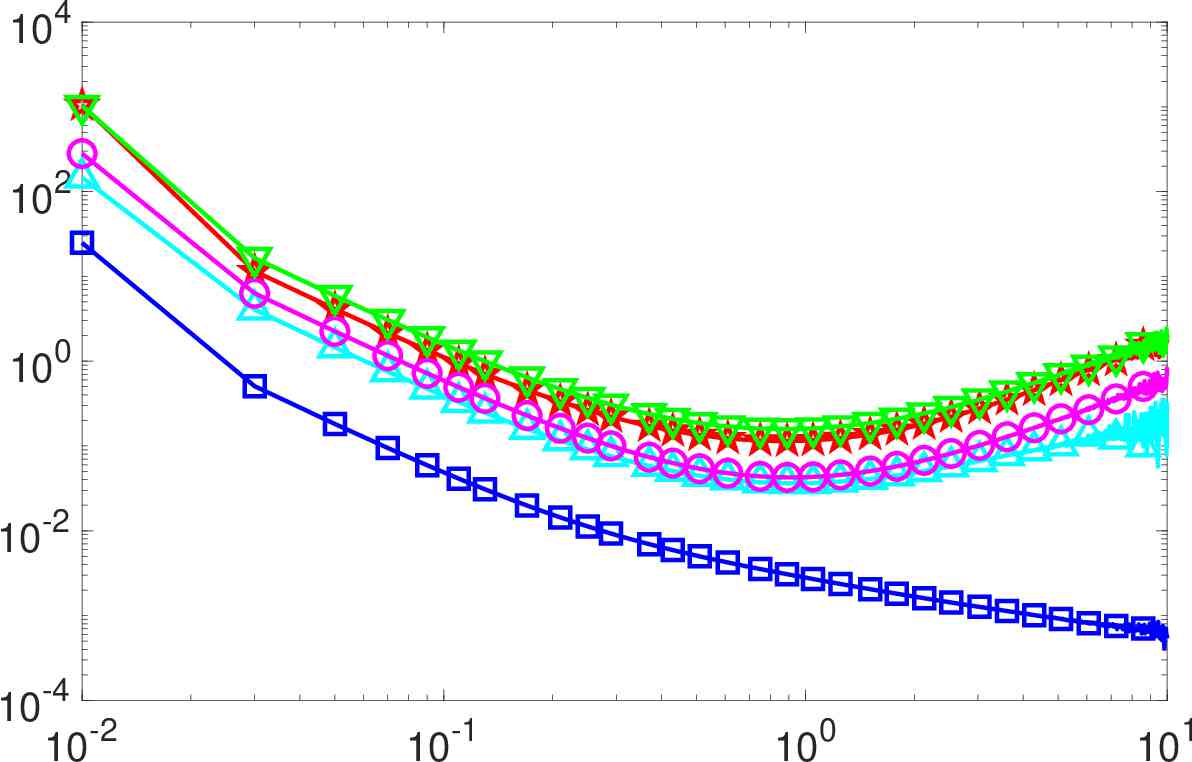}
			\put(42,0){$\widetilde{\tau}(\lambda_j - \lambda_i)$}
			\put(-2,15){\rotatebox{90}{$\condavg{{\widetilde{\tau}}^2 {X_{ij}}^2}{\lambda_j - \lambda_i}$}}
            \begin{tikzpicture}[overlay]
            %\draw [help lines] (0,0) grid (6,6);
            \node[draw,text width=0.70cm, black] at (4.85,3.6) {\scriptsize{{i = 3 j = 2}}};
            %\draw[gray, thick] (-1,-1) -- (2,2);
            \end{tikzpicture}
	\end{overpic}}
    \subfloat[]	
	{\begin{overpic}
	[trim = -30mm -25mm -20mm -1mm,
	scale=0.125,clip,tics=20]{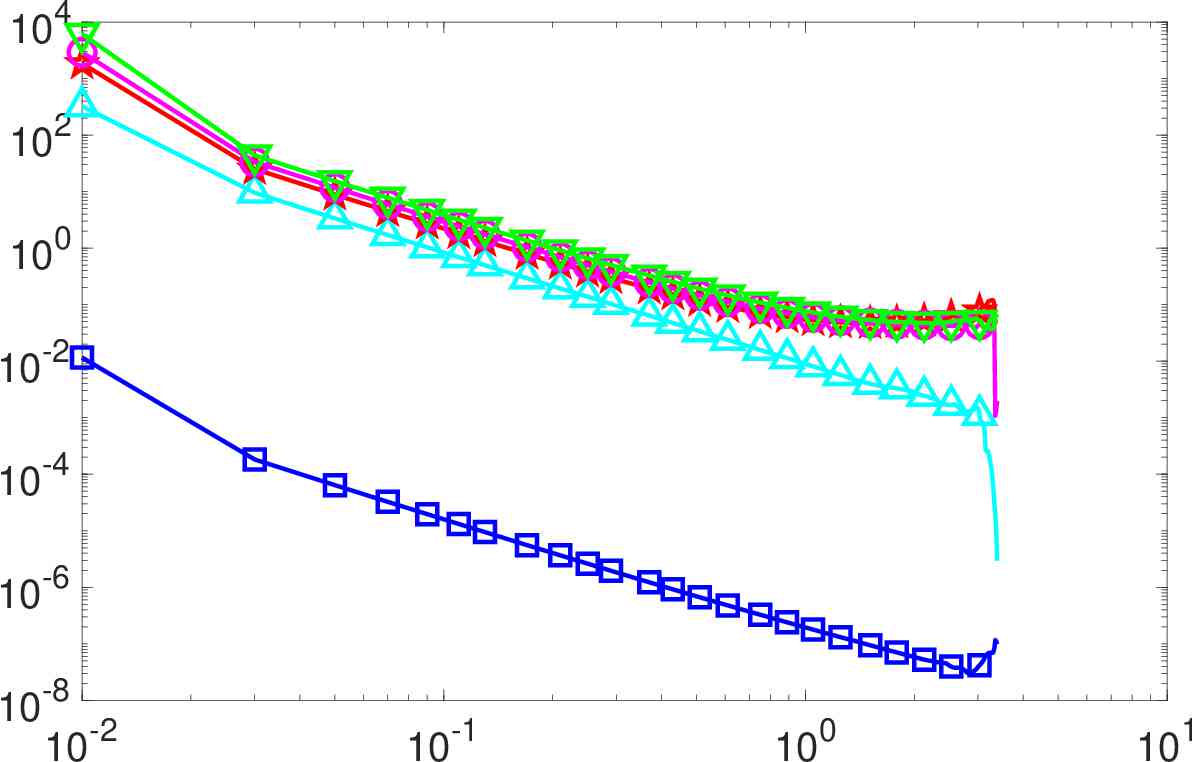}
			\put(43.5,0){$\widetilde{\tau}(\lambda_j - \lambda_i)$}
			\put(-1,15){\rotatebox{90}{$\condavg{{\widetilde{\tau}}^2 {X_{ij}}^2}{\lambda_j - \lambda_i}$}}
			%\put(-6,15){\rotatebox{90}{$\mathbb{E} \big[ X^2 \big| (\widetilde{\tau}\abs{\lambda_1 - \lambda_2}) \big]$}}
            \begin{tikzpicture}[overlay]
            %\draw [help lines] (0,0) grid (6,6);
            \node[draw,text width=0.70cm, black] at (5.0,3.6) {\scriptsize{{i = 3 j = 2}}};
            %\draw[gray, thick] (-1,-1) -- (2,2);
            \end{tikzpicture}
		\end{overpic}}\\
	\subfloat[]
	{\begin{overpic}
	[trim = -20mm -25mm -30mm -1mm,
	scale=0.125,clip,tics=20]{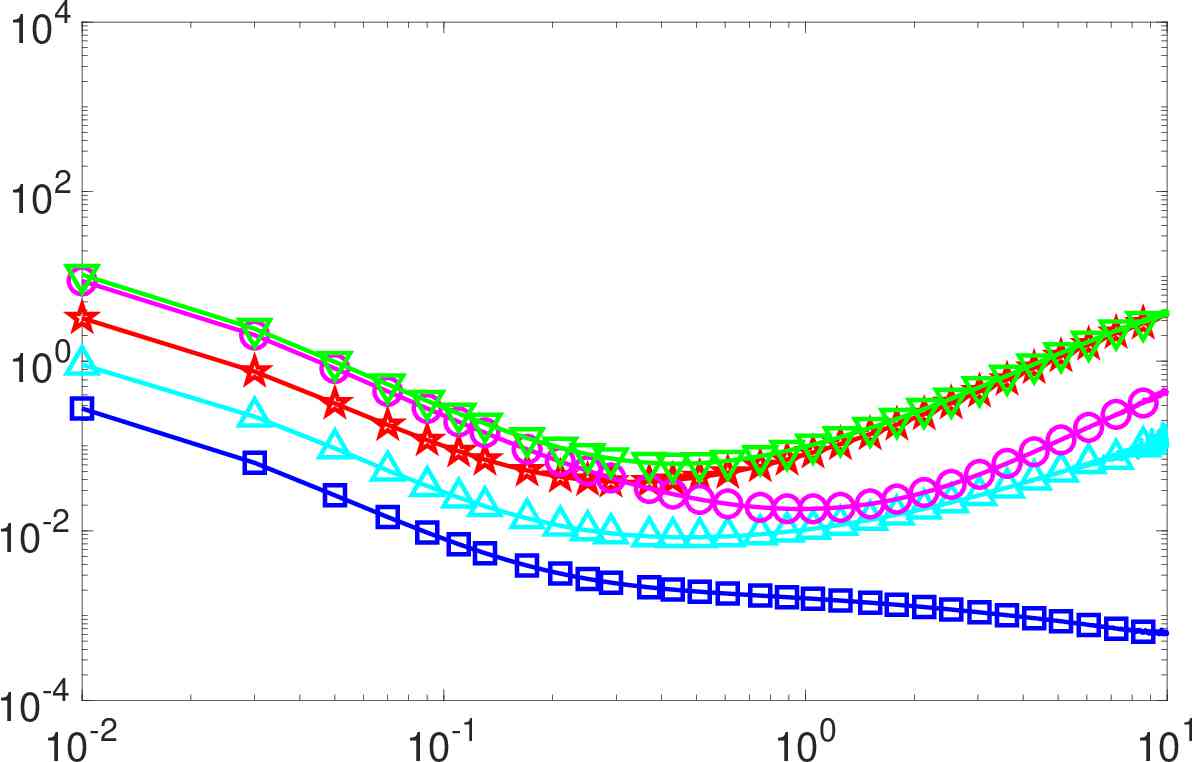}
			\put(42,0){$\widetilde{\tau}(\lambda_j - \lambda_i)$}
			\put(-1,15){\rotatebox{90}{$\condavg{{\widetilde{\tau}}^2 {X_{ij}}^2}{\lambda_j - \lambda_i}$}}
            \begin{tikzpicture}[overlay]
            %\draw [help lines] (0,0) grid (6,6);
            \node[draw,text width=0.70cm, black] at (4.85,3.6) {\scriptsize{{i = 3 j = 1}}};
            %\draw[gray, thick] (-1,-1) -- (2,2);
            \end{tikzpicture}
	\end{overpic}} 
	\subfloat[]
	{\begin{overpic}
	[trim = -30mm -25mm -20mm -1mm,
	scale=0.125,clip,tics=20]{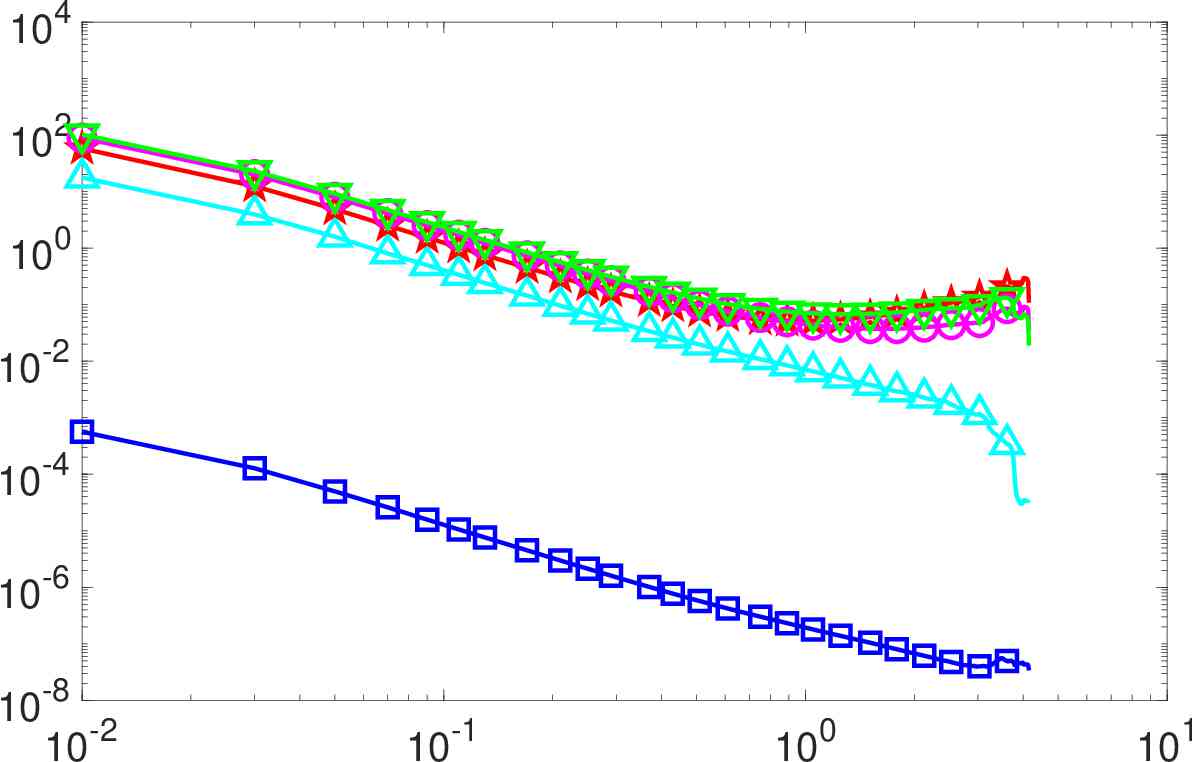}
			\put(43.5,0){$\widetilde{\tau}(\lambda_j - \lambda_i)$}
			\put(-1,15){\rotatebox{90}{$\condavg{{\widetilde{\tau}}^2 {X_{ij}}^2}{\lambda_j - \lambda_i}$}}
            \begin{tikzpicture}[overlay]
            %\draw [help lines] (0,0) grid (6,6);
            \node[draw,text width=0.70cm, black] at (5.0,3.6) {\scriptsize{{i = 3 j = 1}}};
            %\draw[gray, thick] (-1,-1) -- (2,2);
            \end{tikzpicture}
	\end{overpic}}\\
	\subfloat[]
	{\begin{overpic}
	[trim = -20mm -25mm -30mm -1mm,
	scale=0.125,clip,tics=20]{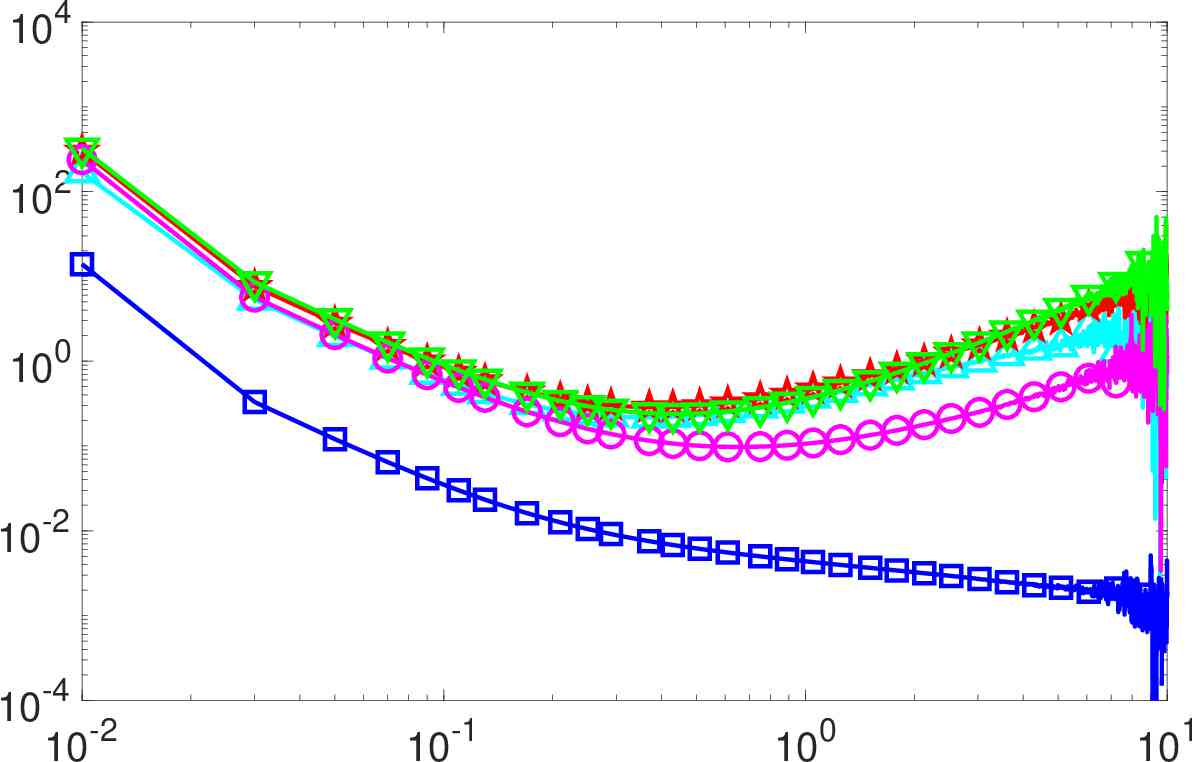}
			\put(42,0){$\widetilde{\tau}(\lambda_j - \lambda_i)$}
			\put(-2,15){\rotatebox{90}{$\condavg{{\widetilde{\tau}}^2 {X_{ij}}^2}{\lambda_j - \lambda_i}$}}
            \begin{tikzpicture}[overlay]
            %\draw [help lines] (0,0) grid (6,6);
            \node[draw,text width=0.70cm, black] at (4.85,3.6) {\scriptsize{{i = 2 j = 1}}};
            %\draw[gray, thick] (-1,-1) -- (2,2);
            \end{tikzpicture}
	\end{overpic}}
	\subfloat[]
	{\begin{overpic}
	[trim = -30mm -25mm -20mm -1mm,
	scale=0.125,clip,tics=20]{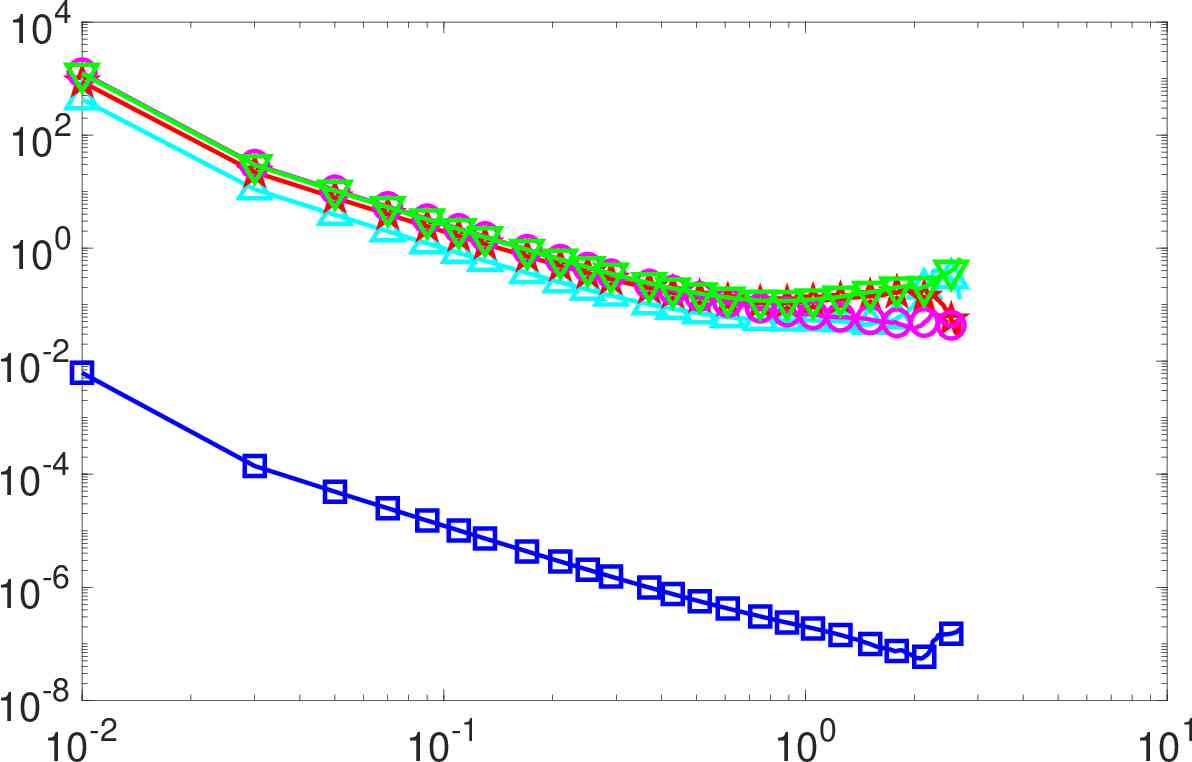}
			\put(43.5,0){$\widetilde{\tau}(\lambda_j - \lambda_i)$}
			\put(-1,15){\rotatebox{90}{$\condavg{{\widetilde{\tau}}^2 {X_{ij}}^2}{\lambda_j - \lambda_i}$}}
            \begin{tikzpicture}[overlay]
            %\draw [help lines] (0,0) grid (6,6);
            \node[draw,text width=0.70cm, black] at (5.0,3.6) {\scriptsize{{i = 2 j = 1}}};
            %\draw[gray, thick] (-1,-1) -- (2,2);
            \end{tikzpicture}
	\end{overpic}}\\
	{\begin{overpic}
	[trim = 35mm 275mm 150mm 20mm,
	scale=0.40,clip,tics=20]{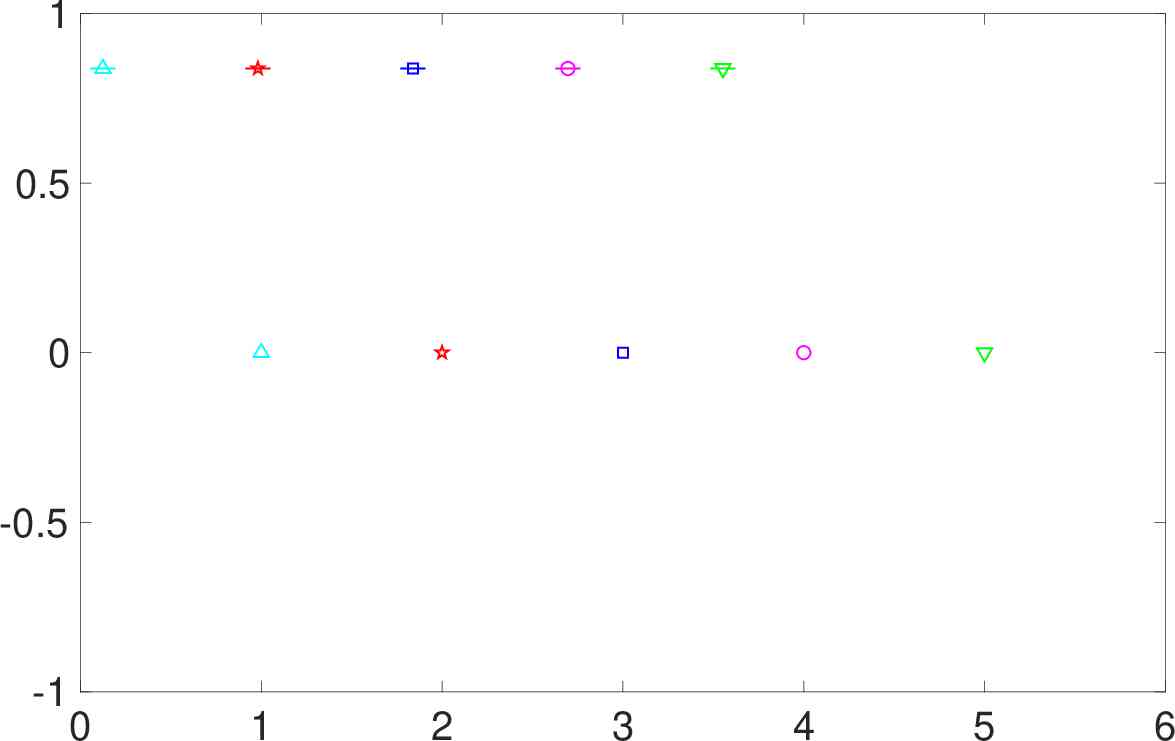}
		\put(5.5,2.5){$X_{ij} = \Pi^{RE\eig}_{ij}$}
	    \put(26.5,2.5){$X_{ij} = \Pi^{P\eig}_{ij}$}
	    \put(47.5,2.5){$X_{ij} = \Pi^{\nu\eig}_{ij}$}
	    \put(68.5,2.5){$X_{ij} = \Pi^{\tau\eig}_{ij}$}
	    \put(89.5,2.5){$X_{ij} = \Pi^\eig_{ij}$}
	\end{overpic}}
	\caption{Second moment of the contributions to the eigenframe rotation-rate conditioned on the difference in the eigenvalues. (a-c-e) correspond to $\ell_F/\eta = 7.0$ and (b-d-f) to $\ell_F/\eta = 209.4$.
	}
	\label{res_strain_rot}
\end{figure}

%%%!
Figures \ref{res_fig_S_rot}(a-b-c) show the second moment of the different contributions to the eigenframe rotation-rate $\Pi_{ij}^*$ (see \eqref{theory_contr_strain_rot}), and figures \ref{res_fig_S_rot}(d-e-f) show the PDF of the square of the difference in the eigenvalues, which represent the resistance of the eigenframe \citep{Vieillefosse1982}.
%%%!
While the local contributions (captured by the RE model) to the eigenframe rotation rate have been investigated in detail \citep{Dresselhaus1992}, together with its impact on the vorticity dynamics, the non-local contributions have received less attention. The second moment of the pressure Hessian contribution is close to the second moment of the overall rotation rate and they are almost independent of the filtering length (when normalized by $\widetilde{\tau}$). This indicates that the non-local effects on the rotation-rate of the eigenframe dominate over the contribution from vorticity that features in the RE model. However, we will observe that the local RE term plays a leading role in the tilting of the vorticity vector.
The sub-grid stress also gives an important contribution to the eigenframe rotation rate at large scales. On the other hand, the viscous contribution is small  with $\inlavg{(\widetilde{\tau} \Pi_{ij}^{\nu\eig})^2 }$ decreasing as $\ell_F^{-\xi}$, with $\xi$ between 2 and 3. The component $\inlavg{(\widetilde{\tau} \Pi_{31}^{\eig})^2 }$, associated with the angular velocity of the eigenframe along $\bm{v}_2$, is much smaller than the other components.
If we consider the PDFs of $(\lambda_j-\lambda_i)^2$ in figures \ref{res_fig_S_rot}(d-e-f) we see that the mode of the PDF is considerably larger for $(\lambda_1-\lambda_3)^2$ than for the other cases. The reason for this is that due to incompressibility, $(\lambda_1-\lambda_3)^2\to 0$ also implies $\lambda_2\to 0$, and the probability of states with strain-rate almost zero is vanishingly small. As a result, the average values of $(\lambda_1-\lambda_3)^2$ are larger than for $(\lambda_2-\lambda_3)^2$ or $(\lambda_1-\lambda_2)^2$, implying on average an increased resistance to rotations of the eigenframe about $\bm{v}_2$, and hence to $\inlavg{(\widetilde{\tau} \Pi_{31}^{\eig})^2 }$ being smaller than the other compoenents.

The results in figures \ref{res_fig_S_rot}(d-e-f) also indicate that the probability of axisymmetric states with $\lambda_2\approx \lambda_1$ and $\lambda_2\approx \lambda_3$ is quite high, with the probability to observe $\lambda_2\approx \lambda_1$ larger, in agreement with previous results that showed that axisymmetric extension occurs more often than axisymmetric compression \citep{Lund1994,Meneveau2011}. However, the probability to observe $\widetilde{\tau}(\lambda_2-\lambda_3)\to 0$ is only three times smaller than that for observing $\widetilde{\tau}(\lambda_2-\lambda_1)\to 0$. A more accurate estimation of the relative occurrence rates of axisymmetric compression and expansion can be achieved by means of a dimensionless parameter \citep{Lund1994} as discussed in section \ref{sec_shape}.

The main effect of filtering on these PDFs is to simply suppress the tails, associated with reduced intermittency at larger scales. However, for the PDF of $(\lambda_1-\lambda_3)^2$, there is also a significant effect of filtering on the behavior for $(\lambda_1-\lambda_3)^2\to 0$, with the probability of states with $(\lambda_1-\lambda_3)^2\to 0$ decreasing as $\ell_F$ is increased.

The quantity $\lambda_j-\lambda_i$ acts as a weight in equation \eqref{theory_eq_strain_rot}, and the behavior of $\Pi_{ij}^\eig$ depends on how the various contributions to $\Pi_{ij}^\eig$ on the right hand side of \eqref{theory_eq_strain_rot} behave as $\lambda_j-\lambda_i$ varies. To explore this, in figure \ref{res_strain_rot} we show results for $\inlavg{X_{ij}^2\vert \lambda_j-\lambda_i }$, with $i>j$ such that $\lambda_j-\lambda_i > 0$, where $X_{ij}$ is either the eigenframe rotation-rate $\Pi_{ij}^*$ or else one of the distinct contributions to $\Pi_{ij}^*$, namely $\Pi^{RE\eig}_{ij}, \Pi^{P\eig}_{ij} , \Pi^{\nu\eig}_{ij} or \Pi^{\tau\eig}_{ij}$ (see equation \eqref{theory_contr_strain_rot}).
The results show that $\inlavg{X_{ij}^2\vert \lambda_j-\lambda_i } \propto(\lambda_j-\lambda_i)^{-2}$ for small $\widetilde{\tau}(\lambda_j-\lambda_i)$, indicating a weak correlation between $(\lambda_j-\lambda_i)$ and the vorticity, pressure Hessian, sub-grid stress and visocus stress in this range. However, for $\widetilde{\tau}(\lambda_j-\lambda_i)\geq \orderof{1}$, $\inlavg{X^2\vert \lambda_j-\lambda_i }$ starts to increase for some of the cases. This non-monotonic behavior is quite intriguing, but the increase cannot persist in the limit $\widetilde{\tau}(\lambda_j-\lambda_i)\to\infty$ if the flow field is to remain regular. Also, the results show that the increase at $\widetilde{\tau}(\lambda_j-\lambda_i)\geq \orderof{1}$ becomes less apparent as the filter length is increased. For the smallest filter scales, the results in figure \ref{res_strain_rot} show that the dominant contribution to the eigenframe rotation-rate comes from the pressure Hessian, with important contributions from the
sub-grid stress and RE term associated with vorticity in some cases (e.g.\ especially for $i=2,j=1$). At larger scales, however, the sub-grid term makes a strong contribution, similar in size to that from the pressure Hessian with the RE playing a smaller role. The viscous contribution is small at all scales and for all components.

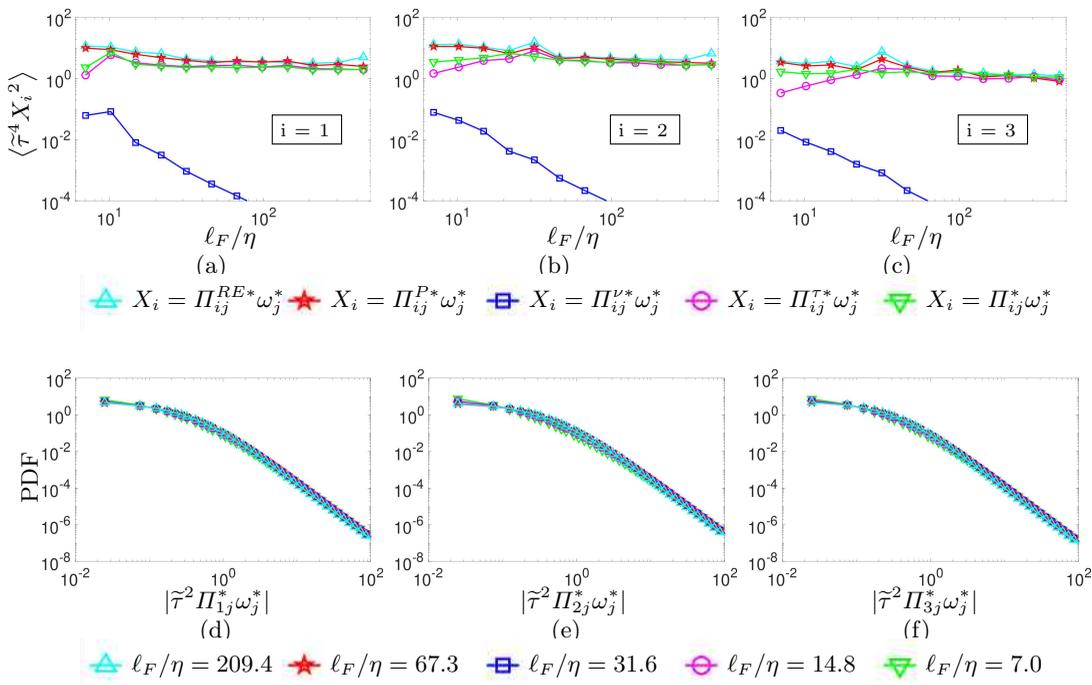
\begin{figure}
\centering
\vspace{0mm}			
    \subfloat[]	
	{\begin{overpic}
	[trim = 0mm -28mm -15mm -1mm,
	scale=0.085,clip,tics=20]{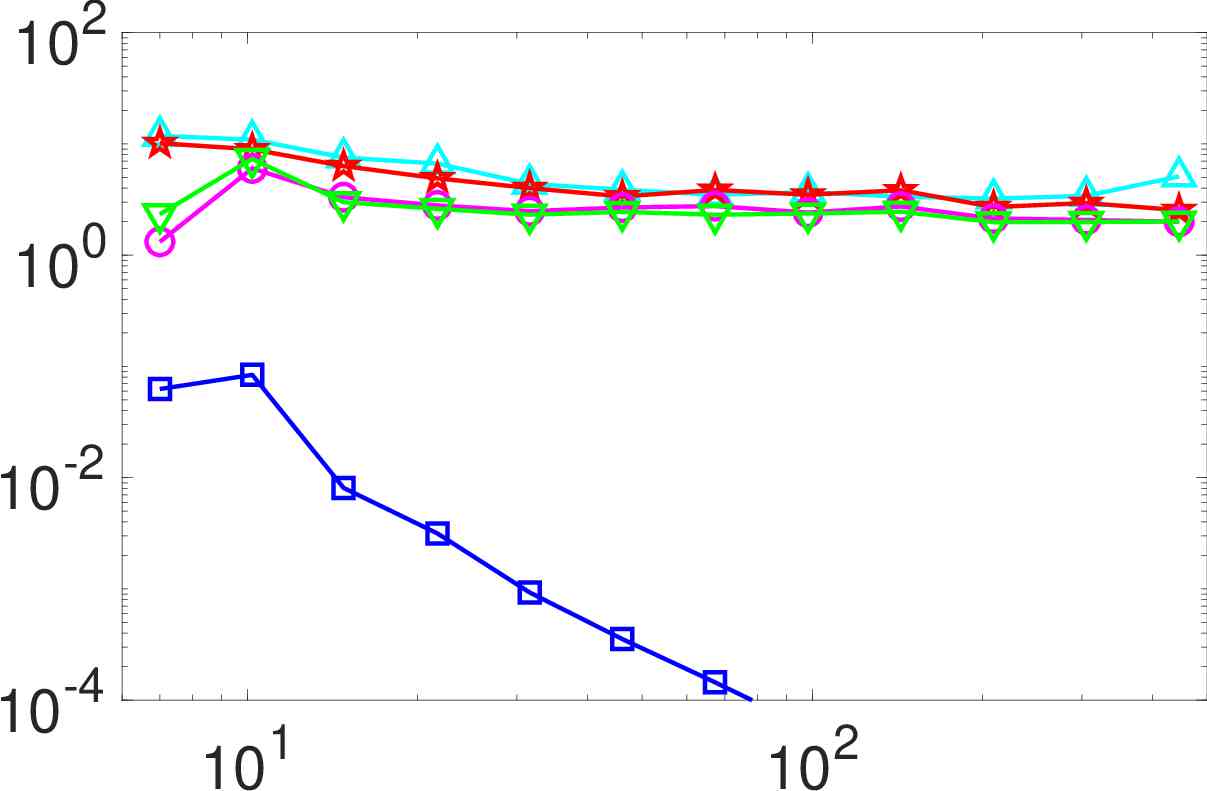}
	\put(-10,25){\rotatebox{90}{$\avg{{\widetilde{\tau}}^4 {X_{i}}^2 }$}}
    \put(48,0){\small{$\ell_F / \eta$}}
        \begin{tikzpicture}[overlay]
            \node[draw,text width=0.70cm, black] at (3.5,1.5) {\scriptsize{{i = 1}}};
        \end{tikzpicture}
	\end{overpic}}
	\subfloat[]
	{\begin{overpic}
	[trim = -15mm -28mm -15mm -1mm,
	scale=0.085,clip,tics=20]{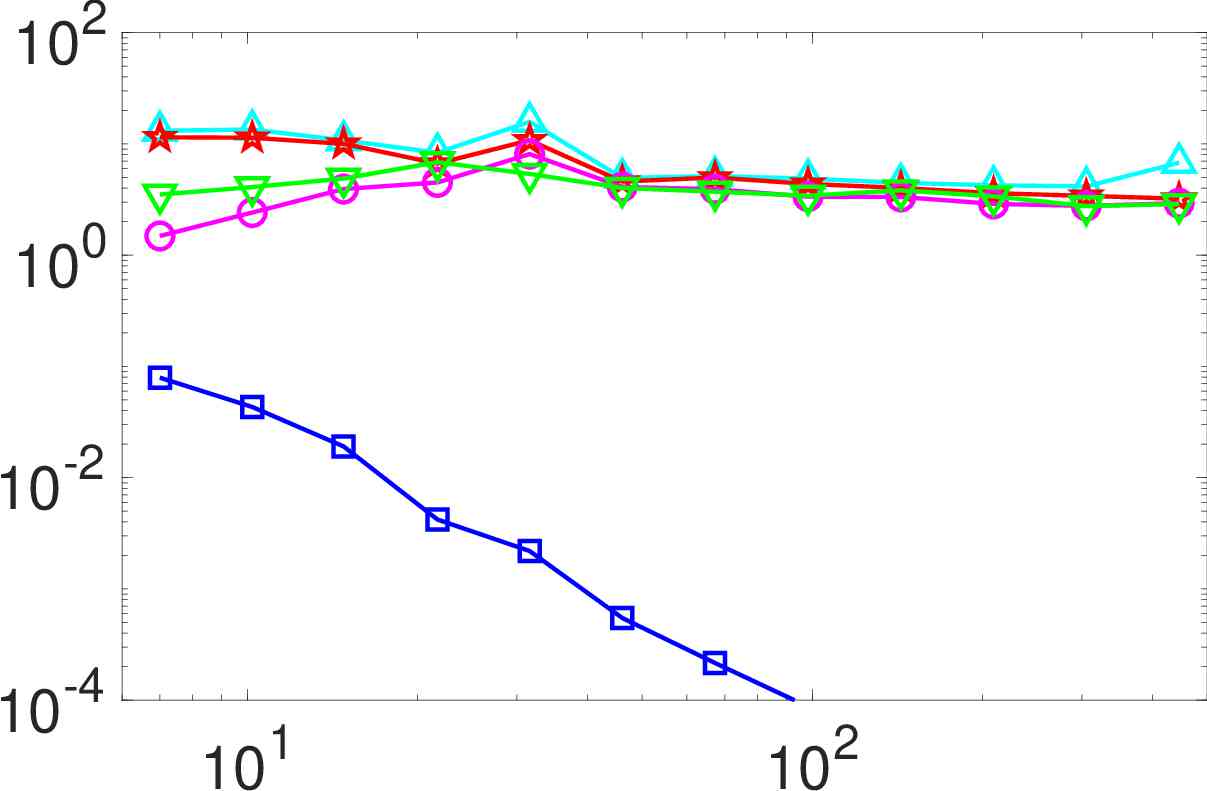}
	%\put(-3,31){\rotatebox{90}{$\avg{{\widetilde{\tau}}^2 {X_{31}}^2 }$}}
    \put(48,0){\small{$\ell_F / \eta$}}
        \begin{tikzpicture}[overlay]
            \node[draw,text width=0.70cm, black] at (3.5,1.5) {\scriptsize{{i = 2}}};
        \end{tikzpicture} 
	\end{overpic}}
	\subfloat[]
    {\begin{overpic}
	[trim = -15mm -28mm 0mm -1mm,
	scale=0.085,clip,tics=20]{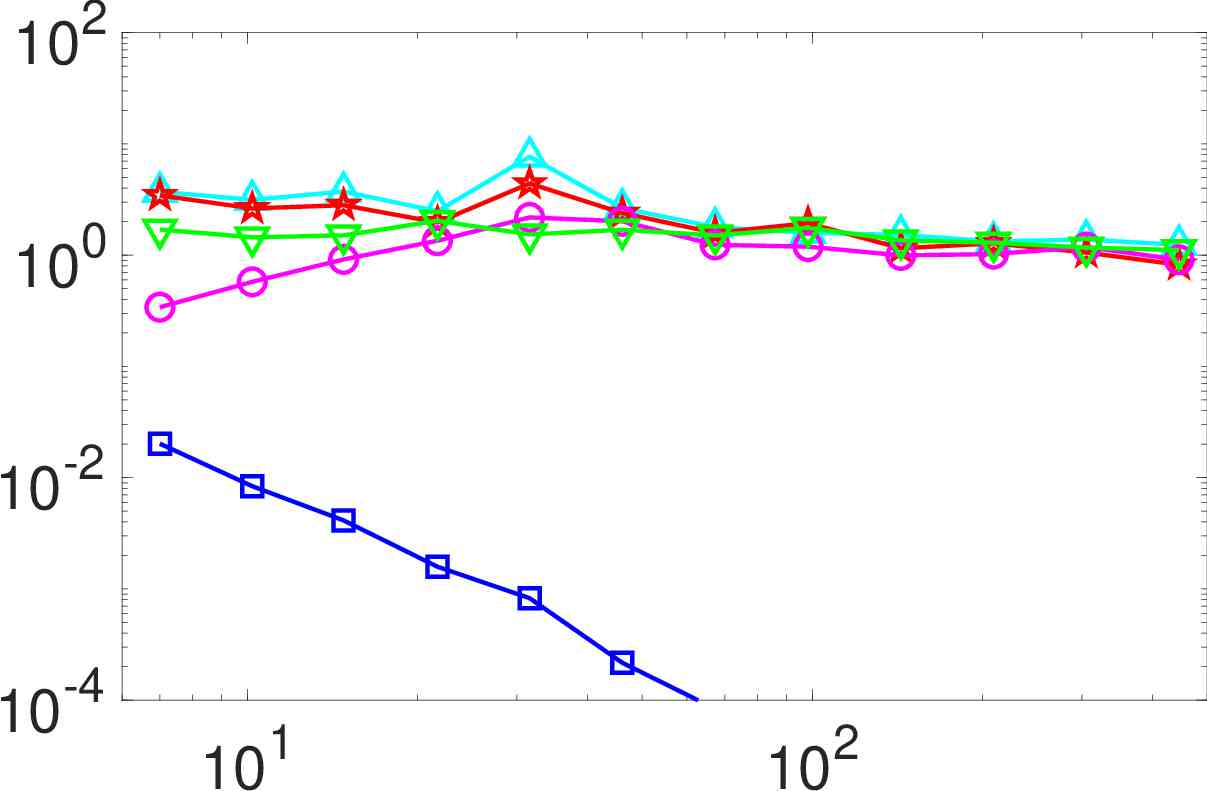}
	%\put(-3,31){\rotatebox{90}{$\avg{{\widetilde{\tau}}^2 {X_{21}}^2 }$}}
    \put(48,0){\small{$\ell_F / \eta$}}
        \begin{tikzpicture}[overlay]
            \node[draw,text width=0.70cm, black] at (3.5,1.5) {\scriptsize{{i = 3}}};
        \end{tikzpicture}     
	\end{overpic}}\\
	{\begin{overpic}
	[trim = 35mm 275mm 150mm 20mm,
	scale=0.40,clip,tics=20]{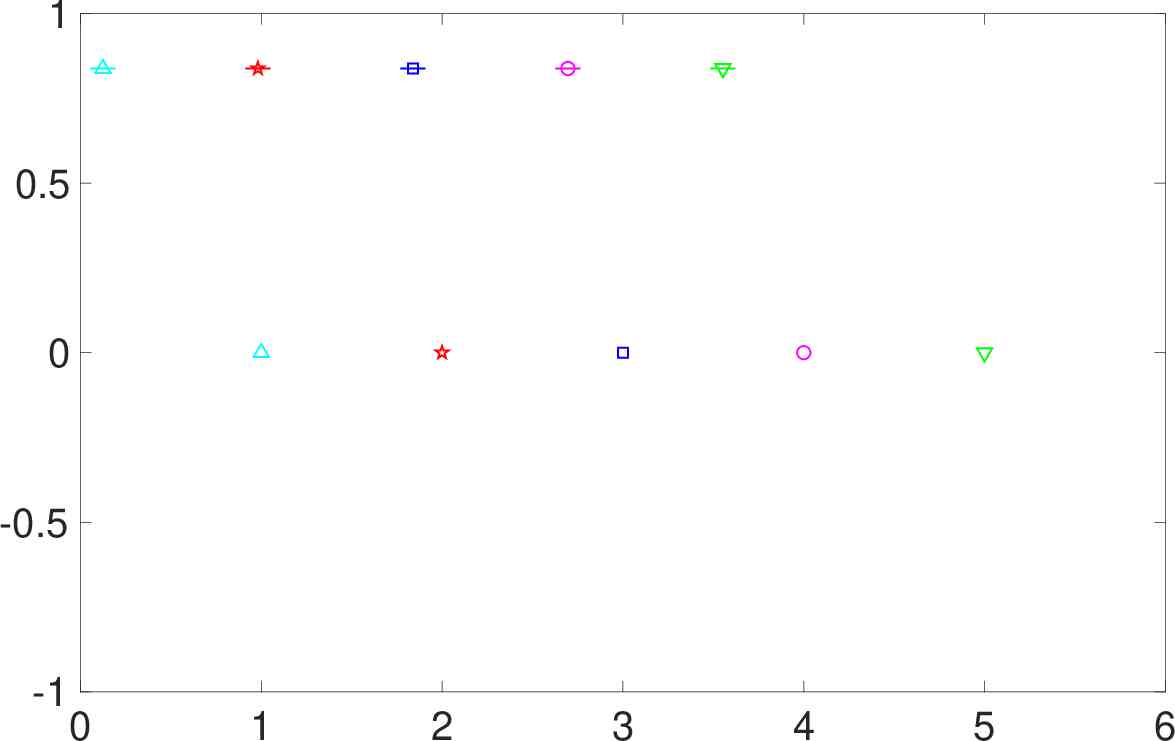}
		\put(5.5,2.5){$X_{i} = \Pi^{RE\eig}_{ij} \omega_j^\eig$}
	    \put(26.5,2.5){$X_{i} = \Pi^{P\eig}_{ij} \omega_j^\eig$}
	    \put(47.5,2.5){$X_{i} = \Pi^{\nu\eig}_{ij} \omega_j^\eig$}
	    \put(68.5,2.5){$X_{i} = \Pi^{\tau\eig}_{ij} \omega_j^\eig$}
	    \put(89.5,2.5){$X_{i} = \Pi_{ij}^\eig \omega_j^\eig$}
	\end{overpic}}\\
	\subfloat[]
    {\begin{overpic}
	[trim = 0mm -34mm -10mm -16mm,
	scale=0.085,clip,tics=20]{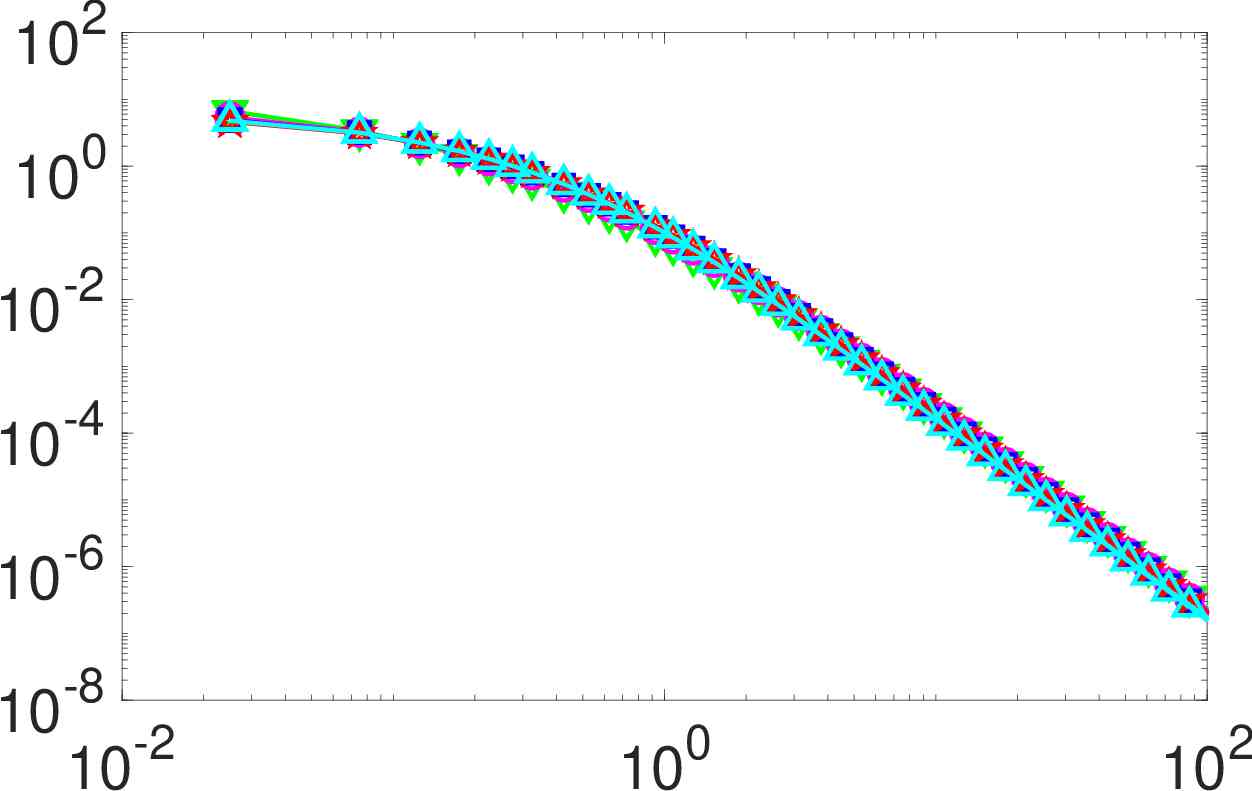}
	\put(-6,30){\rotatebox{90}{PDF}}
    \put(35,0){\small{$|\widetilde{\tau}^2 \Pi^{\eig}_{1j}\omega_j^\eig|$}}
	\end{overpic}}
	\subfloat[]
	{\begin{overpic}
	[trim = -10mm -34mm -10mm -16mm,
	scale=0.085,clip,tics=20]{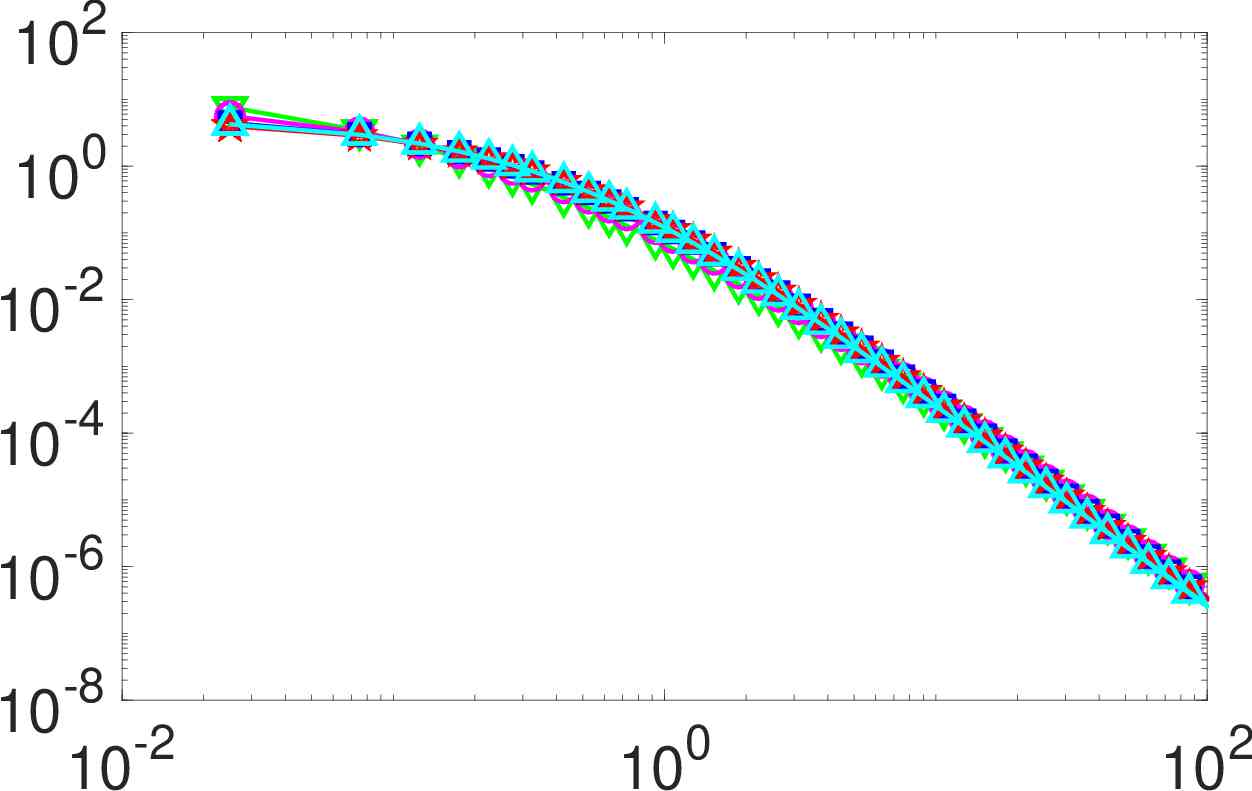}
	%\put(-3,31){\rotatebox{90}{$\avg{{\widetilde{\tau}}^2 {X_{31}}^2 }$}}
    \put(37,0){\small{$|\widetilde{\tau}^2 \Pi^{\eig}_{2j}\omega_j^\eig|$}}
	\end{overpic}}
    \subfloat[]	
	{\begin{overpic}
	[trim = -10mm -34mm 0mm -16mm,
	scale=0.085,clip,tics=20]{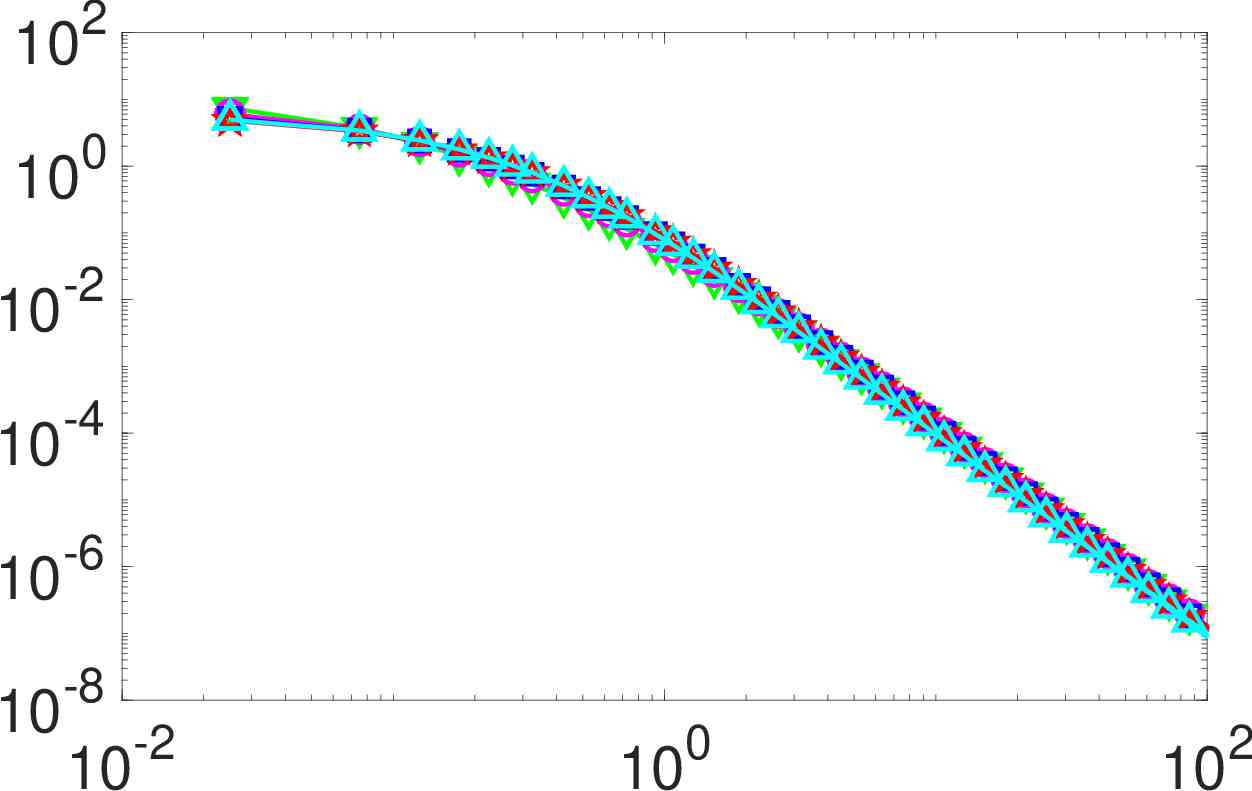}
    \put(37,0){\small{$|\widetilde{\tau}^2 \Pi^{\eig}_{3j}\omega_j^\eig|$}}
	\end{overpic}} \\
	{\begin{overpic}
	[trim = 35mm 275mm 150mm 20mm,
	scale=0.40,clip,tics=20]{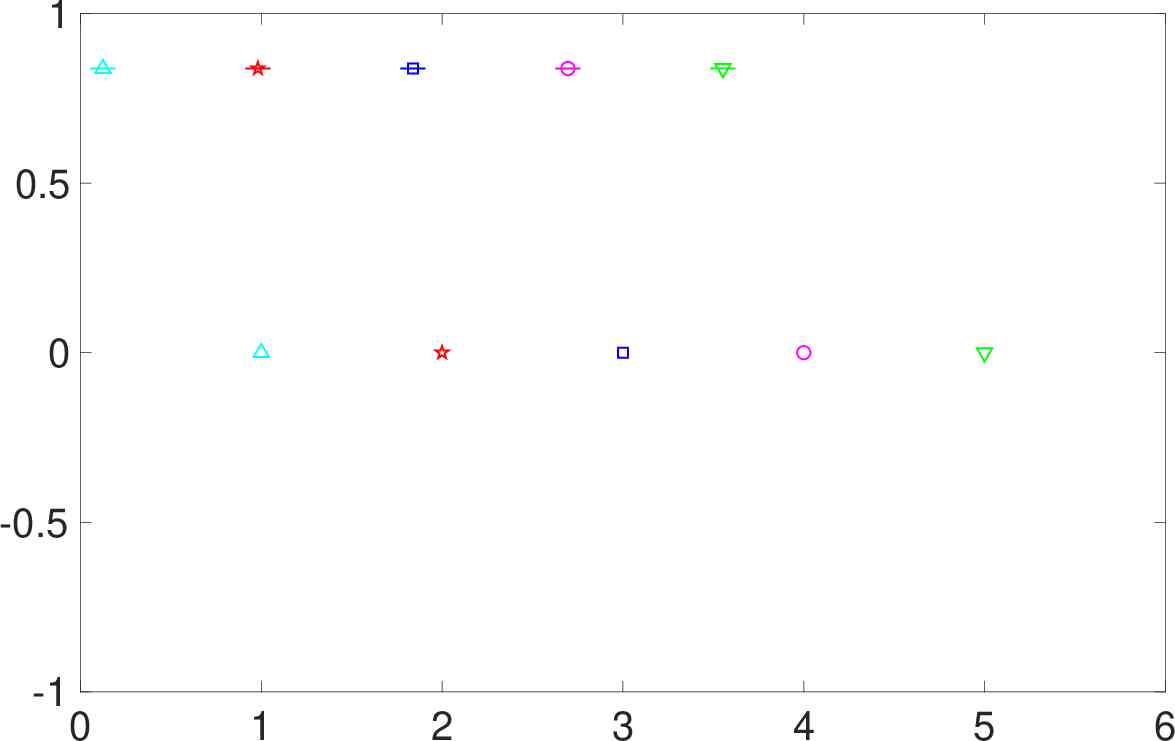}
		\put(5.5,2.5){$\ell_F / \eta = 209.4$}
	    \put(26.5,2.5){$\ell_F / \eta = 67.3$}
	    \put(47.5,2.5){$\ell_F / \eta = 31.6$}
	    \put(68.5,2.5){$\ell_F / \eta = 14.8$}
	    \put(89.5,2.5){$\ell_F / \eta = 7.0$}
	\end{overpic}}
	\caption{ (a-b-c) Second moment of the Restricted Euler, pressure, viscous and sub-grid contributions to the vorticity tilting term, together with the second moment of the total vorticity tilting term.
	(d-e-f) PDF of the vorticity tilting term in equation \eqref{theory_eq_vort_princ}.}
	\label{res_fig_vort_tilt}
\end{figure}

%%%!
The rotation-rate of the eigenframe plays an important role in the vorticity dynamics through the vortex tilting mechanism, described by the term $\Pi_{ij}^\eig\omega_j^\eig$ in equation \eqref{theory_eq_vort_princ}. The second moment of $\Pi_{ij}^\eig\omega_j^\eig$ and the various contributions to it are shown in figures \ref{res_fig_vort_tilt}(a-b-c) together with the PDFs of  $|\Pi_{ij}^\eig\omega_j^\eig|$ in figures \ref{res_fig_vort_tilt}(d-e-f). The Restricted Euler and pressure Hessian contributions to $\langle(\Pi_{ij}^\eig\omega_j^\eig)^2\rangle$ are the largest and are similar in size, showing that the local (i.e.\ that captured by the Restricted Euler model) and non-local contributions to $\langle(\Pi_{ij}^\eig\omega_j^\eig)^2\rangle$ are similar. The reason why the RE term makes a contribution to the vorticity tilting that is similar to that of the pressure Hessian, despite the fact that the former gives a much smaller contribution to the eigenframe rotation-rate, is because of the weak preferential alignment between the vorticity and the pressure Hessian. The viscous contribution is small over the range of $\ell_F$ considered, with $\langle(\widetilde{\tau}\Pi_{ij}^{\nu\eig}\omega_j^\eig)^2\rangle$ decreasing as $\ell_F^{-\xi}$, with $\xi$ between 2 and 3. The vorticity tilting about axis $\bm{v}_2$ is slightly larger than the tilting about the other two axes,
in contrast to the reduced eigenframe rotation rate
about axis $\bm{v}_2$ shown in figure \ref{res_fig_S_rot}.
However, the difference between the components of $\langle(\Pi_{ij}^\eig\omega_j^\eig)^2\rangle$ is smaller than the difference between the components of $\langle(\Pi_{ij}^\eig)^2\rangle$ shown in figure \ref{res_fig_S_rot}. 

The PDFs of $|\Pi_{ij}^\eig\omega_j^\eig|$ are shown in figures \ref{res_fig_vort_tilt}(d-e-f), revealing wide power-law tails and large values of vorticity tilting. Quite remarkably, the PDFs are almost insensitive to the filtering scale $\ell_F$, such that very strong vorticity tilting is a feature that persists beyond just the dissipation range. This behavior, however, is probably kinematic rather than purely dynamical in origin. In particular, equation \eqref{theory_eq_strain_rot} shows that $\Pi_{ij}^\eig$ depends on $(\lambda_j-\lambda_i)^{-1}$, such that small values of $\widetilde{\tau}(\lambda_j-\lambda_i)$, which occur with high probability, can lead to large values of $\widetilde{\tau}^2\Pi_{ij}^\eig \omega_j^\eig$.

\subsection{Characterizing the pressure Hessian}

\begin{figure}
\centering
\vspace{0mm}			
    \subfloat[]	
	{\begin{overpic}
	[trim = 0mm -32mm -40mm -3mm,
	scale=0.125,clip,tics=20]{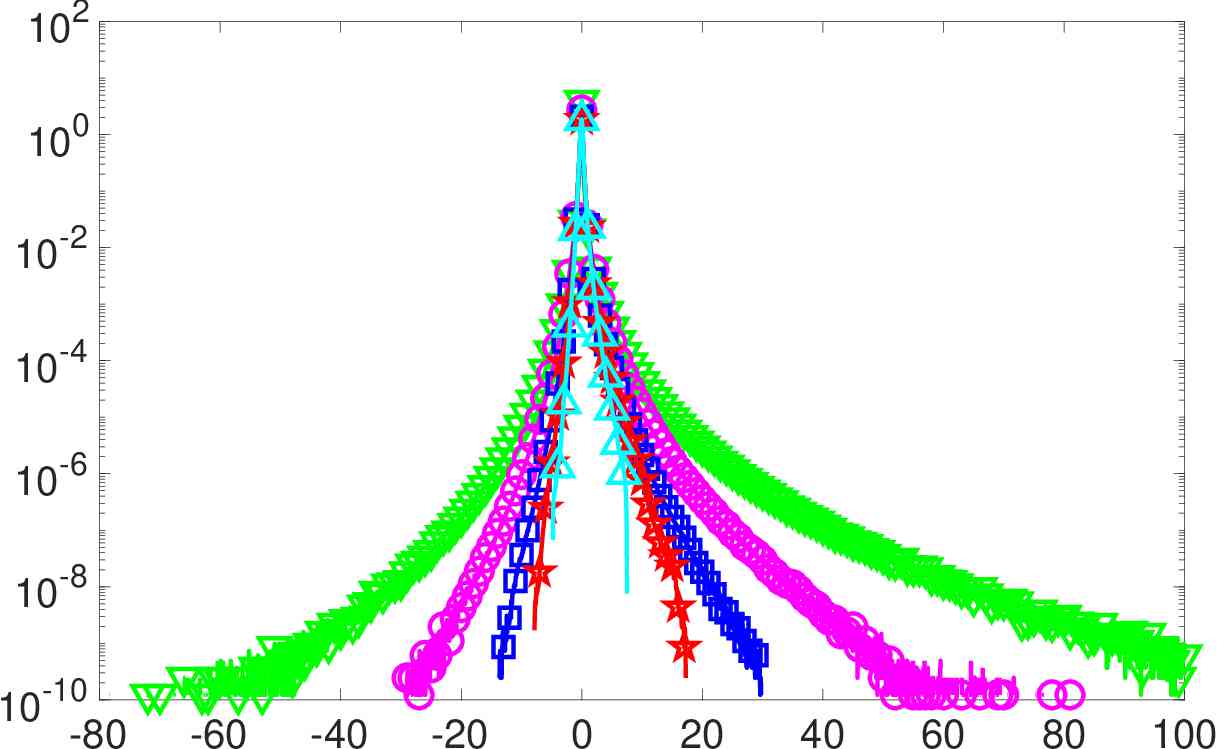}
	\put(40,0){$\widetilde{\tau}^2 {H}_{11}^{P\eig}$}
    \put(-6,30){\rotatebox{90}{PDF}}
	\end{overpic}}
	\subfloat[]
	{\begin{overpic}
	[trim = -40mm -32mm 0mm -3mm,
	scale=0.125,clip,tics=20]{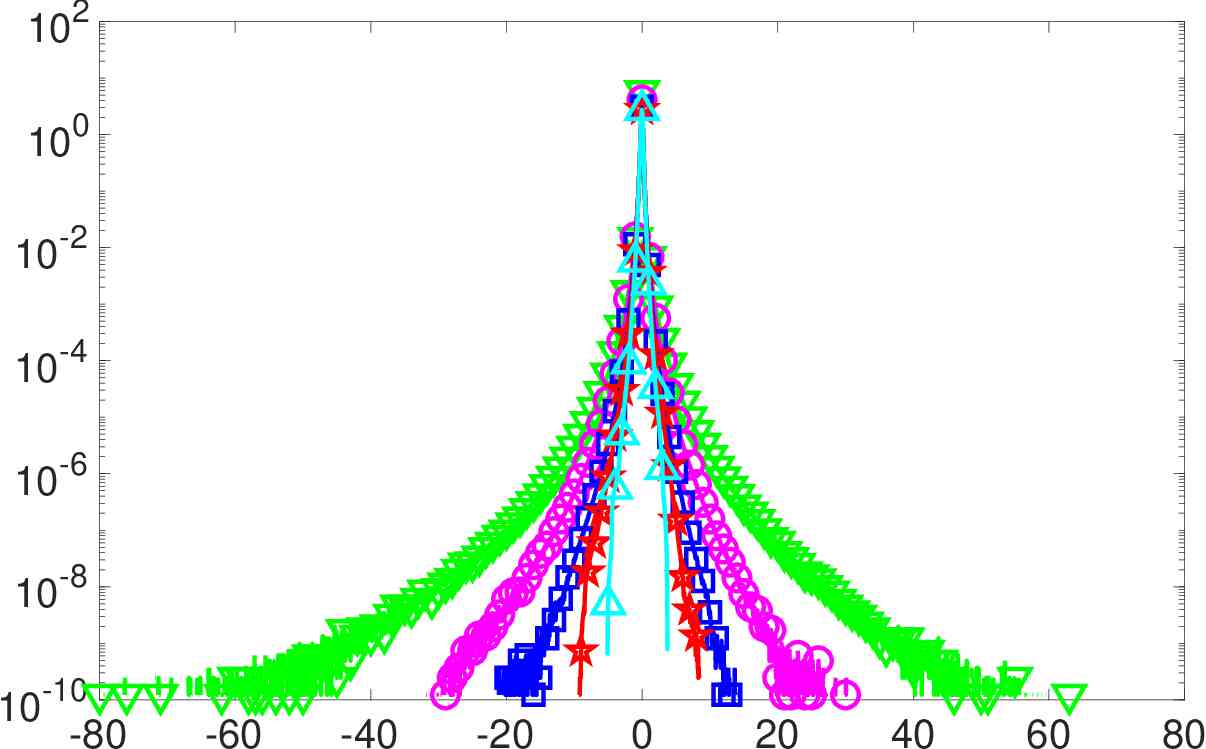}
	\put(50,0){$\widetilde{\tau}^2 \anisotr{H}_{11}^{P\eig}$}
	\put(0,30){\rotatebox{90}{PDF}}
	\end{overpic}}\\
   \subfloat[]	
	{\begin{overpic}
	[trim = 0mm -32mm -40mm -3mm,
	scale=0.125,clip,tics=20]{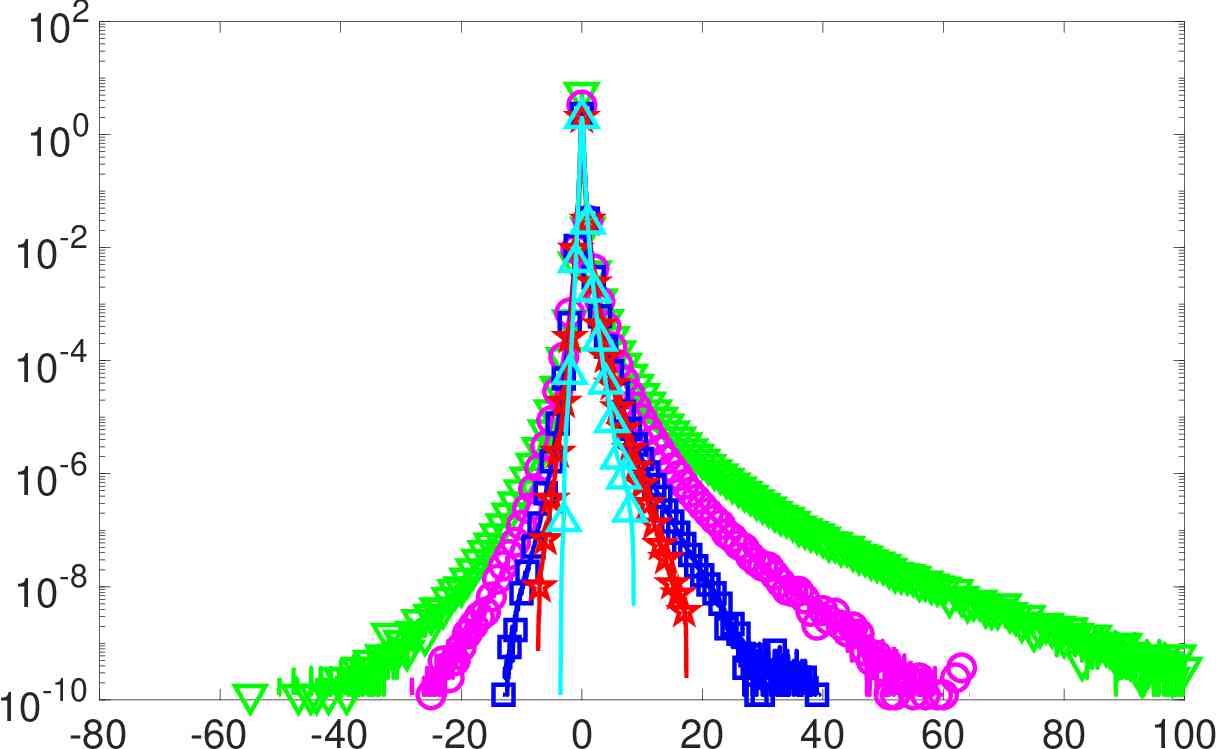}
	\put(40,0){$\widetilde{\tau}^2 {H}_{22}^{P\eig}$}
	\put(-6,30){\rotatebox{90}{PDF}}
	\end{overpic}}
	\subfloat[]
	{\begin{overpic}
	[trim = -40mm -32mm 0mm -3mm,
	scale=0.125,clip,tics=20]{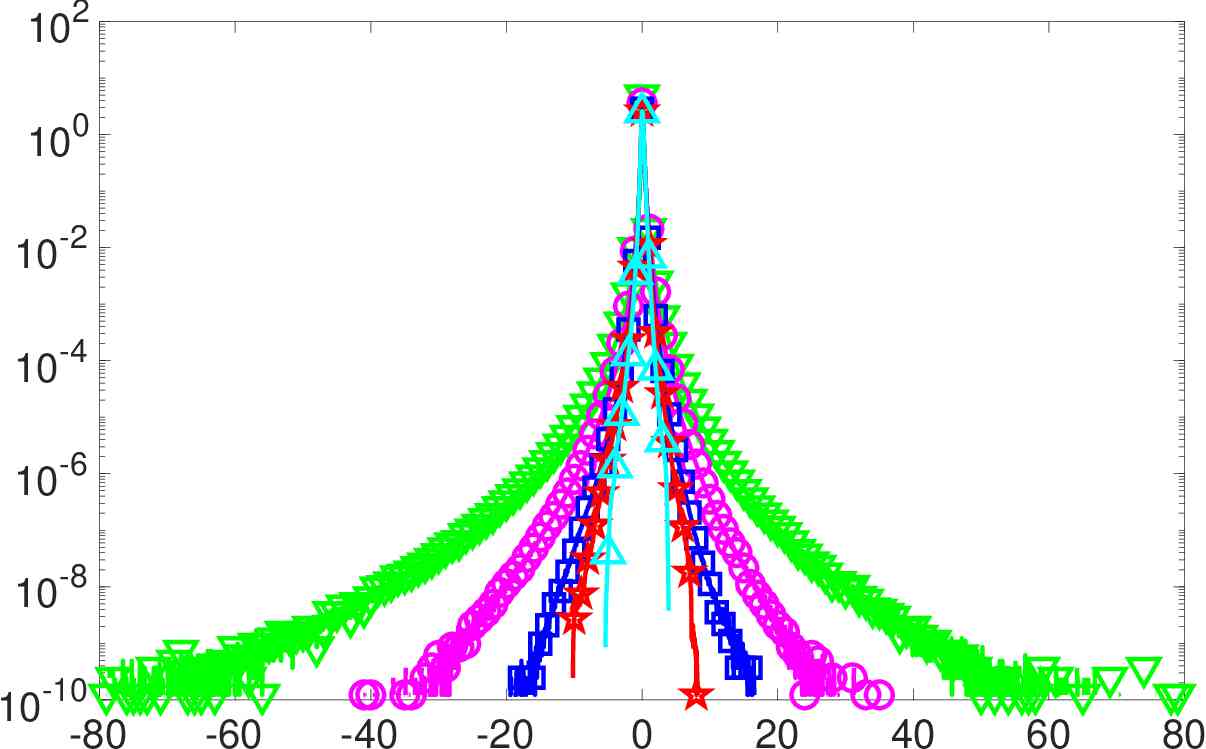}
	\put(50,0){$\widetilde{\tau}^2 \anisotr{H}_{22}^{P\eig}$}
	\put(0,30){\rotatebox{90}{PDF}}
	\end{overpic}}\\
   \subfloat[]	
	{\begin{overpic}
	[trim = 0mm -32mm -40mm -3mm,
	scale=0.125,clip,tics=20]{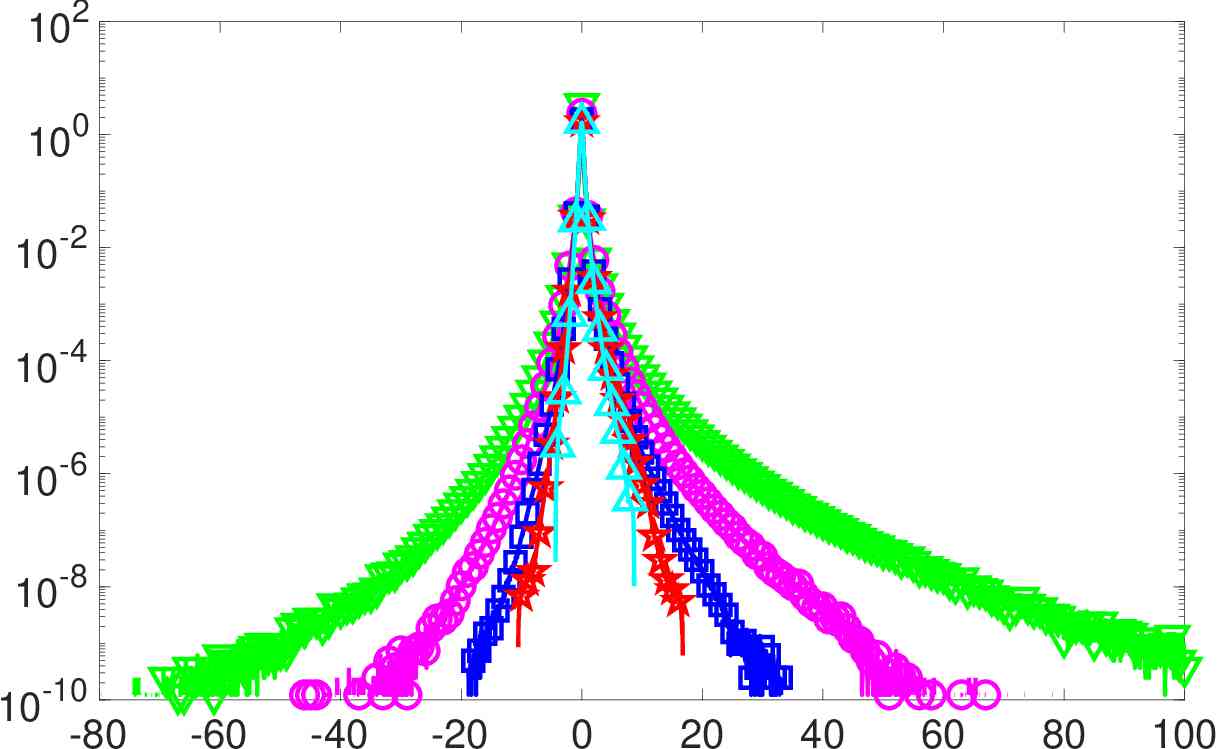}
	\put(40,0){$\widetilde{\tau}^2 {H}_{33}^{P\eig}$}
	\put(-6,30){\rotatebox{90}{PDF}}
	\end{overpic}}
	\subfloat[]
	{\begin{overpic}
	[trim = -40mm -32mm 0mm -3mm,
	scale=0.125,clip,tics=20]{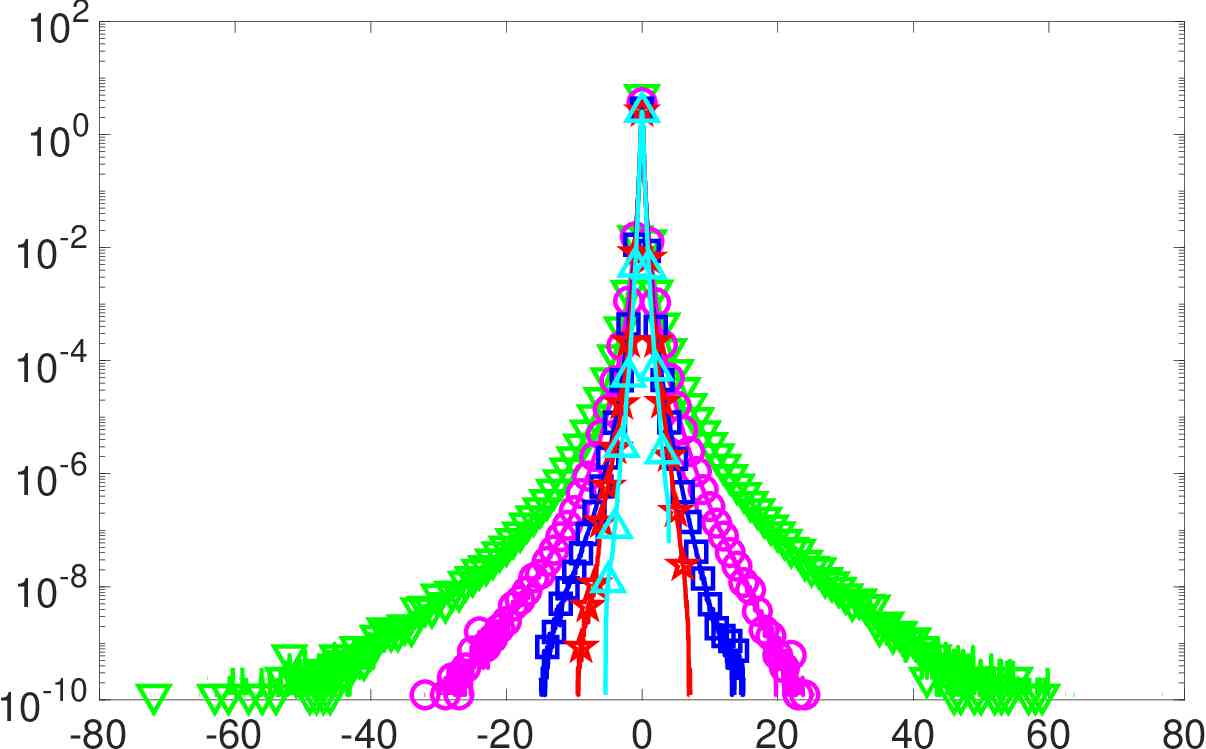}
	\put(50,0){$\widetilde{\tau}^2 \anisotr{H}_{33}^{P\eig}$}
	\put(0,30){\rotatebox{90}{PDF}}
	\end{overpic}}\\
	{\begin{overpic}
	[trim = 35mm 275mm 150mm 20mm,
	scale=0.40,clip,tics=20]{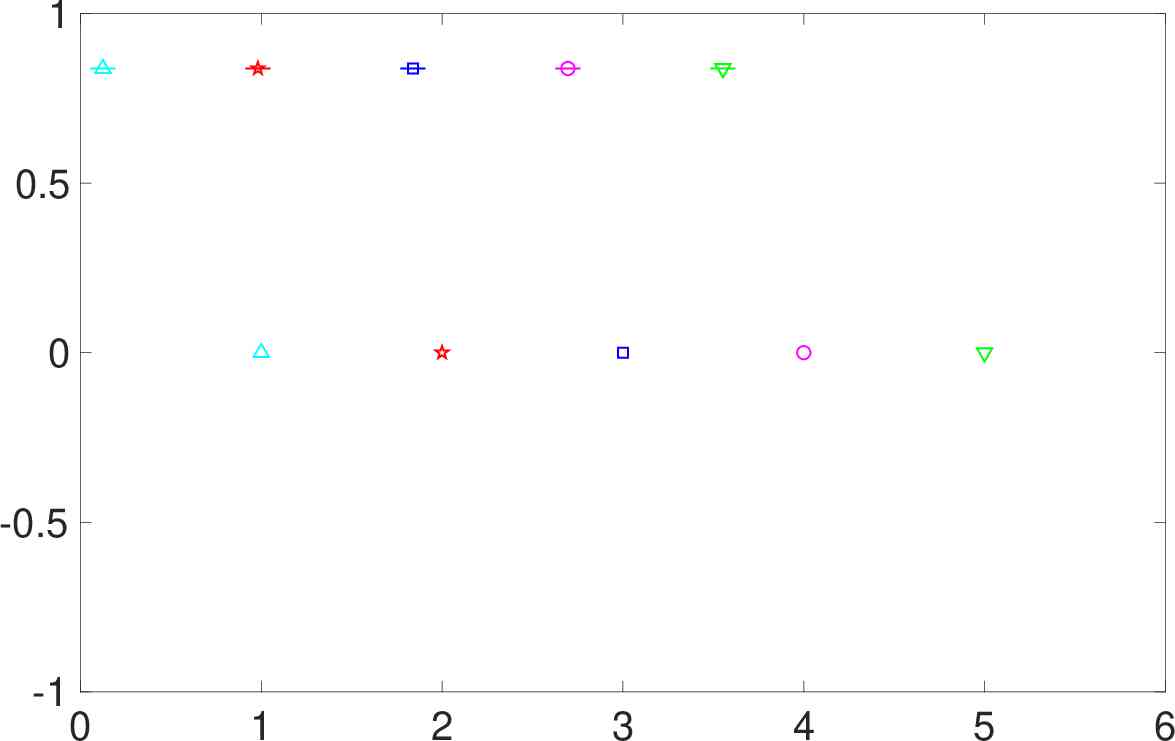}
		\put(5.5,2.5){$\ell_F / \eta = 209.4$}
	    \put(26.5,2.5){$\ell_F / \eta = 67.3$}
	    \put(47.5,2.5){$\ell_F / \eta = 31.6$}
	    \put(68.5,2.5){$\ell_F / \eta = 14.8$}
	    \put(89.5,2.5){$\ell_F / \eta = 7.0$}
	\end{overpic}}
	\caption{
	PDF of the diagonal components of the pressure Hessian in the eigenframe. (a-c-e) show results for the full pressure Hessian, and (b-d-f) are for the anisotopic part of the pressure Hessian.}
\label{res_fig_PDF_H_eigS}
\end{figure}
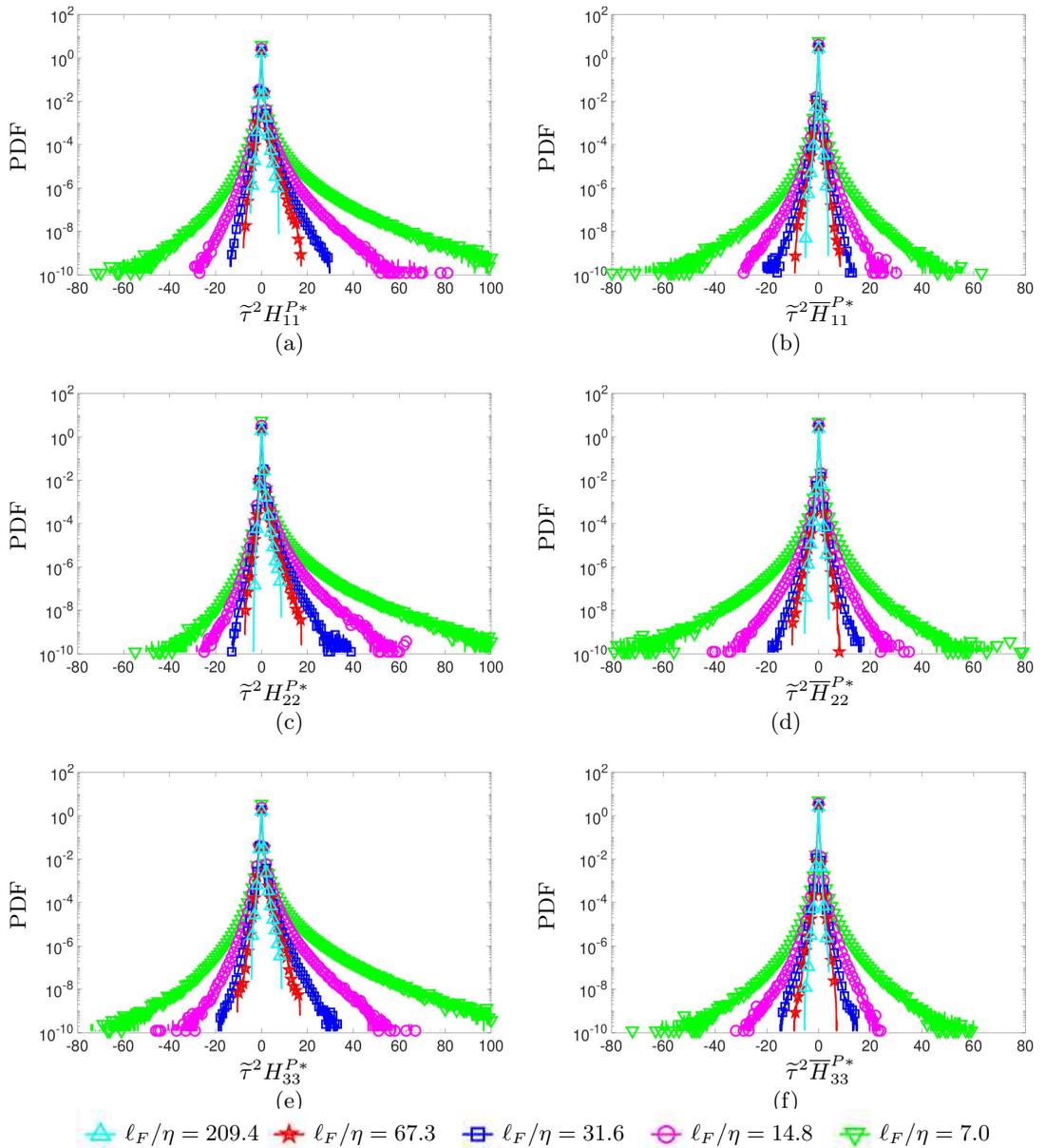

%%%%%PDFs of H
In Figures \ref{res_fig_PDF_H_eigS}(a-c-e) we show the PDFs of the diagonal components of the pressure Hessian in the eigenframe, filtered at various scales. The PDFs show wide tails for the smallest filtering lengths $\ell_F$, i.e.\  with the observation of highly intermittent acceleration statistics in turbulence \citep{Ayyalasomayajula2008}, but the extreme events become much rarer as $\ell_F$ is increased, with the PDFs approaching a Gaussian shape at the largest scales. The PDFs are also strongly positively skewed, with the skewness decreasing as $\ell_F$ is increased. The local/isotropic part of the pressure Hessian is proportional to the invariant $Q=\|\bm{\omega}\|^2/4 - \|\bm{S}\|^2/2$, and since the PDF of $Q$ is positively skewed \citep{Meneveau2011}, so also will be the PDF of the local part of the pressure Hessian. The dependence of the non-local contribution to the pressure Hessian on the local properties of $Q$ is more complicated, however, we note that during large events where $Q\gg \tau_\eta^2$, the local part of the pressure Hessian is expected to dominate over the non-local part (see below), since extreme events in $Q$ are spatially localized.

\subsubsection{Characterising the anisotropic pressure Hessian}

The anisotropic pressure Hessian is defined as
\begin{equation}
\anisotr{H}_{ij}^{P\eig} \equiv H_{ij}^{P\eig} - \frac{1}{3}H_{kk}^{P\eig}\delta_{ij}
\end{equation}
and the PDFs of the diagonal components $\anisotr{H}_{i(i)}^{P\eig}$ are shown in Figures \ref{res_fig_PDF_H_eigS}(b-d-f), for various $\ell_F$. Comparing these results to those in figures \ref{res_fig_PDF_H_eigS}(a-c-e) reveals that the strong positive skewness of the PDF of the full pressure Hessian arises from the dominating contribution of the local pressure Hessian during large events. Indeed, the PDF of $\anisotr{H}_{22}^{P\eig}$ is negatively skewed, indicating that the strongest fluctuations in $\anisotr{H}_{22}^{P\eig}$ tend to help the growth of $\lambda_2$ and hence also vortex stretching. This is in contrast with the average negative value of $-H_{22}^{P\eig}$ observed in figure \ref{res_avg_dlambda}.

%!% H_cond_lam2 and H_cond_omg2 combined
\begin{figure}
\centering
\vspace{0mm}
    \subfloat[]	
	{\begin{overpic}
	[trim = 0mm -32mm -40mm -3mm,
	scale=0.125,clip,tics=20]{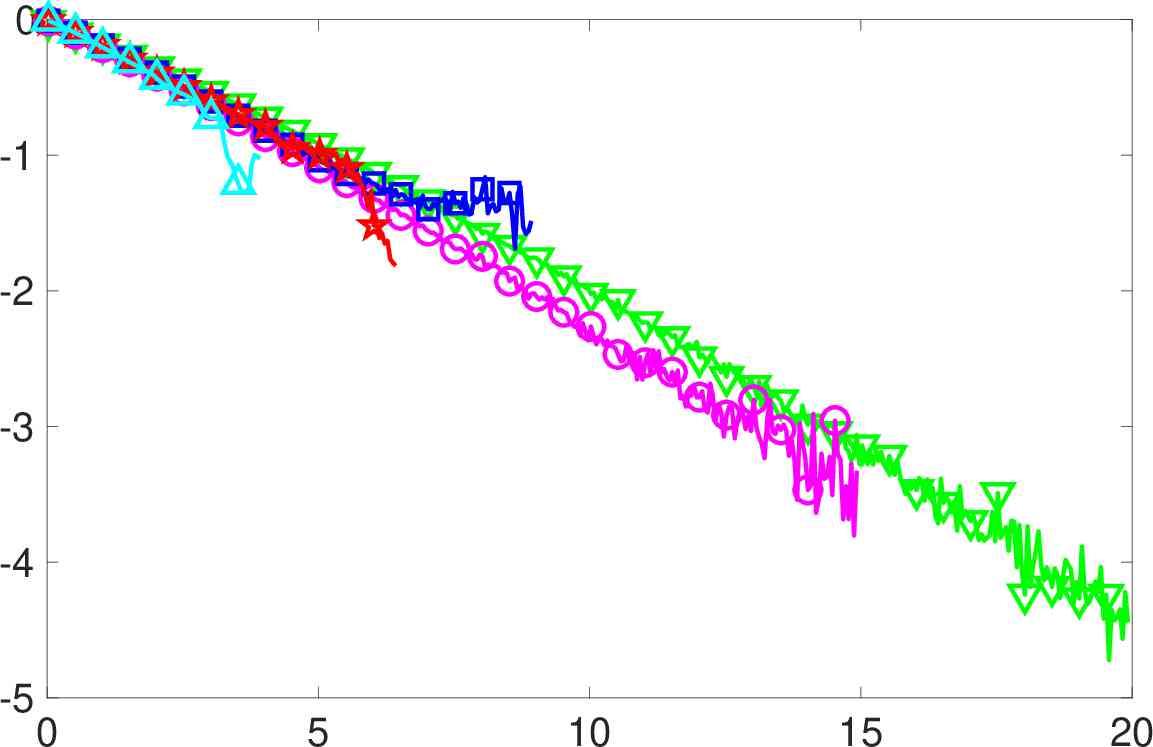}
			\put(42,0){$\widetilde{\tau}^2 {\lambda_1}^2$}
			\put(-10,18){\rotatebox{90}{$\condavg{\widetilde{\tau}^2 \anisotr{H}_{11}^{P\eig}}{\widetilde{\tau}^2 {\lambda_1}^2}$}}
			\put(5,12){
			\begin{overpic}[trim = 0mm 0mm 0mm 0mm, scale=0.05,clip,tics=20]
			{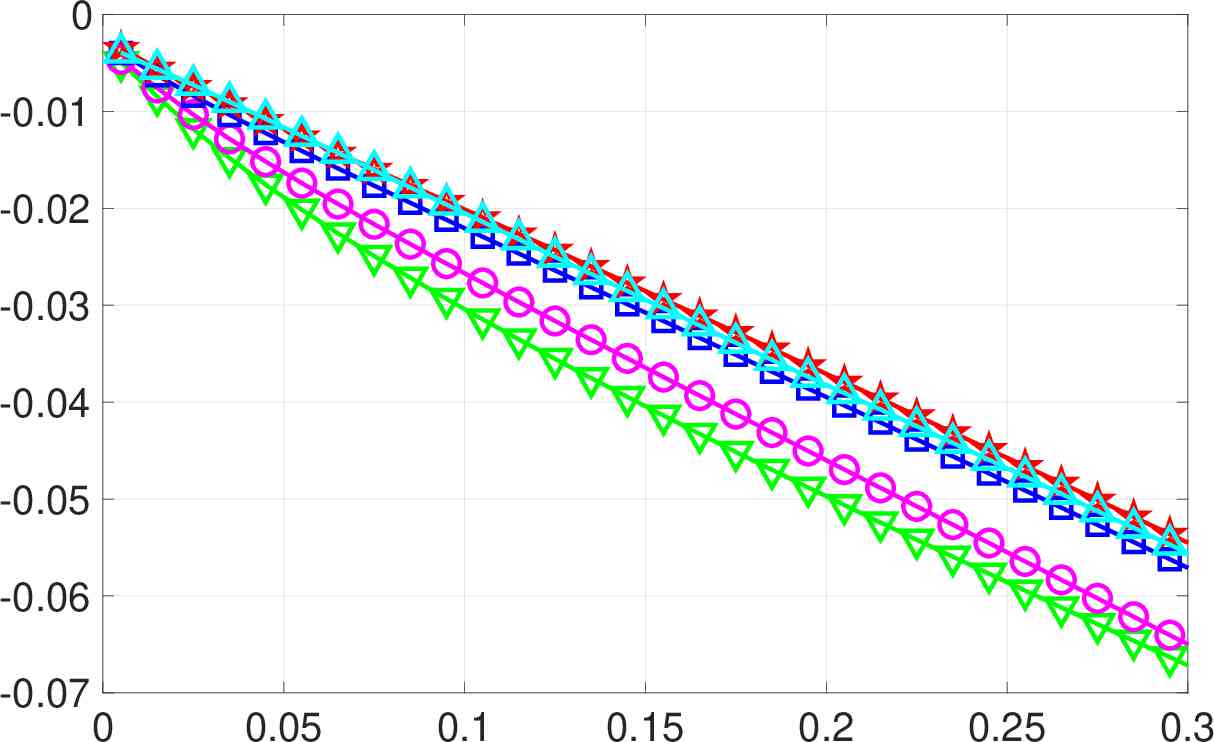}
			\end{overpic}
			}
	\end{overpic}}
	\subfloat[]
	{\begin{overpic}
	[trim = -40mm -32mm 0mm -3mm,
	scale=0.125,clip,tics=20]{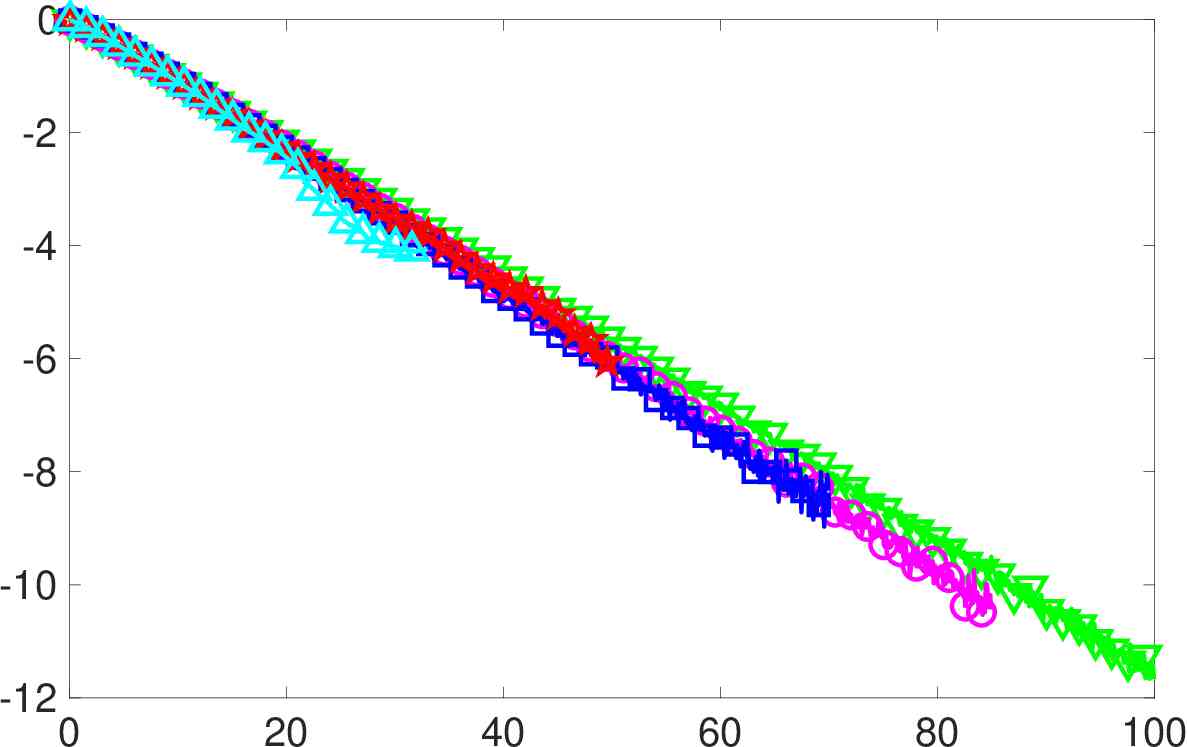}
			\put(53,0){$\widetilde{\tau}^2 {\omega_1^\eig}^2$}
			\put(-2,18){\rotatebox{90}{$\condavg{\widetilde{\tau}^2 \anisotr{H}_{11}^{P\eig}}{\widetilde{\tau}^2 {\omega_1^\eig}^2}$}}
			\put(13,12){
			\begin{overpic}[trim = 0mm 0mm 0mm 0mm, scale=0.05,clip,tics=20]
			{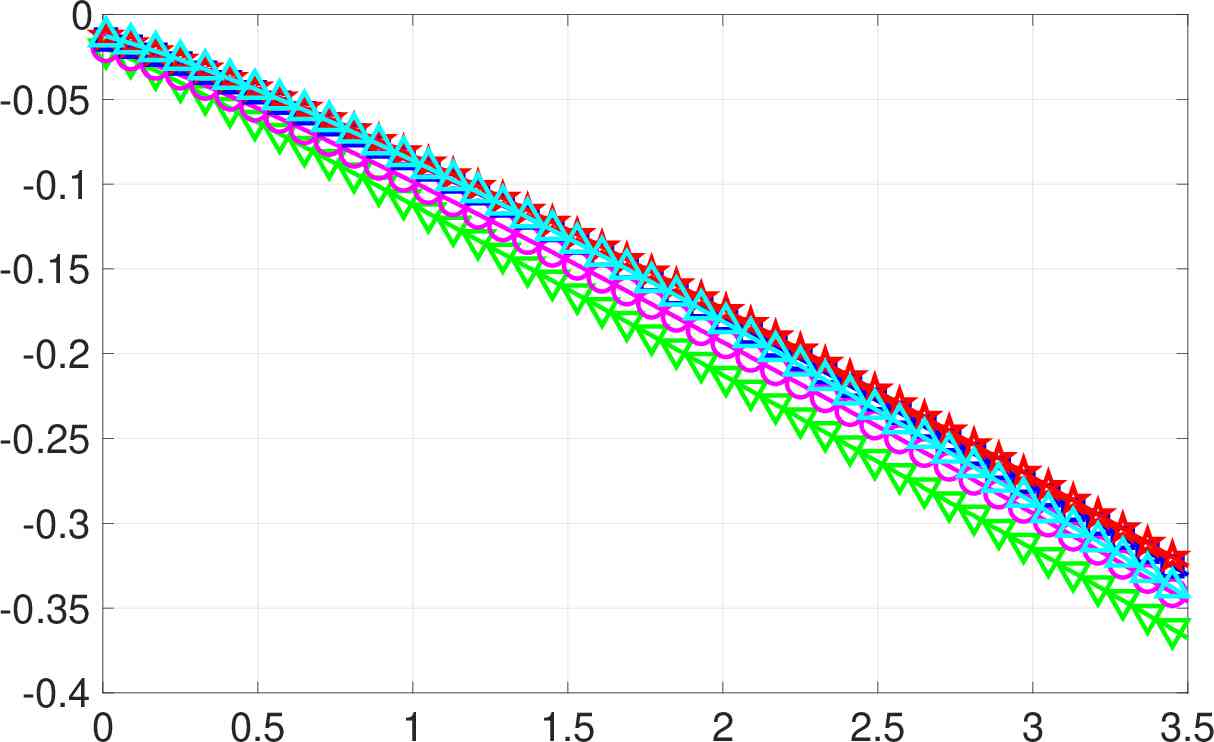}
			\end{overpic}
			}
	\end{overpic}}\\
   \subfloat[]	
	{\begin{overpic}
	[trim = 0mm -32mm -40mm -3mm,
	scale=0.125,clip,tics=20]{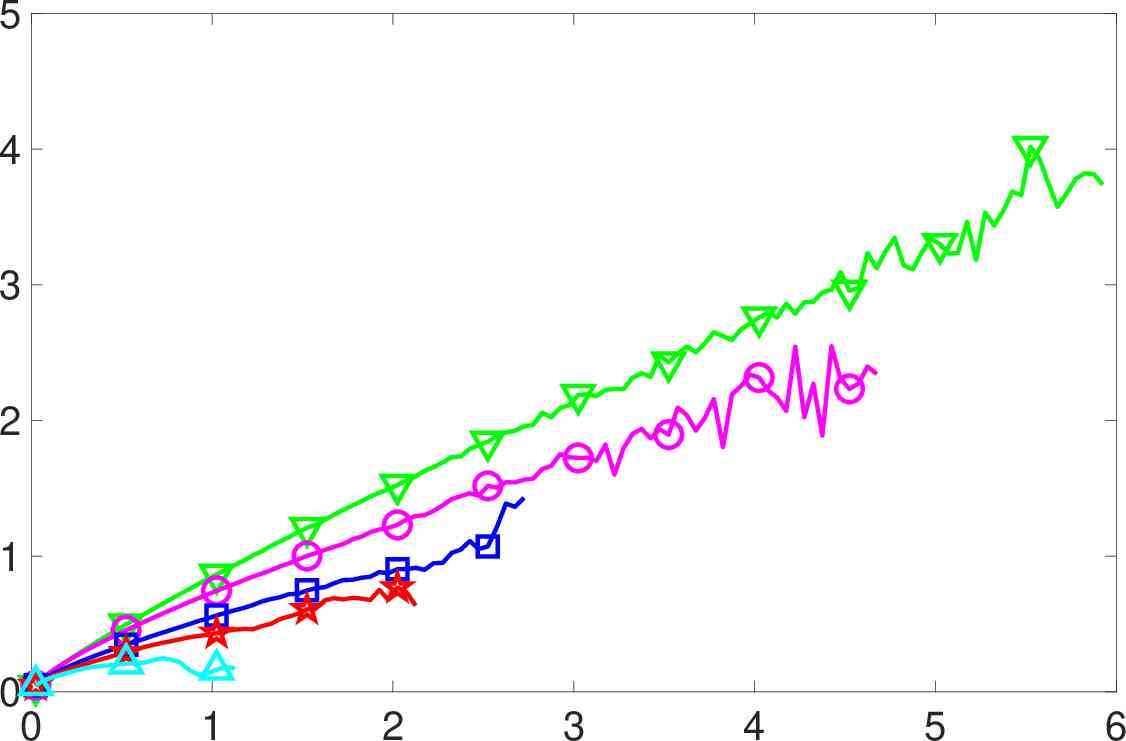}
			\put(42,0){$\widetilde{\tau}^2 {\lambda_2}^2$}
			\put(-10,18){\rotatebox{90}{$\condavg{\widetilde{\tau}^2 \anisotr{H}_{22}^{P\eig}}{\widetilde{\tau}^2 {\lambda_2}^2}$}}
			\put(5,38){
			\begin{overpic}[trim = 0mm 0mm 0mm 0mm, scale=0.05,clip,tics=20]
			{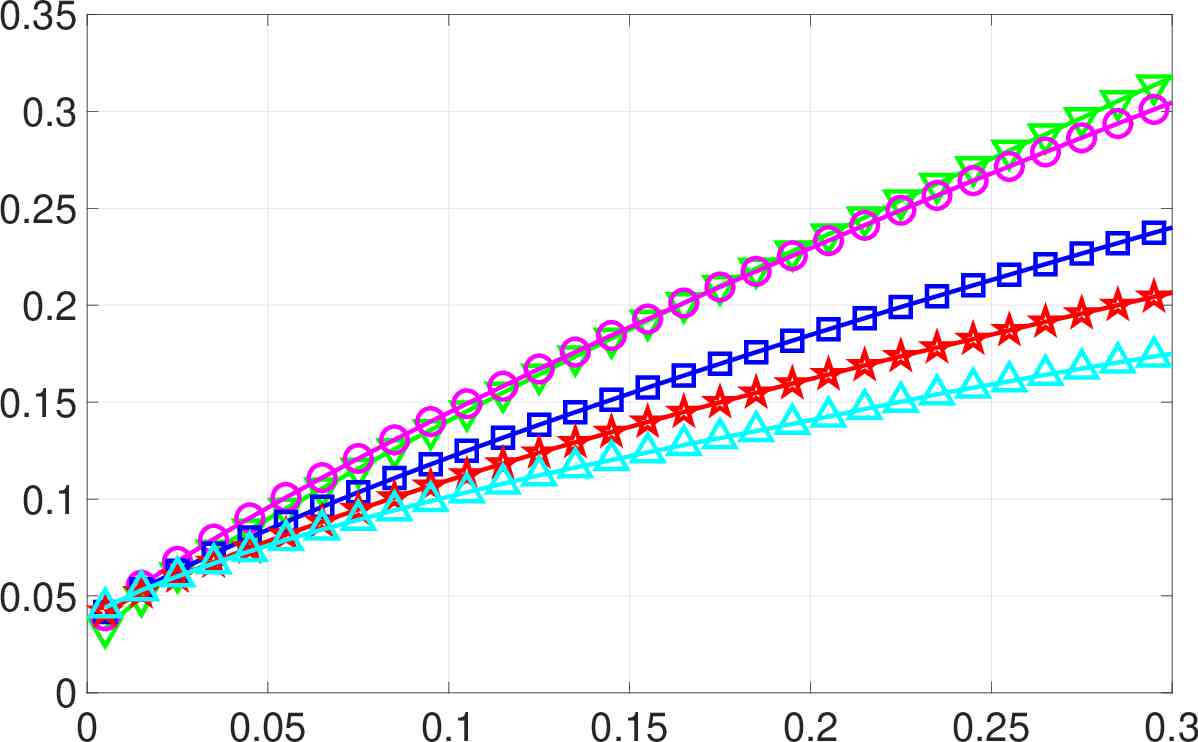}
			\end{overpic}
			}
	\end{overpic}}
	\subfloat[]
	{\begin{overpic}
	[trim = -40mm -32mm 0mm -3mm,
	scale=0.125,clip,tics=20]{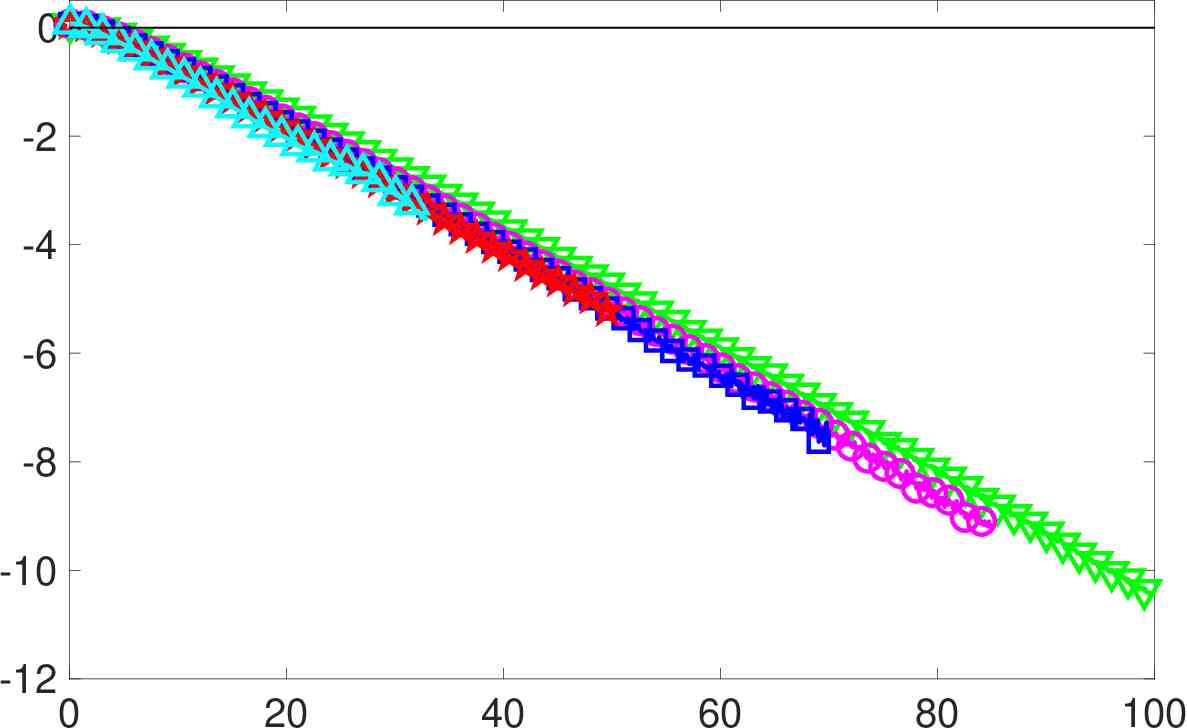}
			\put(53,0){$\widetilde{\tau}^2 {\omega_2^\eig}^2$}
			\put(-2,18){\rotatebox{90}{$\condavg{\widetilde{\tau}^2 \anisotr{H}_{22}^{P\eig}}{\widetilde{\tau}^2 {\omega_2^\eig}^2}$}}
			\put(13,12){
			\begin{overpic}[trim = 0mm 0mm 0mm 0mm, scale=0.05,clip,tics=20]
			{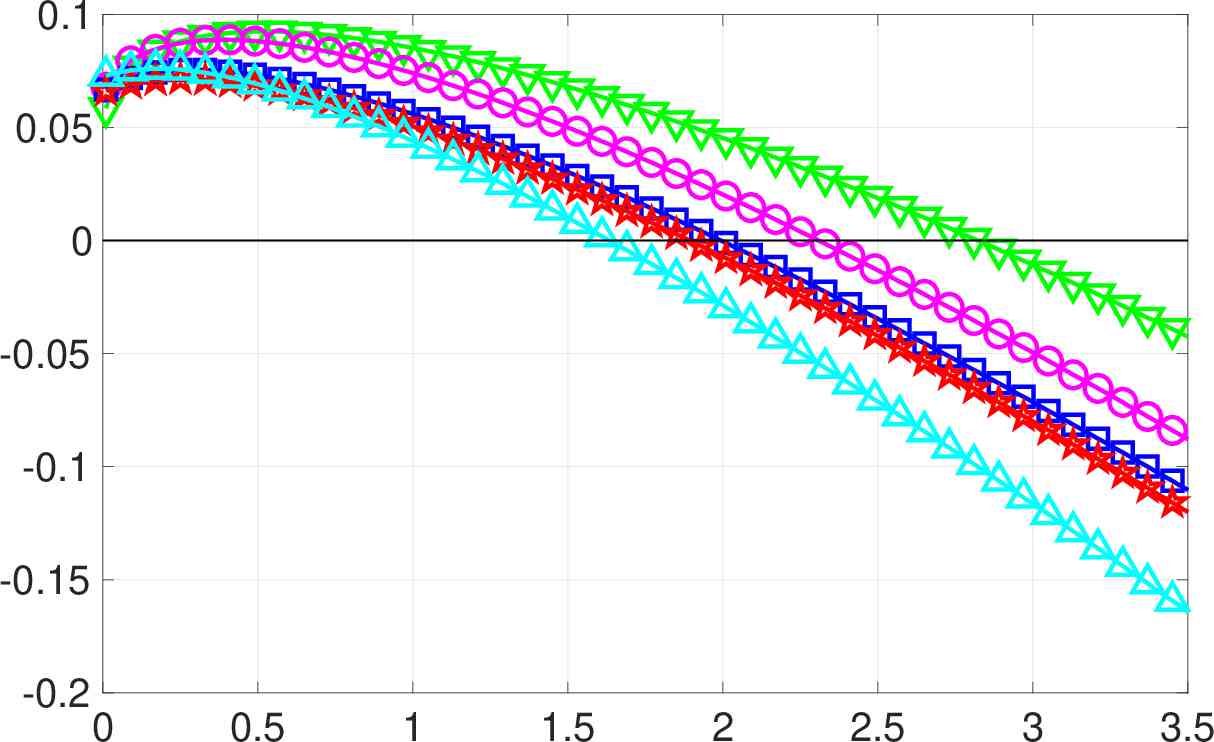}
			\end{overpic}
			}
	\end{overpic}}\\
   \subfloat[]	
	{\begin{overpic}
	[trim = 0mm -32mm -40mm -3mm,
	scale=0.125,clip,tics=20]{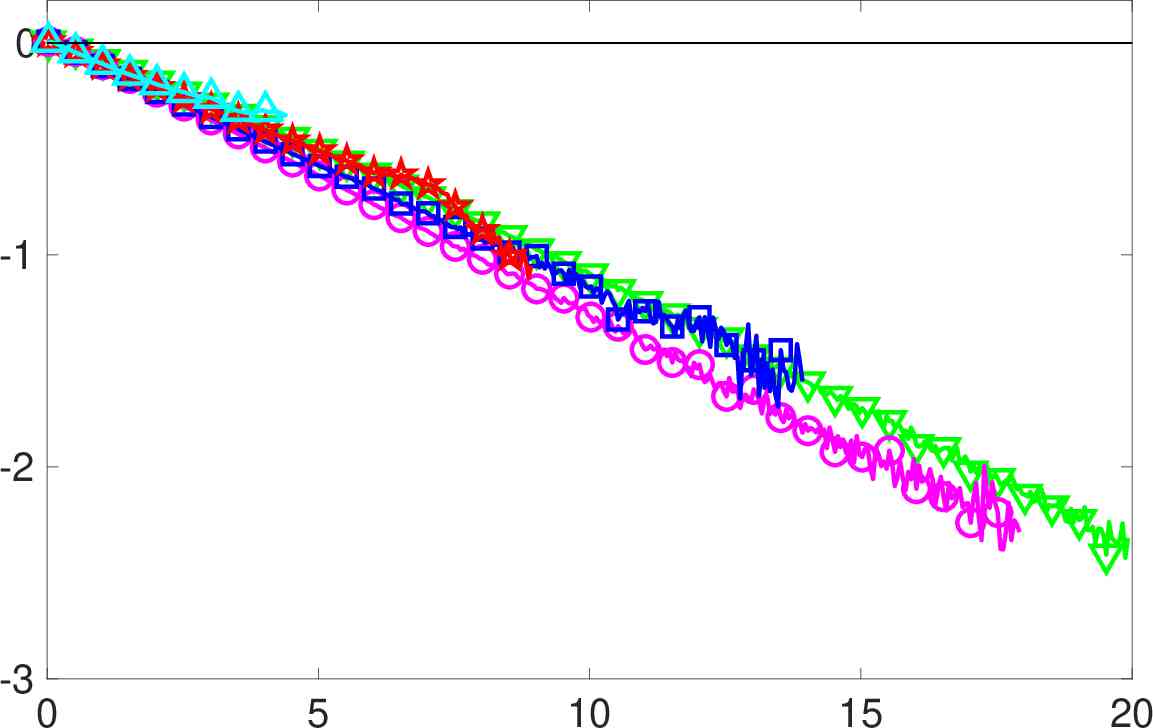}
			\put(42,0){$\widetilde{\tau}^2 {\lambda_3}^2$}
			\put(-10,18){\rotatebox{90}{$\condavg{\widetilde{\tau}^2 \anisotr{H}_{33}^{P\eig}}{\widetilde{\tau}^2 {\lambda_3}^2}$}}
			\put(5,12){
			\begin{overpic}[trim = 0mm 0mm 0mm 0mm, scale=0.05,clip,tics=20]
			{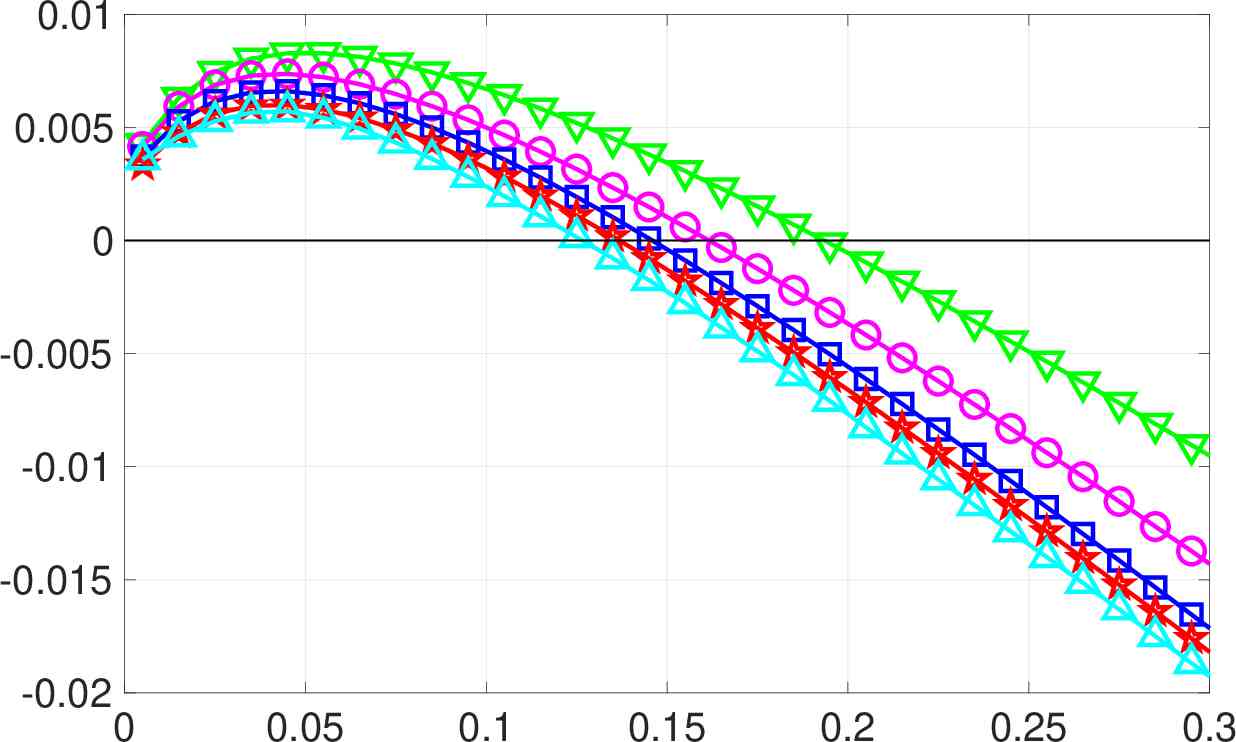}
			\end{overpic}
			}
	\end{overpic}}
	\subfloat[]
	{\begin{overpic}
	[trim = -40mm -32mm 0mm -3mm,
	scale=0.125,clip,tics=20]{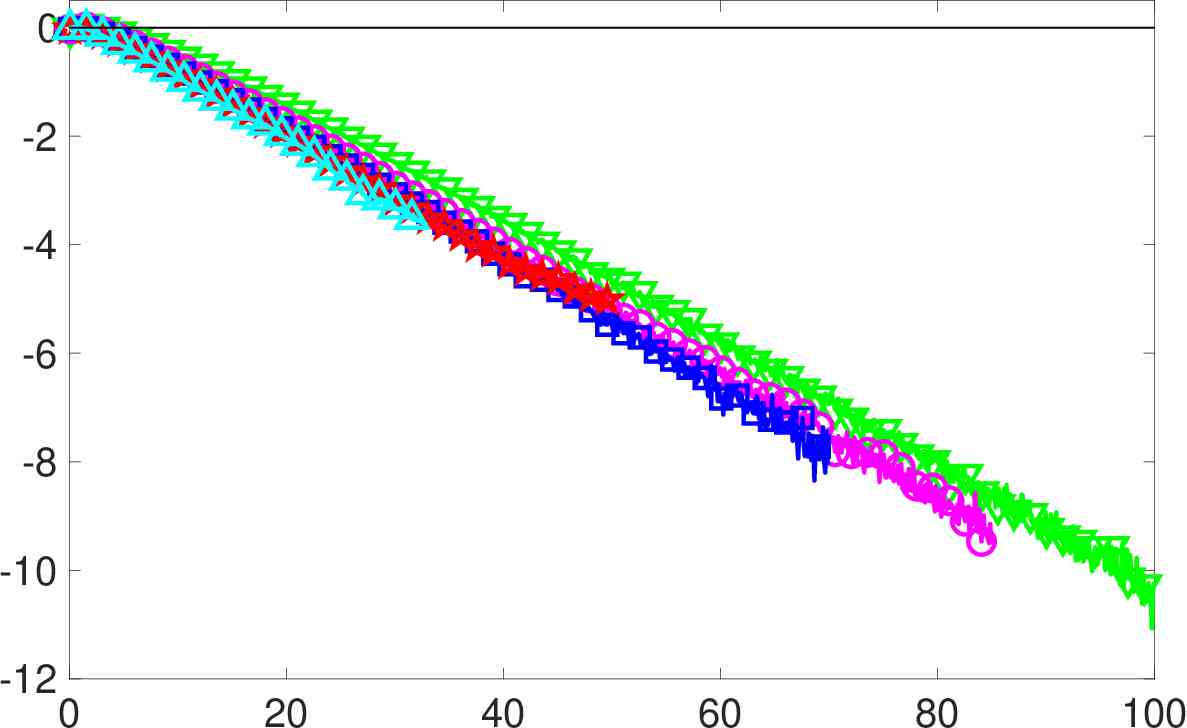}
			\put(53,0){$\widetilde{\tau}^2 {\omega_3^\eig}^2$}
			\put(-2,18){\rotatebox{90}{$\condavg{\widetilde{\tau}^2 \anisotr{H}_{33}^{P\eig}}{\widetilde{\tau}^2 {\omega_3^\eig}^2}$}}
			\put(13,12){
			\begin{overpic}[trim = 0mm 0mm 0mm 0mm, scale=0.05,clip,tics=20]
			{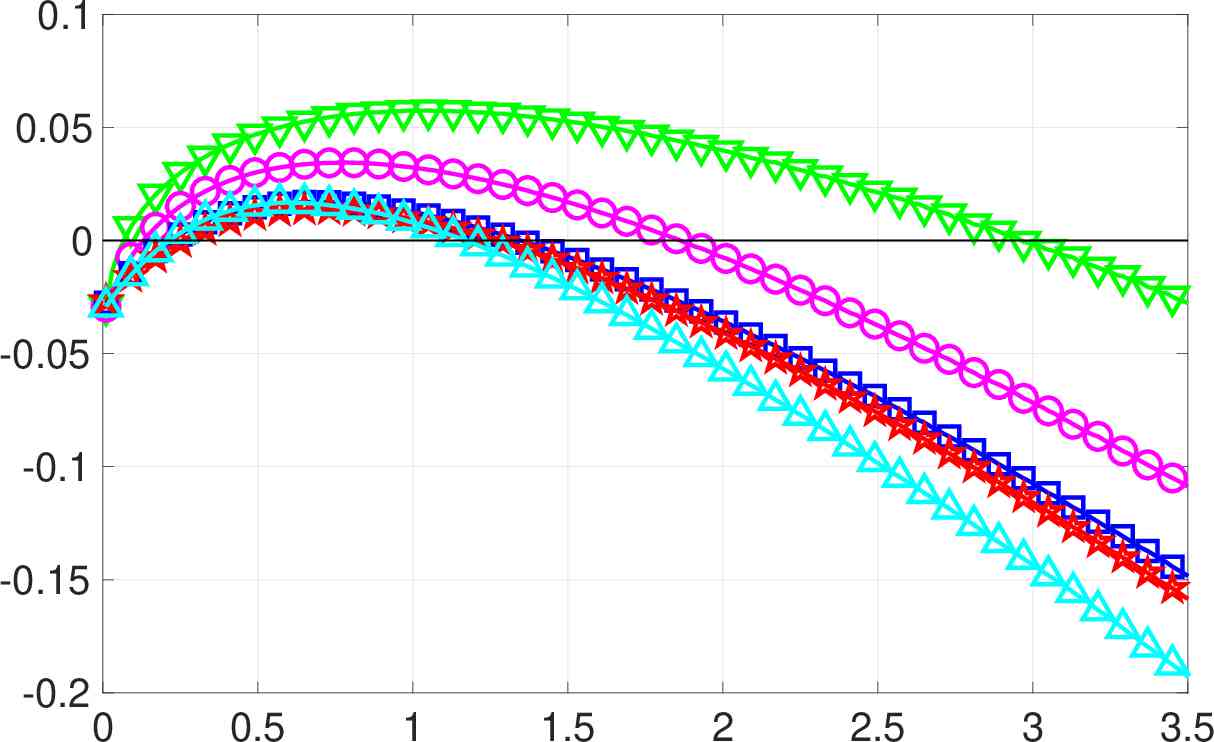}
			\end{overpic}
			}
	\end{overpic}}\\
	{\begin{overpic}
	[trim = 35mm 275mm 160mm 20mm,
	scale=0.40,clip,tics=20]{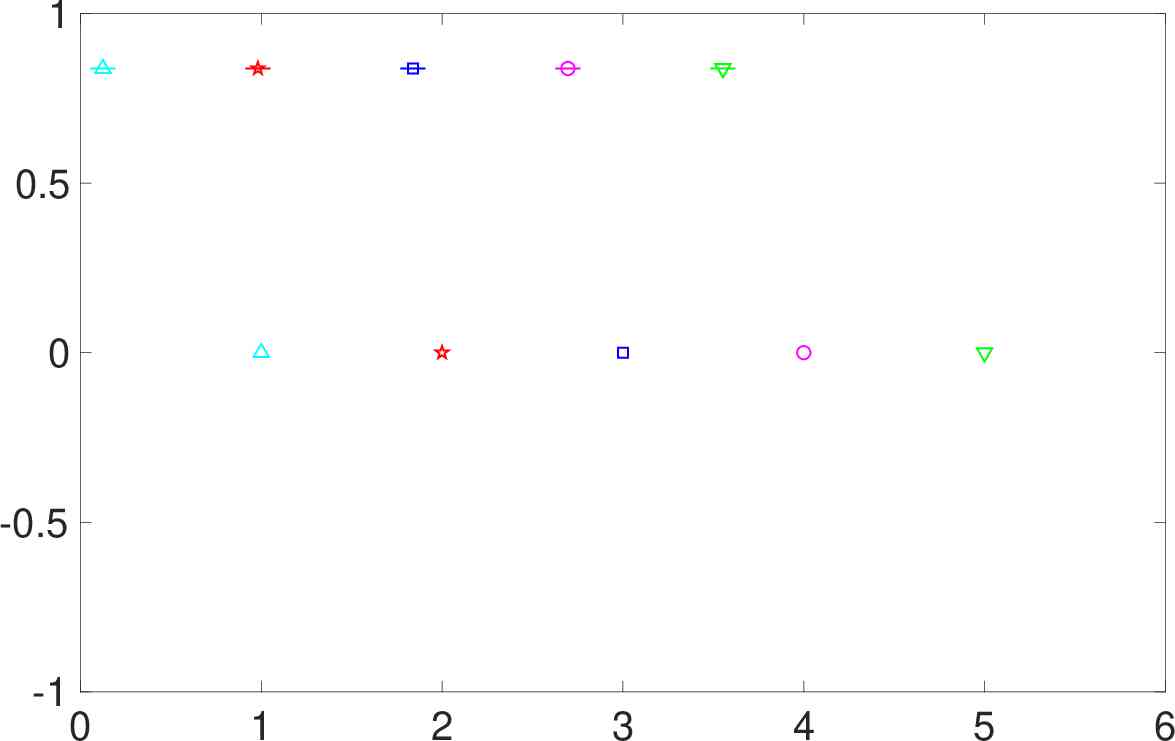}
	    \put(5.5,2.5){$\ell_F/\eta = 209.4$}
	    \put(27.5,2.5){$\ell_F/\eta = 67.3$}
	    \put(49.5,2.5){$\ell_F/\eta = 31.6$}
	    \put(71.5,2.5){$\ell_F/\eta = 14.8$}
	    \put(93.5,2.5){$\ell_F/\eta = 7.0$}
	\end{overpic}}
	\caption{Average of the diagonal components of the anisotropic pressure Hessian in the eigenframe conditioned on (a-c-e) the corresponding eigenvalues squared, and (b-d-f) the corresponding vorticity components squared. Insets highlight the results for small values of $\widetilde{\tau}^2 {\lambda_i}^2$ and $\widetilde{\tau}^2 {\omega_i^\eig}^2$.}
	\label{res_aH_cond_lam2omg2}
\end{figure}
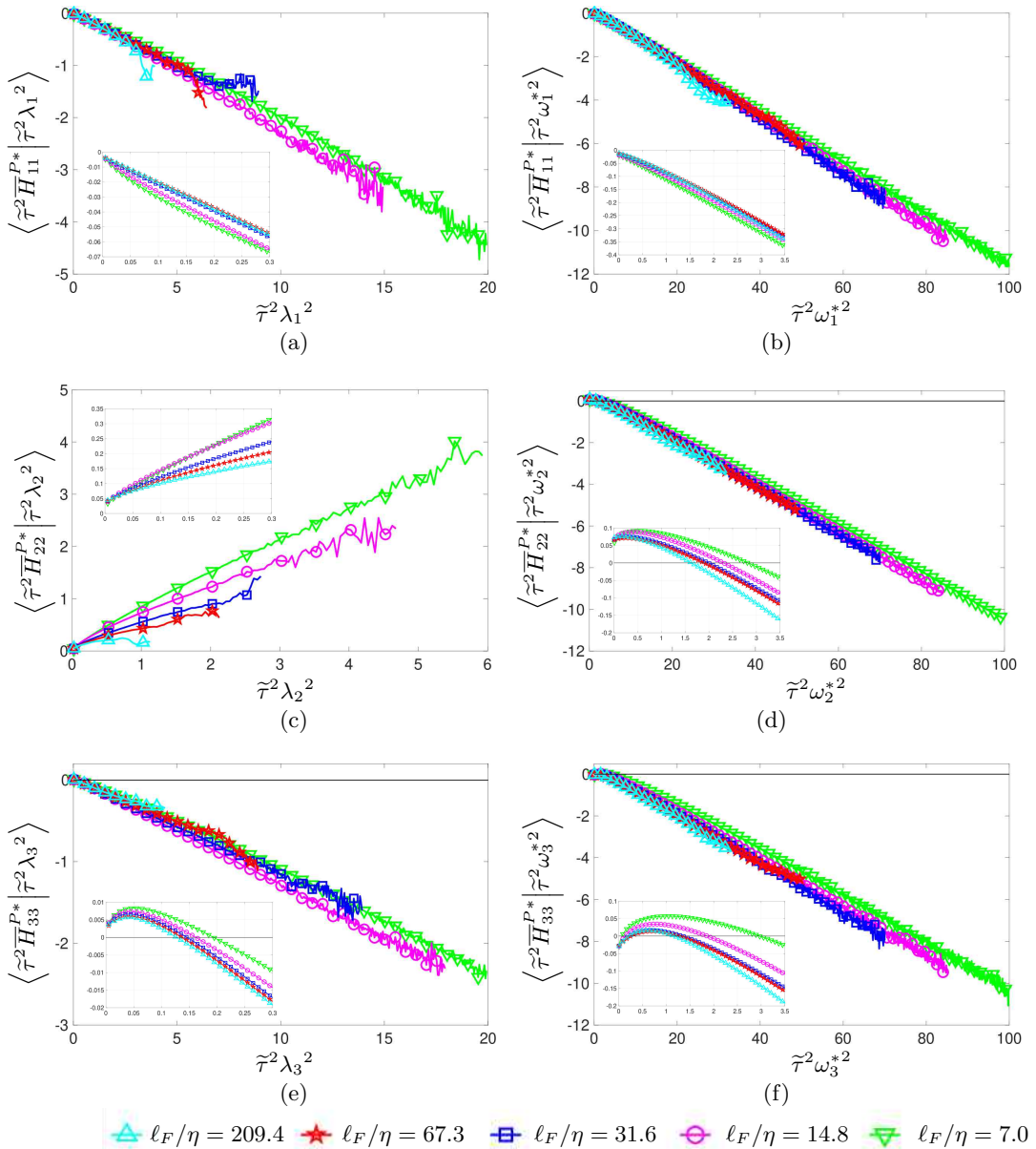

In order to understand more fully the relationship between ${\anisotr{H}_{i(i)}^{P\eig}}$ and the local properties of the flow, in figure \ref{res_aH_cond_lam2omg2} (a-c-e), we consider the results for $\inlavg{\anisotr{H}_{i(i)}^{P\eig}|\lambda_i^2}$ as a function of the filtering scale. For $i=1$, this quantity is always negative, and since the pressure Hessian appears with a negative sign in front of it in the eigenvalue equation \eqref{theory_eq_strain_lambda}, this indicates that on average $\anisotr{H}_{11}^{P\eig}$ acts to amplify $\lambda_1$. For $\widetilde{\tau}\lambda_1\leq O(1)$, $\inlavg{\anisotr{H}_{11}^{P\eig}|\lambda_1^2}$ becomes increasingly negative almost with increasing $\lambda_1^2$, varying linearly with $\lambda_1^2$. The contribution to $\lambda_2$, namely $\inlavg{\anisotr{H}_{22}^{P\eig}|\lambda_2^2}$, is always positive and therefore on average $\anisotr{H}_{22}^{P\eig}$ hinders the growth of positive $\lambda_2$ for all values of $\lambda_2^2$. Furthermore, for $\widetilde{\tau}\lambda_2\leq O(1)$, $\inlavg{\anisotr{H}_{22}^{P\eig}|\lambda_2^2}$ increases almost linearly with increasing $\lambda_2^2$ except for the larger filter scales, for which a nonlinear behavior is apparent even for $\widetilde{\tau}\lambda_2\ll 1$. The behavior of $\inlavg{\anisotr{H}_{33}^{P\eig}|\lambda_3^2}$ is quite peculiar, showing that on average $\anisotr{H}_{33}^{P\eig}$ amplifies $\lambda_3$ for very small $\widetilde{\tau}\lambda_3$, but then hinders its growth outside of this regime. The range of $\widetilde{\tau}\lambda_3$ over which $\anisotr{H}_{33}^{P\eig}$ amplifies $\lambda_3$ on average decreases with increasing filter scale.

Over the entire range of $\widetilde{\tau}\lambda_i$ shown, the effect of filtering appears quite weak (it is more apparent for $i=2$ since in that case the results extend over a smaller region of $\widetilde{\tau}\lambda_i$ owing to the smaller fluctuations of $\lambda_2$ compares with $\lambda_1$ or $\lambda_3$). Nevertheless, the insets to figures \ref{res_aH_cond_lam2omg2} (a-c-e) highlight that for $(\widetilde{\tau}\lambda_i)^2\leq 0.3$, filtering reduces the magnitude of both $\inlavg{\widetilde{\tau}^2\anisotr{H}_{11}^{P\eig}|\lambda_1^2}$ and $\inlavg{\widetilde{\tau}^2\anisotr{H}_{22}^{P\eig}|\lambda_3^2}$, whereas the magnitude of $\inlavg{\widetilde{\tau}^2\anisotr{H}_{33}^{P\eig}|\lambda_3^2}$ is reduced by filtering in the regime where it is positive, but is actually increased in the region where it is negative.

In figure \ref{res_aH_cond_lam2omg2}(b-d-f) we consider the results for $\inlavg{\anisotr{H}_{i(i)}^{P\eig}|\omega_i^{\eig2}}$. For $\widetilde{\tau}^2\omega_i^{\eig2}\gtrsim 3$, $\inlavg{\anisotr{H}_{i(i)}^{P\eig}|\omega_i^{\eig2}}$ is negative, leading to the production of $\lambda_1$ and positive $\lambda_2$, but the suppression of $|\lambda_3|$. In this regime, $\inlavg{\anisotr{H}_{i(i)}^{P\eig}|\omega_i^{\eig2}}\propto \omega_i^{\eig2}$ is a good approximation. For $\widetilde{\tau}^2\omega_i^{\eig2}<3$, however, $\inlavg{\anisotr{H}_{i(i)}^{P\eig}|\omega_i^{\eig2}}$ changes sign for $i=2,3$ and has a highly non-linear behavior, similar to that of $\inlavg{\anisotr{H}_{33}^{P\eig}|\lambda_3^2}$. Taken altogether, the results in figure \ref{res_aH_cond_lam2omg2} indicate that closures such as the tetrad model  \citep{Chertkov1991,Naso2005} and enhanced Gaussian closure \citep{Wilczek2014} which predict that the anisotropic pressure Hessian depends on the square of the local velocity gradient are able to capture several important features, but not all, especially in the regime of small gradients.

%!% H_cond_Q
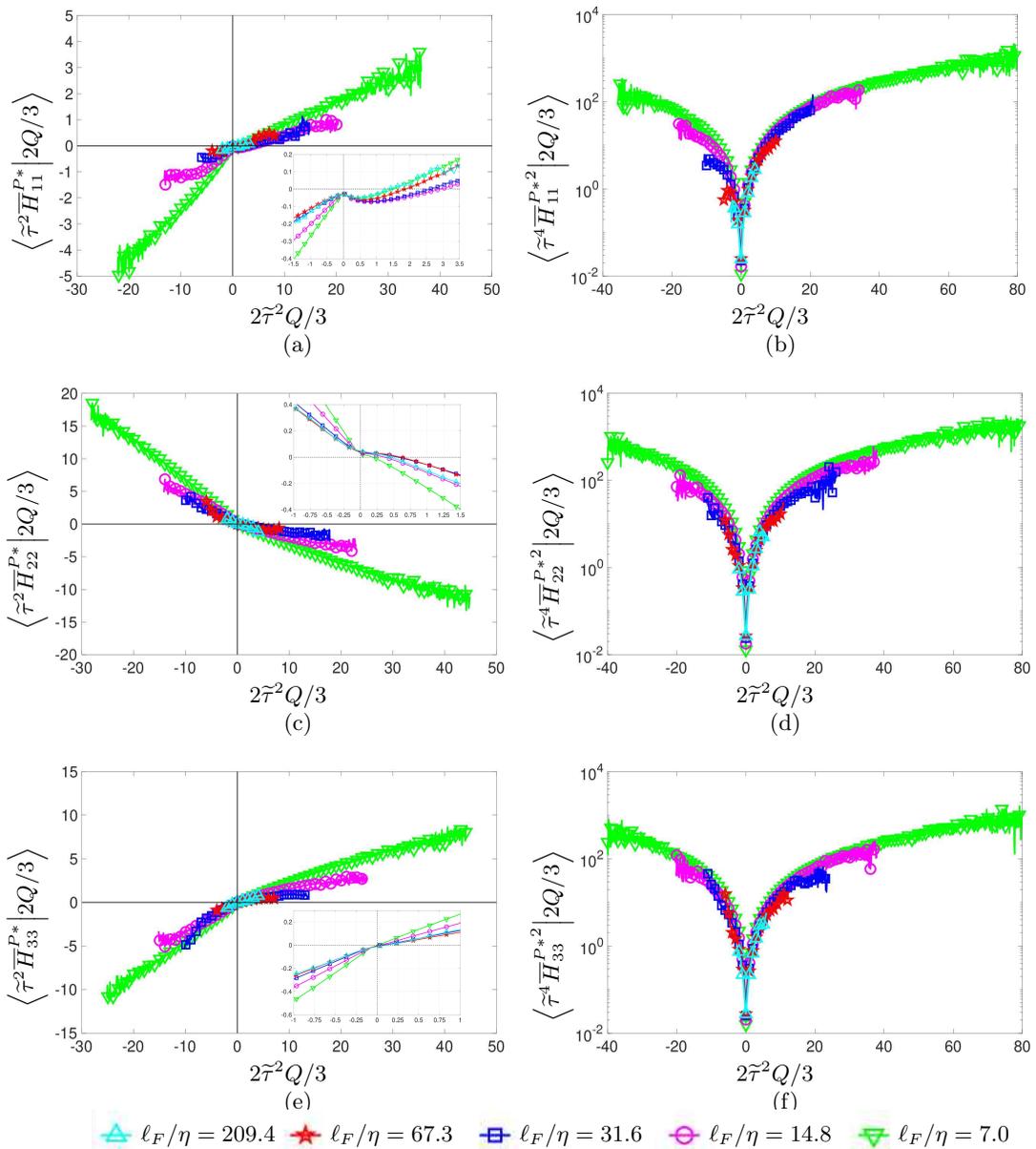
\begin{figure}
\centering
\vspace{0mm}
    \subfloat[]	
	{\begin{overpic}
	[trim = 0mm -32mm -40mm -3mm,
	scale=0.125,clip,tics=20]{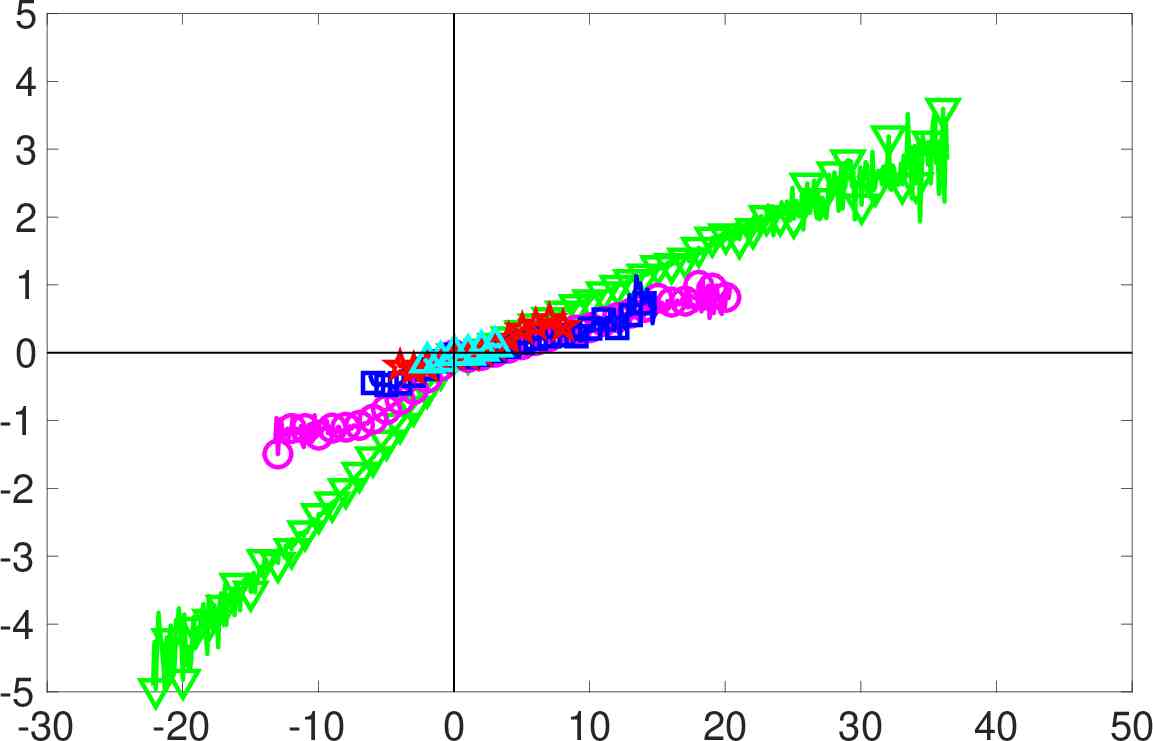}
	\put(40,0){$2\widetilde{\tau}^2 Q/3$}
	\put(-10,15){\rotatebox{90}{$\condavg{\widetilde{\tau}^2  \anisotr{H}_{11}^{P\eig}}{2Q/3}$}}
    \put(45,12){
	    \begin{overpic}[trim = 0mm 0mm 0mm 0mm, scale=0.05,clip,tics=20]
		{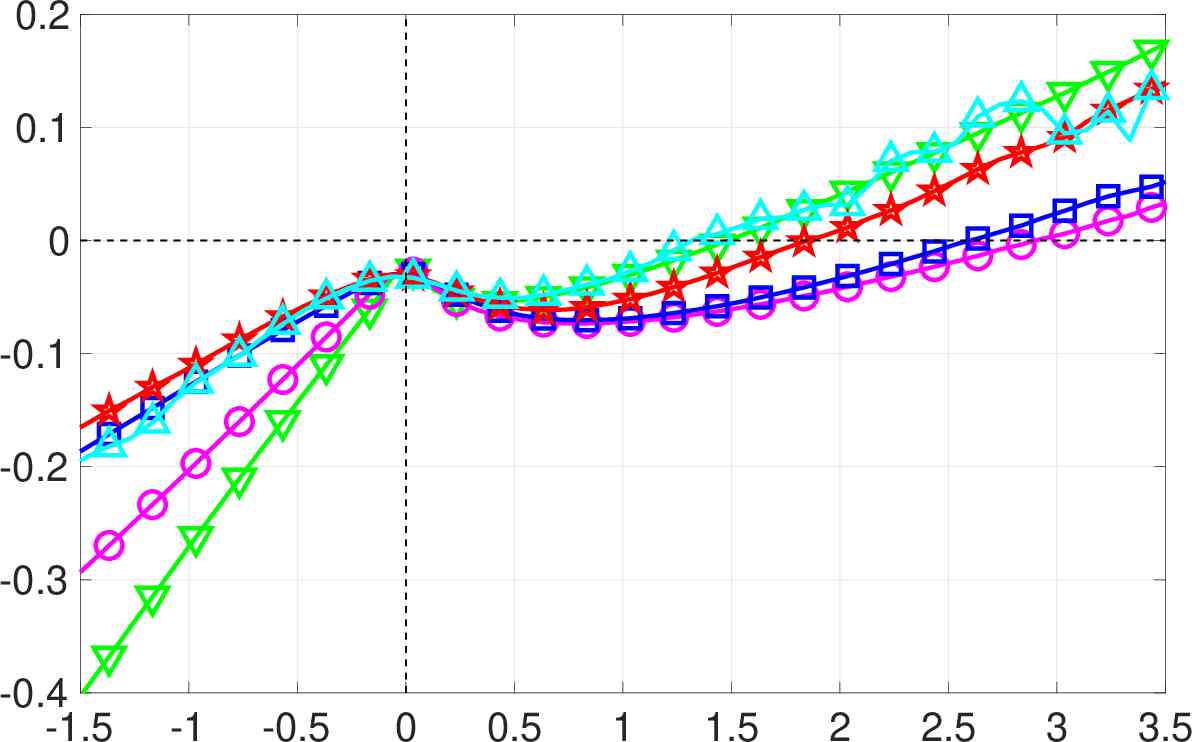}
		\end{overpic}
    }
	\end{overpic}}
	\subfloat[]
	{\begin{overpic}
	[trim = -40mm -32mm 0mm -3mm,
	scale=0.125,clip,tics=20]{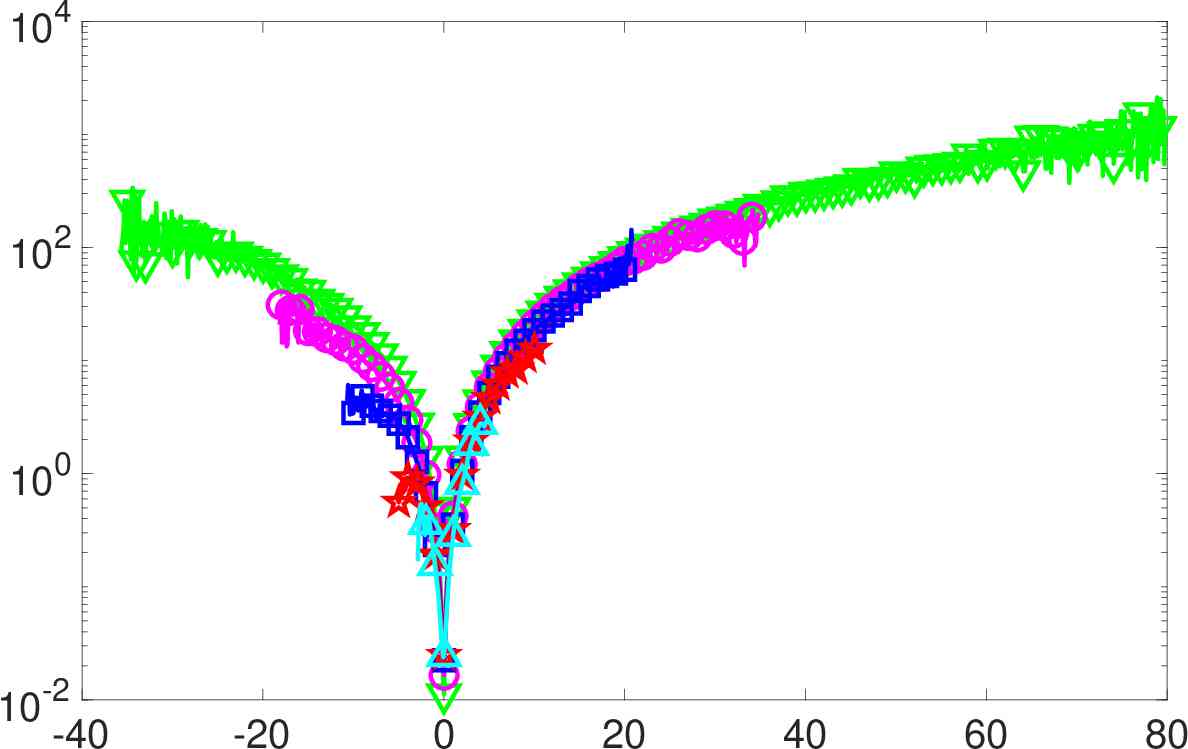}
	\put(40,0){$2\widetilde{\tau}^2 Q/3$}
    \put(-2,12){\rotatebox{90}{$\condavg{\widetilde{\tau}^4  {\anisotr{H}_{11}^{P\eig}}^2}{2Q/3}$}}
	\end{overpic}}\\
   \subfloat[]	
	{\begin{overpic}
	[trim = 0mm -32mm -40mm -3mm,
	scale=0.125,clip,tics=20]{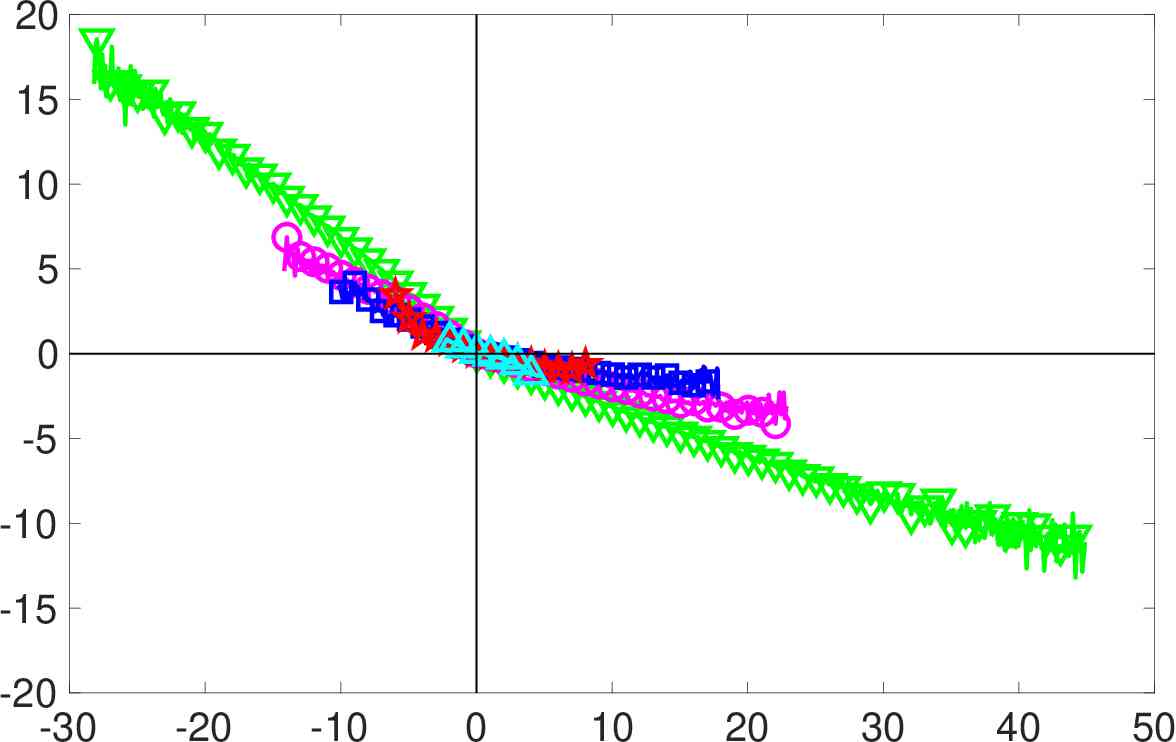}
	\put(40,0){$2\widetilde{\tau}^2 Q/3$}
	\put(-10,15){\rotatebox{90}{$\condavg{\widetilde{\tau}^2  \anisotr{H}_{22}^{P\eig}}{2Q/3}$}}
    \put(45,38){
	    \begin{overpic}[trim = 0mm 0mm 0mm 0mm, scale=0.05,clip,tics=20]
		{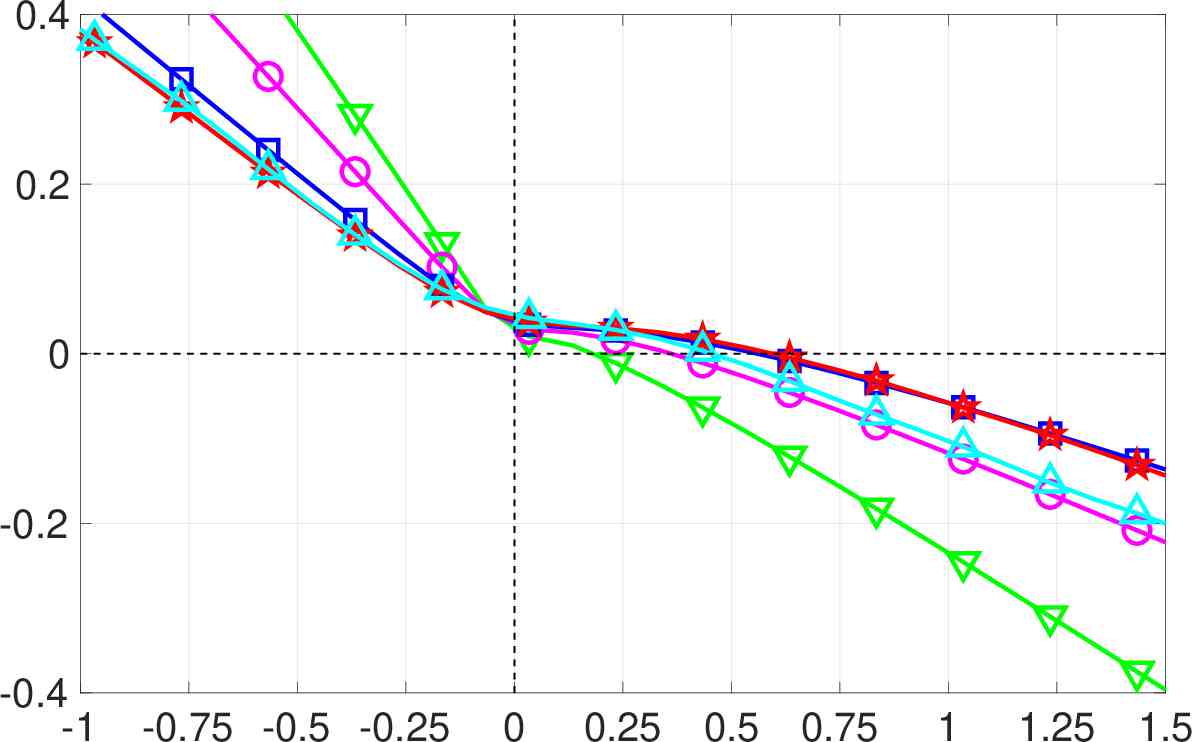}
		\end{overpic}
    }
	\end{overpic}}
	\subfloat[]
	{\begin{overpic}
	[trim = -40mm -32mm 0mm -3mm,
	scale=0.125,clip,tics=20]{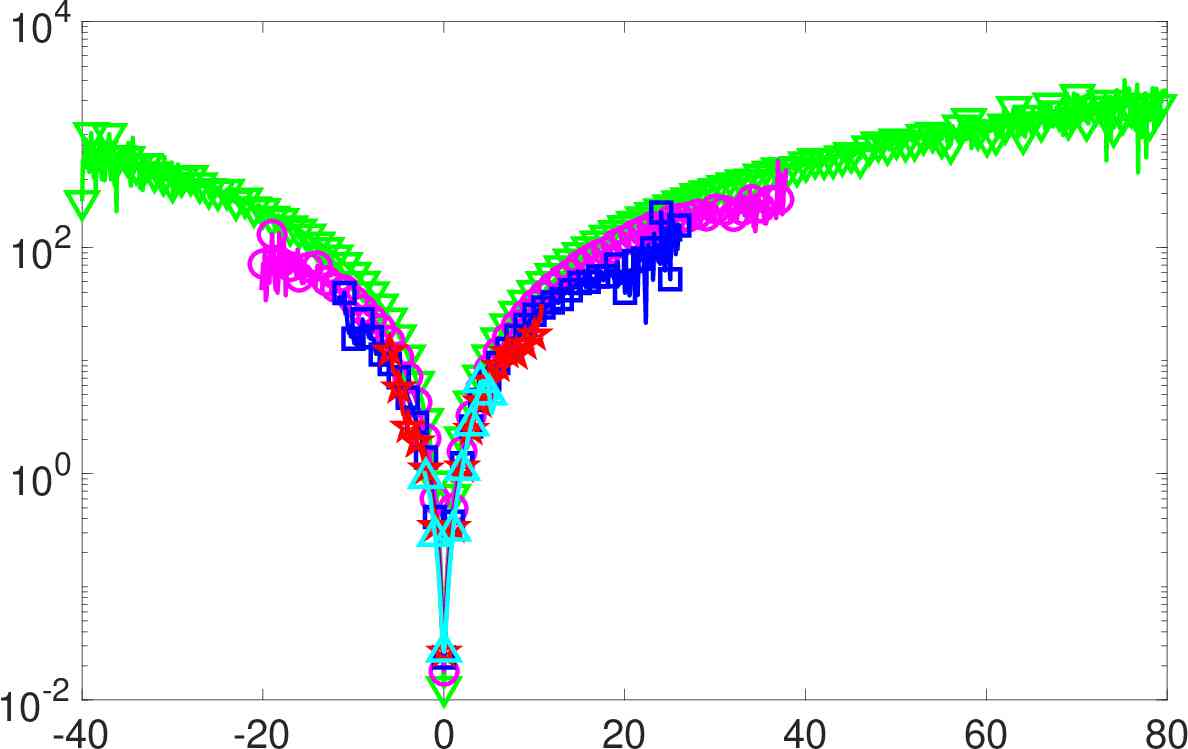}
	\put(40,0){$2\widetilde{\tau}^2 Q/3$}
    \put(-2,12){\rotatebox{90}{$\condavg{\widetilde{\tau}^4  {\anisotr{H}_{22}^{P\eig}}^2}{2Q/3}$}}
	\end{overpic}}\\
   \subfloat[]
	{\begin{overpic}
	[trim = 0mm -32mm -40mm -3mm,
	scale=0.125,clip,tics=20]{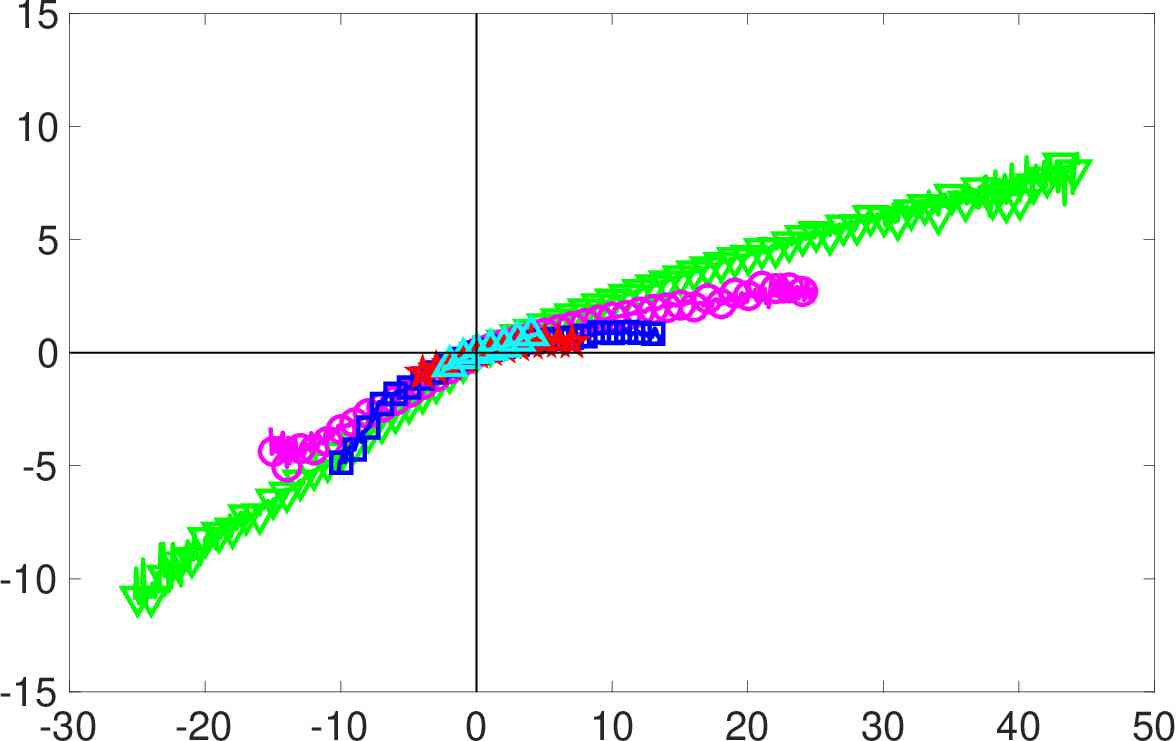}
	\put(40,0){$2\widetilde{\tau}^2 Q/3$}
	\put(-10,15){\rotatebox{90}{$\condavg{\widetilde{\tau}^2  \anisotr{H}_{33}^{P\eig}}{2Q/3}$}}
    \put(45,12){
	    \begin{overpic}[trim = 0mm 0mm 0mm 0mm, scale=0.05,clip,tics=20]
		{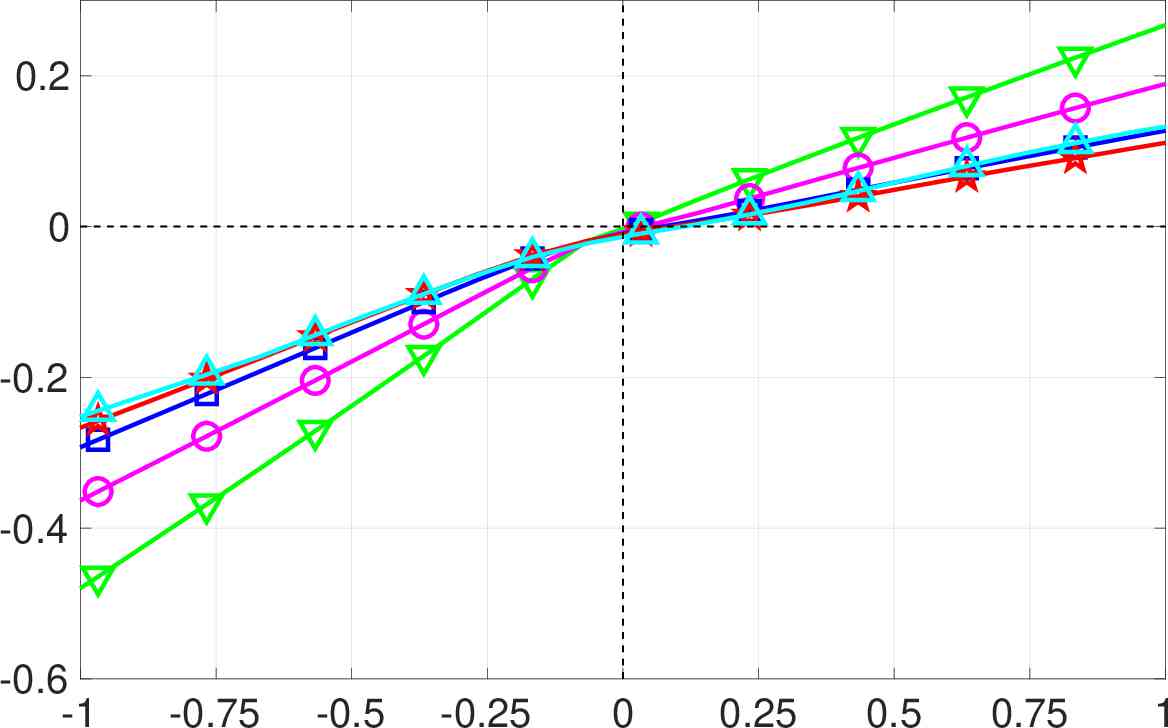}
		\end{overpic}
    }
	\end{overpic}}
	\subfloat[]
	{\begin{overpic}
	[trim = -40mm -32mm 0mm -3mm,
	scale=0.125,clip,tics=20]{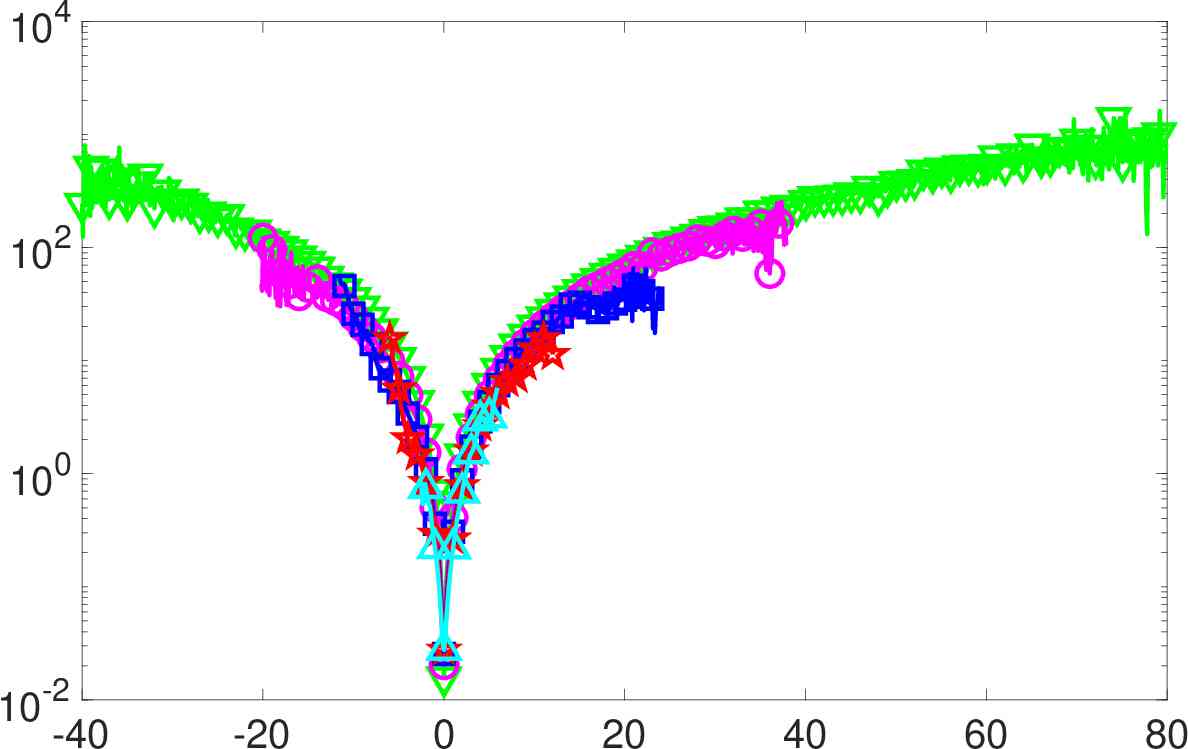}
	\put(40,0){$2\widetilde{\tau}^2 Q/3$}
    \put(-2,12){\rotatebox{90}{$\condavg{\widetilde{\tau}^4  {\anisotr{H}_{33}^{P\eig}}^2}{2Q/3}$}}
	\end{overpic}}\\
	{\begin{overpic}
	[trim = 35mm 275mm 150mm 20mm,
	scale=0.40,clip,tics=20]{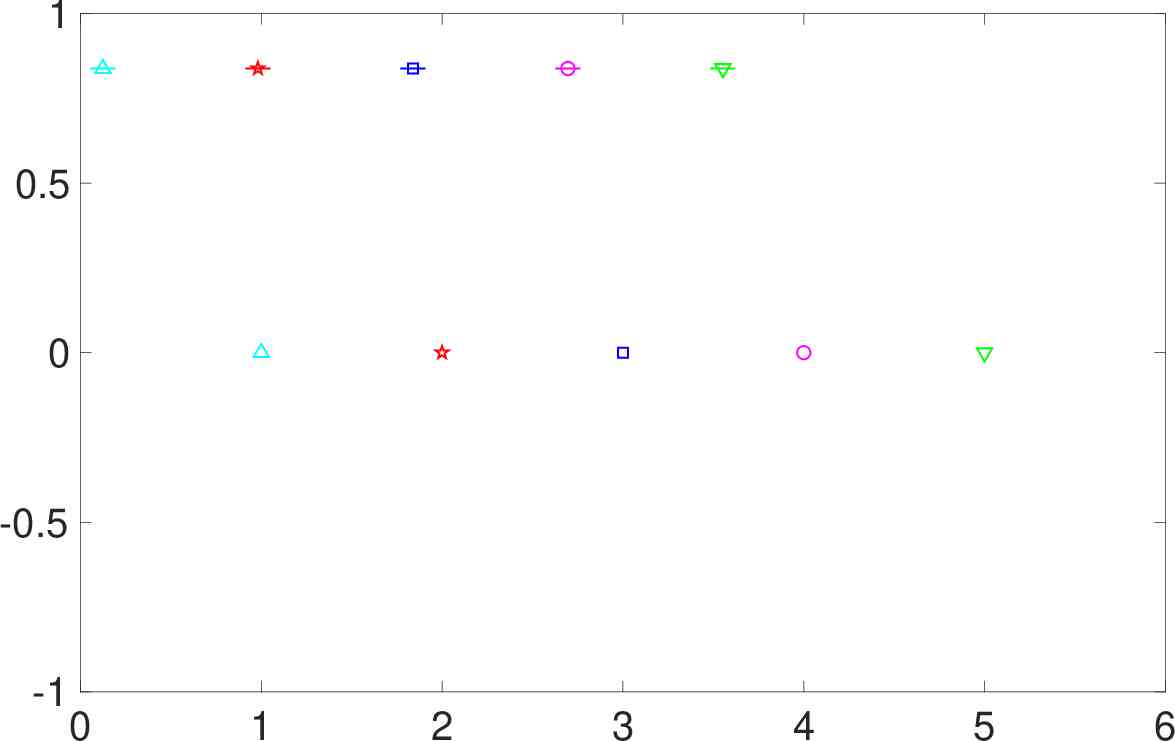}
		\put(5.5,2.5){$\ell_F/\eta = 209.4$}
	    \put(26.5,2.5){$\ell_F/\eta = 67.3$}
	    \put(47.5,2.5){$\ell_F/\eta = 31.6$}
	    \put(68.5,2.5){$\ell_F/\eta = 14.8$}
	    \put(89.5,2.5){$\ell_F/\eta = 7.0$}
	    	\end{overpic}}
	\caption{(a-c-e) the average and (b-d-f) the second moment of the diagonal components of the anisotropic pressure Hessian conditioned on the local part of the filtered pressure Hessian (which is equal to $2/3$ times the second principal invariant of the velocity gradient $Q$), for various filtering lengths $\ell_F/\eta$. Insets in (a-c-e) highlight the results for small values of $2\widetilde{\tau}^2 Q/3$.}
\label{res_fig_avg_aH_cond_Q}
\end{figure}

%%%%%%%cond H on Q
In Figure \ref{res_fig_avg_aH_cond_Q} we show results for the 
%average and variance 
first and second moments of $\anisotr{H}_{i(i)}^{P\eig}$ conditioned on the local part of the pressure Hessian, $H_{kk}^{P\eig}/3 = 2Q/3$.
The results show that $\inlavg{\anisotr{H}_{i(i)}^{P\eig}|2Q/3}$ varies almost linearly with $Q$, but with a different gradient for $Q>0$ and $Q<0$. This provides some support for closure models that assume a linear relationship between the anisotropic pressure Hessian and $Q$ \citep{Chevillard2006,Chevillard2008,Wilczek2014}. However, whether such models correctly describe the change in slope around $Q=0$ should be considered in future work. The results indicate that on average $\anisotr{H}_{11}^{P\eig}$ aids the growth of $\lambda_1$ in strain dominated regions, but counteracts the growth of $\lambda_1$ in vorticity dominated regions. 
The average and the variance of the contribution from $\anisotr{H}_{11}^{P\eig}$ is small compared to the contributions from the other diagonal components, in contrast to what was observed earlier for the full component $H_{11}^{P\eig}$. This differing behavior is due to the contribution of the isotropic pressure Hessian that is proportional to $Q$.
The average contribution from $\anisotr{H}_{22}^{P\eig}$ becomes increasingly negative with increasing $Q$, and it therefore helps the growth of positive $\lambda_2$ in vorticity dominated regions. This opposes the local part of the pressure Hessian that acts to reduce positive $\lambda_2$ events in vorticity dominated regions. The third diagonal component $\anisotr{H}_{33}^{P\eig}$ counteracts the growth of $|\lambda_3|$ in strain-dominated regions, which is critical to stabilize the dynamics. This feature is absent in the RE model where $\anisotr{H}_{ij}^{P\eig}=0$, which is one reason that system blows-up. 

The results in figures \ref{res_fig_avg_aH_cond_Q}(b-d-f)
 for $\inlavg{{\anisotr{H}_{i(i)}^{P\eig}}^2|2Q/3}$, together with the insets of figures \ref{res_fig_avg_aH_cond_Q}(a-c-e), show that the effect of $\anisotr{H}_{i(i)}^{P\eig}$ on the eigenframe dynamics does not vanish when $Q\to 0$, which was also observed in \cite{Chevillard2008}, and is something that is not captured by closure models such as the tetrad model \citep{Chertkov1991}, the Lagrangian linear diffusion model \citep{Jeong2003} or the recent fluid deformation approximation \citep{Chevillard2006}. Moreover, the peculiar behaviour of the anisotropic pressure Hessian components observed in figure \ref{res_aH_cond_lam2omg2} for small $\widetilde{\tau}\lambda_i$ and $\widetilde{\tau}\omega_i$ translates into a non trivial and non monotonic behaviour of the pressure Hessian components at small $Q$, as shown in the insets of figure \ref{res_fig_avg_aH_cond_Q}.

\subsubsection{Preferential states of the pressure Hessian}\label{sec_shape}

\begin{figure}
\centering
\vspace{0mm}			
    \subfloat[]	
	{\begin{overpic}
	[trim = 0mm -25mm -40mm -1mm,
	scale=0.125,clip,tics=20]{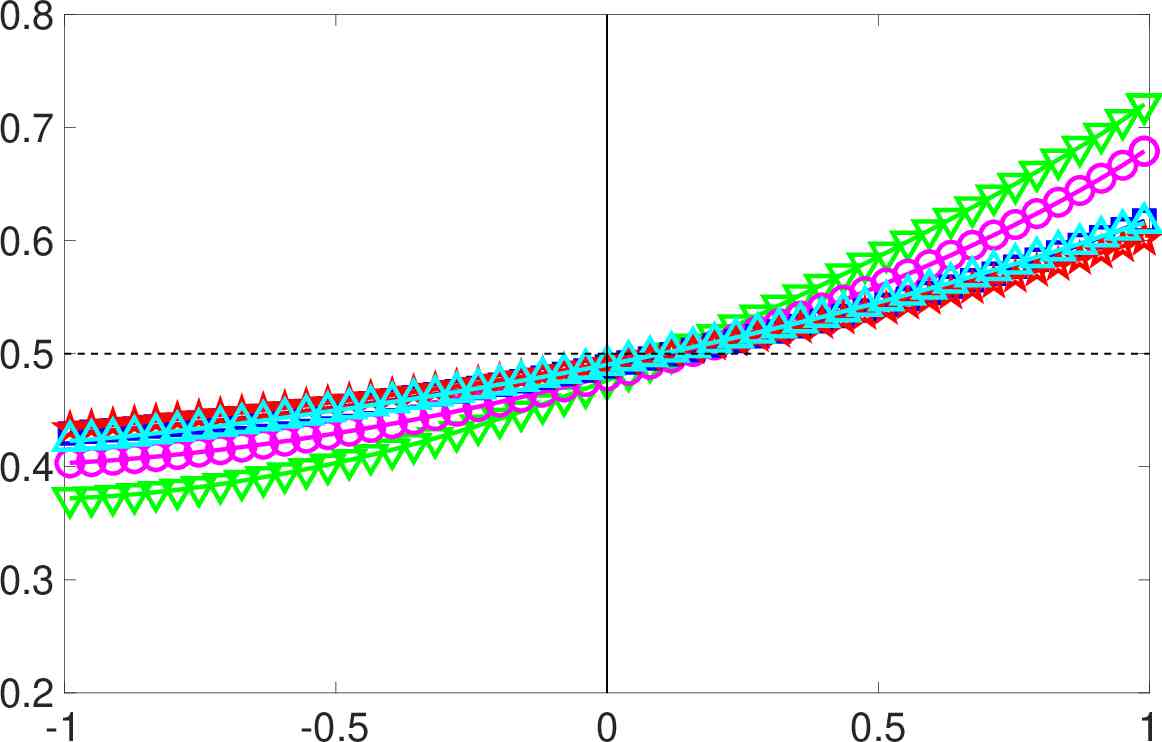}
	    \put(47,0){$s^*$}
        \put(-6,30){\rotatebox{90}{PDF}}
	\end{overpic}}
	\subfloat[]
	{\begin{overpic}
	[trim = -40mm -25mm 0mm -1mm,
	scale=0.125,clip,tics=20]{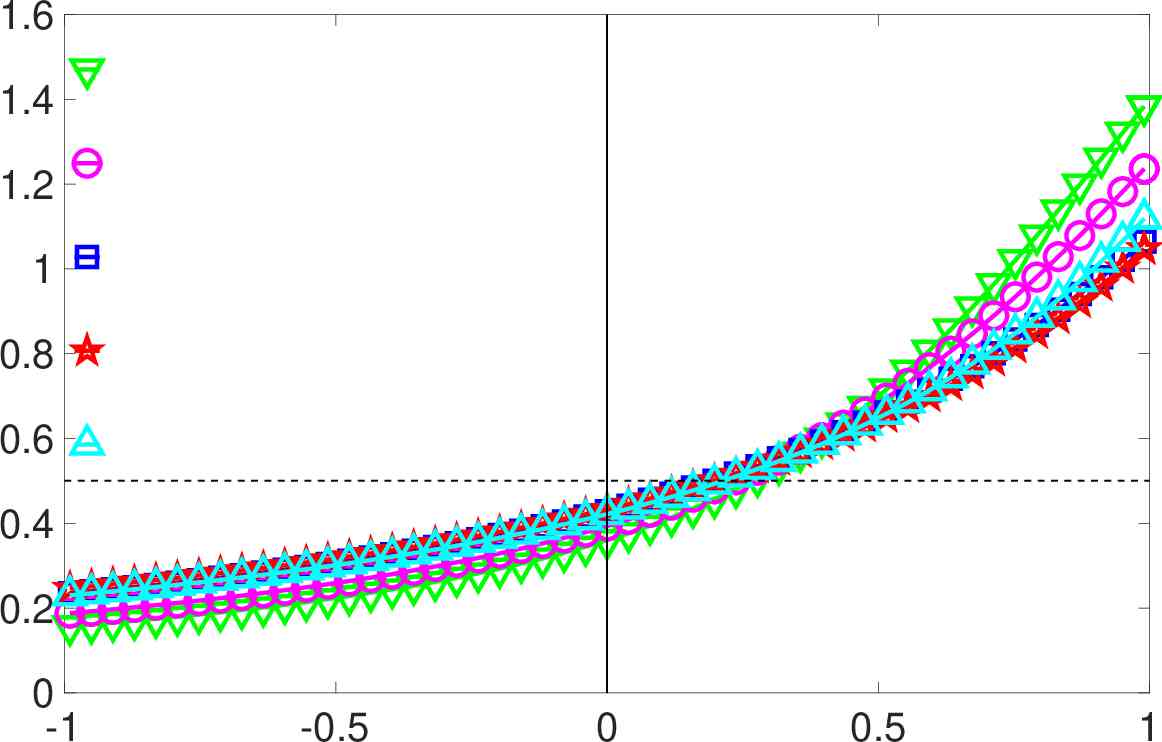}
	    \put(54,0){$s^*$}
        \put(1,30){\rotatebox{90}{PDF}}
        \put(17,56.75){\small{$\ell_F / \eta = 7.0$}} 
        \put(17,49.50){\small{$\ell_F / \eta = 14.8$}} 
        \put(17,42.25){\small{$\ell_F / \eta = 31.6$}}  
        \put(17,35){\small{$\ell_F / \eta = 67.3$}} 
        \put(17,27.75){\small{$\ell_F / \eta = 209.4$}} 
	\end{overpic}} 
	\caption{%\textcolor{red}{EditedFig}
	PDF of $s^*$ for (a) the anisotropic pressure Hessian, and (b) the strain-rate. Dashed lines indicate the probability distribution for a random uncorrelated field.}
    \label{res_fig_s_star}
\end{figure}
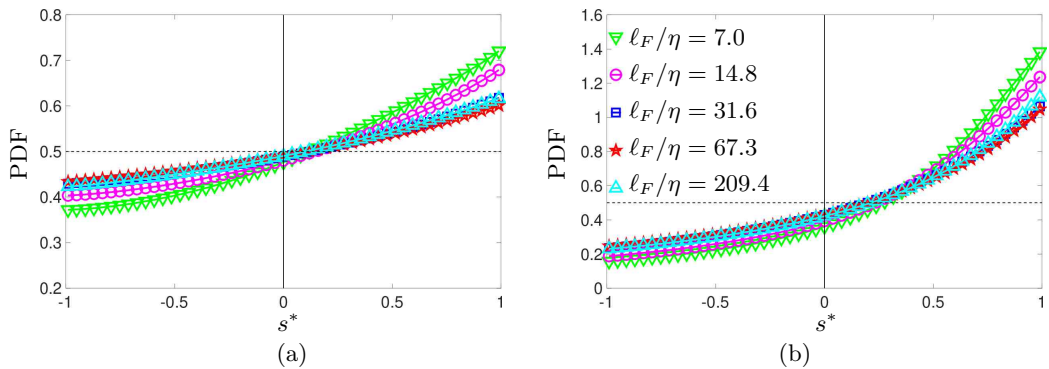

We now turn to characterize the state of the anisotropic pressure Hessian $\bm{\anisotr{H}}^P$ by means of its eigenvalues $\anisotr{\phi}_i$, which are associated with its eigenvectors $\bm{w}_i$. The eigenvalues of the full pressure hessian are denoted by $\phi_i$. The shape of the anisotropic part of the pressure Hessian can be quantified by means of the dimensionless quantity
\begin{equation}
s^* = -\sqrt{6}\frac{\Tr[(\bm{\anisotr{H}^P})^3]}
{\left(\Tr[(\bm{\anisotr{H}^P})^2]\right)^{3/2}} = 
-\frac{3\sqrt{6} \quad  \anisotr{\phi}_1\anisotr{\phi}_2\anisotr{\phi}_3}
{\left(\anisotr{\phi}_1^2+\anisotr{\phi}_2^2+\anisotr{\phi}_3^2\right)^{3/2}}.
\end{equation}
The variable $s^*$ has been proposed in \cite{Lund1994}, and since only $\anisotr{\phi}_2$ is not sign-definite, the sign of $s^*$ is determined by the sign of $\anisotr{\phi}_2$, and the PDFs of $s^*$ can be used to quantify the probability of the tensor being found in axisymmetric states. The PDF of $s^*$ is shown in Figure \ref{res_fig_s_star} (a), and the results show that $s^*$ is preferentially positive, indicating that $\anisotr{\phi}_2$ is also preferentially positive. As $\ell_F$ is increased, the PDF tends to the constant value of $1/2$, corresponding to a uniform random variable \citep{Lund1994}. The preference for positive values of $s^*$ indicates that the anisotropic pressure Hessian exhibits a preference to stretch fluid elements along the direction $\bm{w}_3$ (since $-\bm{H}^P$ appears in equation \eqref{theory_eq_grad}), and to compress them in the plane spanned by $\bm{w}_1$ and $\bm{w}_2$, that is orthogonal to $\bm{w}_3$. In Figure \ref{res_fig_s_star}(b) we show the corresponding results for the strain-rate tensor, for which $s^* = -\sqrt{6}\Tr\left[\bm{S}^3\right]/
\left(\Tr\left[\bm{S}^2\right]\right)^{3/2}$. The results for this quantity in figure \ref{res_fig_s_star}(b) show that the preference for $\bm{S}$ to be in a state of bi-axial extension is much greater than that for $\bm{\anisotr{H}}^P$, though in both cases, this preference for bi-axial extension generally becomes weaker as $\ell_F$ is increased.

In order to gain further insight into the state of the full pressure Hessian $\bm{{H}}^P$, we aim to characterize the full space of the independent dimensionless quantities that can be formed using the invariants of $\bm{{H}}^P$. Since $\bm{{H}}^P$ is symmetric with three real, independent eigenvalues, two dimensionless quantities can be defined. In the context of the Reynolds stresses, the Lumley triangle provides an insightful way to characterize its anisotropic properties \citep{Lumley1979,pope}. However, this cannot be applied to $\bm{{H}}^P$ since it is not positive definite. As an alternative, we seek to construct an invariant triangle, analogous to the Lumley triangle, by employing the invariants of the normalized quantity
\begin{equation}
b_{ij} = \frac{H_{ij}^P-H_{kk}^P\delta_{ij}/3}{\sqrt{H^P_{mn}H^P_{mn}}},
\end{equation}
which simply corresponds to the normalized components of the anisotropic pressure Hessian.
The first two invariants of $\bm{b}$ are $\Tr(\bm{b})=0$ and
\begin{equation}
b^2 \equiv \Tr(\bm{b}^2) = 1-\left(\frac{H_{kk}^P}{\sqrt{3H_{ij}^PH_{ij}^P}}\right)^2.
\label{res_eq_b2}
\end{equation}
Therefore $b^2$ is bounded, $b^2\in[0,1]$. Moreover, the second invariant of $\bm{b}$ is related to the quantity $D^*$ through $b^2 = 1-(D^*)^2$, where
\begin{equation}
D^* = -\frac{\Tr[(\bm{H}^P)]}{\sqrt{3\Tr[(\bm{H}^P)]^2)}} = 
-\frac{\phi_1+\phi_2+\phi_3}{
\sqrt{3}\left( \phi_1^2+\phi_2^2+\phi_3^2\right)^{1/2} },
\end{equation}
which was proposed in \cite{Lund1994} to quantify the relative magnitude of the isotropic dilatation/compression of a tensor.
The constraint on the third invariant is obtained through the discriminant of the characteristic equation of $\bm{b}$
\begin{equation}
\Delta = \frac{1}{2}(\Tr(\bm{b}^2))^3 - 3(\Tr(\bm{b}^3))^2 \ge 0,
\end{equation}
since the eigenvalues of $\bm{b}$ are real and it follows that $\abs{\sqrt{6} \Tr(\bm{b}^3)} \le (\Tr(\bm{b}^2))^{3/2}$.
The zero discriminant case corresponds to
%\begin{equation}
$(\Tr(\bm{b}^2))^3 = 6(\Tr(\bm{b}^3))^2$
%\end{equation}
for which two eigenvalues of $\bm{b}$ coincide, implying that two eigenvalues of $\bm{H}^P$ coincide as well, since the eigenvalues of $\bm{H}^P$ are just shifted and scaled with respect to the eigenvalues of $\bm{b}$.
When $\bm{H}^P$ is traceless, i.e.\ the pressure Hessian is purely non-local, the eigenvalues of $\bm{b}$ and $\bm{H}^P$ are proportional. In that purely anisotropic case we have $\Tr(\bm{b}^2)=1$ and $\sqrt{6}\Tr(\bm{b}^3)=-s^*$.
These relations suggest to employ the following as coordinates on the invariant triangle
\begin{align}
\zeta = -\sqrt{6}\Tr(\bm{b}^3), && \chi =\left(\Tr(\bm{b}^2)\right)^{3/2}.
\label{res_def_T_coords}
\end{align}
With the coordinates defined in equation \eqref{res_def_T_coords}, a triangle is obtained that has straight sides (the original triangle proposed by Lumley had two curved sides), since $0\le \chi\le 1$ and $|\zeta|\le \chi$.

\begin{figure}
\centering
\vspace{0mm}			
    \subfloat[]	
	{\begin{overpic}
	[trim = 0mm -25mm -30mm -1mm,
	scale=0.125,clip,tics=20]{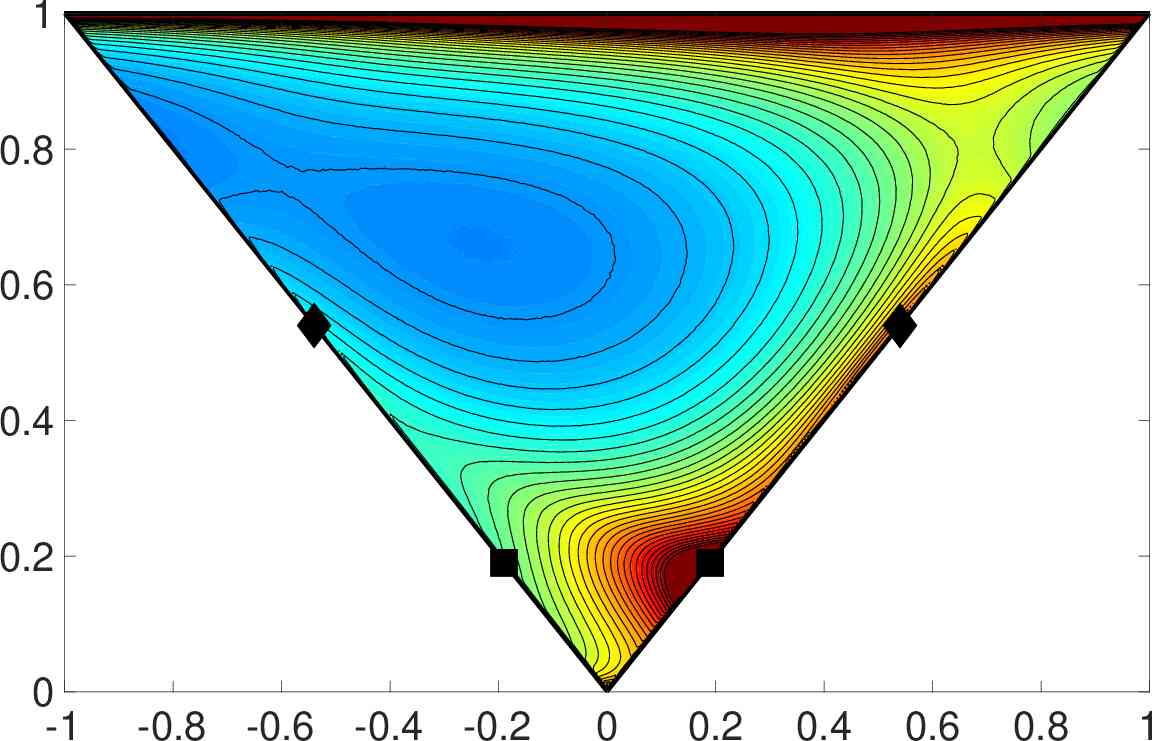}
	\put(-6,35){\rotatebox{90}{$\chi$}}
    \put(48,0){$\zeta$}
	\end{overpic}}
	\subfloat[]
	{\begin{overpic}
	[trim = -30mm -25mm 0mm -1mm,
	scale=0.125,clip,tics=20]{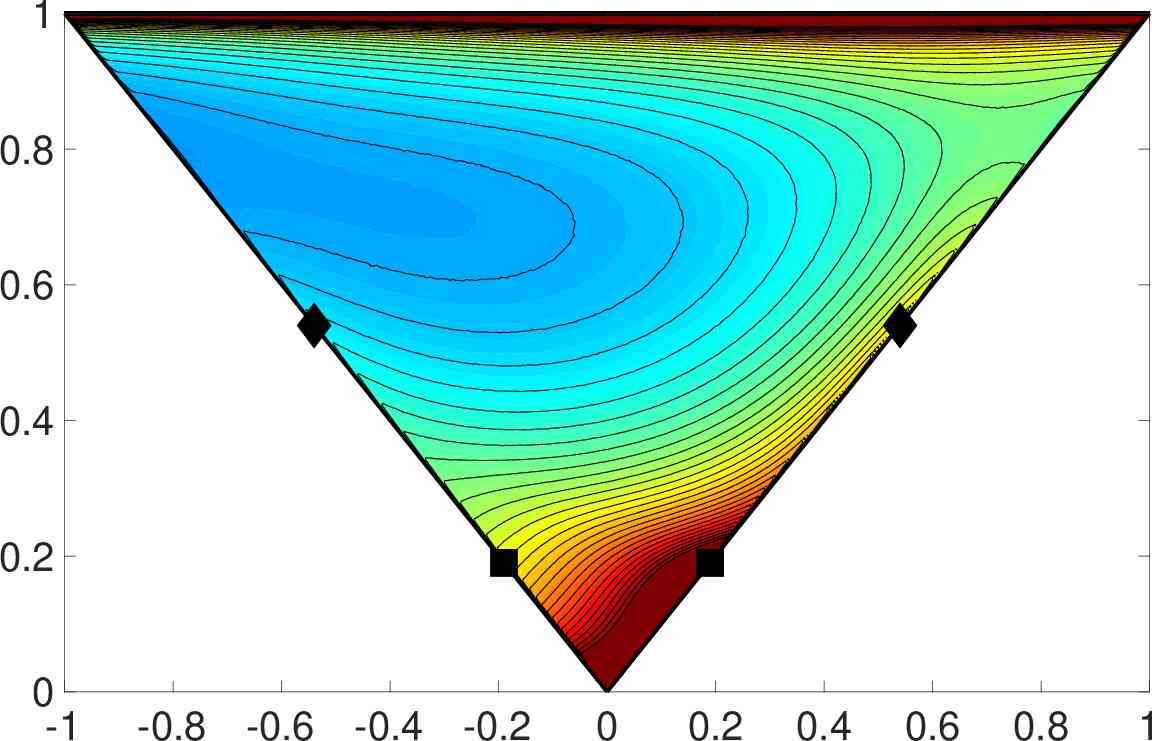}
	\put(0,35){\rotatebox{90}{$\chi$}}
    \put(54,0){$\zeta$}
	\end{overpic}}\\
   \subfloat[]	
	{\begin{overpic}
	[trim = 0mm -25mm -30mm -1mm,
	scale=0.125,clip,tics=20]{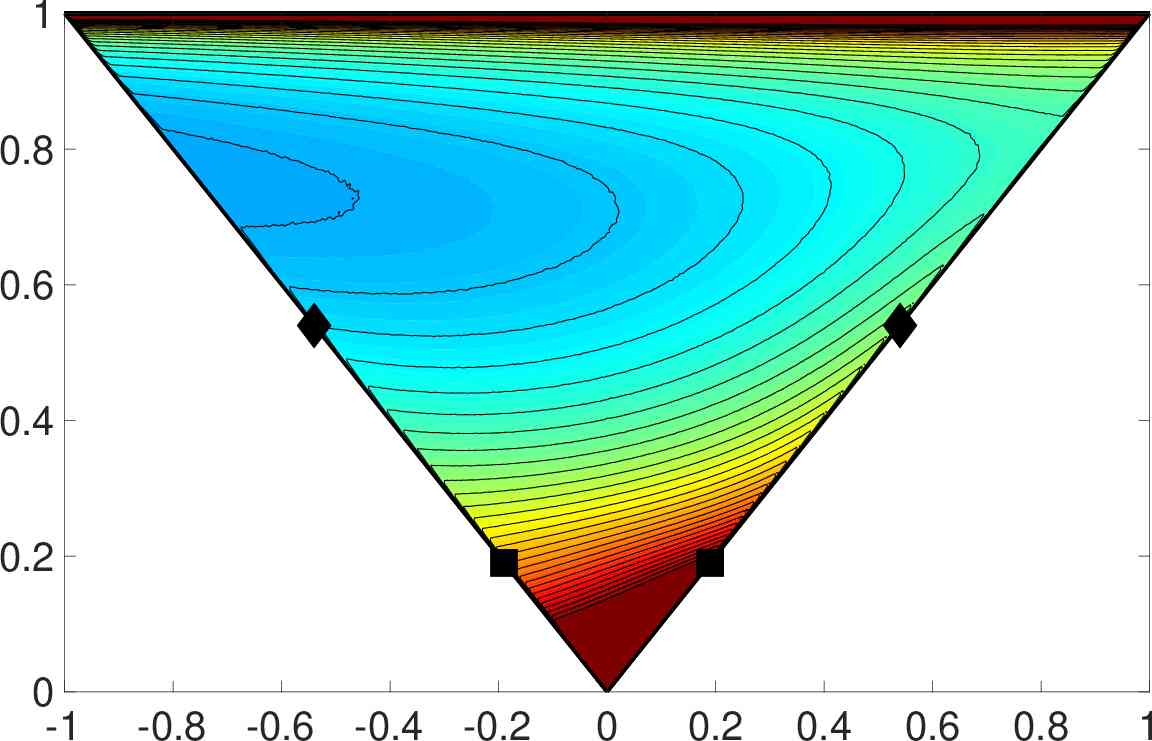}
	\put(-6,35){\rotatebox{90}{$\chi$}}
    \put(48,0){$\zeta$}
	\end{overpic}}
	\subfloat[]
	{\begin{overpic}
	[trim = -30mm -25mm 0mm -1mm,
	scale=0.125,clip,tics=20]{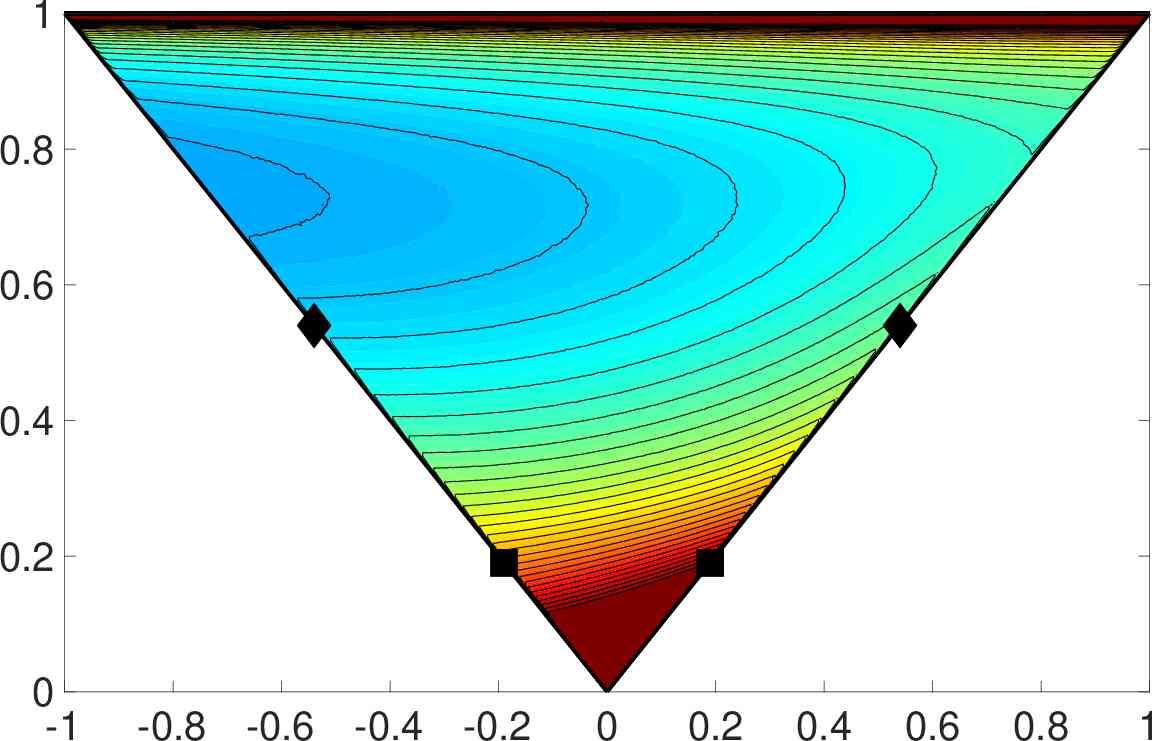}
	\put(0,35){\rotatebox{90}{$\chi$}}
    \put(54,0){$\zeta$}
	\end{overpic}}\\
	{\begin{overpic}
	[trim = 22mm 0mm 0mm 282mm,
	scale=0.25,clip,tics=20]{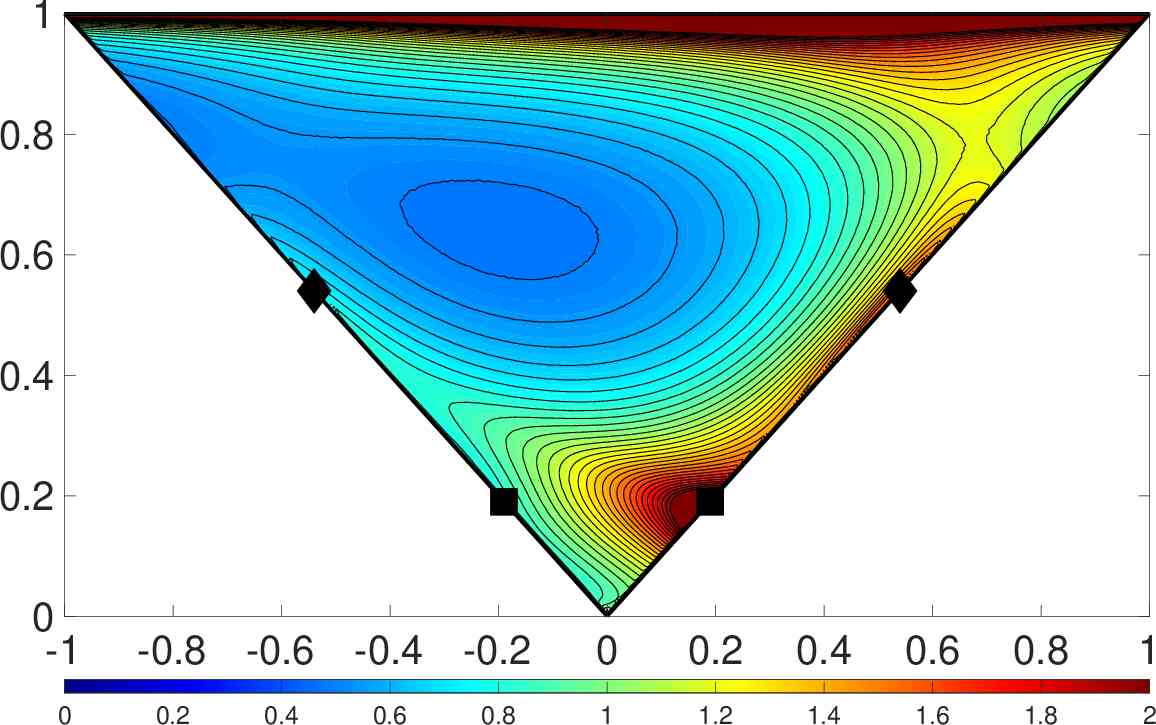}
	\end{overpic}}
	\caption{Joint PDF of the dimensionless invariants $\zeta$ and $\chi$ of the pressure Hessian (defined in \eqref{res_def_T_coords}) at filtering lengths (a) $\ell_F/\eta = 7.0$, (b) $\ell_F/\eta = 14.8$ (c) $\ell_F/\eta = 31.6$ and (d) $\ell_F/\eta = 67.3$. Square markers correspond to two-component axisymmetric configurations, and diamond markers correspond to one component configurations. The colours correspond to the values of the PDF, and the lines are isocontours of PDF shown in increments of 0.05 between 0 and 2.}
\label{res_fig_triangle}
\end{figure}
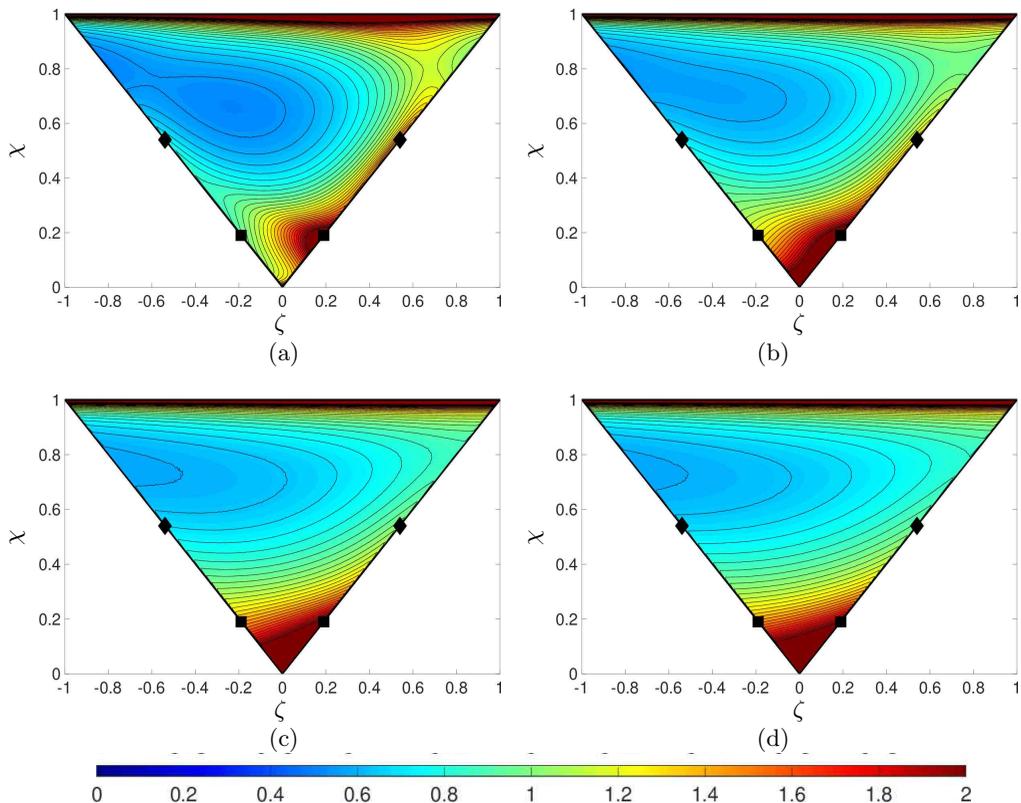

The following states can be observed on the triangle.
The $|\zeta| = \chi$ sides correspond to axisymmetric states ($s^*=1$ on the right and $s^*=-1$ on the left), and the side $\chi=1$ corresponds to purely anisotropic (traceless) state. The point $(\zeta,\chi)=(0,0)$ corresponds to the isotropic state, while the points $(-1,1)$ and $(1,1)$ indicate purely anisotropic states with $s^*=-1$ and $s^*=+1$, respectively.
Moreover, the one component state, $\phi_1=\phi_2 = 0$ and $\phi_3=2Q$, results in $|D^*|=1/\sqrt{3}$, that is $\chi=|\zeta|\simeq 0.54$. The two component axisymmetric state, $\phi_3= 0$ and $\phi_2=\phi_1=Q$, results in $|D^*|=\sqrt{2/3}$, that is $\chi=|\zeta|\simeq 0.19$.

The degree of isotropy is quantified by $\chi$, anisotropy is maximum on the segment $\chi=1$ and decreases towards $\chi=0$ according to equation \eqref{res_eq_b2}. The state of the intermediate eigenvalue of $\bm{b}$, that is its distance from the other two eigenvalues, is measured by the deviation from $\zeta=0$, since $\zeta = s^*\chi$.
If the  tensor considered  (here $\bm{H}^P$) is always traceless, then the support of the proposed triangle reduces to a segment and the joint PDF of $\zeta$ and $\chi$ reduces to the PDF of $s^*$ (with $\chi=1$ fixed). Note that this triangle may be used to quantify the anisotropy of any symmetric second order tensor, and does not require positive definiteness of the tensor. It therefore represents a generalization of the Lumley triangle. 

In figure \ref{res_fig_triangle}, we show results for this invariant triangle of the pressure Hessian for various filtering lengths.
The results show that for the smallest filtering scales there is a high probability region near the two component axisymmetric configuration at $\chi=\zeta =0.19$, and another near the purely anisotropic state close to the edge $\chi=1$. The probability for the pressure Hessian to be in the purely isotropic configuration $\zeta=\chi=0$ is very low, and there is also a low-probability region around the center of the triangle, especially for $\zeta<0$, corresponding to $s^* <0$ and states of bi-axial stretching of the fluid element (since $-\bm{H}^P$ is in equations \eqref{theory_eq_grad}).
As the filtering length is increased, the constant probability lines tend to become parallel to the $\zeta$ axis, associated with the PDF of $s^*$ approaching a uniform distribution as $\ell_F$ is increased. Most interestingly, the probability of observing the purely isotropic state increases significantly as $\ell_F$ increases. Indeed, the peak of the PDF located near $\zeta=\chi\simeq 0.19$ for the smallest filtering scale, shifts towards $\zeta=\chi=0$ as $\ell_F$ is increased. In one sense then, this indicates that the importance of the anisotropic contribution to the pressure Hessian relative to the isotropic contribution reduces as the scale of the flow is increased. However, this is not the whole story since the results also indicate that at all scales there is a significant probability to be close to the purely anisotropic state near the edge $\chi=1$.

%%%VISCOUS
\subsection{Characterization of the viscous stress}

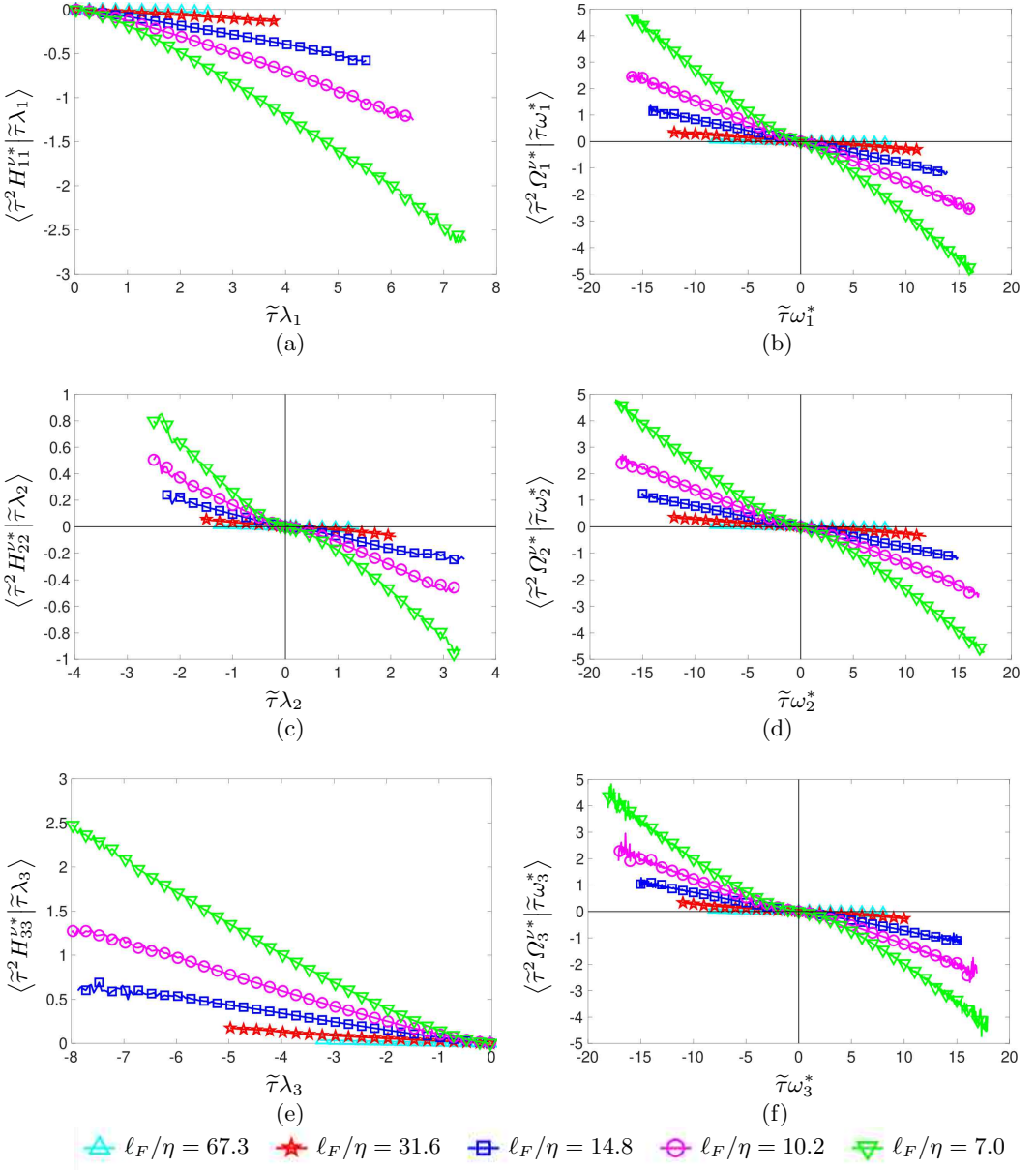
\begin{figure}
\centering
\vspace{0mm}			
    \subfloat[]	
	{\begin{overpic}
	[trim = 0mm -32mm -40mm -3mm,
	scale=0.125,clip,tics=20]{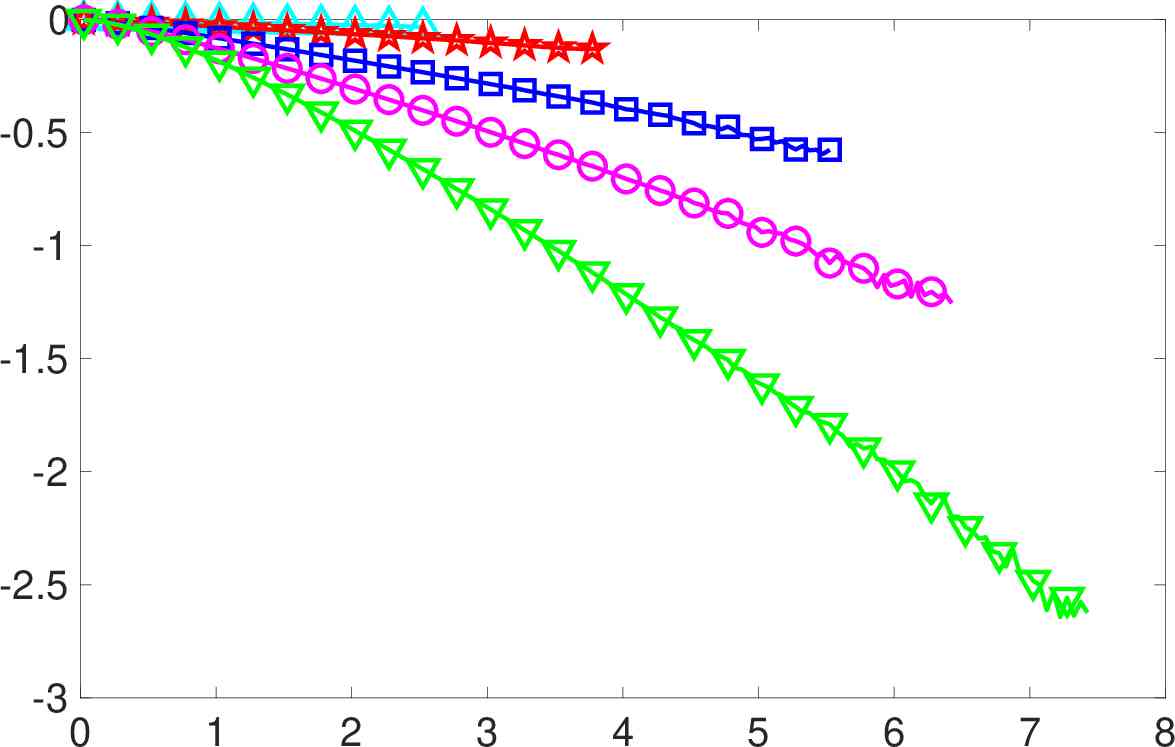}
	\put(45,0){$\widetilde{\tau} \lambda_1$}
	\put(-8,20){\rotatebox{90}{$\condavg{\widetilde{\tau}^2  H_{11}^{\nu\eig}}{\widetilde{\tau} \lambda_1}$}}
    %\put(-8,14){\rotatebox{90}{$\mathbb{E} \big[ \widetilde{\tau}^2  \anisotr{H}_{11}^{P\eig} \Big| 2\widetilde{\tau}^2 Q/3 \Big]$}}
	\end{overpic}}
	\subfloat[]
	{\begin{overpic}
	[trim = -40mm -32mm 0mm -3mm,
	scale=0.125,clip,tics=20]{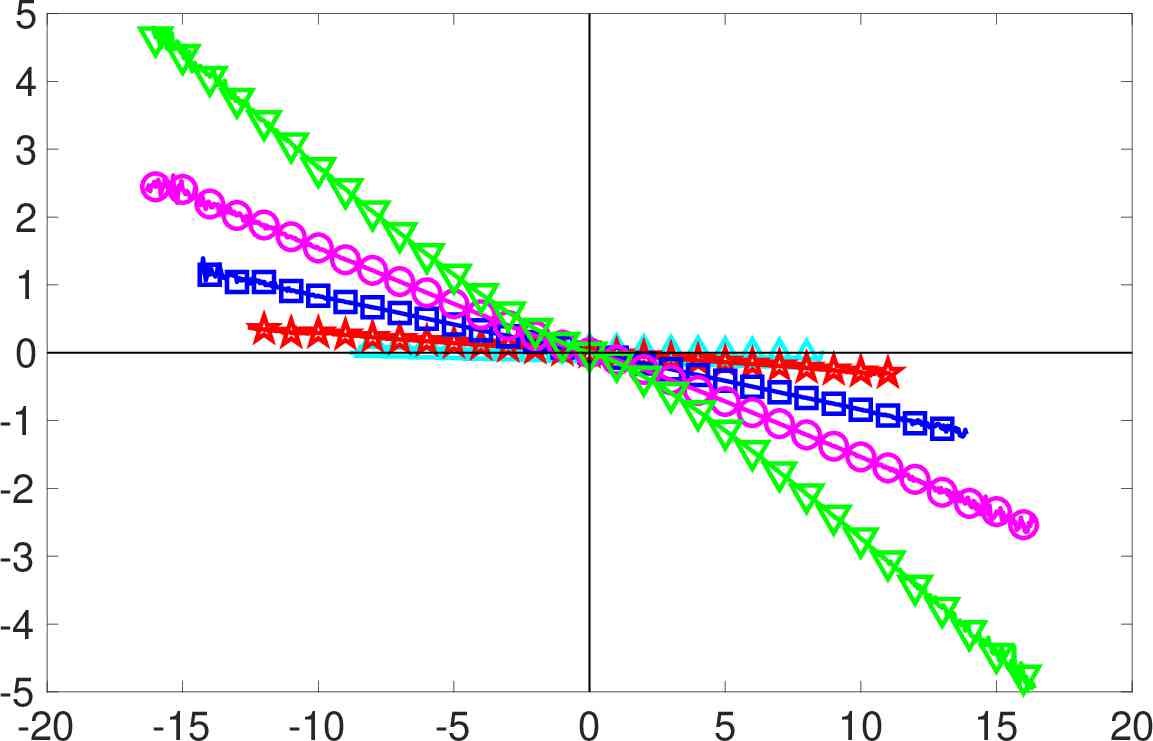}
	\put(50,0){$\widetilde{\tau}\omega_1^\eig $}
    \put(-2,20){\rotatebox{90}{$\condavg{\widetilde{\tau}^2  \Omega_1^{\nu\eig}}{\widetilde{\tau} \omega_1^\eig}$}}
	\end{overpic}}\\
   \subfloat[]	
	{\begin{overpic}
	[trim = 0mm -32mm -40mm -3mm,
	scale=0.125,clip,tics=20]{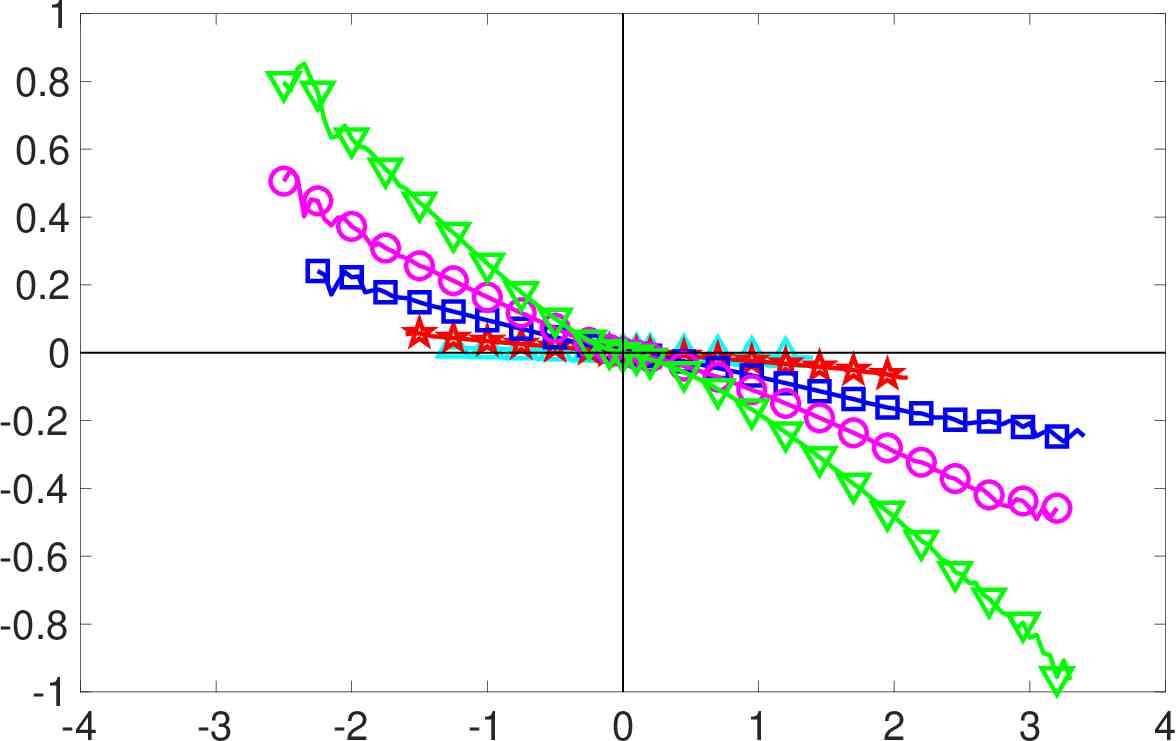}
	\put(45,0){$\widetilde{\tau} \lambda_2$}
	\put(-8,20){\rotatebox{90}{$\condavg{\widetilde{\tau}^2  H_{22}^{\nu\eig}}{\widetilde{\tau} \lambda_2}$}}
	\end{overpic}}
	\subfloat[]
	{\begin{overpic}
	[trim = -40mm -32mm 0mm -3mm,
	scale=0.125,clip,tics=20]{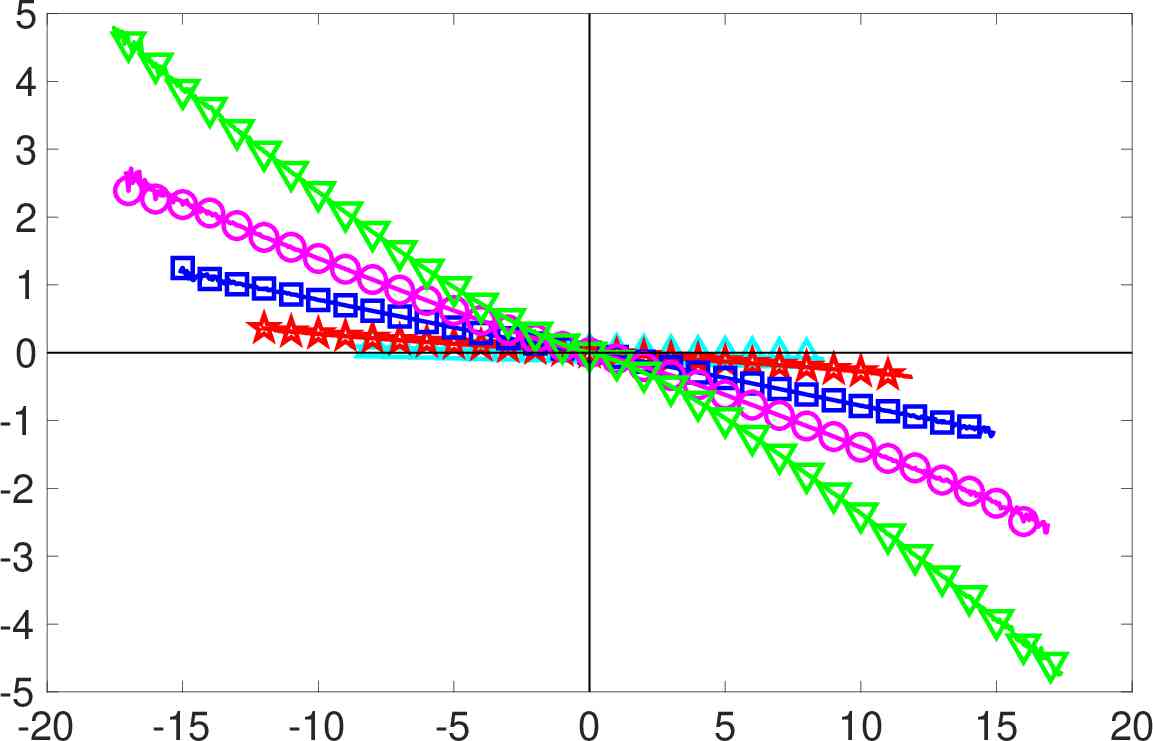}
	\put(50,0){$\widetilde{\tau}  \omega_2^\eig $}
    \put(-2,20){\rotatebox{90}{$\condavg{\widetilde{\tau}^2  \Omega_2^{\nu\eig}}{\widetilde{\tau} \omega_2^\eig}$}}
	\end{overpic}}\\
   \subfloat[]
	{\begin{overpic}
	[trim = 0mm -32mm -40mm -3mm,
	scale=0.125,clip,tics=20]{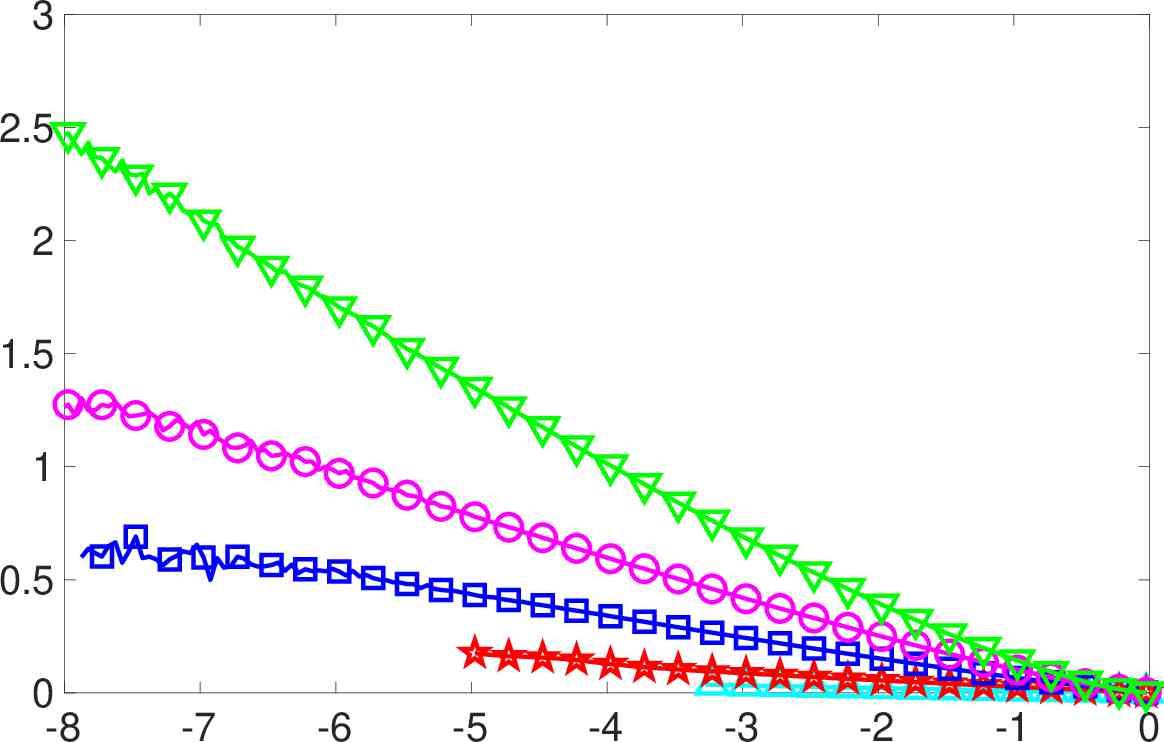}
	\put(45,0){$\widetilde{\tau} \lambda_3$}
	\put(-8,20){\rotatebox{90}{$\condavg{\widetilde{\tau}^2  H_{33}^{\nu\eig}}{ \widetilde{\tau} \lambda_3}$}}
	\end{overpic}}
	\subfloat[]
	{\begin{overpic}
	[trim = -40mm -32mm 0mm -3mm,
	scale=0.125,clip,tics=20]{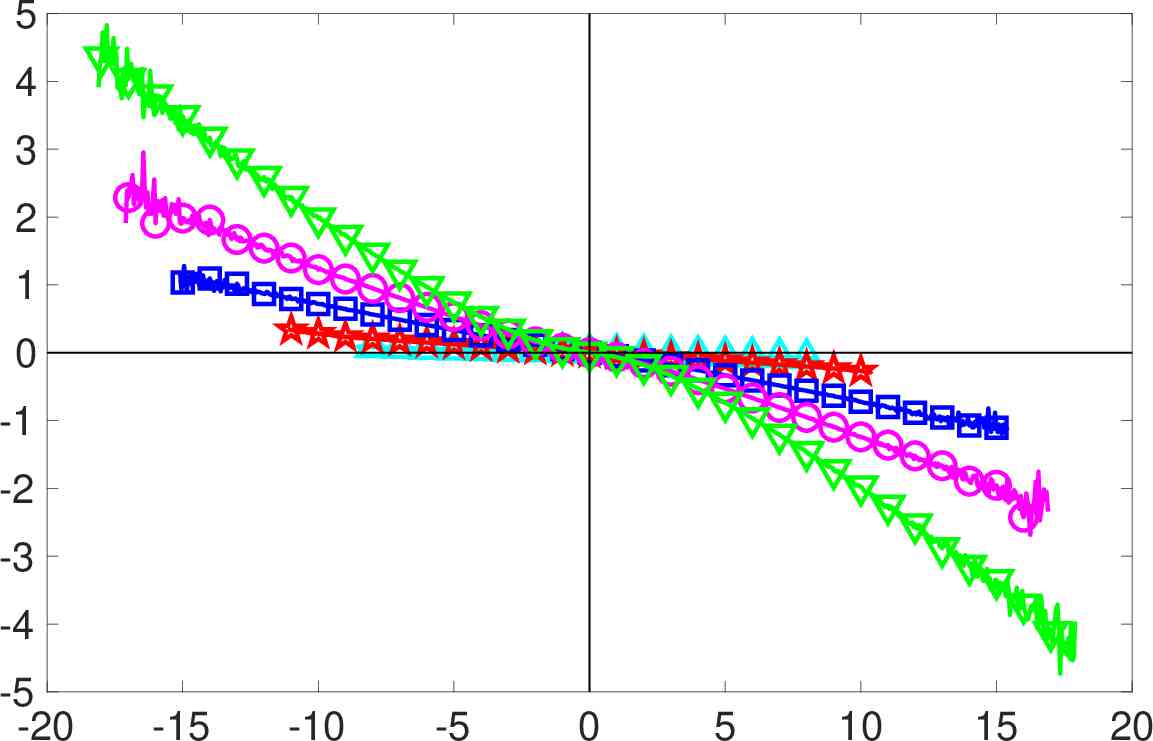}
	\put(50,0){$\widetilde{\tau}  \omega_3^\eig $}
    \put(-2,20){\rotatebox{90}{$\condavg{\widetilde{\tau}^2  \Omega_3^{\nu\eig}}{\widetilde{\tau} \omega_3^\eig}$}}
	\end{overpic}}\\
	{\begin{overpic}
	[trim = 35mm 275mm 150mm 22mm,
	scale=0.40,clip,tics=20]{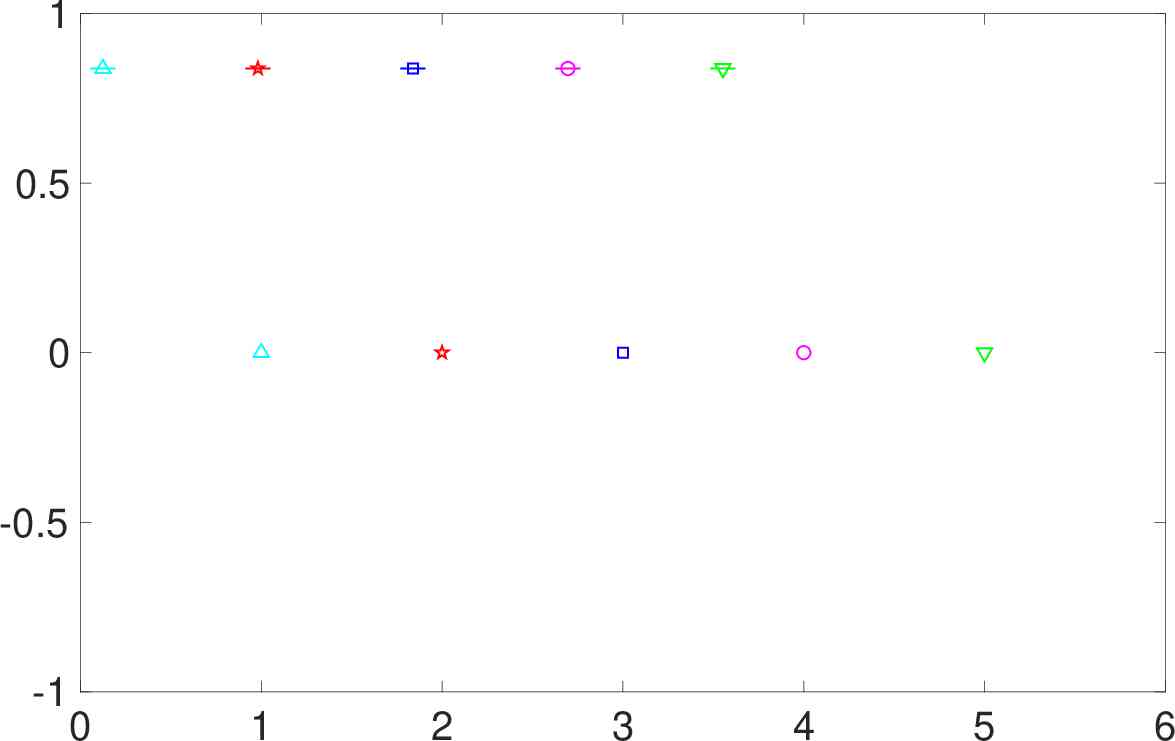}
		\put(5.5,2.5){$\ell_F/\eta = 67.3$}
	    \put(26.5,2.5){$\ell_F/\eta = 31.6$}
	    \put(47.5,2.5){$\ell_F/\eta = 14.8$}
	    \put(68.5,2.5){$\ell_F/\eta = 10.2$}
	    \put(89.5,2.5){$\ell_F/\eta = 7.0$}
	\end{overpic}}
\caption{(a-c-e) Average of the diagonal components of the filtered viscous stress in the eigenframe conditioned on the corresponding eigenvalues (see \eqref{theory_eq_strain_lambda}), for various filtering lengths $\ell_F/\eta$.
(b-d-f) Average of the anti-symetric part of the viscous stress components in the eigenframe conditioned on the corresponding vorticity component (see \eqref{theory_eq_vort_princ}).}
\label{res_fig_avg_lapA_cond_A}
\end{figure}

%%%strain
In figure \ref{res_fig_avg_lapA_cond_A}(a-c-e) we show results for $\inlavg{H_{i(i)}^{\nu\eig}|\lambda_i}$, the average of the diagonal components of the viscous strain-rate term in the eigenframe, conditioned on the corresponding eigenvalue. (Recall that in our notation, $H_{ij}^{\nu\eig}=\nu\bm{v}_i\bm{\cdot}(\nabla^2\bm{S})\bm{\cdot v}_{j}$ is the component of the Laplacian of the strain-rate tensor and not the Laplacian of the strain-rate eigenvalue). The results show that this quantity has the opposite sign to $\lambda_i$, for each $i$. This is in agreement with the results for $\inlavg{H_{i(i)}^{\nu\eig}}$ in figure \ref{res_avg_dlambda}, that revealed a damping effect of the viscous term on the eigenvalue evolution, and indicates a dependence of $\inlavg{H_{i(i)}^{\nu\eig}|\lambda_i}$ on odd powers of $\lambda_i$.
This is consistent with the idea that under time-reversal $t\to -t$, $\nabla^2\bm{A}\to- \nabla^2\bm{A}$ and $\bm{A}\to-\bm{A}$, and therefore a representation of $\nabla^2\bm{A}$ in terms of $\bm{A}$ should satisfy $\nabla^2\bm{A}(\bm{A},\dots) = -\nabla^2\bm{A}(-\bm{A},\dots)$. The results also indicate that $\inlavg{H_{i(i)}^{\nu\eig}|\lambda_i}$ depends non-linearly on $\lambda_i$, but approach a more linear dependence as $\ell_F$ is increased. This would seem to imply that the non-linear dependence at small $\ell_F$ is mainly due to intermittency of the velocity gradient.

\begin{figure}
\centering
\vspace{0mm}			
	{\begin{overpic}
	[trim = 0mm -25mm -70mm -3mm,
	scale=0.175,clip,tics=20]{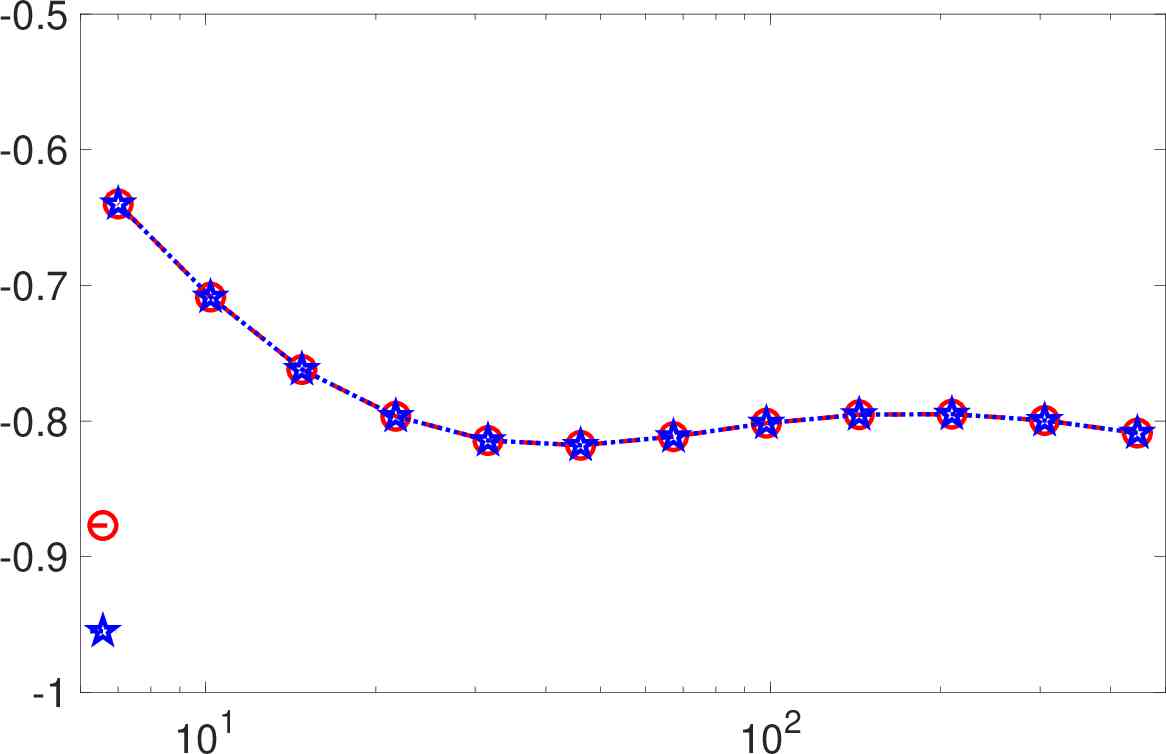}
\put(45,2){$\ell_F / \eta$}
\put(-6,13){\rotatebox{90}{Correlation coefficient, $\psi$}}
\put(10,20.5){$\psi( \bm{H}^{\nu}, \bm{S} )$}
\put(10,12.5){$\psi( \bm{\Omega}^{\nu}, \bm{\omega})$}
	\end{overpic}}
\caption{Correlation coefficient between the filtered strain-rate and the symmetric part of the viscous stress, and
between the filtered vorticity and the anti-symmetric part of the viscous stress. The correlation coefficient $\psi$ is shown as a function of the filtering length $\ell_F/\eta$.}
\label{res_fig_correl_lap_grad}
\end{figure}

In figure \ref{res_fig_avg_lapA_cond_A}(b-d-f) we show results for $\inlavg{\Omega_i^{\nu\eig}|\omega_i^\eig}$, the average of the Laplacian of the vorticity conditioned on the corresponding component of vorticity. Similar to the strain-rate case, we find that $\inlavg{\Omega_i^{\nu\eig}|\omega_i^\eig}$ has the opposite sign to $\omega_i^\eig$, with the curves showing a well defined trend in terms of the first few odd powers of the vorticity components 
\begin{equation}
\avg{\Omega_i^{\nu\eig}|\omega_i^\eig} \simeq a_1\omega_i^\eig + a_3\omega_i^{\eig 3}.
\label{res_exp_lapomg}
\end{equation}
This power law trend partially corroborates linear closures of the form $\nabla^2\bm{A}=-\bm{A}/T$ \citep{Chevillard2006}, where $T$ is a certain time scale, but also shows that higher order terms should be incorporated into the closure in order for it to accurately capture the behavior when $\widetilde{\tau}\omega_i^\eig$ is not small.
The first order term in equation \eqref{res_exp_lapomg} gives a non-zero slope at $\omega_i=0$, while the higher order terms cause departures from the linear trend. As $\ell_F$ is increased, deviations of $\inlavg{\Omega_i^{\nu\eig}|\omega_i^\eig}$ from the linear behavior become less evident, and the linear closure works quite well.
The power law trend \ref{res_exp_lapomg}
is accurately reproduced by the Lagrangian linear diffusion model \citep{Jeong2003}.

The dependence of $\inlavg{H_{i(i)}^{\nu\eig}|\lambda_i}$ on odd powers of $\lambda_i$ and the dependence of $\inlavg{\Omega_i^{\nu\eig}|\omega_i^\eig}$ on odd powers of $\omega_i^\eig$ is reflected by the correlation between the Laplacian of the strain/vorticity and and the strain/vorticity itself, which is shown in Figure \ref{res_fig_correl_lap_grad}, as a function of the filtering length.
Here the correlation coefficients are defined as
\begin{align}
\psi( \bm{H}^{\nu}, \bm{S} )\equiv \frac{ \avg{H_{ij}^{\nu\eig}S_{ij}^{\eig}} }
{ \sqrt{ \avg{H_{ij}^{\nu\eig}H_{ij}^{\nu\eig}}\avg{S_{ij}^\eig S_{ij}^\eig} } }, && 
\psi( \bm{\Omega}^{\nu}, \bm{\omega}) \equiv \frac{ \avg{\Omega_{i}^{\nu\eig}\omega_{i}^{\eig}} }
{ \sqrt{ \avg{\Omega_{i}^{\nu\eig}\Omega_{i}^{\nu\eig}}\avg{\omega_{i}^\eig\omega_{i}^\eig} } }.
\label{res_eq_correl_coeff}
\end{align}
Since $\Omega_i^{\nu\eig}$ decreases on average with $\omega_i^\eig$, as observed above, the correlation coefficient is always negative.
Also, since $\Omega_i^{\nu\eig}$ has a relatively strong linear dependence on $\omega_i^\eig$ when $\widetilde{\tau}\omega_i^\eig$ is not too large, the correlation coefficient is quite large, especially in the inertial range, indicating strong negative proportionality between $\nabla^2\bm{\omega}$ and $\bm{\omega}$. Most striking is that the results show $\psi( \bm{H}^{\nu}, \bm{S} )=\psi( \bm{\Omega}^{\nu}, \bm{\omega})$, such that the correlation between the symmetric and anti-symmetric part of the velocity gradient and their Laplacian is the same. \cite{Betchov1956} proved the relation $\inlavg{\bm{S:S}}=\inlavg{\bm{\omega\cdot\omega}}/2$ for an incompressible and statistically homogeneous flow, and following the same approach it is easily derived that
\begin{equation}
\avg{\nabla^2A_{ij}A_{ji}} = \avg{\partial_j(\nabla^2u_i\partial_iu_j)} = 0.
\end{equation}
Therefore, splitting the velocity gradient into symmetric and anti-symmetricx parts we have
$\avg{S_{ij}\nabla^2S_{ij}} = \avg{W_{ij}\nabla^2W_{ij}}$ and finally
\begin{equation}
\avg{\bm{S :}\nabla^2\bm{S}} = \avg{\bm{\omega\cdot} \nabla^2\bm{\omega}}/2.
\end{equation}
Analogously, it can be derived that $\inlavg{\nabla^2\bm{S :} \nabla^2 \bm{S}} = \inlavg{\nabla^2\bm{\omega\cdot}\nabla^2\bm{\omega}}/2$. As a consequence, the correlations coefficients in equation \eqref{res_eq_correl_coeff} are the same for incompressible, homogeneous flows.

%%%SUB-GRID
\subsection{Characterization of the sub-grid stress}
\begin{figure}
\centering
\vspace{0mm}			
    \subfloat[]
	{\begin{overpic}
	[trim = 0mm -5mm -70mm -3mm,
	scale=0.175,clip,tics=20]{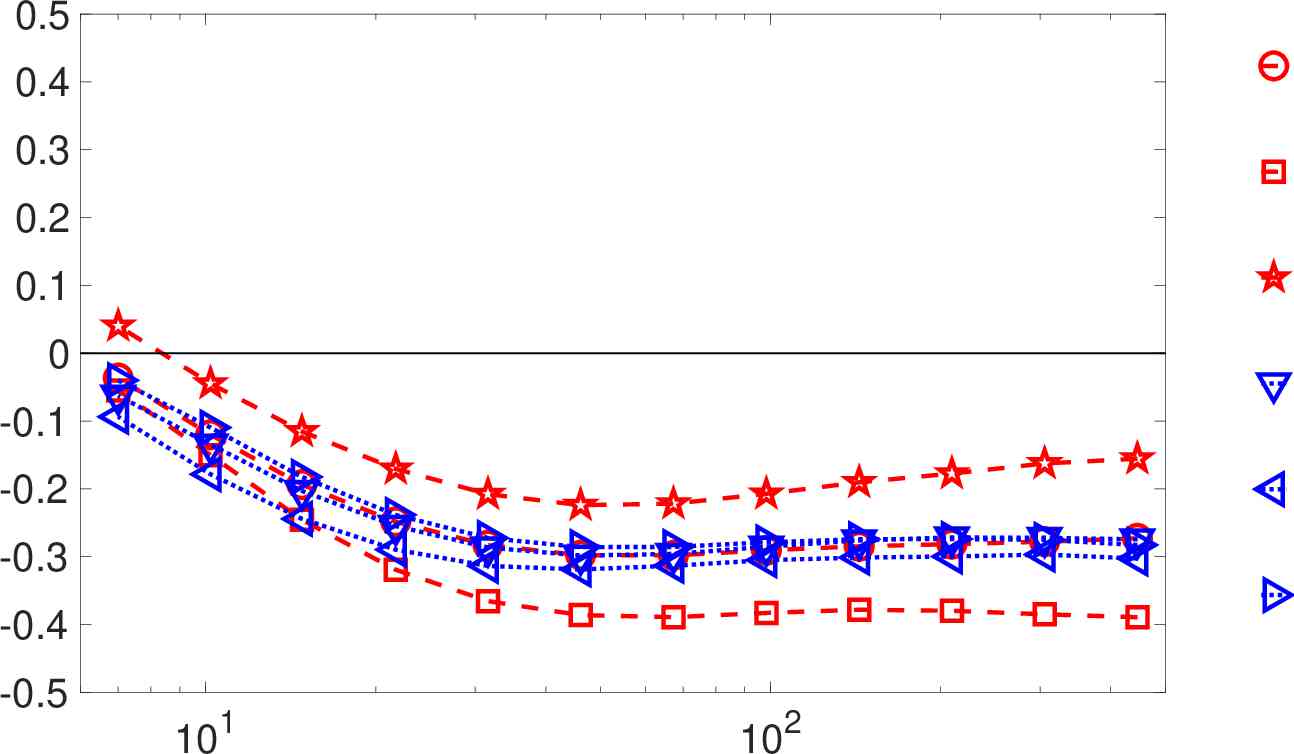}
	\put(40,0){$\ell_F / \eta$}
    \put(-6,13){\rotatebox{90}{Correlation coefficient, $\psi$}}
    \put(90,47.5){$\psi( -H_{11}^{\tau\eig}, -H_{11}^{P\eig} )$}
    \put(90,40.15){$\psi( -H_{22}^{\tau\eig}, -H_{22}^{P\eig} )$}
    \put(90,32.8){$\psi( -H_{33}^{\tau\eig}, -H_{33}^{P\eig} )$}
    \put(90,25.45){$\psi( -H_{32}^{\tau\eig}, -H_{32}^{P\eig} )$}
    \put(90,18.1){$\psi( -H_{31}^{\tau\eig}, -H_{31}^{P\eig} )$}
    \put(90,10.75){$\psi( -H_{21}^{\tau\eig}, -H_{21}^{P\eig} )$}
	\end{overpic}}\\
	\subfloat[]
    {\begin{overpic}
	[trim = 0mm -5mm -70mm -3mm,
	scale=0.175,clip,tics=20]{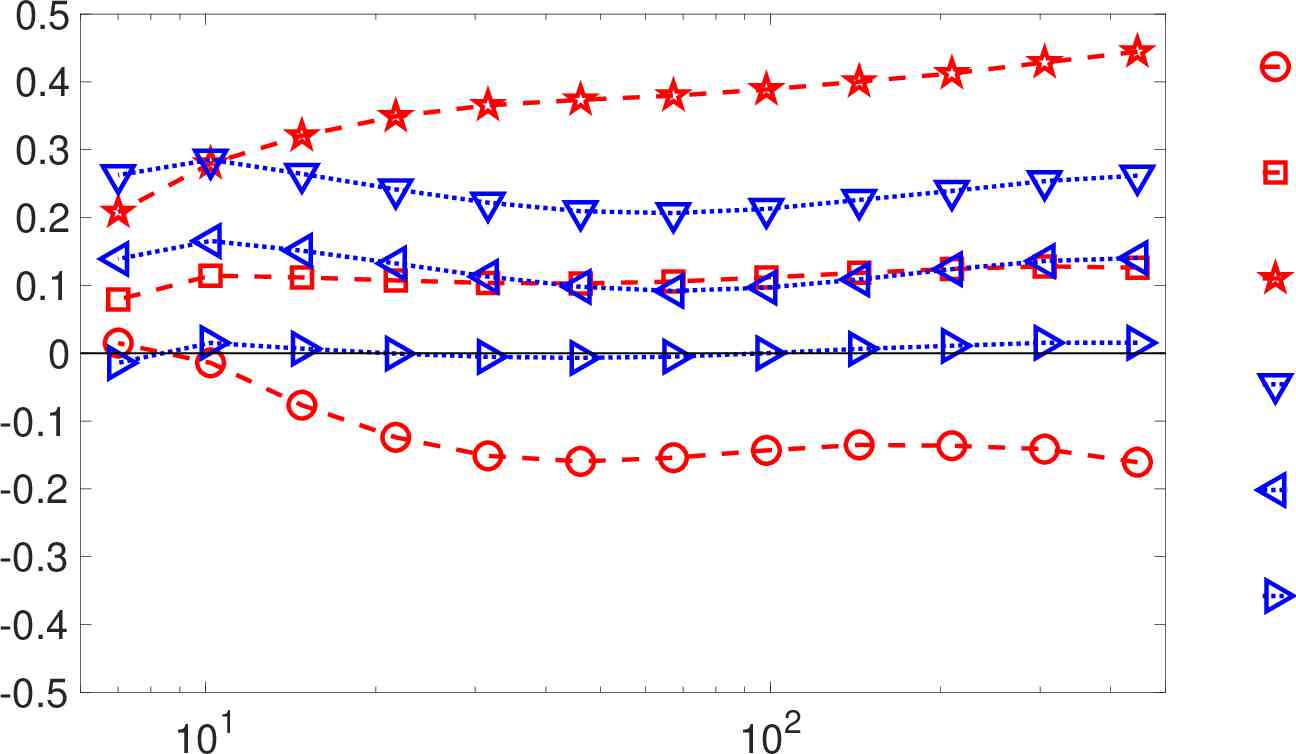}
	\put(40,0){$\ell_F / \eta$}
    \put(-6,13){\rotatebox{90}{Correlation coefficient, $\psi$}}
    \put(90,47.5){$\psi( -H_{11}^{\tau\eig}, H_{11}^{\nu\eig} )$}
    \put(90,40.15){$\psi( -H_{22}^{\tau\eig}, H_{22}^{\nu\eig} )$}
    \put(90,32.8){$\psi( -H_{33}^{\tau\eig}, H_{33}^{\nu\eig} )$}
    \put(90,25.45){$\psi( -H_{32}^{\tau\eig}, H_{32}^{\nu\eig} )$}
    \put(90,18.1){$\psi( -H_{31}^{\tau\eig}, H_{31}^{\nu\eig} )$}
    \put(90,10.75){$\psi( -H_{21}^{\tau\eig}, H_{21}^{\nu\eig} )$}
	\end{overpic}}
	\caption{
	Correlation coefficient, $\psi(X,Y)$ between (a) the sub-grid stress and pressure Hessian and between (b) the sub-grid stress and the symmetric part of the viscous stress, plotted as a function of the filtering length $\ell_F/\eta$.}
\label{res_fig_correl_tau_S}
\end{figure}
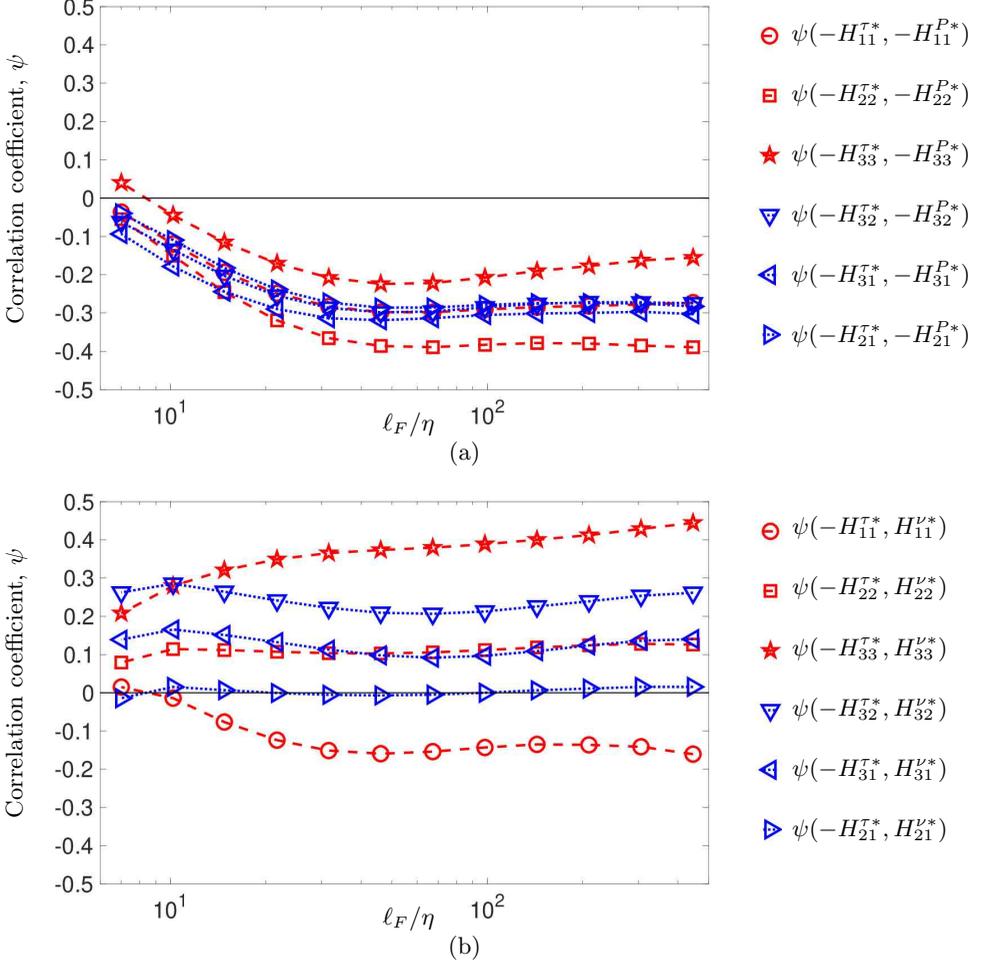

In this subsection, the sub-grid stress is characterized by means of the correlation between its components and the other dynamical terms which contribute to the velocity gradient evolution equations. This can provide insight for closure models in terms of how the sub-grid stress might be related to the filtered quantities in the flow.

We introduce the general correlation coefficient
\begin{equation}
\psi(X,Y) \equiv \frac{
\avg{XY}-\avg{X}\avg{Y}
} {
\sqrt{ \left(\avg{X^2}-\avg{X}^2\right)\left(\avg{Y^2}-\avg{Y}^2\right) }
},
\label{res_def_correl}
\end{equation}
where $X,Y$ are scalar quantities. In figure \ref{res_fig_correl_tau_S}(a) we show results for $\psi(X,Y)$ for the case where $X,Y$ are components of the pressure Hessian and symmetric part of the sub-grid contribution in the eigenframe, both of which contribute to the strain-rate dynamics. The results show that $-H_{ij}^{\tau\eig}$ and $-H_{ij}^{P\eig}$ are negatively correlated (here and throughout this discussion we include in front of the terms the sign with which they appear in the dynamical eqautions), although the correlation is not that strong. A positive correlation is observed only between $-H_{33}^{P\eig}$ and $-H_{33}^{\tau\eig}$ at the smallest scales, implying that in the dissipation range they both tend to hinder the growth of $\lambda_3$.
At larger scales where the correlations are all negative, $-H_{ij}^{\tau\eig}$ and $-H_{ij}^{P\eig}$ have opposite effects on the eigenvalue dynamics and on the angular velocity of the eigenframe.

The correlation coefficient between the viscous $H_{ij}^{\nu\eig}$ and sub-grid $-H_{ij}^{\tau\eig}$ terms as a function of the filtering length is shown in Figure \ref{res_fig_correl_tau_S}(b).
At the smallest scales, the correlations between the diagonal components $H_{i(i)}^{\nu\eig}$ and $-H_{i(i)}^{\tau\eig}$
are positive and the sub-grid stress tends to help the viscous damping effect on $\lambda_i$.  At larger scales, the correlation between $H_{11}^{\nu\eig}$ and $-H_{11}^{\tau\eig}$ becomes negative so that they have opposite effects on the dynamics of $\lambda_1$. The correlations between the off-diagonal components of $H_{ij}^{\nu\eig}$ and $-H_{ij}^{\tau\eig}$ that contribute to the eigenframe rotation-rate are almost independent of $\ell_F$, and the terms $H_{21}^{\nu\eig}$ and $-H_{21}^{\tau\eig}$ are almost entirely uncorrelated. In general, the scale dependence of the correlations between the diagonal and off-diagonal components of $H_{ij}^{\nu\eig}$ and $-H_{ij}^{\tau\eig}$ is quite different, in contrast to the behavior of the correlations between $-H_{ij}^{P\eig}$ and $-H_{ij}^{\tau\eig}$, for which the diagonal and off-diagonal terms behave similarly.

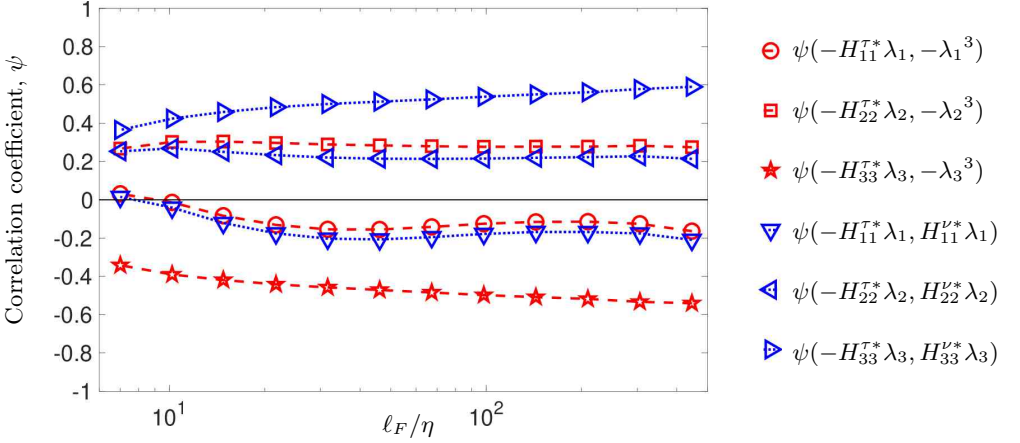
\begin{figure}
\centering
\vspace{0mm}			
	{\begin{overpic}
	[trim = 0mm -5mm -70mm -3mm,
	scale=0.175,clip,tics=20]{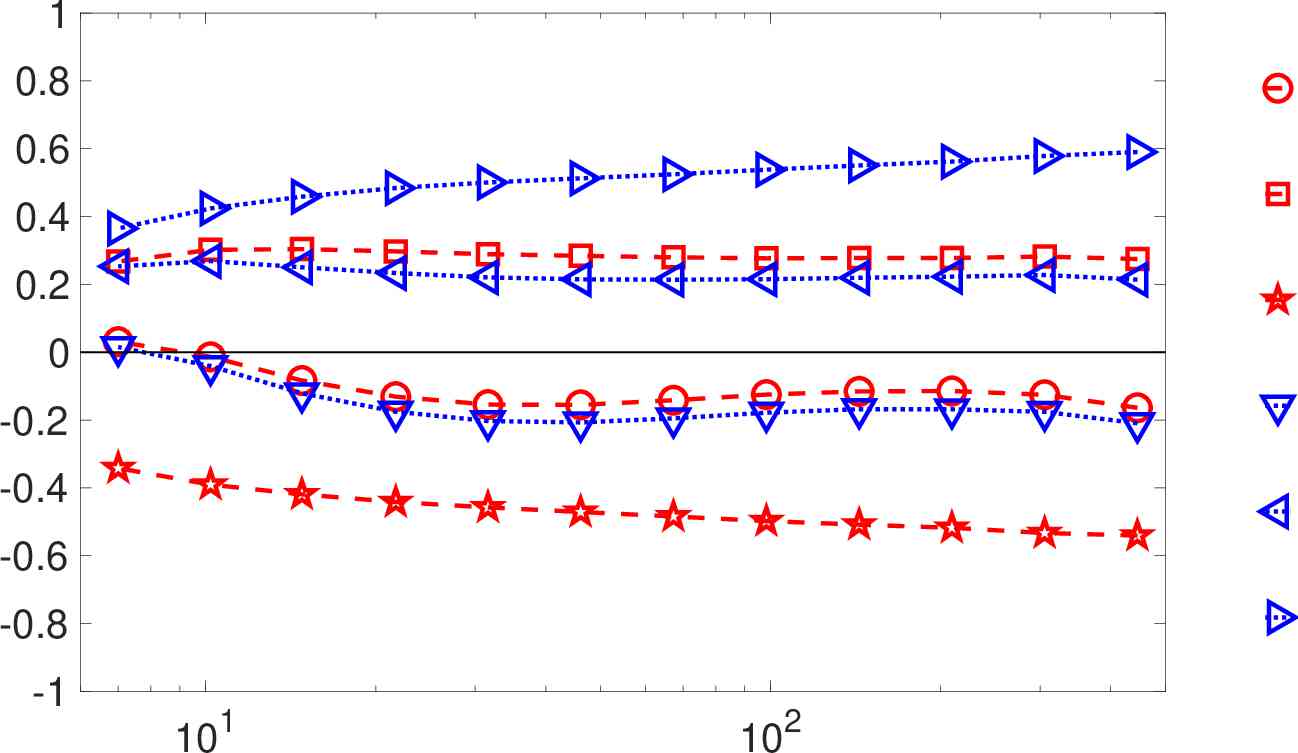}
	\put(40,0){$\ell_F / \eta$}
    \put(-6,13){\rotatebox{90}{Correlation coefficient, $\psi$}}
    \put(90,45.5){$\psi( -H_{11}^{\tau\eig} \lambda_{1}, {-\lambda_{1}}^3 )$}
    \put(90,38.15){$\psi( -H_{22}^{\tau\eig} \lambda_{2}, {-\lambda_{2}}^3 )$}
    \put(90,30.8){$\psi( -H_{33}^{\tau\eig} \lambda_{3}, {-\lambda_{3}}^3 )$}
    \put(90,23.45){$\psi( -H_{11}^{\tau\eig} \lambda_{1}, H_{11}^{\nu\eig} \lambda_{1} )$}
    \put(90,16.1){$\psi( -H_{22}^{\tau\eig} \lambda_{2}, H_{22}^{\nu\eig} \lambda_{2} )$}
    \put(90,8.75){$\psi( -H_{33}^{\tau\eig} \lambda_{3}, H_{33}^{\nu\eig} \lambda_{3} )$}
	\end{overpic}}
	\caption{
	Correlation coefficient, $\psi(X,Y)$ between the sub-grid contribution $-H_{i(i)}^{\tau\eig} \lambda_{(i)}$ and the strain self-amplification term $-\lambda_i^3$, and between the sub-grid contribution $-H_{i(i)}^{\tau\eig} \lambda_{(i)}$ and viscous stress contribution $-H_{i(i)}^{\nu\eig} \lambda_{(i)}$, plotted as a function of the filtering length $\ell_F/\eta$.
	}
\label{res_fig_correl_tau_conv_visc}
\end{figure}

\begin{figure}
\centering
\vspace{0mm}			
	{\begin{overpic}
	[trim = 0mm -5mm -70mm -3mm,
	scale=0.175,clip,tics=20]{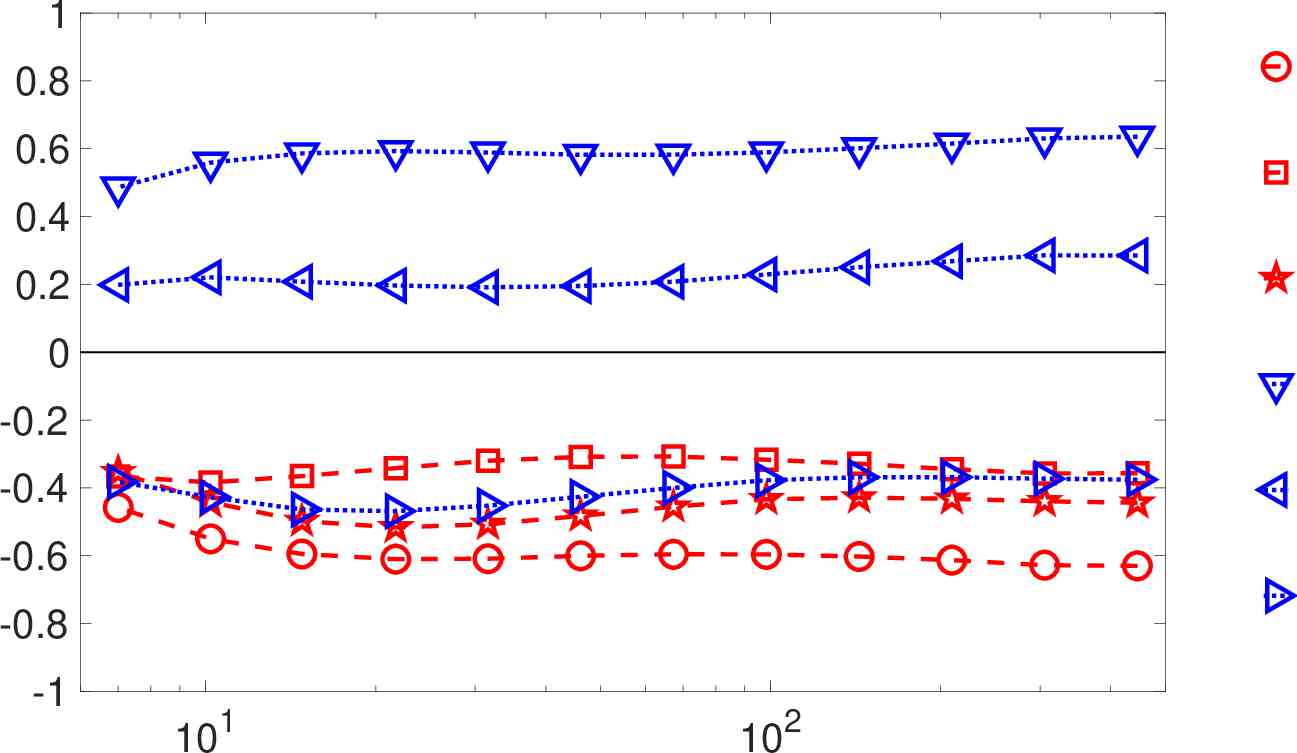}
	\put(40,0){$\ell_F / \eta$}
    \put(-6,13){\rotatebox{90}{Correlation coefficient, $\psi$}}
    \put(90,47.5){$\psi( -\Omega_1^{\tau\eig} \omega_1^\eig, \lambda_{1} {\omega_1^\eig}^2 )$}
    \put(90,40.15){$\psi( -\Omega_2^{\tau\eig} \omega_2^\eig, \lambda_{2} {\omega_2^\eig}^2 )$}
    \put(90,32.8){$\psi( -\Omega_3^{\tau\eig} \omega_3^\eig, \lambda_{3} {\omega_3^\eig}^2  )$}
    \put(90,25.45){$\psi( -\Omega_1^{\tau\eig} \omega_1^\eig, \Omega_1^{\nu\eig} \omega_1^\eig )$}
    \put(90,18.1){$\psi( -\Omega_2^{\tau\eig} \omega_2^\eig, \Omega_2^{\nu\eig} \omega_2^\eig )$}
    \put(90,10.75){$\psi( -\Omega_3^{\tau\eig} \omega_3^\eig, \Omega_3^{\nu\eig} \omega_3^\eig )$}
	\end{overpic}}
	\caption{Correlation coefficient, $\psi(X,Y)$ between the sub-grid contribution $-\Omega_i^{\tau\eig} \omega_{(i)}^\eig$ and the vortex stretching term $\lambda_{(i)} {\omega_i^\eig}^2$, and between the sub-grid contribution $-\Omega_i^{\tau\eig} \omega_{(i)}^\eig$ and the viscous stress contribution $-\Omega_i^{\nu\eig} \omega_{(i)}^\eig$ .}
\label{res_fig_correl_tau_omg}
\end{figure}
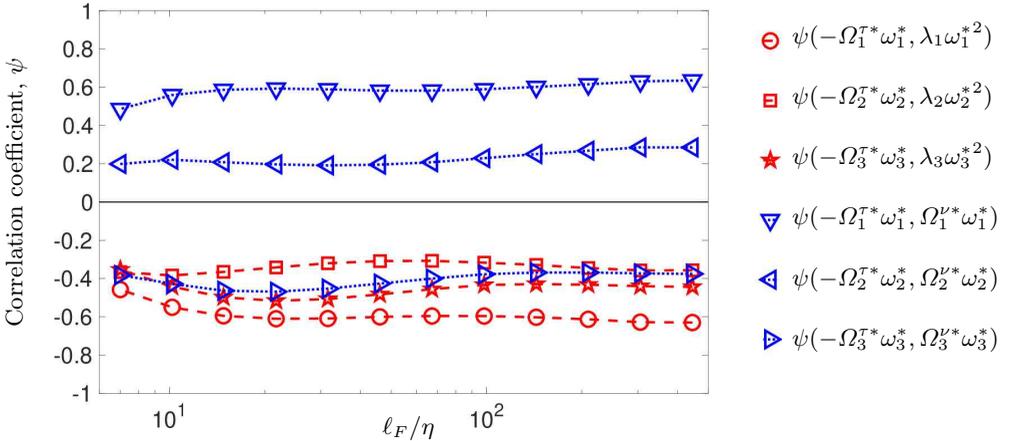

Figure \ref{res_fig_correl_tau_conv_visc} shows the 
correlation between $-H_{i(i)}^{\tau\eig}\lambda_{(i)}$ and both the strain self-amplification term $-\lambda_i^3$
and $H_{i(i)}^{\nu\eig}\lambda_{(i)}$, terms which appear in the equation governing $\lambda_{i}^2$.
Concerning the correlation between $-H_{i(i)}^{\tau\eig}\lambda_{(i)}$ and $-\lambda_i^3$, the correlations are positive for $i=2$, and negative for $i=1,3$ at all scales (although the data may indicate that the $i=1$ component becomes positive for $\ell_F/\eta\to 0$). Therefore, the sub-grid stress tends to oppose the growth of $\lambda_1$ and positive $\lambda_2$, although the correlation is weak. The correlation is strongest for $i=3$, and the negativity of the correlation indicates that the sub-grid stress act to stabilize the dynamics by opposing the growth of $|\lambda_3|$. Interestingly, the correlations between
$-H_{i(i)}^{\tau\eig}\lambda_{(i)}$ and $H_{i(i)}^{\nu\eig}\lambda_{(i)}$ are similar in magnitude to those between $-H_{i(i)}^{\tau\eig}\lambda_{(i)}$ and $-\lambda_i^3$. The correlation between $-H_{i(i)}^{\tau\eig}\lambda_{(i)}$ and $H_{i(i)}^{\nu\eig}\lambda_{(i)}$ is positive for $i=3$, indicating that the sub-grid stress also acts to help the viscous stress in reducing $|\lambda_3|$.

Figure \ref{res_fig_correl_tau_omg} shows the correlation between $-\Omega_i^{\tau\eig} \omega_{(i)}^\eig$ and both $\lambda_i{\omega_{(i)}^\eig}^2$ and $\Omega_i^{\nu\eig} \omega_{(i)}^\eig $, terms which appear in the enstrophy equation \eqref{res_eq_vort2_princ}. The correlations between  $-\Omega_i^{\tau\eig} \omega_{(i)}^\eig$ and $\lambda_i{\omega_{(i)}^\eig}^2$ are negative for all $i$ and at all scales, showing that the sub-grid stress acts on average to counteract the amplification of $\omega_1^\eig$ and the reduction of $\omega_3^\eig$, and also to hinder vortex stretching along the direction $\bm{v}_2$. 
The correlation between $-\Omega_i^{\tau\eig} \omega_{(i)}^\eig$ and $\Omega_i^{\nu\eig} \omega_{(i)}^\eig $ is positive for $i=2$, showing that the sub-grid stress acts together with the viscous stress to hinder the growth of $\omega_2^{\eig 2}$, at all scales, though the correlation is weak. The correlation is also positive for $i=1$, showing that the sub-grid and viscous terms act together to reduce $\omega_1^{\eig 2}$, and in this case, the correlation is quite strong, reaching values around $0.6$ in the inertial range. In contrast to the behavior for $i=1,2$, the sub-grid stress opposes the viscous effect on $\omega_3^{\eig 2}$ at all scales, just as it also opposes the vortex compression that occurs for the $i=3$ component.

%!%
%%%
\section{Conclusions}\label{sec:Conclusions}
%%%\subfile{Sections/Sec05_Conclusions.tex}
%!%
In this paper, we have explored the properties of the dynamical equations governing the filtered velocity gradient tensor written in the eigenframe of the strain-rate tensor. The mean contributions to the evolution of the square of the eigenframe vorticity components are dominated by vortex stretching and vorticity tilting at smallest scales, while the sub-grid stress contribution grows with increasing filtering length and makes an important contributions at large scales.
The eigenvalues and vorticity components conditioned on the principal invariants of the filtered velocity gradient tensor reveal a significant misalignment of the vorticity with the compressional eigendirection. This misalignment gives rise to a finite centrifugal force associated with the local spinning  of the fluid that opposes the self-amplification of the eigenvalues and helps to prevent the blow-up of the compressional eigenvalue along the right Vieillefosse tail. This supports ``reduction of non-linearity'' models that assume that the vorticity is effective in stabilizing dynamics.

The vorticity tilting term exhibits very large fluctuations and striking intermittency, indicating rapid rotations of the vorticity vector with respect to the strain-rate eigenframe. This intermittency persists even at large filtering scales, a significant cause of which is kinematic, being related to the large probability to have very small differences between the eigenvalues. In particular, the differences in the eigenvalues act as a moment of inertia in the eigenframe rotation-rate equation, and so small differences in the eigenvalues allows for large rotation-rates. The non-local anisotropic pressure Hessian is the dominating contribution to the eigenframe rotation-rate, being much larger than the contribution from the local spinning of the fluid due to vorticity (the sub-grid stress also makes and important contribution outside the dissipation range). However, these local and non-local contributions to the eigenframe rotation-rate give comparable contributions to the vorticity tilting term, due to the fact that the vorticity exhibits weak preferential alignment with respect to the anisotropic pressure Hessian.

The average of the diagonal components of the pressure Hessian conditioned on both the square of the eigenvalues and vorticity components exhibit an almost linear relationship to the variables on which they are conditioned. This supports models such as the Gaussian closure \citep{Wilczek2014} that expresses the pressure Hessian as the square of the velocity gradient. However, the results also show that for relatively small values (compared to the eddy turnover timescale based on the filtering scale) of the eigenvalues and vorticity the dependence is highly non-linear for some of the components, features that are very challenging to replicate in models. Corresponding conditional averages show that the symmetric part of the viscous stress behaves as an odd function of eigenvalue. The anti-symmetric part of the viscous stress shows a similar dependence on the vorticity components in the eigenframe. While velocity gradient models such as \cite{Chevillard2008} describe these odd functions with a linear behavior, our results showed that cubic terms should also be included. This is confirmed by the correlation coefficient between the filtered velocity gradient and its Laplacian, which is negative but differs from $-1$, especially at small scales.

In order to characterize the state of the pressure Hessian we developed a generalization of the classical Lumley triangle that does not require positive definiteness of the tensor. The results using this triangle revealed a preference for the intermediate eigenvalue of the pressure Hessian to be positive, but the preference is much weaker than for the strain-rate. Moreover, the pressure Hessian filtered at small scales is rarely in the isotropic configuration, while the most probable states are complete anisotropy and two-dimensional axisymmetric expansion. Therefore, the pressure Hessian preferentially exerts a two-dimensional axisymmetric compression on the fluid element along two of its eigendirections. As the filtering scale is increased, the probability of two-dimensional axisymmetric expansion reduces and the peak in probability that occurred there at small scales shifts towards the isotropic state. However, there still exists a significant probability for configurations where the non-local contribution dominates the pressure Hessian.

Finally, the sub-grid stress has been characterized by means of its component-wise correlation with the other terms in the eigenframe equations for the filtered velocity gradient. Concerning the correlations between sub-grid stress and filtered pressure Hessian, the off-diagonal and diagonal elements of the tensors show similar behavior. This is in contrast to the correlations between the sub-grid stress and the symmetric part of the viscous stress, for which the diagonal and off-diagonal components show considerably different behavior. The intermediate components of the strain self-amplification and the sub-grid stress are positively correlated in contrast to the correlations between vortex stretching and the sub-grid stress, which are always negative.

Taken together, the results presented provide a comprehensive statistical description of the filtered velocity gradient dynamics from the perspective of the strain-rate eigenframe. Lagrangian models for the velocity gradient tensor can be tested against the data, but predicting the non-trivial behaviour of the non-local terms highlighted throughout the paper represents a challenge for those models. The new results on the preferential state of the pressure Hessian, viscous and sub-grid stress constitute a reference for modelling those unclosed parts of the equations and they can enhance our understanding of the non-linear and non-local evolution of the filtered velocity gradient in turbulence.
%!%
%%%
\section*{Acknowledgements}

This work used the Extreme Science and Engineering Discovery Environment (XSEDE) under allocation CTS170009, which is supported by National Science Foundation (NSF) grant number ACI-1548562 \citep{xsede}. Specifically, Stampede2 cluster operated by Texas Advanced Computing Center (TACC) was used to obtain the results in this work. The Comet cluster operated by the San Diego Supercomputer Center (SDSC) and Duke Computing Cluster (DCC) operated by Duke University Research Computing was used to obtain some of the preliminary results for this study. The authors gratefully acknowledge discussions and input from Mohammadreza Momenifar on the issue discussed in Appendix A.

\section*{Declaration of Interests}
The authors report no conflict of interest.

\appendix
%%% \subfile{Appendix/Appendix_dealiasing.tex}
%!%
\section{Preserving incompressibility and positive definiteness in de-aliased computation of the pressure Hessian}
\label{app_dealiasing}

In this Appendix, we discuss an issue that arises when computing the pressure Hessian $\bm{H}^P$, which, if not handled correctly, leads to a violation of the incompressibility constraint
\begin{equation}
\Tr(\bm{H}^P + \bm{A\cdot A})=0.
\label{app_incompr}
\end{equation}
This issue was is part discussed in \cite{Cheng1996}. However, there are additional complications not discussed in that paper which, to the best of our knowledge, have been overlooked in previous work. We discuss the issues in one spatial dimension for simplicity, however the conclusions are easily extended to three dimensions by separation of variables.

In a spectral representation, aliasing errors arise when one attempts to represent a field that has active Fourier modes with wavenumbers larger than the maximum wavenumber that is resolved by the discrete numerical grid \citep{Canuto1988}. In the NSE, the non-linear term naturally tends to violate this constraint, since even if the initial field is resolved by the grid, the non-linear term can excite Fourier modes with wavenumbers larger than what the grid can resolve. The maximum wavenumber that can be represented on the grid is the Nyquist wavenumber $N_\textrm{Nyq}=N/2$ (here $N$ is the number of grid points and the grid resolution is $\Delta x = 2\pi/N$) and wavenumbers exceeding the Nyquist wavenumber, $|k|>N_\textrm{Nyq}$, are aliased to wavenumbers
$\textrm{mod}(k, N/2)\in (-N/2,N/2)$.
The convective non-linear term requires the computation of products of the velocity field with itself
and, since the velocity field is represented as the superposition of $N$ Fourier modes, then the non-linear convective term involves $2N$ active Fourier modes, which can not be represented on the grid.
In order to remove the aliasing error, the Fourier transform of the velocity field is set to zero at wavenumbers $|k|\ge N/3$ and it is easily shown that the resulting non-linear convective term is not aliased at wavenumbers $|k|< N/3$ \citep{Orszag1971}.
However, the non-linear convective term still involves aliased wavenumbers in the range $N/3 <|k|\le N/2$, and these are removed once the convolution sum is transformed back to Fourier space by setting to zero the amplitudes associated with $|k|\ge N/3$. The algorithm sketched above, the $2/3$ rule, is very well known, and is the basis of the majority of pseudo-spectral codes.

Having summarized the de-aliasing procedure for the $2/3$ rule, we may now present the issue that arises when computing the pressure Hessian. Using the Fourier transform of the velocity field, namely
$\widehat{u}_i=\mathcal{F}[u_i]$, the de-aliased pressure Hessian is computed as
\begin{subequations}
\label{app_H_2third}
\begin{align}
{C_{pq}} &= \mathcal{F}^{-1}\left[\widehat{u}_pB_{1/3}\right] \mathcal{F}^{-1}\left[\widehat{u}_qB_{1/3}\right]\label{app_H_2third_a},\\
H_{ij}^P &= \mathcal{F}^{-1}\left[k_ik_j \frac{k_pk_q}{k^2}\mathcal{F}\left[{C_{pq}}\right]B_{1/3} \right],
\label{app_H_2third_b}
\end{align}
\end{subequations}
where $\mathcal{F}$ indicates the Fourier transform and $B_{1/3}$ is the box function, $B_{1/3}(k)=1$ for $|k|<N/3$ and zero otherwise. The velocity gradient $A_{ij}$ is constructed from $\widehat{{u}}_i$ as
%
%\begin{subequations}
%\label{app_A2_2third}
\begin{align}
{A_{ij}} &= \mathcal{F}^{-1}\left[k_j \widehat{u}_i B_{1/3} \right]  \label{app_A2_2third_a}.
%A_{il}A_{lj} &= \mathcal{F}^{-1}\left[\mathcal{F}\left[{A_{ij}^2}\right] B_{1/3} \right]. \label{app_A2_2third_b}
\end{align}
%\end{subequations}
%
However, computing $H_{ij}^P$ and $A_{ij}$ in this way violates equation \eqref{app_incompr}, and our data indicates that it may be significantly violated. This issue was pointed out in \cite{Cheng1996}, who noted that the violation arises because when computed in this way, $\bm{A\cdot A}$ effectively contains information at higher wavenumbers than does $\bm{H}^P$. They did not, however, provide a method to address the issue, but simply noted that their results were affected by it.

One way to ensure that \eqref{app_incompr} is satisfied is to apply an additional truncation step when computing $\bm{A\cdot A}$, namely
\begin{align}
A_{ij}A_{lm} &= \mathcal{F}^{-1}\left[\mathcal{F}\left[{A_{ij}A_{lm}}\right] B_{1/3} \right], \label{app_A2_2third_b}
\end{align}
with $A_{ij}$ on the right hand side constructed using \eqref{app_A2_2third_a}. Using \eqref{app_H_2third} and \eqref{app_A2_2third_b} satisfies equation \eqref{app_incompr}. However, \eqref{app_A2_2third_b} violates the fundamental constraint $A_{ij}A_{ij}\geq 0$. This is because the additional truncation in Fourier space described by \eqref{app_A2_2third_b} corresponds to the convolution of $\bm{AA}$ with the sinc function in physical space \citep{Beylkin1995}, and this function takes on negative values. Our data indicates that violations of $A_{ij}A_{ij}\geq 0$ through the use of \eqref{app_A2_2third_b} can be significant, and therefore this method should be rejected.

In order to satisfy $A_{ij}A_{ij}\geq 0$ and equation \eqref{app_incompr}, we truncate the Fourier transform of the velocity field at $|k|\le N/4$ so that quadratic products such as $A_{ij}A_{lm}$ are resolved on the grid, and no additional truncation or de-aliasing is required. Therefore, in our paper, the pressure Hessian is computed using
\begin{subequations}
\label{app_H_our}
\begin{align}
C_{pq} &= \mathcal{F}^{-1}\left[\widehat{u}_pB_{1/4}\right] \mathcal{F}^{-1}\left[\widehat{u}_qB_{1/4}\right], \label{app_H_our_a}
\\
H_{ij}^P &= \mathcal{F}^{-1}\left[k_ik_j \frac{k_pk_q}{k^2}\mathcal{F}\left[C_{pq}\right]\right],
\label{app_H_our_b}
\end{align}
\end{subequations}
together with
\label{app_A2_our}
\begin{align}
A_{ij}A_{lm}
%A_{ij}^{2}
&= -\mathcal{F}^{-1}\left[k_j \widehat{u}_i B_{1/4} \right] \mathcal{F}^{-1}\left[k_m \widehat{u}_l B_{1/4} \right],
\label{app_A2_our_a}
%\\
%A_{il}A_{lj} &= \mathcal{F}^{-1}\left[\mathcal{F}\left[A_{ij}^{2}\right] \right],
%\label{app_A2_our_b}
\end{align}
%\end{subequations}
%
where $B_{1/4}$ is the box function, $B_{1/4}(k)=1$ for $|k|\le N/4$ and zero otherwise. Moreover, for consistency, all quantities were computed from the $N/4$ truncated velocity field. A consequence of this is that the smallest filtering scale that can be considered in our analysis is $\ell_F=2\pi/(N/4)\approx 7\eta$.
%!%
%%% 

\bibliographystyle{jfm}
\bibliography{bib_multiscale_grad}

\end{document}